\documentclass{ucetd}
\usepackage{amsfonts,amssymb}
\usepackage{mathtools}
\usepackage{graphicx}
\usepackage{wrapfig}
\usepackage{caption}
\usepackage{feynmp-auto}
\usepackage{comment}
\usepackage{setspace}
\usepackage{multirow}

\makeatletter
\newcommand*\wt[1]{\mathpalette\wthelper{#1}}
\newcommand*\wthelper[2]{%
        \hbox{\dimen@\accentfontxheight#1%
                \accentfontxheight#11.15\dimen@
                $\m@th#1\widetilde{#2}$%
                \accentfontxheight#1\dimen@
        }%
}

\newcommand*\accentfontxheight[1]{%
       \fontdimen5\ifx#1\displaystyle
                \textfont
        \else\ifx#1\textstyle
                \textfont
        \else\ifx#1\scriptstyle
                \scriptfont
        \else
                \scriptscriptfont
        \fi\fi\fi3
}
\makeatother

\newcommand*{\momentumarrow}[4]{%
    \fmfcmd{style_def marrow#1
    expr p = drawarrow subpath (0.3, 0.7) of p shifted #3 #2 withpen pencircle scaled 0.4;
    enddef;}
    \fmf{marrow#1,tension=0}{#4}}

\newcommand*{\tr}{\ensuremath{\mathbf{tr}}}
\newcommand*{\EoM}{\ensuremath{{\mbox{\tiny EoM}}}}
\newcommand*{\z}{\ensuremath{{\vec{z}}}}
\newcommand*{\x}{\ensuremath{{\vec{x}}}}
\newcommand*{\p}{\ensuremath{{\vec{p}}}}
\newcommand*{\q}{\ensuremath{{\vec{q}}}}
\newcommand*{\kk}{\ensuremath{{\vec{k}}}}
\renewcommand*{\l}{\ensuremath{{\vec{l}}}}

\newcommand*{\sgn}{\ensuremath{\mathrm{sgn}}}
\renewcommand*{\Im}{\ensuremath{\mathrm{Im}}}
\renewcommand*{\Re}{\ensuremath{\mathrm{Re}}}

\newcommand*{\Ical}{\ensuremath{{\mathcal{I}}}}
\newcommand*{\Jcal}{\ensuremath{{\mathcal{J}}}}

\begin{document}
\unitlength = 1mm

\title{Berry Phase Physics in Free and Interacting Fermionic Systems}
\author{Jingyuan Chen}
\department{Physics}
\division{Physical Sciences}
\degree{Doctor of Philosophy}
\date{June 2016}
\maketitle

%

\tableofcontents
\listoffigures
\listoftables

\acknowledgments{
Slightly more than five and a half years ago, when I was applying to the University of Chicago for graduate study in physics, I was far more ignorant than I should about physics research: at the time I had not heard of condensed matter physics, I barely had any idea that there is an important subject called quantum field theory to be learned\textellipsis Despite such blindness, I had the great luck to be admitted by the Physics Department at the University of Chicago. During my five years here, the Department provided an incredible environment for learning physics knowledge, for exploring different research areas in physics, and for fostering stimulating discussions. I have to thank all of those who have worked towards establishing, sustaining and developing such a wonderful academic environment.

I would like to express my deepest gratitude towards Dam Thanh Son, my Ph.D. advisor and thesis committee chair. His broad interest over different fields of physics, his insight to catch interesting problems, his clarity in presenting ideas\textellipsis have all deeply influenced me during these years, and will continue to exert larger impact on me in the future. He is also very nice and interesting to get along with. I feel privileged to have him as my advisor, and I really enjoy these years working with him.

I am honored to have William Irvine, Michael Levin and Lian-Tao Wang as my thesis committee members. I am interested in their research works and I am looking forward to more discussions with them in the future.

For a period of time, Misha Stephanov was almost like my secondary advisor. I had many long and fruitful discussions with him, and received many helpful advices from him. Some of my works presented in this thesis were done in collaboration with him, as well as with Ho-Ung Yee and Yi Yin.

I am also grateful to Lian-Tao Wang and Rocky Kolb, who were my Ph.D. advisors at an early stage and gave me valuable first guidance into physics research, although my works on dark matter done with them, as well as with Michael Fedderke, are not presented in this thesis.

I benefited a lot from the lectures and discussions with the faculty members, most thankfully Cheng Chin, Jeffery Harvey, Leo Kadanoff, Kathryn Levin, Michael Levin, Emil Martinec, Savdeep Sethi, Dam Thanh Son, Robert Wald, Lian-Tao Wang and Paul Weigmann. I have had valuable discussions with many postdoc fellows and students, most thankfully Jo\~{a}o Caldeira, Chien-Hung Lin, Eun-Gook Moon, Matthew Roberts, Chengjie Wang, Yizhi You and Hao Zhang; I especially appreciate Caner Nazaroglu, Michael Geracie and Dung Xuan Nguyen for many stimulating discussions over the years. The Kadanoff Center Journal Club, initiated by Siavash Golkar, has played an extremely important role in broadening my perspective in physics. I should also thank some friends I knew from on-line through discussing physics: Zhen Gao, Wei Gu, Jiajun Li, Jin-Bo Yang and Yi-Zhuang You. Tracing back further, I would like to thank Andreas Blass for being my undergraduate research advisor on set theory, and Dragan Huterer on cosmology.

Finally, I want to thank my father Shao-Gang Chen for his initiation, and my mother Yue-Lan Mo for her long-term support, of my interest in science, and more generally for their shaping of me as an individual person. And I owe everything to my beloved Ying Zhao, for her wit, affection and understanding. Without her my five years here could not have been so colorful.
}

\abstract{
Berry phase plays an important role in many non-trivial phenomena over a broad range of many-body systems. In this thesis we focus on the Berry phase due to the change of the particles' momenta, and study its effects in free and interacting fermionic systems. We start with reviewing the semi-classical kinetic theory with Berry phase for a non-interacting ensemble of fermions -- a Berry Fermi gas -- which might be far-from-equilibrium. We particularly review the famous Berry phase contribution to the anomalous Hall current. We then provide a concrete and general path integral derivation for the semi-classical theory. Then we turn to the specific example of Weyl fermion, which exhibits the profound quantum phenomenon of chiral anomaly; we review how this quantum effect, and its closely related chiral magnetic effect and chiral vortical effect, arise from Berry phase in the semi-classical kinetic theory. We also discuss how Lorentz symmetry in the kinetic theory of Weyl fermion, seemly violated by the Berry phase term, is realized non-trivially; we provide a physical interpretation for this non-trivial realization, and discuss its mathematical foundation in Wigner translation. Next, we turn towards interacting fermionic systems. We consider Fermi liquid near equilibrium, and propose the Berry Fermi liquid theory -- the extension to Landau Fermi liquid theory incorporating Berry phase (and other) effects. In our proposed Berry Fermi liquid theory, we can show the Berry phase is a Fermi surface property, qualitatively unmodified by interactions. But there also arise new effects from interactions, most notably the emergent electric dipole moment which contributes to the anomalous Hall current in addition to the usual Berry phase contribution. We prove our proposed Berry Fermi liquid theory from quantum field theory to all orders in Feynman diagram expansion under very general assumptions.

The discussion of Berry Fermi gas is based on the previous literature and the author's works \cite{Chen:2014cla} and \cite{Chen:2015gta}. The Berry Fermi liquid theory follows from the author's work \cite{chen2016berry}.
}

\mainmatter

\chapter{Introduction}

\section{Background, Problems and Results}

As we learn to do quantum mechanics, we learn to diagonalize the Hamiltonian:
\begin{eqnarray}
H = U \, D \: U^\dagger.
\end{eqnarray}
The diagonal matrix $D$ is the spectrum of energy eigenvalues, while the unitary matrix $U$ has its columns as energy eigenstates. We have learned to pay much attention to $D$ when considering the dynamics, or time evolution, of the system. But it would be hard to believe if the $U$ part has no effect on dynamics.

Indeed, the $U$ part has effects that people have noticed in different problems. But it was not until 1984 that a unified, general picture appeared. This is Berry's picture of geometric phase~\cite{berry1984quantal}, or \emph{Berry phase} as we call it. Suppose the Hamiltonian depends on a few adjustable parameters $Q_I$. For simplicity, we let $|u(Q)\rangle$ be an energy eigenstate (a column of $U(Q)$) whose corresponding energy eigenvalue $E(Q)$ (an entry of $D(Q)$) is well gapped with other eigenvalues. Now we adjust the parameters $Q_I$ in time, $Q_I=Q_I(t)$; the variation in time is so slow compared to the energy gap over $\hbar$. Then, according to the quantum adiabatic theorem, if the initial state is prepared in $u(Q(t=0))$, the state of the system will stay in the instantaneous eigenstate $|u(Q(t))\rangle$ of the Hamiltonian, up to a complex phase. That is, the state at time $t$ is $|\psi_u(t)\rangle = e^{i\theta(t)} |u(Q(t))\rangle$ with $\theta(t=0)=0$. Now let's compute the phase $\theta(t)$. Using the Schr\"{o}dinger's equation $i\hbar \partial_t |\psi_u(t)\rangle = H(Q(t))\: |\psi_u(t)\rangle$, we have
\begin{eqnarray}
i\hbar\, \partial_t |u(Q(t))\rangle - \hbar\,\partial_t \theta(t) \: |u(Q(t))\rangle = E(Q(t)) \: |u(Q(t))\rangle
\end{eqnarray}
Contracting with $\langle u(Q(t))|$ on the left, and integrating over time, we find the phase is
\begin{eqnarray}
\theta(t) = -\int_0^t dt' \: \frac{E(Q(t'))}{\hbar} - \int_{Q(0)}^{Q(t)} dQ_I \: a^I(Q), \ \ \ \ \ \ a^I(Q) \equiv -i\, \langle u(Q) | \partial_{Q_I} |u(Q)\rangle
\end{eqnarray}
The first term is the usual dynamical phase from the diagonal $D$ part of the Hamiltonian. The second term is the geometric phase due to the $U$ part, and $a^I$ is called the \emph{Berry connection} in the space of $Q_I$. But we are not done yet. The definition of $|u(Q)\rangle$ is ambiguous. We can redefine it by $|u(Q)\rangle \rightarrow e^{i\phi(Q)} |u(Q)\rangle$ (corresponding to redefining $U$ by multiplying a unitary diagonal matrix to its right), and the Berry connection will shift by $a^I \rightarrow a^I + \partial_{Q_I} \phi$, so its contribution to $\theta(t)$ will be shifted by the boundary term $\phi(Q(t))-\phi(Q(0))$. Thus, it seems the Berry connection contribution is just arbitrary. It indeed is, except if the we adjust the system back to its initial setup $Q(t)=Q(0)$. In that case, the phase
\begin{eqnarray}
\theta_{Berry} = \int_{Q(0)}^{Q(t)=Q(0)} dQ_I \: a^I(Q)
\end{eqnarray}
is independent of the choice of $\phi$, but only depends on the loop $Q_I(t)$ traces out (due to our physical adjustment) in the $Q$ space. This unambiguous phase is the \emph{Berry phase}. Note that the Berry phase does not depend on how $Q_I(t)$ changes with time, but only on the loop traced out. A closely related concept is the \emph{Berry curvature}
\begin{eqnarray}
b^{IJ}(Q) \equiv \partial_{Q_I} a^J(Q) - \partial_{Q_J} a^I(Q)
\end{eqnarray}
which represents ``the density of Berry phase per unit area in the $Q$ space'', and is unambiguously independent of the choice of phase $\phi(Q)$. If we compare the Berry phase $\theta_{Berry}$ to the magnetic flux $\Phi$, then the Berry connection $a^I(Q)$ is analogous to the vector potential $A_i(\x)$, while the Berry curvature $b^{IJ}(Q)$ is analogous to the magnetic field $F_{ij}(\x)$.

The phenomenon of Berry phase is rooted in the mismatching between the inner product structure of the Hilbert space (``dagger-ing'') and the detailed way the Hamiltonian depends on $Q$; this mismatching is captured in the definition of $a^I$. This is analogous to another phenomenon of geometric phase -- the Foucault pendulum, where the precession angle is due to the mismatching between the pendulum's tendency to move under inertia and the transport of the pendulum as the Earth rotates. In both cases adiabaticity plays a crucial role. In the Berry phase case, the time variation of $Q$ must be slow enough for the state $|\psi\rangle$ to stay proportional to the instantaneous eigenvector $|u(Q(t))\rangle$. In the Foucault pendulum case, the transport of the pendulum must be slow enough that it does not exert force on the pendulum's motion.

In the above we have considered the parameters $Q$ being externally adjustable parameters. But they can also be quantum numbers, and that is the scenario we focus on in this thesis. Consider the Weyl Hamiltonian as an example: $\hat{H}^\alpha_{\ \beta} = \hat{p}_i (\sigma^i)^\alpha_{\ \beta}$. The Hilbert space has an infinite dimensional subspace, on which the momentum operator acts, as well as a two dimensional subspace, the internal space, on which the Pauli matrices act. If we diagonalize the momentum operators first, we can use the quantum numbers $p_i$ to label the left-over part of the Hamiltonian, which now acts only on the two dimensional internal space (spinor space) $H^\alpha_{\ \beta}(p) = p_i (\sigma^i)^\alpha_{\ \beta}$. Another example is the Bloch wave in a lattice. We can diagonalize the lattice momentum part first, and then the Hamiltonian becomes a matrix, labelled by the lattice momentum, acting on the discrete band index. In this thesis, the $Q_I$ we consider are the spatial momentum $p_i$, so the Berry curvature is a curvature in the momentum space.

Of course, if $p_i$ are good quantum numbers, the momentum will be constant in time and there will be no Berry phase effect. Therefore, to see the effects of the momentum space Berry curvature, we must perturb the system so that the particle's momentum changes slowly in time.

As we can see from the above, Berry curvature effects are very common; they generally exist if there is an internal Hilbert space of more than one dimension (i.e. more than one band). However, their effects are order $\hbar$ suppressed compared to the usual phase due to the energy eigenvalue. Therefore, Berry curvature effects are only visible if we consider the next-to-leading order effects in $\hbar$ expansion, or equivalently low frequency / long wavelength expansion. This is why they are usually overlooked in more traditional studies of bands.

\begin{table}
\centering
\begin{tabular}{|c||c|c|}
\hline
Fermionic systems & Neglect Berry Phase & Include Berry Phase \\ \hline\hline
Free & simple Fermi gas & Berry Fermi gas \\ \hline
\multirow{2}{*}{Interacting} & Landau's Fermi liquid & Berry Fermi liquid \\[.1cm] \cline{2-3}
                                           & \multicolumn{2}{c|}{non-Fermi liquid\textellipsis} \\ \hline                 
\end{tabular}
\caption{Free and interacting fermionic systems, with and without Berry phase.}
\label{table_fermionic_systems}
\end{table}

In this thesis, we consider multi-band fermionic systems, both free and interacting ones, in which the particles' momenta change in time due to external electromagnetic field, manifesting the effects of the momentum space Berry curvature. Let's organize the different cases as Table \ref{table_fermionic_systems}. In this thesis we study the two cell on the right, the non-interacting Berry Fermi gas, and the interacting Berry Fermi liquid. Below we introduce the contents in this table in some details.

The upper left cell of simple Fermi gas is the most elementary case. The key concept is Pauli's exclusion principle, leading to phenomena such as degeneracy pressure, the existence of Fermi surface\textellipsis which are fundamental to our present day understanding of the physical world.

In simple Fermi gas, Pauli exclusion is crucial to the macroscopic ensemble behavior; however, the particles' dynamics is completely classical. If we go one order higher in $\hbar$ expansion, we will see the quantum effects on the dynamics due to the momentum space Berry curvature. This brings us into the regime of Berry Fermi gas theory. Berry phase effect in fermionic system made its first appearance as early as 1950s, when Karplus, Luttinger and Kohn studied the anomalous Hall effect in ferromagnets and attributed the effect to an ``anomalous velocity'' (explained below) of the electrons in the lattice~\cite{KarplusLuttinger, kohn1957quantum, luttinger1958quantum, luttinger1958theory}. Of course, this was before the formulation of the general concept of Berry phase~\cite{berry1984quantal}. The formal connection between the anomalous velocity and the Berry phase was realized by Chang, Niu and Sundaram in the 1990s~\cite{chang1996berry,chang1995berry,Sundaram:1999zz}. The anomalous velocity is a quantum correction to the particle's velocity perpendicular to the external force (electric field), much like the Lorentz force is a force perpendicular to the particle's velocity. Chang, Niu and Sundaram made this analogy mathematically concrete; in particular, the role played by the magnetic field (a curvature of magnetic flux in position space) in Lorentz force is played by the momentum space Berry curvature in the anomalous velocity.

Berry curvature in Fermi gas leads to other interesting phenomena besides anomalous Hall effect. Most amazingly, the profound phenomenon of chiral anomaly can be reproduced from Berry curvature. The chiral anomaly is the phenomenon that, in a system of Weyl fermions, the right-handed fermion number and the left-handed fermion number are separately conserved classically, but quantum mechanically they are not conserved, only their sum, the total fermion number (total electric charge) is conserved. This phenomenon has important experimental consequences (e.g. the pion decay rate) and deep mathematical structure. In recently years it was shown this phenomenon can be reproduced from the Berry curvature effect in the kinetic theory of Weyl fermions \cite{Son:2012wh,Son:2012zy,Stephanov:2012ki,Gao:2012ix,Chen:2012ca}. Two closely associated phenomena, chiral magnetic effect (CME)~\cite{PhysRevD.22.3080,Fukushima:2008xe} and the chiral vortical effect (CVE)~\cite{PhysRevD.20.1807,Son:2009tf}, can also be reproduced from the Berry curvature in kinetic theory.

The above aspects of Berry Fermi gas will be reviewed in Chapter \ref{chap_BFG} of this thesis.

At the same time as Berry Fermi gas theory achieves such success in describing Weyl fermions, a sharp problem arises. Weyl fermion is a relativistic particle respecting Lorentz invariance. On the other hand, the entire story of Berry curvature explicitly breaks Lorentz invariance, since it is a curvature in the space of spatial momentum $\p$, i.e. it depends on the choice of a reference frame. In Chapter \ref{chap_LI} of this thesis we provide the resolution to this problem~\cite{Chen:2014cla}. For the semi-classical theory with Berry curvature to be Lorentz invariant, or equivalently frame independent, it turns out the ``spacetime position'' $x^\mu$ and the ``four momentum'' $p_\mu$ of the Weyl fermion must be frame dependent. This sounds odd, but it is a consequence of the fact that the semi-classical Weyl fermion has non-point-like feature due to its spin, and therefore the physical meaning of $x^\mu$ and $p_\mu$ must be interpreted carefully. It was also known that the frame dependence of $x$ and $p$ is related to Wigner translation~\cite{Duval:2014cfa,Stone:2015kla}, the non-compact part of the little group of a massless spinning particle; in this thesis we will present the relation in a way somewhat different from the presentation in the literature.

We will take one step further, and consider how collisions of Weyl fermions can be included in the semi-classical kinetic theory in compatible with Lorentz invariance~\cite{Chen:2015gta}. Collisions will relax a far-from-equilibrium ensemble, described by our chiral kinetic theory, to local equilibrium, described by chiral hydrodynamics. This is an important step, establishing the bridge from microscopic quantum mechanics to locally-equilibrium hydrodynamics, via the far-from-equilibrium kinetic theory.

The above explains the first row in Table \ref{table_fermionic_systems}. But in real systems interactions are generally present, and usually not small. Therefore we need to turn to the second row.

Landau's theory of Fermi liquid~\cite{landau1957theory,landau1957oscillations}, established in the 1950s, is one of the cornerstones of condensed matter physics. A \emph{Fermi liquid} is an interacting fermionic system satisfying two conditions: its ground state is described by a Fermi surface much like that of a non-interacting Fermi gas, and moreover its low energy spectrum is described by ``quasiparticle'' excitations which are qualitatively similar to non-interacting particles. With his superb insight, Landau realized there is only one effect of interaction at low energy -- a local interaction potential energy between quasiparticles. Quasiparticle decay and collision arising from interactions are suppressed by the low energy, due to the limited availability of decay / collision channels. Thus, the low energy behavior of a Fermi liquid can be described by a kinetic theory very similar to that of a Fermi gas, with the addition of a local interaction potential energy. This simple picture, the Landau Fermi liquid theory, has been extremely successful in describing a large class of interacting fermionic systems, most notably Helium-3~\cite{baym1978physics} and electrons in normal metal.

Landau's intuitive theory was soon proven by matching with Feynman diagram expansion to all orders in perturbations theory, by Landau himself and others~\cite{landau1959theory,Nozieres:1962zz,Luttinger:1962zz,abrikosov1975methods}. In the 1990s, Landau's original insight was finally concretely casted in the language of low energy effective field theory~\cite{Polchinski:1992ed,Shankar:1993pf}. Except for a possible instability in the Bardeen-Cooper-Schrieffer channel, Landau's Fermi liquid theory provides a truly universal low-energy effective description of Fermi liquid systems with short-ranged interactions.

We shall now turn to the motivation of our Berry Fermi liquid theory \cite{chen2016berry}. As mentioned above, people have derived many interesting results from Berry Fermi gas theory, but how much of those survives once interaction is included (as in real systems)? Also, are there any new effects arising from interactions that are not present in Berry Fermi gas theory? To answer these two questions, we have to extend Landau's Fermi liquid theory into next-to-leading order in low energy / long wavelength expansion. This extension, we call \emph{Berry Fermi liquid theory}, is presented in Chapter \ref{chap_BFL}. We will also provide a proof to the Berry Fermi liquid theory by resumming Feynman diagram expansion to all orders. Our main results are the followings. First, it is known that for a Fermi gas, the Berry curvature effects can be written as Fermi surface integral~\cite{Haldane:2004zz}. It would be important if the same can be done for a Fermi liquid, because for a Fermi liquid the system's properties far from the Fermi surface are generally complicated and non-universal. We show this indeed can be done in Fermi liquid, and the form of the Fermi surface integral is almost the same as in Fermi gas. Second, in addition to the usual Berry curvature contribution, there is a new contribution to the anomalous Hall effect from the quasiparticles' emergent electric dipole moment due to interactions. Third, at next-to-leading order in low energy / long wavelength expansion, we have to consider quasiparticle decay and collisions; however, they have no contribution to the ``interesting physics'' such as anomalous Hall effect, chiral magnetic effect, etc.

In the previous literature, the work that has the most overlap with our Berry Fermi liquid theory is Ref.~\cite{shindou2008gradient} where the interplay between Berry curvature and interaction has been studied in a very general context. The authors of Ref.~\cite{shindou2008gradient} showed, via the Keldysh formalism, that the quasiparticles' motion has an anomalous velocity due to the Berry curvature, as in the non-interacting case, but the content of the Berry curvature is modified by interactions. There are four main differences between Ref.~\cite{shindou2008gradient} and our Chapter \ref{chap_BFL}. First, in contrast to Ref.~\cite{shindou2008gradient}, we only study linear response, which does not see the effect of the anomalous velocity. Second, in our theory the Berry curvature effects show up in (the non-quasiparticle contribution to) the current, which is not computed in Ref.~\cite{shindou2008gradient}. Third, we are able to take into account the effect of the quasiparticle collisions and the finite quasiparticle lifetime. Last, we are able to answer the question whether interesting transport phenomena such as the anomalous Hall effect involve Fermi surface contribution only, or involve Fermi sea contribution as well.

One can notice at the bottom of Table \ref{table_fermionic_systems}, there is also ``non-Fermi liquid''. Fermi liquid describes a large class of interacting fermionic systems, but there are also many non-Fermi liquid systems, not satisfying one or both assumptions about Fermi liquid. The study of these systems is an important subject, but far beyond the scope of this thesis.

Finally, we emphasize that impurities / disorders are completely neglected in this thesis. This is a purely theoretical idealization. There are extra non-trivial effects if disorders are included. For instance, there will be extra contribution to anomalous Hall effect from skew-scattering and side-jump; see Ref.\cite{nagaosa2010anomalous} for review.

\

The chapters of this thesis are related to the author's research works as the following. Chapter \ref{chap_BFG} is mostly a summarization of previous literature, except for Section \ref{sect_BFG_PIderivation}, which is a generalized and refined version of the path integral derivation in the author and collaborators' work Ref.\cite{Chen:2014cla}. The first two sections of Chapter \ref{chap_LI} are also based on \cite{Chen:2014cla}; the third section is from some unpublished work of the author; Section \ref{sect_collisions} is an elaboration of the author and collaborators' work \cite{Chen:2015gta}. Chapter \ref{chap_BFL} is mainly a reproduction of the author and advisor's recent work~\cite{chen2016berry}, with the example in Section \ref{sect_Invitation} added for demonstration purpose.

Needless to say, the references included in this thesis are far from a complete list on this rich subject. Here the author only included those which were more familiar to the author during the course of study. There must be a lot of important works missing from the references due to the author's limitation.

\section{Organization of the Thesis}

Chapters \ref{chap_BFG} and \ref{chap_LI} are devoted to Berry Fermi gas system without interaction. In Chapter \ref{chap_BFG}, we start with a review of the symplectic formulation of classical mechanics. Then we present the semi-classical action of a single particle with Berry curvature in external electromagnetic field, and extract the implications from the action using the symplectic formulation. We then consider a non-interacting ensemble of such particles -- we focus on fermions in particular -- and discuss the general effects of Berry curvature on the ensemble's macroscopic behavior; we will pay particular attention to the anomalous Hall current due to Berry curvature. To support the validity of the semi-classical action, in Section \ref{sect_BFG_PIderivation} we derive it from path integral under very general setting, and get to understand the microscopic origin of the particle's Berry curvature as well as its magnetic dipole moment. We then turn from the general case to the specific example of Weyl fermion as a concrete demonstration of the general formalism. The example of Weyl fermion is important in its own. It has band touching, and exhibits chiral anomaly as well as the associated chiral magnetic effect and chiral vortical effect. We will present the computation of chiral magnetic effect and chiral anomaly from the semi-classical theory of Berry Fermi gas, and discuss the physical picture.

The semi-classical theory of Weyl fermion seems to violate Lorentz invariance, but it must not. In Chapter \ref{chap_LI} we resolve this puzzle. We will first provide a physical argument arguing that the notion of ``position'' of a semi-classical massless spinning particle must be frame dependent. Then we present in detail how Lorentz invariance is realized non-trivially in the semi-classical theory -- both the notion of ``position'' and ``momentum'' are frame dependent. We will show that, despite the frame dependence of position and momentum, the current and stress-energy tensor are frame independent (Lorentz covariant), and therefore they are the legitimate basic physical observables. We will give a physical interpretation to the non-trivial realization of Lorentz invariance based on the helical (spinning) feature of the current. In Section \ref{sect_Wigner_transl} we will explore the mathematical foundation of the non-trivial realization of Lorentz invariance -- it is due to Wigner translation, the non-compact part of the little group of a massless spinning particle. Finally, in Section \ref{sect_collisions} we will consider one step beyond Fermi gas theory. We assume the Weyl fermions collide, and study the restrictions on the form of the collisionful kinetic theory due to Lorentz invariance. We show that such collisionful kinetic theory of Weyl fermions will relax to the familiar chiral hydrodynamical limit, in which we will compute the chiral vortical effect.

Chapter \ref{chap_BFL} is devoted to interacting fermionic systems with Berry curvature. We start with a simple example, Dirac fermion with weak contact interaction in 2 spatial dimensions, to demonstrate the appearance of new contribution to anomalous Hall current not present in non-interacting systems. This example shows the inclusion of interaction must be a non-trivial story. To proceed, we first provide a quick review on the computation of linear response in Landau's Fermi liquid theory. Then we present our proposed Berry Fermi liquid theory, which incorporates Berry curvature effects (as well as other effects) in extension to Landau's theory. In Section \ref{sect_QFT} we derive the Berry Fermi liquid theory from perturbative quantum field theory under very general assumptions. Our derivation is valid to all orders in Feynman diagram expansion, as long as we stay in the long wavelength limit. The quantum field theoretic derivation itself is quite technical and length, involving heavy use of Cutkosky cut, Ward-Takahashi identity and combinatorial diagrammatic techniques; the organization of the derivation is provided at the beginning of Section \ref{sect_QFT}.

At the end of each of these three chapters there is a ``Summary and Outlook'' section, summarizing the main results and ideas in that chapter, and discussing unresolved problems and future directions of study. These discussions will not be repeated in the final Conclusion chapter. The final Conclusion chapter will contain brief discussions in broader perspective.

\section{Conventions in the Thesis}

To make the computations readable, it is important to clarify the conventions.

We generally consider systems in $d$ spatial dimensions, i.e. $(d+1)$ spacetime dimensions, with $d\geq 2$.

We denote spacetime (relativistic or not) coordinate as $x^\mu$, with $x^0$ being time $t$ (instead of $ct$ in usual relativistic context) and $x^i \ \ (i=1, \dots, d)$ being spatial position. Spatial momentum is denoted as $p^i=p_i$, while kinetic energy is denoted as $p^0 = -p_0$. For derivatives with respect to momentum, we denote $\partial_p^0 \equiv \partial / \partial p_0 = -\partial / \partial p^0 = \partial_{p_0} = - \partial_{p^0}$ while $\partial_p^i \equiv \partial / \partial p_i = \partial / \partial p^i  = \partial_p^i= \partial_{p^i}$. This rule of raising and lowering indices on momentum is compatible with, but does not rely on the Minkowski metric; it follows from the canonical structure of classical / quantum mechanics.

For relativistic systems, our convention of Minkowski metric is $\eta_{\mu\nu} = \mathrm{diag}(-1, 1, \dots, 1)$, and our convention of spacetime Levi-Civita symbol is $\epsilon_{012\dots d} = -\epsilon^{012\dots d} = 1$.

We set the Boltzmann constant to $k_B=1$. For relativistic systems the speed of light $c=1$ is understood. However, we will generally leave $\hbar$ explicit to keep track of the quantum effects in the semi-classical Berry curvature framework. We will set $\hbar=1$ in the quantum field theory computations in Sections \ref{sect_Invitation} and \ref{sect_QFT}.

The validity of the semi-classical picture relies on low frequency / long wavelength expansion, i.e. the expansion of $q_\mu \sim \hbar\, \partial_{x^\mu}$ over the typical energy / momentum scale of the system. It is equivalent to keep track of $\hbar$ or $q$; in Chapters \ref{chap_BFG} and \ref{chap_LI} we keep track of the former, and in Chapter \ref{chap_BFL} the latter.

The notation $X^{[i_1\dots i_n]}$ means $X^{i_1\dots i_n}$ with all the indices antisymmetrized; there are $n!$ terms in total, and in our convention there is an overall factor of $1/n!$. The notation $X^{\{i_1\dots i_n\}}$ means $X^{i_1\dots i_n}$ with all the indices symmetrized; again there are $n!$ terms in total and in our convention there is an overall factor of $1/n!$.

\chapter{Berry Fermi Gas}
\label{chap_BFG}

In this chapter, we discuss the semi-classical theory of a single particle with Berry curvature, and extract its implications on an ensemble of non-interacting fermions -- a Berry Fermi gas. The particular strength of the Berry Fermi gas theory, as a Boltzmann type kinetic theory, is that it describes macroscopic systems far-from-equilibrium, and therefore has broader application (when interactions are small) than the commonly used hydrodynamics theory which assumes local equilibrium. We will see how the Berry phase-related interesting physics of anomalous Hall effect, chiral magnetic effect and chiral anomaly are computed in the Berry Fermi gas theory.

This field has been very well-developed, therefore this chapter is mostly a summarization of previous works. Section \ref{sect_BFG_PIderivation} contains a path integral derivation of the semi-classical theory with full generality, which is not written down explicitly in the previous literature.

\section{Review of Symplectic Formulation of Classical Mechanics}
\label{Rev_Classical_Mech}

To most efficiently convey the Berry phase physics, we use the symplectic form formulation of classical mechanics. Below is a brief review of the formulation.

\subsection{Worldline Parametrized by Time}

Consider the action of a single particle moving in $d$-dimensional space under Hamiltonian $H(\p, \x, t)$:
\begin{eqnarray}
S[\p(t), \x(t)] = \int dS, \ \ \ \ \ \ \ dS= p_i dx^i - H(\p, \x, t) dt
\label{general_usual_action}
\end{eqnarray}
where the integration is along an arbitrary worldline of the particle in the $2d$-dimensional phase space. We regard the first term of $dS$ as the \emph{symplectic part}, and the second term as the \emph{Hamiltonian part}. Now we denote $(x^i, p_i)$ collectively as $\xi_I$, so the action can be written as
\begin{eqnarray}
dS=\omega_I d\xi^I - H(\xi, t) dt
\label{general_action}
\end{eqnarray}
where $\omega_I$ is called the \emph{symplectic 1-form}, $\omega_{x^i}=p_i, \omega_{p_i}=0$. Adding to $\omega_I$ a total $\xi^I$ derivative changes the action by a boundary term and affects no dynamics.

Consider arbitrary infinitesimal variations $\delta\xi^I$:
\begin{eqnarray}
\delta dS = \delta \xi^I \omega_{IJ} d\xi^J - \delta \xi^I \left(\partial_{\xi^I} H + \partial_t \omega_I \right) dt,
\label{general_action_variation}
\end{eqnarray}
where dropping total derivative terms is always implicitly understood. Here $\omega_{IJ}\equiv 2\partial_{[\xi^I} \omega_{J]}$ is the \emph{symplectic 2-form}. Explicitly,
\begin{eqnarray}
\omega_{IJ} = \left[ \begin{array}{c|c} \omega_{x^i x^j} = 0 & \omega_{x^i p_j} = -\delta_i^j \\ \hline \omega_{p_i x^j} = \delta^i_j & \omega_{p_i p_j} = 0 \end{array} \right].
\label{standard_symplectic}
\end{eqnarray}
On the other hand, $\partial_t \omega_I=0$ in the present case; we will see below why we formally keep it.

The equation of motion (EoM) is derived by requiring $d\xi$ to be such that $\delta dS$ is total derivative for arbitrary $\delta \xi$. This requires
\begin{eqnarray}
\omega_{IJ} d\xi^J = \left(\partial_{\xi^I} H + \partial_t \omega_I\right) \ dt, \ \ \ \ \ \mbox{i.e.} \ \left. d\xi^I/dt \right|_\EoM = \omega^{IJ} \left(\partial_{\xi^J} H + \partial_t \omega_J\right),
\label{general_EoM}
\end{eqnarray}
where $\omega^{IJ} \equiv (\omega^{-1})^{IJ}$ is the inverse symplectic 2-form whose components are those of the \emph{Poisson brackets}:
\begin{eqnarray}
\omega^{IJ} \equiv (\omega^{-1})^{IJ} = \left[ \begin{array}{c|c} \{x^i, x^j\} = 0 & \{x^i, p_j\} = \delta^i_j \\ \hline \{p_i, x^j\} =-\delta_i^j & \{p_i, p_j\} = 0 \end{array} \right].
\end{eqnarray}
Similar, the symmetries of an action can be derived by requiring $\delta\xi$ to be such that $\delta dS$ is total derivative for arbitrary $d\xi$. This requires the existence of some function $Q(\x, \p, t)$, called the \emph{Noether charge} of the corresponding symmetry, such that
\begin{eqnarray}
\delta \xi^I \omega_{IJ} = -\partial_{\xi^J} Q, \ \ \ \ \ \ \delta \xi^I \left(\partial_{\xi^I} H + \partial_t \omega_J\right) = \partial_t Q.
\label{general_Noether_charge}
\end{eqnarray}
It is easy to see when the EoM \eqref{general_EoM} is satisfied, $dQ/dt|_\EoM = -\delta dS/dt = 0$, i.e. the Noether charge is a conserved quantity at EoM. Finally, if the symmetry transformation involves the transformation of some non-dynamical parameter $\alpha$ on which $H$ depends, then the second equation in \eqref{general_Noether_charge} should have an extra term $\delta\alpha \partial_\alpha H$ on the left-hand-side.

All these seem quite trivial -- all we have done is to state the textbook Hamiltonian mechanics~\cite{goldstein1980classical} in some fancy language. But the key point is the following: Now it is no longer necessary to use $(x^i, p_i)$ to parametrize the phase space; $\xi^I$ can be arbitrary parametrization of the phase space, and \eqref{general_action_variation}, \eqref{general_EoM} and \eqref{general_Noether_charge} still hold, as long as we require $\omega_{IJ}$ to transform as a (0, 2)-tensor on the phase space. That is, under reparametrization $\xi^I \rightarrow \xi'^{I'}$,
\begin{eqnarray}
\omega_{I'J'} = \omega_{IJ} \partial_{\xi'^{I'}} \xi^I \partial_{\xi'^{J'}} \xi^J.
\label{symplectic_as_2-form}
\end{eqnarray}
If $\omega_{IJ}$ in a parametrization $\xi^I$ still takes the ``standard form'' \eqref{standard_symplectic}, then the parametrization $\xi^I$ are called \emph{canonical variables}, otherwise $\xi^I$ are non-canonical; for the study of Berry curvature physics it is convenient to use non-canonical variables, as we will see later. Clearly, in any parametrization, $\omega$ is always a non-degenerate, closed 2-form -- in fact, these two properties are the defining properties of a symplectic 2-form. Non-degeneracy means the phase space is physical and involves no redundant gauge degree of freedom; the dynamics is completely determined by the EoM \eqref{general_EoM}. The meaning of closed-ness will be discussed later.

The reparametrization of $\xi^I \rightarrow \xi'^{I'}$ can even be time dependent, as long as we require the Hamiltonian to transform accordingly:
\begin{eqnarray}
H' =  H + \omega_J \partial_t \xi^J
\label{H'_H_transf}
\end{eqnarray}
so that the action \eqref{general_action}, or equivalently its variation \eqref{general_action_variation}, remains invariant up to total derivative.

Let's see a simple example of non-canonical variables. A particle moving in external electromagnetic field has action
\begin{eqnarray}
dS=p_i dx^i + A_i(\x, t) dx^i - H(\p, \x, t) dt
\label{action_in_EM}
\end{eqnarray}
(we have absorbed the electric potential $-A_0$ into $H$). Here $p_i$ is the physical momentum and is not canonical to $x^i$. Explicitly,
\begin{eqnarray}
\omega_{IJ} = \left[ \begin{array}{c|c} \omega_{x^i x^j} = F_{ij} & \omega_{x^i p_j} = -\delta_i^j \\ \hline \omega_{p_i x^j} = \delta^i_j & \omega_{p_i p_j} = 0 \end{array} \right]
\label{EM_symplectic}
\end{eqnarray}
where $F_{ij}\equiv 2\partial_{[x^i} A_{j]}$ is the magnetic field, and its corresponding block $\omega_{x^ix^j}$ gives rise to the Lorentz force. One can define the canonical momentum
\begin{eqnarray}
P_i(t) \equiv p_i(t) + A_i(\x(t)),
\label{EM_canonical_P}
\end{eqnarray}
so that $(x^i, P_i)$ are canonical variables and $\omega$ is brought back into the standard form \eqref{standard_symplectic}, at the price that $H$ would now depend on $A$ through $p_i=P_i-A_i$.

The transformation law \eqref{symplectic_as_2-form} implies the volume element
\begin{eqnarray}
d^{2d} \xi \: \sqrt{\det \omega_{IJ}} = d^{2d} \xi' \: \sqrt{\det \omega_{I'J'}}
\end{eqnarray}
is the natural volume element in the phase space -- this is called the Liouville volume element. For canonical variables clearly it is just $d^{2d} \xi$. An important property of the volume element is the Liouville's theorem:
\begin{eqnarray}
\partial_t \sqrt{\det\omega} + \partial_{\xi^I} \left(\sqrt{\det\omega} \left.\frac{d\xi^I}{dt}\right|_\EoM\right) &=& -\frac{\sqrt{\det\omega}}{2} \ 3\partial_{[\xi^I}\omega_{JK]} \ \omega^{IJ} \omega^{KL} \ \left(\partial_{\xi_L} H + \partial_t \omega_L\right) \nonumber \\[.2cm]
&=& 0
\label{Liouville_Thm}
\end{eqnarray}
where the second equality follows from the closed-ness $\partial_{[\xi^I}\omega_{JK]}=0$. One can understand this theorem as: a fluid (ensemble of independent particles) is flowing in the phase space in time according to the EoM, but the Liouville volume of this fluid remain the same over time.

A question arises. In general relativity, we know not all metrics are locally reparametrization-equivalent to the Minkowski metric, and the local inequivalence is measured by the Riemann curvature. Are all symplectic 2-forms locally reparametrization-equivalent to the standard one \eqref{standard_symplectic}? The answer is Yes: All non-degenerate closed 2-form is locally reparametrization-equivalent to the form \eqref{standard_symplectic}. This is the Darboux theorem and can be easily proven by constructing a reparametrization using interpolation~\cite{da2001lectures}. Thus, in classical mechanics, locally there is no non-trivial background in the phase space. The most familiar example is \eqref{EM_canonical_P}.

But globally there is non-trivial information in the symplectic 2-form. While closed 2-forms are locally exact, globally they may not be if the phase space is not simply-connected; if so, $\omega_I$ cannot be a globally continuous 1-form over the phase space. The failure of $\omega_{IJ}$ being globally exact is measured by its de Rham cohomology. In our Berry phase story, this issue is associated with the semi-classical picture of chiral anomaly in Section \ref{sect_chiral_anomaly}, where the classical phase space has non-trivial topology because of band-touching points in the momentum space.

\subsection{Worldline Parametrized by Intrinsic Parameter}

In the above, we have reviewed the symplectic formulation of single particle classical mechanics. While the formalism is mostly beautiful, it has one ugliness: Time $t$, which we use to parametrize the worldline, is a physical quantity, so the system may have time dependence, and the phase space reparametrization $\xi\rightarrow \xi'$ might also have time dependence. These time dependence makes the symplectic 1-form appear explicitly in \eqref{general_EoM} and \eqref{H'_H_transf}. But the symplectic 1-form is contains non-physical information -- the phase space gauge freedom of adding total $\xi$ derivative. Is there a formulation that can get rid all these troubles due to time dependences?

There is another motivation for an alternative formulation. If we consider relativistic systems (which we will), Lorentz invariance puts time and space at equal footing. But using time to parametrize the worldline makes time special.

With these motivations, it is now clear how we shall reformulate the physics. Instead of using the physical time $t$ to parametrize the particle's worldline, we use some intrinsic time parameter $\tau$. Since $\tau$ is intrinsic to the worldline only, the physical system shall not depend on $\tau$, and the any physically meaningful transformation should not depend on $\tau$. The price we pay is we promote the physical time and the associated energy to dynamical quantities. For example, the action \eqref{action_in_EM} is reformulated as
\begin{eqnarray}
S = \int dS, \ \ \ \ \ \ dS=p_\mu dx^\mu + A_\mu(x) dx^\mu - \mathcal{H}(p, x) \lambda d\tau
\end{eqnarray}
where $x^0\equiv t$ is time and $-p_0=p^0$ is energy. $\lambda$ is the einbein on the worldline such that $\lambda d\tau = \lambda' d\tau'$ under worldline reparametrization. $\lambda$ is also a Lagrange multiplier -- the variation with respect to it demands $\mathcal{H}(p, x)=0$, from which the energy $p^0$ can be solved in the form
\begin{eqnarray}
p^0 = H(\p, \x, t)
\end{eqnarray}
where $H$ is the Hamiltonian in our old formulation. The EoM for $x^0$ relates the intrinsic $\lambda d\tau$ to the physical $dx^0$, while the EoM for $p^0$ will be compatible with $dp^0/\lambda d\tau=dH/\lambda d\tau$. For non-relativistic particle, we can choose $\mathcal{H}=p^0-|\p|^2/2m$. For relativistic particle, a nice choice is $\mathcal{H}=p_\mu p^\mu+m^2$ so that the entire action is manifestly Lorentz invariant (demanding $p^0>0$ in the solution is implicitly understood). 

Now the phase space is $2(d+1)$-dimensional, and instead of the $(x^\mu, p_\mu)$ parametrization, we can consider arbitrary parametrization $\xi^\Ical$; any parametrization of the phase space should be independent of the worldline parameter $\tau$. The formal results in the old formulation are straightforwardly carried over, except $dt$ are replaced by $\lambda d\tau$, and now nothing has the $\partial_\tau$ dependence. The only unobvious modification is the Liouville phase space volume now becomes
\begin{eqnarray}
\int d^{2d}\xi \sqrt{\det\omega_{IJ}} = \int d^{2d+2}\xi \sqrt{\det\omega_{\Ical\Jcal}} \ \delta(\mathcal{H}) \: \frac{d}{{\lambda d\tau}}.
\label{Liouville_volume_intrinsic}
\end{eqnarray}
(This can be proven by invoking the EoM for $dx^0/\lambda d\tau$.) In particular, the right-hand-side is invariant under reparametrization $\xi\rightarrow \xi'$, because the reparametrization must be $\tau$ independent and hence $\mathcal{H}'=\mathcal{H}$. For relativistic systems, this expression of Liouville phase space volume is very useful because it is manifestly Lorentz invariant as long as we have chosen a Lorentz invariant expression for $\mathcal{H}$ (e.g. $\mathcal{H}=p_\mu p^\mu+m^2$).

\section{Semi-Classical Particle with Berry Phase}

Having had the formalism of symplectic 2-form introduced, we are ready to present the semi-classical mechanics of a single particle with Berry curvature. The quantum mechanical justification of this formalism is left to Section \ref{sect_BFG_PIderivation}. To be consistent with the literature, we will use the physical time as the worldline parameterization in this chapter. The intrinsic parametrization will be used in the next chapter when we discuss the Lorentz invariance of semi-classical Weyl fermions.

The action is~\cite{Chang:2008zza,Xiao:2009rm}
\begin{eqnarray}
S=\int dS, \ \ \ \ \ \ dS=p_i dx^i + A_i(\x, t) dx^i - \hbar \: a^i(\p) dp_i - H_0(\p, \x, t) dt + A_0(\x, t) dt
\label{action_Berry}
\end{eqnarray}
where we have separated $A_0$ from $H$, so that $H=H_0-A_0$. $a^i(\p)$ is the Berry connection in the momentum space. Microscopically it is given by
\begin{eqnarray}
\hbar \: a^i(\p) = -i\hbar \: u_\alpha^\dagger(\p) \: \partial_p^i u^\alpha(\p) = i\hbar \: \partial_p^i u_\alpha^\dagger(\p) \: u^\alpha(\p)
\end{eqnarray}
where $u^\alpha(\p)$ is the normalized Bloch state or spinor of the quantum mechanical particle when $A_\mu=0$. The Berry curvature is given by
\begin{eqnarray}
\hbar \: b^{ij}(\p) \equiv \hbar\, 2\partial_p^{[i} a^{j]}(\p) = -2i\hbar \: \partial_p^{[i} u_\alpha^\dagger(\p) \: \partial_p^{j]} u^\alpha(\p).
\label{Berry_curv_def}
\end{eqnarray}
We leave $\hbar$ explicit to keep track of semi-classical effects. Quantum mechanically, we can multiply $u^\alpha$ by a $\p$-dependent complex phase, and no physics should change; indeed, such a transformation adds to $a^i$ a total $p_i$-derivative, under which the action and the Berry curvature are left invariant. 

What is the range of applicability of this action? As we will see in the derivation in Section \ref{sect_BFG_PIderivation}, the semi-classical dynamics is valid to first order in $\hbar$ (compared to the typical action of a worldline). This means $\hbar\partial_x/|\p|$ is kept to first order, so the expansion over $\hbar$ is equivalent to the low frequency / long wavelength expansion over $\partial_{x^\mu}$.

The symplectic 2-form is given by
\begin{eqnarray}
\omega_{IJ} = \left[ \begin{array}{c|c} \omega_{x^i x^j} = F_{ij} & \omega_{x^i p_j} = -\delta_i^j \\ \hline \omega_{p_i x^j} = \delta^i_j & \omega_{p_i p_j} = -\hbar \: b^{ij} \end{array} \right].
\label{Berry_symplectic}
\end{eqnarray}
Since the semi-classical theory is valid to order $\hbar$, we only need to find the determinant and inverse of $\omega_{IJ}$ to order $\hbar$. The determinant, and hence Liouville volume, is modified by the Berry curvature~\cite{Xiao:2005qw,Duval:2005vn}
\begin{eqnarray}
\sqrt{\det\omega} = 1+ \hbar F_{ij} b^{ij} /2.
\label{Liouville_with_Berry}
\end{eqnarray}
The inverse, or Poisson bracket, is given by
\begin{eqnarray}
\omega^{IJ} = \left[ \begin{array}{c|c} \{x^i, x^j\} = -\hbar\, b^{ij} & \{x^i, p_j\} = \delta^i_j + \hbar\, b^{ik} F_{kj} \\ \hline \{p_i, x^j\} =-\delta_i^j - \hbar\, F_{ik} b^{kj} & \{p_i, p_j\} = F_{ij} + \hbar\, F_{ik} b^{kl} F_{lj} \end{array} \right].
\label{Berry_Poisson}
\end{eqnarray}
The EoM \eqref{general_EoM} written explicitly to order $\hbar$ will read
\begin{eqnarray}
\left.\frac{dx^i}{dt}\right|_\EoM &=& \partial_p^i H_0 + \hbar\, b^{ij} \left( F_{j0} + F_{jk} \partial_p^k H_0 - \partial_{x^j} H_0 \right),  \nonumber \\[.3cm]
\left.\frac{dp_i}{dt}\right|_\EoM &=& F_{i0} + F_{ij} \left.\frac{dx^j}{dt}\right|_\EoM - \partial_{x^i} H_0 \nonumber \\[.2cm]
&=& \left(\delta_i^j + \hbar\, F_{il} b^{lj}\right) \left( F_{j0} + F_{jk} \partial_p^k H_0 - \partial_{x^j} H_0 \right)
\label{BFG_EoM}
\end{eqnarray}
(recall that $F_{ij}$ is the magnetic field and $F_{i0}=-F_{0i}$ is the electric field). The $\hbar$ term in $dx^i/dt$, given by the Berry curvature contracting with the force, is the famous \emph{anomalous velocity}~\cite{KarplusLuttinger,Sundaram:1999zz}.

Here we work with Berry curvature in the momentum space only. Of course, if there is some external adjustable parameter, on which the Hamiltonian depends, that varies slowly in space and time, there would be extra Berry curvature components $b_{p_i x^j}$, $b_{x^i p^j}$ and $b_{x^i x^j}$~\cite{Sundaram:1999zz,Chang:2008zza,Xiao:2009rm}. In this thesis we would not consider such generality.

Although the Darboux theorem mentioned in Section \ref{Rev_Classical_Mech} guarantees the existence of canonical variables, they are rarely used in the semi-classical Berry phase physics literature. Here we mention them for completeness. If Berry curvature is absent, $x^i$ and $P_i \equiv p_i+A_i$ are canonical variables. In the presence of Berry curvature, a convenient choice of canonical variables $(\mathcal{X}^i, \mathcal{P}_i)$ valid to order $\hbar$ is
\begin{eqnarray}
&& \hspace{-.5cm} x^i = \mathcal{X}^i - \hbar\, a^i(\vec{\mathcal{P}}-\vec{A}(\vec{\mathcal{X}}, t)), \ \ \ \ p_i = \mathcal{P}_i - A_i(\vec{\mathcal{X}}, t) - \hbar\, a^j(\vec{\mathcal{P}}-\vec{A}(\vec{\mathcal{X}}, t)) \: F_{ij}(\vec{\mathcal{X}}, t); \nonumber \\[.2cm]
&& \hspace{-.5cm} \mathcal{X}^i = x^i + \hbar\, a^i(\p), \ \ \ \ \mathcal{P}_i = p_i + A_i(\x, t) + \hbar \, a^j(\p) \: \partial_{x^i} A_j(\x, t).
\end{eqnarray}
(Here we expand in $\hbar$ which is the natural expansion parameter in semi-classical physics; in \cite{Xiao:2009rm} the authors expanded in $A$.) According to \eqref{H'_H_transf}, the Hamiltonian should shift by
\begin{eqnarray}
H_{cano.}(\mathcal{P}, \vec{\mathcal{X}}, t) &=& H(\p, \x, t) - \hbar \, a^i(\p) \: \partial_t A_i(\x, t) \nonumber \\[.2cm]
&=& H_0(\vec{\mathcal{P}}-\vec{A}, \vec{\mathcal{X}}, t) - A_0(\vec{\mathcal{X}}, t) \ + \ \hbar \, a^i(\vec{\mathcal{P}}-\vec{A}) \: \times \nonumber \\[.2cm]
&& \hspace{.5cm} \left(F_{i0} +F_{ij}\, \partial_{\mathcal{P}_j} H_0(\vec{\mathcal{P}}-\vec{A}, \vec{\mathcal{X}}, t) - \partial_{\vec{\mathcal{X}}^i} H_0(\vec{\mathcal{P}}-\vec{A}, \vec{\mathcal{X}}, t)\right)
\end{eqnarray}
where $H=H_0-A_0$, and the argument of $A_\mu$ is $(\vec{\mathcal{X}}, t)$. The Liouville volume in canonical variables is the usual $d^d \mathcal{P} \: d^d \mathcal{X}$.

\section{Ensemble of Berry Fermi Gas}

In the previous section we considered the Berry curvature effects for a semi-classical single particle. Now we study the effects in an ensemble of non-interacting particles -- in this thesis we focus on fermions. One of the important consequence is the relation between the anomalous Hall current and the anomalous velocity due to Berry curvature~\cite{KarplusLuttinger,Sundaram:1999zz}.

\subsection{Boltzmann Equation and Current}

We let $\rho$ denote the particle density per unit $d^{2d}\xi = d^d x \: d^dp$:
\begin{eqnarray}
\rho(\p, \x, t) \equiv \frac{dN}{d^d x \: d^dp}.
\end{eqnarray}
If particles are moving independently and not being created or annihilated, the continuity equation
\begin{eqnarray}
\partial_t \rho + \partial_{\xi^I} \left(\rho \ d\xi^I/dt \right) = 0
\label{cont_Eq}
\end{eqnarray}
must hold whether or not EoM is satisfied. Now we define the distribution function $f(\p, \x, t)$ to be the particle density per Liouville volume:
\begin{eqnarray}
f \equiv \frac{\rho}{\sqrt{\det \omega}} = \frac{dN}{d^{2d}\xi \sqrt{\det \omega}}.
\end{eqnarray}
With the aid of the Liouville's theorem \eqref{Liouville_Thm}, the continuity equation implies the collisionless Boltzmann equation
\begin{eqnarray}
\partial_t f + \left.\left(d\xi^I/dt\right)\right|_\EoM \ \partial_{\xi^I} f \ = \ \partial_t f + \partial_{\xi^I} f \: \omega^{IJ} \left(\partial_{\xi^J} H+\partial_t \omega_J\right)= 0.
\label{Boltzmann_Eq}
\end{eqnarray}
For fermions, Pauli exclusion requires $0\leq f \leq 1$.

A system is in equilibrium, i.e. $\partial_f=0$, if $\omega_{IJ}, H$ are time independent and $f=f(H)$. However, in reality there is always some small collision (whose relaxation time is perhaps much longer compared to our time scale of interest), therefore really the only equilibrium distribution is the Fermi Dirac distribution $f=f_{FD}(H)$.

The usual physical observables for an ensemble are the current and the stress-energy tensor. First consider the current of a single particle with worldline $(z^i(t), p_i(t))$ in the phase space satisfying the EoM:
\begin{eqnarray}
J^\mu_{sp}(\x, t)[\p, \z] &\equiv& \left.\frac{\delta S[\p, \z]}{\delta A_\mu(\x, t)}\right|_\EoM \nonumber \\[.3cm]
&=& \int dt' \left(\left.\frac{dz^\mu}{dt'}\right|_\EoM - 2\frac{\partial H(\p, \z, t')}{\partial F_{\nu\mu}(\z, t')} \: \partial_{z^\nu}\right) \delta^d(\z-\x) \delta(t'-t)
\end{eqnarray}
where the integral is along the physical worldline (satisfying EoM) of the particle in the phase space, and $z^0\equiv t'$, $x^0\equiv t$. For an ensemble of non-interacting particles, the total current is just the integral of the single particle currents over the particle density in the phase space:
\begin{eqnarray}
&& J^\mu(\x, t) \nonumber \\[.2cm]
&\equiv& \int\frac{d^d p \: d^d z \: \sqrt{\det\omega(\p,\z, t)}}{(2\pi \hbar)^d} \: f(\p,\z, t) \: J^\mu_{sp}(\x, t)[\p, \z] \nonumber \\[.2cm]
&=& \int \frac{d^d p \: \sqrt{\det\omega(\p,\x, t)}}{(2\pi \hbar)^d} \left[\left.\frac{dx^\mu}{dt}\right|_\EoM f(\p,\x, t) + \partial_{x^\nu} \left(-2\frac{\partial H(\p, \x, t)}{\partial F_{\mu\nu}(\x, t)} f(\p,\x, t)\right) \right].
\label{J_expression}
\end{eqnarray}
The $J^0$ component is interpreted as the particle density at $(\x, t)$, while the $J^i$ components are the current density. The conservation $\partial_{x^\mu} J^\mu=0$ just follows from the $U(1)$ gauge invariance as usual.

Clearly the first term in \eqref{J_expression} is the transport current. What is the nature of the second term? Note that $-2\partial H/\partial F_{ij}$ can be understood as the magnetic dipole moment and $-2\partial H/\partial F_{i0}$ the electric dipole moment, so the second term really corresponds to the electric polarization density (when $\mu=0$) and magnetization / electric polarization current (when $\mu=i$). This term plays an important role in, for instance, defining the chiral vortical current for Weyl fermions, as we will see in Section \ref{ssect_CVE}.

We emphasize that although in the definition of the current, we made use of the electromagnetic field, the result also holds for neutral particles, as we can take the charge of the particle (which has been absorbed in $A$) to the zero limit.

If we couple the action to a spacetime structure (general relativity geometry for relativistic particle or Newton-Cartan geometry for non-relativisitic particle), we can also derive the stress-energy tensor. This is beyond the scope of this thesis. We just comment the followings. The energy density $T^{00}$ should take the form
\begin{eqnarray}
\int\frac{d^d p \: \sqrt{\det\omega(\p,\x, t)}}{(2\pi \hbar)^d} \left[ H(\p, \x, t) f(\p,\x, t) + \mbox{(terms with $\partial_x f$)} \right],
\end{eqnarray}
while the momentum density $T^{i0}$ should take the form
\begin{eqnarray}
\int\frac{d^d p \: \sqrt{\det\omega(\p,\x, t)}}{(2\pi \hbar)^d} \left[ p_i f(\p,\x, t) + \mbox{(terms with $\partial_x f$)} \right],
\end{eqnarray}
as one would expect. Moreover, the conservation $\partial_{x^\nu} T^{\mu\nu}=F^\mu_{\ \lambda} J^\lambda$ must hold. When we discuss the specific example of Weyl fermion in Section \ref{sect_BFG_Weyl}, we will present the full form of its stress-energy tensor, omitting the derivation which requires either coupling to spacetime metric or performing the Belinfante procedure to the Noether stress-energy tensor.

All the discussion so far has been completely general. Now we focus on Berry fermion gas in particular. We impose some additional restrictions:
\begin{itemize}
\item
The electric dipole moment vanishes. This is a reasonable assumption for non-interacting fermions; in particular this is true for non-interacting electrons in a lattice. However, as we will consider in Chapter \ref{chap_BFL}, interactions in general lead to emergent electric dipole moment.

\item
The magnetic dipole moment of a single fermion is order $\hbar$, and depends on $\p$ only but not on $\x$. This is the case because the only characteristic scale of a particle that has dimension $[x]$ (needed to cancel the $[x]^{-1}$ from $\partial_x$ in electromagnetic field) should be $\hbar/|\p|$.

\item
The only $x$ dependence in $H$ is the electric potential and the magnetic dipole term. Physically, this corresponds to no force other than electromagnetic force is acting on the particle, or alternatively, we have effectively absorbed all other forces acting on the particle into the electric potential $A_0$.
\end{itemize}
With these assumptions, we can write
\begin{eqnarray}
H(\p, \x, t) = H'(\p, \x, t) - A_0(\x, t) = E(\p) - \hbar\, \mu^{ij}(\p) F_{ij}(\x, t)/2 - A_0(\x, t)
\label{H_for_BFG}
\end{eqnarray}
where $\hbar\, \mu^{ij}(\p)$ is the magnetic dipole moment; its microscopic relation to the Bloch state or spinor $u^\alpha$ will be given in the next section when we perform the path integral derivation. Now, the current \eqref{J_expression} satisfying the EoM \eqref{BFG_EoM} reads
\begin{eqnarray}
J^0(\x, t) = \int\frac{d^d p}{(2\pi\hbar)^d} \left(1+\hbar \frac{F_{ij} b^{ij}}{2}\right) f,
\end{eqnarray}
\begin{eqnarray}
J^i(\x, t) = \int\frac{d^d p}{(2\pi\hbar)^d} \left[ \partial_p^i H \: f + \hbar\, b^{ij} F_{j0} \: f + \hbar\, F_{jk} \frac{3b^{[ij} \partial_p^{k]} H}{2} \: f + \hbar\: \mu^{ij} \partial_{x^j}f \right].
\label{BFG_current}
\end{eqnarray}
Below we look at some interesting implications from these expressions. It is these implications that drove people into the study of Berry Fermi gas systems.

\subsection{Anomalous Hall Effect}

First we consider the anomalous Hall current in a spatially uniform electric field oscillating in time (so that the particles do not keep accelerating), without magnetic field. The anomalous Hall current, i.e. that perpendicular to the electric field, is given by~\cite{chang1996berry,chang1995berry,Sundaram:1999zz}
\begin{eqnarray}
\delta J_H^i(\x, t) = F_{j0}(\x, t) \int\frac{d^d p}{(2\pi\hbar)^d} \hbar \, b^{ij}(\p) \: f(\p, \x, t).
\end{eqnarray}
This is correct in the semi-classical framework presented so far. However, recall that when we defined the Berry curvature, the spinor / Bloch state $u^\alpha$ is multi-component, which means there exists bands besides the $u$-band, and each of these bands has its own distribution $f$ and Berry curvature $b^{ij}$. To get the full anomalous Hall current, we must sum up \eqref{semi-classical_AHE} for all bands.

More specifically, suppose $u$ is the conducting band in a lattice system, there should also be some valence band(s) and some empty bands. Let's further assume the system is oscillating around in thermal equilibrium, i.e. $f(\p, \x, t)$ equals $f_{FD}(E(\p))$ plus oscillation due to the electric field. If we consider linear response, in \eqref{semi-classical_AHE} we can approximate $f=f_{FD}$. Finally, we assume the temperature is very small compared to the gap between the $u$-band and the other bands, so we can approximately take $f=1$ for all valence bands and $f=0$ for all empty bands. The total physical anomalous Hall current in linear response
\begin{eqnarray}
{\delta J_H^i}_{\ tot.} = F_{j0} \left(\int\frac{d^d p}{(2\pi\hbar)^d} \hbar \, b^{ij} \: f_{FD}(E) + \sum_w \int\frac{d^d p}{(2\pi\hbar)^d} \hbar \, b_w^{ij} \right)
\label{semi-classical_AHE}
\end{eqnarray}
where $w$ runs over the valences bands, and the integration is over the Brillouin zone (BZ) for a lattice system. 

For $d=2$ spatial dimensions, each valence band $w$ contributes to the Hall conductivity $\epsilon^{ij}C_w/(2\pi \hbar)$, where $C_w$ is an integer called the \emph{Chern number} of the $w$-band. The proof is the following~\cite{Thouless:1982zz,kohmoto1985topological}. When $b_w^{ij}(\p)$ is an exact 2-form over the BZ, by Stoke's Theorem the valence band contribution vanishes. $b_w^{ij}$ is non-exact when the complex phase of the spinor / Bloch state $w^\alpha(\p)$ cannot be chosen smoothly over the BZ. In such case, we can divide the BZ into multiple patches. The complex phase of $u^\alpha(\p)$ is smooth within each patch, but discontinuous over the boundaries (which are topologically $S^1$ circles) between the patches. By Stoke's Theorem,
\begin{eqnarray}
\int\frac{d^d p}{(2\pi\hbar)^d} \hbar \, b_w^{ij} = \frac{\epsilon^{ij}\hbar}{(2\pi\hbar)^2}\sum_{\mathcal{B}} \int_0^{2\pi} d\theta_\mathcal{B} \frac{d \Delta\phi^w_\mathcal{B}(\theta_\mathcal{B})}{d\theta_\mathcal{B}}
\end{eqnarray}
where $\mathcal{B}$ labels the boundaries between patches, and $\theta_\mathcal{B}$ parametrizes the boundary $\mathcal{B}$ (topologically an $S^1$ circle), and $\Delta\phi^w_\mathcal{B}(\theta_\mathcal{B})$ is the discontinuity of the complex phase of $w^\alpha$ over $\mathcal{B}$. Since $w^\alpha(\p)$ is single valued in each patch, $\Delta\phi^w_\mathcal{B}$ must change by a multiple of $2\pi$ as $\theta_{\mathcal{B}}$ goes from $0$ to $2\pi$. Therefore each $\mathcal{B}$ contributes an integer to $C_w$, and hence $C_w$ is an integer. Thus, in $d=2$ each valence band contributes a quantized Hall conductivity, whilst the conducting band contributes a non-quantized part.

\subsection{Equilibrium Current}

Now we look at the $\hbar$ corrections to the equilibrium current. We have $f=f_{FD}(H)$, and we can choose the gauge so that $A_\mu$ is time independent. Using the expression \eqref{H_for_BFG} for the Hamiltonian, we find
\begin{eqnarray}
J^0(\x, t) &=& \int\frac{d^d p}{(2\pi\hbar)^d} \left[ \: f_{FD}(E-A_0) + \hbar \frac{F_{ij}}{2} \left(b^{ij} - \mu^{ij} \partial_E \right) f_{FD}(E-A_0) \right],
\end{eqnarray}
\begin{eqnarray}
J^i(\x, t) &=& \int\frac{d^d p}{(2\pi\hbar)^d} \left[\: v^i \: f_{FD}(E-A_0) + \hbar\, F_{j0} \left(b^{ij} - \mu^{ij} \partial_E \right) f_{FD}(E-A_0) \right. \nonumber \\[.2cm]
&& \hspace{5cm} \left. + \ \hbar\, F_{jk} \frac{3b^{[ij} v^{k]}}{2} \: f_{FD}(\epsilon-A_0) \right]
\label{equilibrium_current}
\end{eqnarray}
where $v^i(\p)\equiv \partial_p^i E(\p)$ is the band velocity. In $J^0$, we see both the Berry curvature correction~\cite{Xiao:2005qw, Duval:2005vn} and the magnetic dipole moment correction to the fermion density. In $J^i$ (whose first term is zero is usual systems), the anomalous Hall current receives an additional magnetic dipole moment contribution compared to to the spatially uniform case \eqref{semi-classical_AHE} (note that the equilibrium current depends on $\mu^{ij}$ only through this anomalous Hall current term). The last term of $J^i$ is a magnetic field induced equilibrium current; as we will see in Section \ref{sect_BFG_Weyl}, in the case of Weyl fermion, this term gives rise to the chiral magnetic effect.

Again, since $u^\alpha$ is multi-component, there exist other bands. We have to add up the contributions to $J^\mu$ from all bands. The total equilibrium Hall current is (not assuming small temperature)
\begin{eqnarray}
{\delta J_H^i}_{\ tot.} = F_{j0} \sum_u \int\frac{d^d p}{(2\pi\hbar)^d} \hbar \left(b_u^{ij} - \mu_u^{ij} \partial_E \right) f_{FD}(E_u-A_0).
\label{equilibrium_current_Hall}
\end{eqnarray}
Again, in $d=2$, if for any band $w$ we can approximately take $f_{FD}(E_w-A_0)=1$ (valence band), then its contribution is quantized. The total magnetic field induced equilibrium current is
\begin{eqnarray}
{\delta J_M^i}_{\ tot.} = F_{jk} \sum_u \int\frac{d^d p}{(2\pi\hbar)^d} \hbar \frac{3b_u^{[ij} v_u^{k]}}{2} \: f_{FD}(E_u-A_0).
\label{equilibrium_current_mag}
\end{eqnarray}
We may write $v_u^k f_{FD}=\partial_{p^k} \int_{const}^{E_u-A_0} d\varepsilon \: f_{FD}(\varepsilon)$ and integrate $\p$ by parts, so that the above would be proportional to $\partial_p^{[k} b_w^{ij]}$ which vanishes. But there is an important caveat, as we will discuss through the example of Weyl fermion in Section \ref{sect_BFG_Weyl}.

\section{Derivation from Path Integral}
\label{sect_BFG_PIderivation}

By now we have presented the semi-classical formalism of particle with Berry curvature; the formalism is valid to first order in $\hbar\partial_x/|\p|$. We also derived some interesting implications in non-interacting fermionic systems. In this section we provide the quantum mechanical derivation of the semi-classical formalism. The original derivation is to evolution a wave packet in time, see \cite{Chang:2008zza} for review. This method has an intuitive physical picture; however, its drawback is the steps in the derivation are not so straightforward, and the role of the assumed shape of the original wave packet is not entirely clear. Alternatively, there are derivations from field theory, which have the strength of being generalizable to interacting systems (\cite{shindou2008gradient} and Section \ref{sect_QFT} of this thesis). But such derivations have a significant drawback: they are limited to the near-equilibrium situation, i.e. $f=f_{FD}+\mbox{perturbations}$, while for non-interacting systems the semi-classical formalism should hold far-from-equilibrium. There is also a non-commutative coordinate method~\cite{Gosselin:2006bv} which is quite general, but may appear unfamiliar at first sight. The method we want to present here is a path integral derivation, based on \cite{Stephanov:2012ki,Chen:2014cla} and refined to be more systematic and generalized. This method is very general, straightforward and familiar compared to other methods.

The idea of the method is the following~\cite{Stephanov:2012ki}. The single particle time-ordered propagator (i.e. the unitary time evolution) from $t=0$ to $t=T$ is given by
\begin{eqnarray}
\hat{G}^\alpha_{\ \beta}(T, 0) = \mathcal{T} \exp\left(-\frac{i}{\hbar} \int_0^T dt \: \hat{H}(\hat{P}, \hat{x}, t)\right)^\alpha_{ \ \ \beta}
\end{eqnarray}
where $\hat{x}^i$ is the position operator and $\hat{P}_i$ is the canonical momentum operator, and $\hat{H}^\alpha_{\ \beta}(\hat{P}, \hat{x}, t)$ is the multi-component Hamiltonian operator and we assume its $(\hat{x}, t)$ dependence is entirely due to $A_\mu$. Let $u^\alpha$ be an eigenvector (band) of $\hat{H}^\alpha_{\ \beta}$ when $A_\mu=0$, well gapped with other bands. We want to evaluate $\hat{G}^\alpha_{\ \beta}$ in between initial and final states in the $u$-band, and express the amplitude as
\begin{eqnarray}
&& \left\langle \ u_\alpha^\dagger (\vec{p}_T, T) \ \left| \ \hat{G}^\alpha_{\ \beta}(T, 0) \ \right| \ u^\beta (\vec{p}_0, 0) \ \right\rangle \nonumber \\[.3cm]
&=& \int \mathcal{D}(\p, \x) \ \exp\left[\frac{i}{\hbar} \left(S[\p, \x] + \mathcal{O}(\hbar^2)\right)\right] + \mbox{(band hopping contributions)}.
\end{eqnarray}
We want to show:
\begin{itemize}
\item
The action $S[\p, \x]$, to order $\hbar$, is given by \eqref{action_Berry}, with Hamiltonian \eqref{H_for_BFG}.

\item
The path integral measure $\mathcal{D}(\p, \x)$ is the product of Liouville volume measure over infinitely many time slices.

\item
The band hopping contributions are order $\mathcal{O}(\hbar^2 F^2/\Delta^4)$ suppressed, where $\Delta$ is the energy gap between the $u$-band and other bands. This is analogous to the quantum adiabatic theorem.
\end{itemize}
Below we present the derivation in full details.

Microscopically the Hamiltonian should take the form of the ``quantum version'' of \eqref{H_for_BFG}.
\begin{eqnarray}
\hat{H}^\alpha_{\ \beta}(\hat{P}, \hat{x}, t) &=& \hat{E}^\alpha_{\ \beta}(\hat{P}-A(\hat{x}, t)) - \delta^\alpha_\beta \: A_0(\hat{x}, t) \nonumber \\[.2cm]
&& - \hbar\: (\hat{\mu}^\alpha_{\ \beta})^{ij}(\hat{P}-A(\hat{x}, t)) \ \frac{F_{ij}(\hat{x}, t)}{2} + \mathcal{O}(\hbar^2)
\label{H_op_for_BFG}
\end{eqnarray}
Here $(\hat{\mu}^\alpha_{\ \beta})^{ij}$ is the intrinsic (bare) magnetic dipole matrix that the fermion might have. Now we use the commutation between $\hat{x}$ and $\hat{P}$ to rewrite the operators in ``anti-commutator ordering'': every term is ordered as the anti-commutator of a purely $\hat{P}$-dependent operator and a purely $\hat{x}$-dependent operator. That is
\begin{eqnarray}
\hat{H}^\alpha_{\ \beta}(\hat{P}, \hat{x}, t) &=& \left(\hat{E}^\alpha_{\ \beta}(\hat{P}) - \frac{1}{2}\left\{\partial_{\hat{P}_i} \hat{E}^\alpha_{\ \beta}(\hat{P}), A_i(\hat{x}, t)\right\} + \mathcal{O}(\vec{A}^2) \right) \ - \ \delta^\alpha_\beta \: A_0(\hat{x}, t) \nonumber \\[.2cm]
&& - \left(\frac{1}{2} \left\{\hbar(\hat{\mu}^\alpha_{\ \beta})^{ij}(\hat{P}), \frac{F_{ij}(\hat{x}, t)}{2} \right\} + \mathcal{O}(\hbar \vec{A}^2) \right)  + \mathcal{O}(\hbar^2)
\end{eqnarray}
(all the higher order terms are also in anti-commutator ordering). We use the anti-commutator ordering because it is the simplest ordering rule that preserves EM $U(1)$ gauge invariance to order $\hbar$ in all intermediate steps.

For the purpose of deriving semi-classical physics, the $\mathcal{O}(\hbar^2)$ terms can be dropped. For now, let's also neglect the order $A^2$ terms. They will be restored easily at the end using $U(1)$ gauge invariance.

As usual, we discretize the period from $t=0$ to $t=T$ into many time slices of duration $\delta t$, so small that $\hat{H}^2 \, \delta t^2 / \hbar^2$ can be neglected. The time slice labeled by $t$ lasts from $t-\delta t/2$ to $t+\delta t/2$. At the beginning time $t-\delta t/2$ of the time slice, we insert
\begin{eqnarray}
\mathbf{1} = \int \frac{d^d x_{t-\delta t/2}}{(2\pi\hbar)^{d/2}} \ | x_{t-\delta t/2} \rangle \ \langle x_{t-\delta t/2} |.
\end{eqnarray}
Then we are left with $\hat{P}$ operators residing in the time slice, and we decompose them into
\begin{eqnarray}
\hat{P}_i = \int \frac{d^d P_t}{(2\pi\hbar)^{d/2}} \ | P_t \rangle \ (P_t)_i \ \langle P_t |.
\end{eqnarray}
This procedure keeps the EM $U(1)$ gauge invariance 
\begin{eqnarray}
A_i(\z, t) \rightarrow A_i(\z, t) + \partial_{z^i} f(\z), \ \ \ \ (P_t)_i \rightarrow (P_t)_i +\frac{\partial_{X^i} f(x_{t+\delta t/2}^i) + \partial_{X^i} f(x_{t-\delta t/2}^i)}{2}
\label{discrete_EM_gauge_inv}
\end{eqnarray}
manifest. The propagator now becomes
\begin{eqnarray}
\hat{G}(T, 0)&=& \mathcal{T} \prod_t \left[\hat{1} - \frac{i\, \delta t}{\hbar} \hat{H}(\hat{P}, \hat{x}, t)\right] \nonumber \\[.2cm]
&=& \mathcal{T} \prod_t \frac{d^d P_t \: d^d x_{t-\delta t/2}}{(2\pi\hbar)^d} \: \exp\left(\frac{i}{\hbar} (P_t)_i \left(x_{t+\delta t/2} - x_{t-\delta t/2} \right)^i \right) \ \times \nonumber \\[.2cm]
&& \hspace{0cm} \left[\hat{1} - \frac{i\, \delta t}{\hbar} \: \hat{E}(P_t) + \frac{i\, \delta t}{\hbar} \: \partial_P^i \hat{E}(P_t)\: \frac{A_i(x_{t+\delta t/2}, t) + A_i(x_{t-\delta t/2}, t)}{2}\right. \nonumber\\[.2cm]
&& \hspace{.5cm} \left. + \ \frac{i\, \delta t}{\hbar} \: \hat{1} \: \frac{A_0(x_{t+\delta t/2}, t) + A_0(x_{t-\delta t/2}, t)}{2} \right. \nonumber\\[.2cm]
&& \hspace{.5cm} \left. + \ \frac{i\, \delta t}{\hbar} \: \frac{\hbar \, \hat{\mu}^{ij}(P_t)}{2} \: \frac{F_{ij}(x_{t+\delta t/2}, t) +F_{ij}(x_{t-\delta t/2}, t)}{2}\right]
\label{G_XP_PI}
\end{eqnarray}
where the $\alpha, \beta$ indices are hidden in the matrix multiplication, with $\hat{1}$ being $\delta^\alpha_\beta$.

Here we remind that the canonical commutation relation is realized in path integral as the following~\cite{feynman1942principle}. Denote $\delta P_{t+\delta t/2} \equiv P_{t+\delta t} - P_t$. It is important that $\delta P_{t+\delta t/2}$ should not be regarded as of order $\delta t$. Rather, we have
\begin{eqnarray}
&& \int d^d x_{t+\delta t/2} \: \exp\left(-\frac{i}{\hbar} x_{t+\delta t/2}^i \: (\delta P_{t+\delta t/2})_i\right) \ f(x_{t+\delta t/2}) \ (\delta P_{t+\delta t/2})_k \nonumber \\[.2cm]
&=& \int d^d x_{t+\delta t/2} \: \exp\left(-\frac{i}{\hbar} x_{t+\delta t/2}^i \: (\delta P_{t+\delta t/2})_i\right) \ (-i\hbar) \partial_{x_{t+\delta t/2}^k} f(x_{t+\delta t/2})
\end{eqnarray}
via integration of $x_{t+\delta t/2}$ by parts. Therefore, effectively $-\delta P_{t+\delta t/2} \rightarrow i\hbar \partial_{x_{t+\delta t/2}}$. Similarly, denote $\delta x_t \equiv x_{t+\delta t/2} - x_{t-\delta t/2}$, effectively we have $-\delta x_t \rightarrow -i\hbar \partial_{P_t}$.

We need to reduce the multi-band problem to a single band problem. Recall we are regarding $A$ as a perturbation such that each $A_\mu(x_{t+\delta t/2}, t)$ can be kept to linear order (we will easily restore $\mathcal{O}(A^2)$ terms by gauge invariance at the end). We decompose the ``unperturbed part'' $\hat{1} - (i\delta t/\hbar) \hat{E}(P_t)$ into eigenvectors of $\hat{E}(P_t)$:
\begin{eqnarray}
\delta^\alpha_\beta - \frac{i\delta t}{\hbar} \hat{E}^\alpha_{\ \beta}(P_t) = \left(1 - \frac{i\delta t}{\hbar} E(P_t) \right) \ u^\alpha(P_t) \ u^\dagger_\beta(P_t) \ + \ \mbox{(other bands)}
\end{eqnarray}
where $u(P)$ is an eigenvector with energy eigenvalue $E(P)$, and in the $u$ band lies the initial and final states. The crucial step towards the semi-classical picture is, in the decomposition above we can keep the $uu^\dagger$ term and drop all other bands. The reason is the following. It is not hard to see, after the decomposition, any matrix element of hopping between the $u$-band and another band is of order
\begin{eqnarray}
A \ \delta P / \Delta^2 \sim \hbar F / \Delta^2
\end{eqnarray}
where $\Delta$ is the energy gap between the $u$-band and the other hand; this is completely analogous to the quantum adiabatic theorem mentioned in the Introduction, except now there is no externally adjustable parameter, but a quantum number $P$ which is not strictly conserved due to the EM field. Our initial and final states are both in the $u$-band, so if hopping occurs, it must occur at least twice -- hop out and eventually hop back, yielding an amplitude of order $\hbar^2 F^2 / \Delta^4$, which we can neglect. Therefore, it is legitimate to order $\hbar$ to move towards the semi-classical picture by keeping $uu^\dagger$ in the decomposition above. This also explains why the semi-classical formalism is usually not applicable to higher order in $\hbar$ -- because then we will have to consider hopping which is a non-classical behavior.

After the projection to the $u$-band, the projected propagator $G_{uu}$ reads
\begin{eqnarray}
&& G_{uu}(0, T) \nonumber\\[.2cm]
&=& \prod_t \frac{d^d P_t \: d^d x_{t-\delta t/2}}{(2\pi\hbar)^d} \: \exp\left(\frac{i}{\hbar} (P_t)_i \left(x_{t+\delta t/2} - x_{t-\delta t/2} \right)^i \right) \ \times \nonumber \\[.2cm]
&& \hspace{0cm} \left[ \ u_\alpha^\dagger(P_t) \ u^\alpha(P_{t-\delta t}) \ \left(1 - \frac{i\, \delta t}{\hbar} \: E(P_t)\right) \phantom{\frac{1^\beta}{1_\beta}} \right. \nonumber \\[.2cm]
&& \left. \hspace{.5cm} + \ \frac{i\, \delta t}{\hbar} \ u_\alpha^\dagger(P_{t+\delta t}) \partial_P^i \hat{E}^\alpha_{\ \beta}(P_t) u^\beta(P_{t-\delta t}) \ \frac{A_i(x_{t+\delta t/2}, t) + A_i(x_{t-\delta t/2}, t)}{2}\right. \nonumber\\[.2cm]
&& \hspace{.5cm} \left. + \ \frac{i\, \delta t}{\hbar} \ u_\alpha^\dagger(P_{t+\delta t}) \ u^\alpha(P_{t-\delta t}) \ \frac{A_0(x_{t+\delta t/2}, t) + A_0(x_{t-\delta t/2}, t)}{2} \right. \nonumber\\[.2cm]
&& \hspace{.5cm} \left.  - \ \frac{i\, \delta t}{\hbar} \: \frac{u_\alpha^\dagger(P_{t+\delta t}) \: \hbar\, (\hat{\mu}^\alpha_{\ \beta})^{ij}(P_t) \ u^\beta(P_{t-\delta t})}{2} \: \frac{F_{ij}(x_{t+\delta t/2}, t) +F_{ij}(x_{t-\delta t/2}, t)}{2}\right] \nonumber\\[.2cm]
&& \phantom{1}
\end{eqnarray}
Now, we expand those EM perturbation terms in the square bracket in powers of $\delta P_{t+\delta t/2}$ and $\delta P_{t-\delta t/2}$, and then, as mentioned before, replace them by $-i\hbar \partial_{x_{t+\delta t/2}}$ and $-i\hbar \partial_{x_{t-\delta t/2}}$. We define the ``band magnetic dipole moment''
\begin{eqnarray}
\mu_{band}^{ij} &\equiv& -i \ \partial_P^{[i} u^\dagger_\alpha \left(\delta^\alpha_\beta E - \hat{E}^\alpha_{\ \beta}\right) \partial_P^{j]}  u^\beta \nonumber \\[.2cm]
&=& \frac{i}{2} \left(\partial_P^j u^\dagger_\alpha \: \partial_P^i \hat{E}^\alpha_{\ \beta} \: u^\beta - u^\dagger_\alpha \: \partial_P^i \hat{E}^\alpha_{\ \beta} \: \partial_P^j u^\beta" \right) - a^j \ \partial_P^i E
\label{BFG_mu_band}
\end{eqnarray}
and the ``bare magnetic dipole moment''
\begin{eqnarray}
\mu_{bare}^{ij} \equiv u^\dagger_\alpha \: \left(\hat{\mu}^\alpha_{\ \beta}\right)^{ij} \: u^\beta.
\label{BFG_mu_bare}
\end{eqnarray}
The total magnetic dipole moment is defined as $\mu^{ij} \equiv \mu_{band}^{ij} + \mu_{bare}^{ij}$. We find, to linear order in $A$ and in $\hbar$,
\begin{eqnarray}
&& G_{uu}(0, T) \nonumber\\[.2cm]
&=& \prod_t \frac{d^d P_t \: d^d x_{t-\delta t/2}}{(2\pi\hbar)^d} \: \exp\left(\frac{i}{\hbar} (P_t)_i \left(x_{t+\delta t/2} - x_{t-\delta t/2} \right)^i \right) \ \times \nonumber \\[.2cm]
&& \hspace{0cm} \left[ \ u_\alpha^\dagger(P_t) \ u^\alpha(P_{t-\delta t}) \ \left(1 - \frac{i\, \delta t}{\hbar} \: E(P_t)\right) + \frac{i\, \delta t}{\hbar} \frac{\hbar\, \mu^{ij}(P_t)}{2}  \frac{F_{ij}(x_{t+\delta t/2}, t) + F_{ij}(x_{t-\delta t/2}, t)}{2} \right. \nonumber \\[.2cm]
&& \left. \hspace{.5cm} + \ \frac{i\, \delta t}{\hbar} \ \partial_P^i E(P_t) \left(1 - \hbar\, a^j(P_t) \: \partial_{x^j}\right) \frac{A_i(x_{t+\delta t/2}, t) + A_i(x_{t-\delta t/2}, t)}{2}\right. \nonumber\\[.2cm]
&& \hspace{.5cm} \left. + \ \frac{i\, \delta t}{\hbar} \ \left(1 - \hbar\, a^j(P_t) \: \partial_{x^j}\right) \frac{A_0(x_{t+\delta t/2}, t) + A_0(x_{t-\delta t/2}, t)}{2} \right] \nonumber \\[.3cm]
&=& \prod_t \frac{d^d P_t \: d^d x_{t-\delta t/2}}{(2\pi\hbar)^d} \: \exp\left(\frac{i}{\hbar} (P_t)_i \left(x_{t+\delta t/2} - x_{t-\delta t/2} \right)^i \right) \ \times \nonumber \\[.2cm]
&& \hspace{0cm} u_\alpha^\dagger(P_t) \ u^\alpha(P_{t-\delta t}) \left[ \ 1 - \frac{i\, \delta t}{\hbar} \left(E(P_t) - \frac{\hbar\, \mu^{ij}(P_t)}{2}  \frac{F_{ij}(x_{t+\delta t/2}, t) + F_{ij}(x_{t-\delta t/2}, t)}{2} \right. \right. \nonumber \\[.2cm]
&& \left. \hspace{0cm} - \left. \partial_P^i E(P_t) \ \frac{A_i(x_{t+\delta t/2}, t) + A_i(x_{t-\delta t/2}, t)}{2} - \frac{A_0(x_{t+\delta t/2}, t) + A_0(x_{t-\delta t/2}, t)}{2} \right) \right]
\end{eqnarray}
Note that $\mu_{band}^{ij}$ has no direct counter-part in the original Hamiltonian operator \eqref{H_op_for_BFG}; its emergence is a purely quantum effect. For example, in particle physics, the famous $g=2$ of a Dirac fermion (say electron) is its $\mu_{band}^{ij}$, and there is no $\mu_{bare}^{ij}$ as the usual Dirac equation has no Pauli term. On the other hand, as we move on to solid state physics, the electron's $g=2$ serves as its $\mu_{bare}^{ij}$ in Bloch's band theory, and the observed magnetic dipole moment usually involves an additional $\mu_{band}^{ij}$ due to the lattice.

Finally we recognize the consecutive product of $uu^\dagger$ as the momentum space Wilson line with connection $a^i$, and exponentiate the $i\delta t/\hbar$ terms. Passing over to the  continuum limit, we find
\begin{eqnarray}
G_{uu}(0, T) &=& \prod_t \frac{d^d P_t \: d^d x_{t-\delta t/2}}{(2\pi\hbar)^d} \: \times \nonumber \\[.2cm]
&& \hspace{1cm} \exp\left(\frac{i}{\hbar}\int_0^T \left(P_i dx^i - \hbar\, a^i(\vec{P}) dP_i - H(\vec{P}-\vec{A}, \x, t)\, dt\right)\right)
\end{eqnarray}
where the Hamiltonian takes the form \eqref{H_for_BFG}, with $\vec{P}-\vec{A}(\x, t)$ in place of $\p$. To connect towards \eqref{action_Berry}, we define the physical momentum
\begin{eqnarray}
(p_t)_i = (P_t)_i - \frac{A_i(X_{t+\delta t/2}) + A_i(X_{t-\delta t/2})}{2}
\end{eqnarray}
which is invariant under \eqref{discrete_EM_gauge_inv}. Note that $d^d P \: d^d x = d^dp \: d^d x$, however
\begin{eqnarray}
\exp\left(-\frac{i}{\hbar} \hbar\, a^i(\vec{P}) dP_i \right) &=& \exp\left(-\frac{i}{\hbar} \hbar\, a^i(\p) dp_i - \frac{i}{\hbar} \hbar\, b^{ij}(\p) A_i(\x, t) dp_j \right) \nonumber \\[.2cm]
&=& \left(1 - \frac{i}{\hbar} \hbar\, b^{ij}(\p) A_i(\x, t) dp_j\right) \exp\left(-\frac{i}{\hbar} \hbar\, a^i(\p) dp_i\right)
\end{eqnarray}
(total derivative dropped). The second term is of order $\hbar A$, so we can view it as a perturbation, and replace $A_i dp_j \rightarrow -i\hbar \partial_{x^i} A^j$ without affecting other terms (to order $\hbar$ and order $A$). This gives the Berry curvature correction to the Liouville volume element \eqref{Liouville_with_Berry}.

In summary, we have proven the $u$-band propagator
\begin{eqnarray}
G_{uu}(0, T) = \int \mathcal{D}\left(\p, \x\right) \: \exp\left(\frac{i}{\hbar} S[\p, \x] \right)
\end{eqnarray}
where $S$ is given by \eqref{action_Berry} with Hamiltonian \eqref{H_for_BFG}, and the path integral measure is
\begin{eqnarray}
\mathcal{D}(\p, \x) \equiv \prod_t \frac{d^d p_t \: d^d x_{t-\delta t/2}}{(2\pi\hbar)^d} \left(1+ \hbar\, \frac{b^{ij}(p_t)}{2} \frac{F_{ij}(x_{t+\delta t/2}, t) + F_{ij}(x_{t-\delta t/2}, t)}{2} \right),
\end{eqnarray}
where the second term is the Berry curvature correction to the Liouville volume element. Band hopping contributions are order $\hbar^2 F^2/\Delta^4$ suppressed compared to $G_{uu}$, therefore we drop them; on the other hand, this shows one cannot improve the semi-classical action to order $\hbar^2$, because then one has to include hopping, which is a non-classical behavior.

Our derivation above assumed linear response and kept each $A_\mu(x_{t-\delta t/2}, t)$ to leading order. However, it is now obvious that our conclusion remains unchanged when we include higher orders in $A$. This is because the EM $U(1)$ gauge invariance must be respected. Any order $A^n$ ($n\geq 2$) term will just enter the higher order expansion of $E(\vec{P}-\vec{A})$, any order $A^n \delta P$ or order $\hbar F A^{n-1}$ term will enter the higher order expansion of $\mu^{ij}(\vec{P}-\vec{A})$, and any non-trivial new effects (similar to the emergence of $\mu_{band}^{ij}$) are at least order $A^2 \delta P^2 \sim \hbar^2 F^2$ which are negligible at order $\hbar$ anyways.

\section{Example: Chiral Kinetic Theory of Weyl Fermion}
\label{sect_BFG_Weyl}

So far we have introduced the semi-classical formalism of Berry Fermi gas, and established its microscopic foundation from path integral. Here we demonstrate the formalism in a concrete example. The example we choose is that of Weyl fermion, which have received much attention in recent years. This example is a two band system that is easy to solve, but it has very rich physics due to the band touching point (Weyl node) at $\p=0$ -- most notably the chiral anomaly and its associated effects. In high energy physics, in particular in the study of hot, dense quark gluon plasma, light quarks can be approximately viewed as Weyl fermions, and the semi-classical theory can be useful in describing its far-from-equilibrium state~\cite{Son:2012wh, Stephanov:2012ki, Chen:2012ca}.

Moreover, thanks to its simplicity of Weyl Hamiltonian, its appearance is very general -- if a multi-band system has two bands touching, then almost always it can be approximated (unless there is some prohibition due to symmetries) by the Weyl fermion band structure for energy and momenta near the Weyl node. If such band touching occurs in a solid state system, such material is called a \emph{Weyl semimetal}~\cite{wan2011topological, yang2011quantum, turner2013beyond}, which is one of the most focused area of experimental study in recent years~\cite{huang2015weyl, xu2015discovery, lv2015experimental}. A closedly related solid state system, in which two Weyl nodes of opposite chiralities reside at the same point in momentum space (usually due to symmetry; band degeneracy near the Weyl node required), called \emph{Dirac semimetal}~\cite{young2012dirac}, is also under extensive experimental study~\cite{liu2014discovery,liu2014stable,neupane2014observation,li2014chiral}.

\subsection{Single Weyl Fermion}

We consider a right-handed Weyl fermion. The quantum mechanical Hamiltonian is
\begin{eqnarray}
\hat{H}^\alpha_{\ \beta} = (P_i - A_i) \: (\sigma^i)^\alpha_{\ \beta} - A_0 \: \delta^\alpha_\beta
\label{Weyl_Hamiltonian}
\end{eqnarray}
for $i=1, 2, 3$ (for left-handed Weyl fermion, we have $-\sigma^i$ in place of $\sigma^i$). When $A_\mu=0$, the eigenvalues are $\pm|\p|$, whose associated eigenvectors (spinors) are respectively
\begin{eqnarray}
u^\alpha(\p)=\left[\begin{array}{c} \cos(\theta/2) \\ e^{i\phi}\sin(\theta/2) \end{array}\right], \ \ \ \ \ w^\alpha(\p)=\left[\begin{array}{c} -e^{-i\phi}\sin(\theta/2) \\ \cos(\theta/2) \end{array}\right]
\end{eqnarray}
where $(|\p|, \theta, \phi)$ are the spherical coordinates in the momentum space. We view the negative energy $w$-band as the Dirac sea; a hole in the Dirac sea is recognized an anti-particle. We can compute the Berry curvature in the $u$-band according to \eqref{Berry_curv_def}
\begin{eqnarray}
\epsilon^{ijk} \: \hbar\, b_{k}(\p) \equiv \hbar\, b^{ij}(\p) = \frac{\hbar}{2} \epsilon^{ijk} \frac{p_k}{|\p|^3},
\label{Weyl_Berry_curv}
\end{eqnarray}
and the magnetic dipole moment in the $u$-band according to \eqref{BFG_mu_band}
\begin{eqnarray}
\epsilon^{ijk} \: \hbar\, \mu_k(\p) \equiv \hbar\, \mu^{ij}(\p) = \frac{\hbar}{2} \epsilon^{ijk} \frac{p_k}{|\p|^2}.
\end{eqnarray}
In both expressions, the $\hbar/2$ factor is the helicity of the right-handed Weyl fermion; for a right-handed Weyl anti-fermion (hole in the $w$-band), it will be $-\hbar/2$. (For left-handed, the $\pm \hbar/2$ would be exchanged.) In particular, $\mu_k$ being pointed along $p_k$ is related to the the well known fact that the direction of spin of a relativistic massless spinning particle is locked with the direction of momentum~\cite{weinberg1995quantum}.

The semi-classical action and Hamiltonian are
\begin{eqnarray}
dS = p_i dx^i + A_\mu dx^\mu - a^i dp_i - H_0 dt, \ \ \ \ \ \ H_0 = H + A_0 = |\p| - \hbar \: B^k p_k/|\p|^2
\label{Weyl_Action}
\end{eqnarray}
where $\epsilon_{ijk} B^k\equiv F_{ij}$ is the magnetic field and $E_k=F_{k0}$ is the electric field. (Note that accidentally one can write $H_0 = |\p|/\sqrt{\det\omega}$.) A puzzle arises immediately. The microscopic theory of Weyl fermion is Lorentz invariant, but the semi-classical action is not; more broadly, the entire Berry curvature formalism is not Lorentz invariant because Berry curvature by definition is purely spatial. Where did the Lorentz invariance go? We put this aside for now. The entire next chapter will be devoted on this problem.

Needless to say, the expressions of $b_k$ and $\mu_k$ become problematic as $\p\rightarrow 0$. But they should, because $\p\rightarrow 0$ is where the $u$-band and the $w$-band meet, and according to Section \ref{sect_BFG_PIderivation}, the semi-classical single particle picture is legitimate only when $(\hbar F/\Delta)^2$ negligible, where in the present case $\Delta = 2|\p|$. Therefore, physically we should place some infrared (IR) cutoff $|\p|_{IR}$ such that $\hbar F / |\p|_{IR}^2 \lesssim 1$, and the semi-classical theory is only valid if the particle momentum $|\p| \gg |\p|_{IR}$. The most important fact associated with this band touching is the Berry curvature obeys the ``inverse square law'' as if there is Berry curvature monopole residing at the band touching point (Weyl node) $\p=0$. Mathematically
\begin{eqnarray}
\hbar\, 3\partial_p^{[k} b^{ij]} = \hbar\, \epsilon^{ijk} \partial_p^l b_l = 4\pi \epsilon^{ijk} \ \frac{\hbar}{2} \delta^3(\p).
\label{Berry_monopole}
\end{eqnarray}
(This is related to the fact there is no way to choose $u$ and $w$ so that their phases are continuously well-defined all over the momentum space. For example, in the phases we have chosen above, $u$ and $w$ are undefined along $\theta=\pi \ (|\p|=-p_3)$.) Consequently the symplectic 2-form has Liouville flow monopole
\begin{eqnarray}
\partial_p^{[i} \omega_{p_i p_j]} = - 4\pi \epsilon^{ijk} \ \frac{\hbar}{2} \delta^3(\p).
\label{symplectic_monopole}
\end{eqnarray}
This does not mean the symplectic 2-form is not closed, because $\p=0$ hidden under the IR cutoff is not part of the classical phase space anyways. Rather, this means the symplectic 2-form is not exact in the classical phase space with the $|\p| < |\p|_{IR}$ region removed (and hence becomes topologically non-trivial); there can be symplectic flow (Liouville flow) flowing in and out through the $|\p|=|\p|_{IR}$ classical-quantum interface. This is closely related to Nielsen and Ninomiya's famous spectral flow interpretation of chiral anomaly~\cite{Nielsen:1983rb}, as we will elaborate on in the next section.

The EoM \eqref{BFG_EoM} for Weyl fermions are explicitly
\begin{eqnarray}
\frac{dx^i}{dt} &=& \frac{p^i}{|\p|} \left(1+ \hbar\, B^k b_k\right) + \epsilon^{ijk} \hbar\, b_k E_j, \\[0.2cm]
\frac{dp_i}{dt} &=& \left(1+ \hbar\, B^l b_l\right) \epsilon_{ijk} B^k \frac{p_j}{|\p|} + \left(1- \hbar\, B^k b_k\right) E_i+ \hbar\, \mu_k \frac{\partial B^k}{\partial x^i} + \hbar\, b_i E^j B_j.
\label{Weyl_EoM}
\end{eqnarray}
Notably, since $b_i=p_i/2|\p|^3$, the last term in $dp_i/dt$ is radial in the momentum space when $E^j B_j \neq 0$. If we have an ensemble of particles, collectively there will be a total number of particles flowing through the $|\p| = |\p|_{IR}$ the interface, so that they appear up in / disappear from the classical phase space. Of course this is again related to the chiral anomaly to be explored in the next section.

Before we move on, we comment about the generality of the Weyl Hamiltonian \eqref{Weyl_Hamiltonian}. In a multi-band system, if we are interested in two bands which are touching, we can always write the Hamiltonian about this two bands as
\begin{eqnarray}
\hat{H}^\alpha_{\ \beta} = h_i (\p) (\sigma^i)^\alpha_{\ \beta} + h_0(\p) \delta^\alpha_\beta.
\end{eqnarray}
Suppose the two bands are touching at some momentum $\p$. Clearly $h_0$ has no business to do with the touching. Moreover, since there are three $h_i$'s and at the same time three $p_j$ components, generically the Jacobian $\det(\partial_p^j h_i)$ is non-zero near the Weyl node (unless dictated by some symmetry), that is, the mapping of the 3-dimensional momentum space into the 3-dimensional space of $h_i$ is generically non-degenerate. Therefore, up to redefinition of variables, the Weyl Hamiltonian is very generic in band touching systems. This is why this example is very useful. In a lattice system, there is the famous Weyl fermion doubling theorem~\cite{Nielsen:1980rz, Nielsen:1981xu} stating that the number of Weyl nodes with $\det(\partial_p^j h_i)>0$ (right-handed) and $\det(\partial_p^j h_i)<0$ (left-handed) must be equal; we will provide a simple proof to it later, based on Berry curvature.

\subsection{Ensemble}
\label{ssect_BFG_Weyl_ensemble}

The above discussion focuses on a single particle. Now we consider the behavior of an ensemble of non-interacting Weyl fermion gas. The kinetic theory is called \emph{chiral kinetic theory}~\cite{Stephanov:2012ki}. The chiral kinetic theory enables us to study macroscopic ensemble of Weyl fermions far-from-equilibrium, and hence (in systems where interactions effects are small) has broader application then the chiral hydrodynamics. In particular, the chiral anomaly computation in the next section is legitimate even when the system is far-from-equilibrium.

The density and current \eqref{BFG_current} are explicitly
\begin{eqnarray}
J^0 = \int \frac{d^3 p}{(2\pi\hbar)^3} \left(1+ \frac{\hbar}{2} \, B^k \frac{p^k}{|p|^3}\right) \: f,
\end{eqnarray}
\begin{eqnarray}
J^i = \int\frac{d^3 p}{(2\pi\hbar)^3} \left[ \left(1+ \hbar\, B^k \frac{p^k}{|\p|^3}\right) \frac{p^i}{|\p|} f + \frac{\hbar}{2}\, \epsilon^{ijk} E_j \frac{p_k}{|\p|^3} f + \frac{\hbar}{2}\, \epsilon^{ijk} \frac{p_k}{|\p|^2} \: \partial_{x^j} f \right].
\label{BFG_Weyl_current}
\end{eqnarray}
The anomalous Hall current term will be non-vanishing only if the distribution function is in a rotationally non-symmetric configuration.

\begin{figure}
\centering
\includegraphics[width=0.5\textwidth]{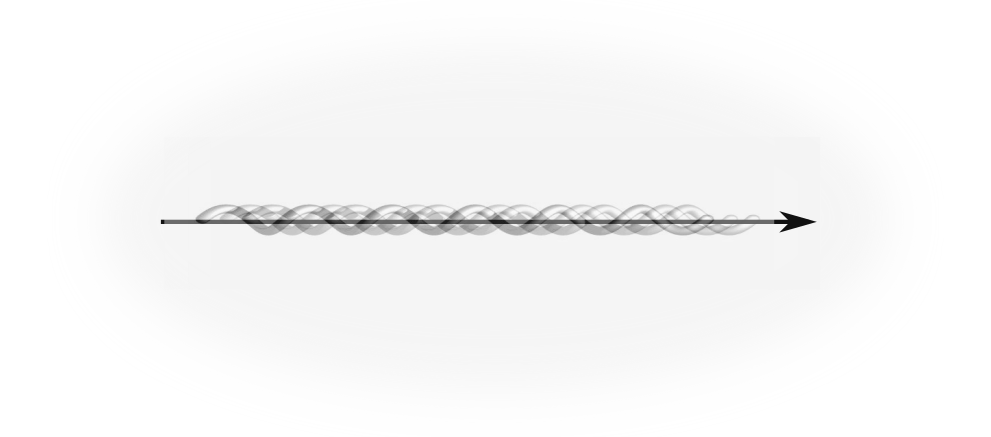}
\caption{As the narrow wave packet of a Weyl fermion moves along a path in the space, its spatial current, imagined as a bunch of arrows on the wave packet, will trace out a bunch of helical trajectories round the path over time.}
\label{Helical_Current_Pic}
\end{figure}

With this expression for $J^i$, we can picture the single particle current. Let $f(\p, \x, t)$ be a narrow distribution localized in position space and momentum space (can be thought of as representing a Gaussian wave packet). If the particle is spinless, the current $J^i$ can be viewed as a narrow bunch of arrows pointing along its direction of velocity, and over time these arrows will trace out a narrow bunch of parallel lines. However, with spin, and in particular for massless spinning particle whose direction of spin is locked to the direction of momentum~\cite{weinberg1995quantum}, the magnetization current term will make the bunch arrows wind around the direction of velocity, and over time the arrows trace out a narrow bunch of helices, as illustrated by Figure \ref{Helical_Current_Pic}. Therefore, in the sense of the ``shape'' of the single particle current, we really can view the quantum mechanical spin as the particle's current ``physically'' spinning around its direction of velocity. This helical feature of the current is closely related to the Lorentz invariance of the semi-classical Weyl fermion, as we will see in the next chapter.

We will write down the stress-energy tensor without derivation, as the derivation requires either coupling to spacetime metric or performing the Belinfante procedure to the Noether stress-energy tensor, both of which are beyond the scope of this thesis. The energy density is
\begin{eqnarray}
T^{00} = \int \frac{d^3 p}{(2\pi\hbar)^3} |\p| f
\end{eqnarray}
which is the usual integration of the Hamiltonian (without the $A_0$ part), with the magnetic dipole correction to the energy and the Berry curvature correction to the Liouville measure cancelled out. The momentum density and energy flux is
\begin{eqnarray}
T^{i0} = T^{0i} = \int \frac{d^3 p}{(2\pi\hbar)^3} \left[ \left(1+ \frac{\hbar}{2}\, B^k \frac{p^k}{|p|^3}\right) p^i f + \frac{\hbar}{2}\, \epsilon^{ijk} \frac{p_k}{2|\p|} \: \partial_{x^j} f \right].
\end{eqnarray}
The spatial stress tensor is
\begin{eqnarray}
T^{ij} \!\!\! &=& \!\!\! \int\frac{d^3 p}{(2\pi\hbar)^3} \left[ \left(1+ \hbar\, B^k \frac{p^k}{|p|^3}\right) \frac{p^i p^j}{|\p|} f + \frac{\hbar}{2} \left(p^{\{i}\epsilon^{j\}kl} E_k \frac{p_l}{|\p|^3} + \frac{B^{\{i} p^{j\}}}{|\p|^2} - \delta^{ij} \frac{B^k p_k}{|\p|^2}\right) f \right. \nonumber \\[.2cm]
&& \hspace{2cm} \left. + \ \frac{\hbar}{2} p^{\{i} \epsilon^{j\}kl} \frac{p_l}{|\p|^2} \partial_{x^k} f \right].
\end{eqnarray}
The terms in $T^{i0}$ and $T^{ij}$ with $\partial_x f$ arise from the spin $1/2$ of the Weyl fermion coupling to the spacetime spin connection, they can be viewed as the momentum and the stress carried in the ``spinning motion'' of the particle. The stress-energy tensor is traceless $-T^{00}+T^{ij}\delta_{ij}=0$, as the Weyl fermion is massless.

With some work one can verify the energy and momentum conservation
\begin{eqnarray}
\partial_{x^\mu} T^{\mu\nu} = F^\nu_{\ \lambda} J^\lambda
\end{eqnarray}
is satisfied. On the other hand, the charge conservation $\partial_{x^\mu} J^\mu$ is violated due to the chiral anomaly proportional to $\hbar E_k B^k$, as we will see in the next section.

\subsection{Chiral Magnetic Effect}
\label{sect_BFG_CME}

Although the chiral kinetic theory is originally purposed for describing far-from-equilibrium systems, from the theory we can learn interesting effects in an equilibrium state too. Most notable are the chiral magnetic effect (CME)~\cite{PhysRevD.22.3080,Fukushima:2008xe} and the chiral vortical effect (CVE)~\cite{PhysRevD.20.1807,Son:2009tf}. Both of them are tightly related to the chiral anomaly~\cite{Son:2009tf,Neiman:2010zi,Loganayagam:2012pz}, and equivalently, the Berry curvature monopole~\cite{Son:2012wh,Stephanov:2012ki,Chen:2012ca,Chen:2014cla}. Here we will discuss the CME, which has been observed in a recent experiment~\cite{li2014chiral} on Dirac semimetal. On the other hand, the CVE will be discussed later in Section \ref{ssect_CVE} when we find the most general form of equilibrium distribution in the absence of external field.

Consider a Weyl fermion ensemble in a static (but might be spatially non-uniform) magnetic field. Assume the system is in Fermi-Dirac distribution. By \eqref{equilibrium_current}, we find the equilibrium current from the $u$-band is
\begin{eqnarray}
J^i &=& \hbar\, F_{jk} \int \frac{d^d p}{(2\pi\hbar)^d} \frac{3b^{[ij} \partial_p^{k]} E}{2} f_{FD}(E) \nonumber \\[.2cm]
&=& \hbar\, F_{jk} \int \frac{d^d p}{(2\pi\hbar)^d} \frac{3}{2} b^{[ij} \partial_p^{k]} \ \int_{\varepsilon_{min}}^{E} d\varepsilon \ f_{FD}(\varepsilon)
\end{eqnarray}
where the lower limit $\varepsilon_{min}$ of the $\varepsilon$ integral is arbitrary. Now we integrate $p_k$ by parts, which yields~\cite{Son:2012wh}
\begin{eqnarray}
J^i = \hbar\, F_{jk} \int \frac{d^d p}{(2\pi\hbar)^d} \frac{3}{2} \partial_p^{[i} b^{jk]} \ \int_E^{\varepsilon_{surf}} d\varepsilon \ f_{FD}(\varepsilon).
\end{eqnarray}
In particle physics, $\varepsilon_{surf} \equiv E(\p\rightarrow\infty) = \infty$ arises as the boundary term from the integration by part. In solid state physics, the momentum space a is boundary-less BZ, so we can let $\varepsilon_{surf}$ be our arbitrarily chosen $\varepsilon_{min}$. The current $J^i$ is physical, so it must be independent of our arbitrarily chosen $\varepsilon_{min}$. Indeed, this is due to
\begin{eqnarray}
\int \frac{d^d p}{(2\pi\hbar)^d} \frac{3}{2} \partial_p^{[i} b^{jk]} = 0
\end{eqnarray}
being a total derivative in a boundary-less BZ. For $d=3$ this is just the fermion double theorem, since right-handed and left-handed Weyl nodes have opposite Berry curvature charge~\cite{Nielsen:1980rz, Nielsen:1981xu}.

A question about the general computation above is, we know the semi-classical picture fails near the band touching region where $\partial_p^{[i} b^{jk]}\neq 0$, so how can we trust the integral of $f_{FD}$ near such region? Recall the system is in equilibrium, so there shall be no net band hopping, and we can extrapolate the validity of semi-classical theory arbitrarily close to the Weyl nodes. Even in non-equilibrium state, as long as a sufficiently large neighborhood of the band touching region is approximately in equilibrium, we should still be able to extrapolate.

The discussion above is completely general. Specifically for our system of Weyl fermion, $\varepsilon_{surf}=\infty$, and the monopole is given by \eqref{Berry_monopole} occurring at $E=0$. To get the total current, we also need to first add the $w$-band contribution, and then subtract the contribution when the $w$-band is full. The subtraction is due to the regularization of the definition of $J^i$ against the infinite Dirac sea. This is equivalent to subtracting the contribution of the anti-particles, which are under the same temperature as the particles but at opposite chemical potential, and have opposite Berry curvature and opposite electric charge (couples oppositely to magnetic field). Therefore
\begin{eqnarray}
J_{tot}^i &=& \frac{4\pi}{(2\pi\hbar)^3} \frac{\hbar}{2} B^i \ \int_0^\infty d\varepsilon \left(f_{FD}(\varepsilon; \mu, T) - f_{FD}(\varepsilon; -\mu, T)\right) \nonumber \\[.2cm]
&=& \frac{4\pi}{(2\pi\hbar)^3} \frac{\hbar}{2} B^i \ \mu
\end{eqnarray}
regardless of temperature. One can also compute the momentum density / energy flux and find
\begin{eqnarray}
T_{tot}^{i0} &=& \frac{4\pi}{(2\pi\hbar)^3} \frac{\hbar}{2} B^i \ \left(\frac{\mu^2}{2} + \frac{\pi^2 T^2}{6} \right).
\end{eqnarray}
This is the well-known chiral magnetic effect in current and in momentum density (for sure, the two are related~\cite{Son:2009tf,Neiman:2010zi,Loganayagam:2012pz}).

On the other hand, let's consider a two band Weyl semimetal instead. In Weyl semimetal, there is no regularization against any Dirac sea, so we just add up the contributions from both bands -- the fermions in both bands have the same electric charge and are subjected to the same chemical potential and energy, but see opposite Berry curvature monopoles charge at each Weyl node. Therefore
\begin{eqnarray}
J_{tot}^i = \hbar\, F_{jk} \int \frac{d^d p}{(2\pi\hbar)^d} \frac{3}{2} \partial_p^{[i} b^{jk]} \ \int_{\varepsilon^w_{surf}}^{\varepsilon^u_{surf}} d\varepsilon \ f_{FD}(\varepsilon) = 0
\end{eqnarray}
since the $\varepsilon$ integral is independent of $\p$ and the $\p$ integral is a total derivative integral over a boundary-less BZ. This proves that there is no CME for any Weyl semimetal in equilibrium~\cite{Basar:2013iaa}. In order to observe CME in a Weyl semimetal, the regions near the right-handed Weyl node(s) and the left-handed Weyl node(s) must be prepared to fill up to different energies; the simplest way to do so is by the chiral anomaly~\cite{Son:2012bg} introduced in the next section. This is how the recent experiment on the observation of CME~\cite{li2014chiral} (in Dirac semimetal) is performed.

\section{Chiral Anomaly in Chiral Kinetic Theory}
\label{sect_chiral_anomaly}

Anomaly is one of the most fascinating subjects of quantum mechanics. It refers to the fact that a symmetry in classical mechanics may not be consistent with quantum mechanics, and hence must be broken by order $\hbar$. Anomalies have significant experimental consequences, and also profound mathematical structure. Here we will not review the subject; see e.g. \cite{weinberg1996quantum} for good pedagogical introduction.

The earliest and simplest example of anomaly is the chiral anomaly~\cite{Adler:1969gk, bell1969pcac}, originally studied in the context of quarks and successfully explained the experimentally observed pion lifetime. This anomaly can be stated as the following. Consider a left-handed Weyl fermion species and a right-handed Weyl fermion species, uncoupled with each other, and have the same coupling to the electromagnetic (EM) field. One would expect both of them have their own fermion number conservation $U(1)_L$ and $U(1)_R$; the sum of their fermion numbers is the total EM $U(1)$ charge, and the difference is called the axial $U(1)_A$ charge. The statement of chiral anomaly is, quantum mechanically it is inconsistent to have $U(1)_L$, $U(1)_R$ and the left- right-handed parity all in one theory. In reality the spacetime parity (this is built-in if we view the left- and right-handed Weyl fermions as a single Dirac fermion) and the EM $U(1)$ gauge invariance should be respected, but then the axial $U(1)_A$ must be broken by the anomaly, i.e. the left- and right-handed fermion numbers are not separately conserved, although their sum, the total physical electric charge, is conserved. More precisely,
\begin{eqnarray}
\partial_{x^\mu} J^\mu_R = -\partial_{x^\mu} J^\mu_L = \frac{4\pi}{(2\pi\hbar)^3} \frac{\hbar}{2} \frac{\epsilon^{\mu\nu\rho\sigma}}{8} F_{\mu\nu} F_{\rho\sigma} = \frac{4\pi}{(2\pi\hbar)^3} \frac{\hbar}{2} E_k B^k
\label{ABJ_chiral_aom}
\end{eqnarray}
(remarkably, this is the exact result \cite{adler1969absence}) so that their sum vanishes but their difference does not. Can we avoid this if we have only right-handed Weyl fermion to start with? As long as we want to it to behave as ``right-handed'' at the quantum mechanical level, its fermion number must be non-conserved as the above.

But where does the fermion number go? In solid state system, Nielsen and Ninomiya provided a very intuitive spectral flow interpretation with concrete computation~\cite{Nielsen:1983rb}. Suppose we have an electron in the valence band. If we turn on $E_k B^k>0$, the electron rises into the conducting band through the right-handed Weyl node so that $\partial_{x^\mu} J^\mu_R >0$. Eventually, in the conducting band, the electron must move towards in the left-handed Weyl node (whose existence is guaranteed the doubling theorem mentioned before), through which it sinks back into the valence band, and hence $\partial_{x^\mu} J^\mu_L<0$, but the total electron number is of course conserved. In high energy physics, there is no doubling theorem and no path connecting the right-handed Weyl mode to the left-handed Weyl mode. However, there are infinite Dirac seas. With $E_k B^k>0$, right-handed Weyl fermions in their Dirac sea will increase in energy so that the top ones will pop up as right-handed fermion excitations, while left-handed Weyl fermions in their Dirac sea will decrease in energy, leaving some holes in the Dirac sea, which are left-handed anti-fermion excitations (there is some language ambiguity here -- this is an anti-fermion of the Weyl mode of left-handed chirality, but the anti-fermion itself has right-handed helicity). Of course, this statement requires ultra-violet (UV) regularization, and for the regularization to be consistent with parity and EM $U(1)$ gauge invariance, the rate of creating right-handed fermion excitations and left-handed anti-fermion excitations must be equal.

\begin{figure}
\centering
\includegraphics[width=0.98\textwidth]{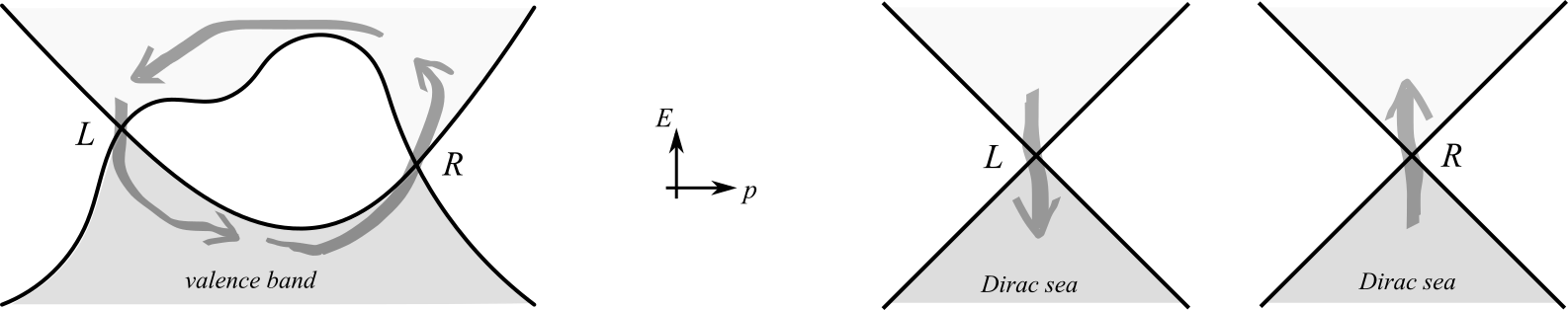}
\caption{The spectral flow under chiral anomaly with $\vec{E}\cdot\vec{B}>0$ for Weyl semimetal (left) and for Weyl fermions (right).}
\label{spectral_flow}
\end{figure}

The remarkable success of chiral kinetic theory is that such spectral flow can even to computed in the (semi-)classical framework \cite{Son:2012wh,Son:2012zy,Stephanov:2012ki} (closed related ideas also in \cite{Gao:2012ix,Chen:2012ca}). The spectral flow is illustrated in Figure \eqref{spectral_flow}.

We already have all the recipes for the computation. To expedite the computation, for now let's pretend we can trust the semi-classical theory even in the vicinity of Weyl node and the Boltzmann equation \eqref{Boltzmann_Eq} holds in this vicinity. Clearly this assumption is unphysical, but we will justify the result of our computation later. We take $\partial_{x^\mu}$ on the current \eqref{J_expression}. We can ignore the magnetization current term because $\partial H /\partial F$ is order $\hbar$ and $\partial_x \sqrt{\det\omega}$ is also order $\hbar$. We are left with the velocity term:
\begin{eqnarray}
\partial_{x^\mu} J^\mu(x) &=& \int\frac{d^d p}{(2\pi\hbar)^d} \: \partial_{x^\mu} \left(\sqrt{\det\omega} \: \left.\frac{dx^\mu}{dt}\right|_\EoM \: f\right) \nonumber \\[.2cm]
&=& \int\frac{d^d p}{(2\pi\hbar)^d} \left[ \partial_t \left(\sqrt{\det\omega} \: f\right) + \partial_{\xi^I} \left(\sqrt{\det\omega} \: \left.\frac{d\xi^I}{dt}\right|_\EoM \: f\right) \right].
\label{chiral_anomaly_first_step}
\end{eqnarray}
In the second line we added a total $\p$ derivative which equals zero, because there is no boundary term as we assumed we can include the vicinity of the Weyl node as part of the classical phase space. The integrand of the second line is just the left-hand-side of the continuity equation \eqref{Liouville_Thm}, which usually vanishes. However, in the spirit of spectral flow, we allow it to be violated at the Weyl node, but we assume the Boltzmann equation \eqref{Boltzmann_Eq} still holds. The difference between the continuity equation and the Boltzmann equation is given by \eqref{Liouville_Thm} which is non-zero at the Weyl node where the symplectic 2-form fails to be closed. Therefore
\begin{eqnarray}
\partial_{x^\mu} J^\mu(x) = -\int\frac{d^d p}{(2\pi\hbar)^d} f \: \frac{\sqrt{\det\omega}}{2} \ 3\partial_{[\xi^I}\omega_{JK]} \ \omega^{IJ} \omega^{KL} \ \left(\partial_{\xi_L} H + \partial_t \omega_L \right).
\end{eqnarray}
This is the general formula relating the chiral anomaly to the non-closedness of the symplectic 2-form at band touching. For a right-handed Weyl fermion, we apply \eqref{symplectic_monopole}. Since the non-closedness is order $\hbar$, we can ignore any $\hbar$ term in \eqref{Berry_Poisson} for $\omega^{IJ}$, and also take $\partial_{x^i} H + \partial_t \omega_L \simeq -E_i$. Thus, for a right-handed Weyl fermion,
\begin{eqnarray}
\partial_{x^\mu} J^\mu = \frac{4\pi}{(2\pi\hbar)^3} \ \frac{\hbar}{2} E_k B^k \ f(\p=0, \x, t).
\label{chiral_anomaly_second_last_step}
\end{eqnarray}
But can we make clear sense of $f$ at the Weyl node? We do not need to worry about this, because to get $\partial_{x^\mu} J_R^\mu$, we also need to subtract the anti-fermion contribution, which has opposite Berry curvature. Therefore
\begin{eqnarray}
\partial_{x^\mu} J_R^\mu = \partial_{x^\mu} J^\mu - \partial_{x^\mu} \wt{J}^\mu = \frac{4\pi}{(2\pi\hbar)^3} \ \frac{\hbar}{2} E_k B^k \left(f(\p=0, \x, t) + \wt{f}(\p=0, \x, t) \right).
\end{eqnarray}
The factor in the parenthesis is $1$ because anti-fermion density is just hole density in the $w$-band, and at $\p=0$ the $w$-band meets the $u$-band. Thus, we have computed the chiral anomaly \eqref{ABJ_chiral_aom} in semi-classical (order $\hbar$) chiral kinetic theory. No assumption of near equilibrium is needed.

Finally we justify the left-over subtleties. When we add the total $\p$ derivative term in \eqref{chiral_anomaly_first_step}, there really should be a boundary term due to the $|\p|_{IR}$ cutoff. By the Stoke's Theorem, clearly \eqref{chiral_anomaly_second_last_step} is just computing this boundary term, but instead of $f(\p=0, \x, t)$, we should integrate about $f(\p, \x, t)$ over the $|\p|_{IR}$ interface; similarly for the anti-fermions. However, $f$ and $\wt{f}$ at the $|\p|_{IR}$ interface not necessarily add up to $1$ because they are in different bands. How shall we interpret our computation then? With the spectral flow picture, it is easy to see that, in the presence of the chiral anomaly flow, $f+\wt{f}$ around the $|\p|_{IR}$ interfaces (one in the $u$ band and the other in the $w$ band) cannot stay away from $1$ because the momentum space below these interfaces has limited volume of order $\sqrt{\hbar F}^3$. Instead, $f+\wt{f}$ around the $|\p|_{IR}$ interfaces must average to $1$ over the spacetime scale $\gtrsim\sqrt{\hbar/F}$. Therefore, our computation must be interpreted as the effect averaged over a spacetime scale $\gtrsim\sqrt{\hbar/F}$. But this indeed is the regime of validity of our semi-classical $\hbar \partial_{x} / |\p|$ expansion. Therefore the computation is justified within our framework.

\section{Summary and Outlook}

In this chapter, we have reviewed formal aspects of the Berry Fermi gas theory, the semi-classical theory of non-interacting Fermi gas with Berry curvature, and presented a general and conceptually simple derivation of the semi-classical theory from single particle path integral. We demonstrated the idea by the concrete example of Weyl fermion, which, despite its computational simplicity, has rich and deep physical implications and broad applications. Most notably, the chiral anomaly and the associated chiral magnetic effect can be computation from the Berry curvature monopole residing at the band touching point. We kept our computations in the most formal expressions, so that it is clear that these effects are robust and do not depend on the details of the system. We also presented the relation between anomalous Hall effect and Berry curvature, which was the earliest relation that drew people's attention toward Berry curvature in fermionic systems.

For sure, the formalism we presented here does not encompass the full story of Berry phase physics in non-interacting fermionic system. We considered momentum space Berry phase only, but clearly there can also be other components of Berry phase~\cite{Sundaram:1999zz,Chang:2008zza,Xiao:2009rm} due to, e.g. extra externally adjustable parameter in the system. More importantly, even for momentum space Berry phase only, we restricted to particles that have no internal degrees of freedom. That is, we assumed our particle is microscopically described by a single band $u$. We made this assumption for simplicity. In general physical systems there might be band degeneracy so that the particle must be described by multiple degenerate bands, and therefore has internal degrees of freedom. In such case, the Berry curvature would have internal indices and becomes a non-abelian curvature~\cite{culcer2005coherent,shindou2005noncommutative,Xiao:2009rm}, and leads to some new interesting implications. The path integral and the semi-classical formalism we presented here can be straightforwardly generalized to the case with internal degrees of freedom.

There are not much formal, general, outstanding puzzles lingering around the Berry Fermi gas theory, because the whole formalism can be derived by the quantum mechanics of single particle (e.g. our path integral derivation) due to the absence of interactions. But for applications to specific systems, there certainly are remaining problems.

The next problem we will discuss is, in the particular system of Weyl fermions, how is the Lorentz symmetry realized in the Berry curvature formalism, which intrinsically separates time and space. This is clearly a sharp problem in chiral kinetic theory. The resolution to this problem, however, also deepens our understanding of the Berry curvature formalism in general.

\chapter{Lorentz Symmetry in Chiral Kinetic Theory}
\label{chap_LI}

In this chapter we study how Lorentz symmetry is realized in the semi-classical theory of Weyl fermion~\eqref{Weyl_Action}. Our conclusion is that, there is nothing wrong with the semi-classical theory, but the realization of Lorentz symmetry is non-trivial. More particularly, we should not regard the spacetime position $(t, \x)$ and energy-momentum $(H_0, \p)$ as Lorentz covariant physical observables; their definitions are frame dependent. Rather, the frame independent observables are (as one might expect), the current and stress-energy tensor in Section \ref{ssect_BFG_Weyl_ensemble}.

We start with stating a puzzle motivating the idea that the notion of position must be frame dependent. Then we use the intrinsic-time parameter formulation of the action introduced in Section \ref{Rev_Classical_Mech} and see how the frame independence of the action is realized. We provide a physical explanation based on the single particle current Figure \ref{Helical_Current_Pic}, and a mathematical explanation relating the non-trivial frame independence with the little group of a massless spinning particle. Finally, we discuss if the Weyl fermions have a collision term in the Boltzmann equation, how the collision should be realized to be compatible with the non-trivial frame independence. We will see this colliding theory relaxes to the well-known chiral hydrodynamics, and we compute the famous chiral vortical effect (CVE)~\cite{Son:2009tf}.

This chapter resolves the puzzle of Lorentz symmetry in the semi-classical Berry curvature formalism. However, resolving this problem also sheds light on the interpretation of the formalism in general, as we will see along the text.

\section{Puzzle}

The most direct puzzle we see is that the action \eqref{Weyl_Action} for a semi-classical Weyl fermion manifestly breaks Lorentz invariance. In particular, the Berry curvature depends on the spatial momentum only, how can the action possibly be Lorentz invariant at all?

A resolution would be that, in addition to the usual Lorentz boost, $\x$ and $\p$ must transform with additional order $\hbar$ corrections; this is indeed how the problem is resolved. However, such modification to the Lorentz boost would mean the four vectors $x^\mu$ and $p_\mu$ are frame dependent quantities, which sounds strange. Now we present a physical argument why this has to be the case~\cite{Chen:2014cla}.

\begin{figure}
\centering
\includegraphics[width=0.318\textwidth]{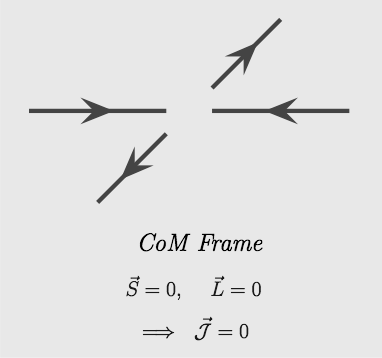}
\hspace{2cm}
\includegraphics[width=0.38\textwidth]{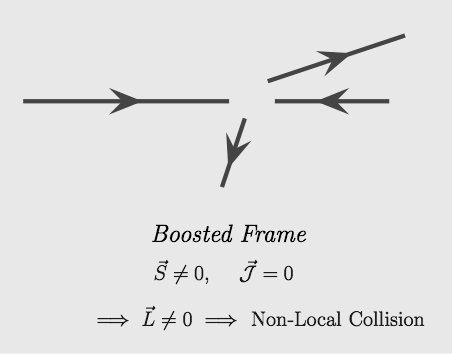}
\caption{Consider a head-on collision between two massless spinning particles. In the center-of-mass frame, the incoming particles have no orbital angular momentum $\vec{L}$ and no total spin angular momentum $\vec{S}$, and hence no total angular momentum $\vec{\mathcal{J}}$; so are the outgoing particles. If we boost along the direction of the incoming particles, then the incoming particles still have no $\vec{L}$ and no $\vec{S}$ and hence $\vec{\mathcal{J}}=0$; however, the out-going particles have $\vec{S}\neq 0$, and in order for $\vec{\mathcal{J}}$ to be conserved, they must have non-zero $\vec{L}=-\vec{S}$.}
\label{head-on_collision}
\end{figure}

\begin{figure}
\centering
\includegraphics[width=0.3\textwidth]{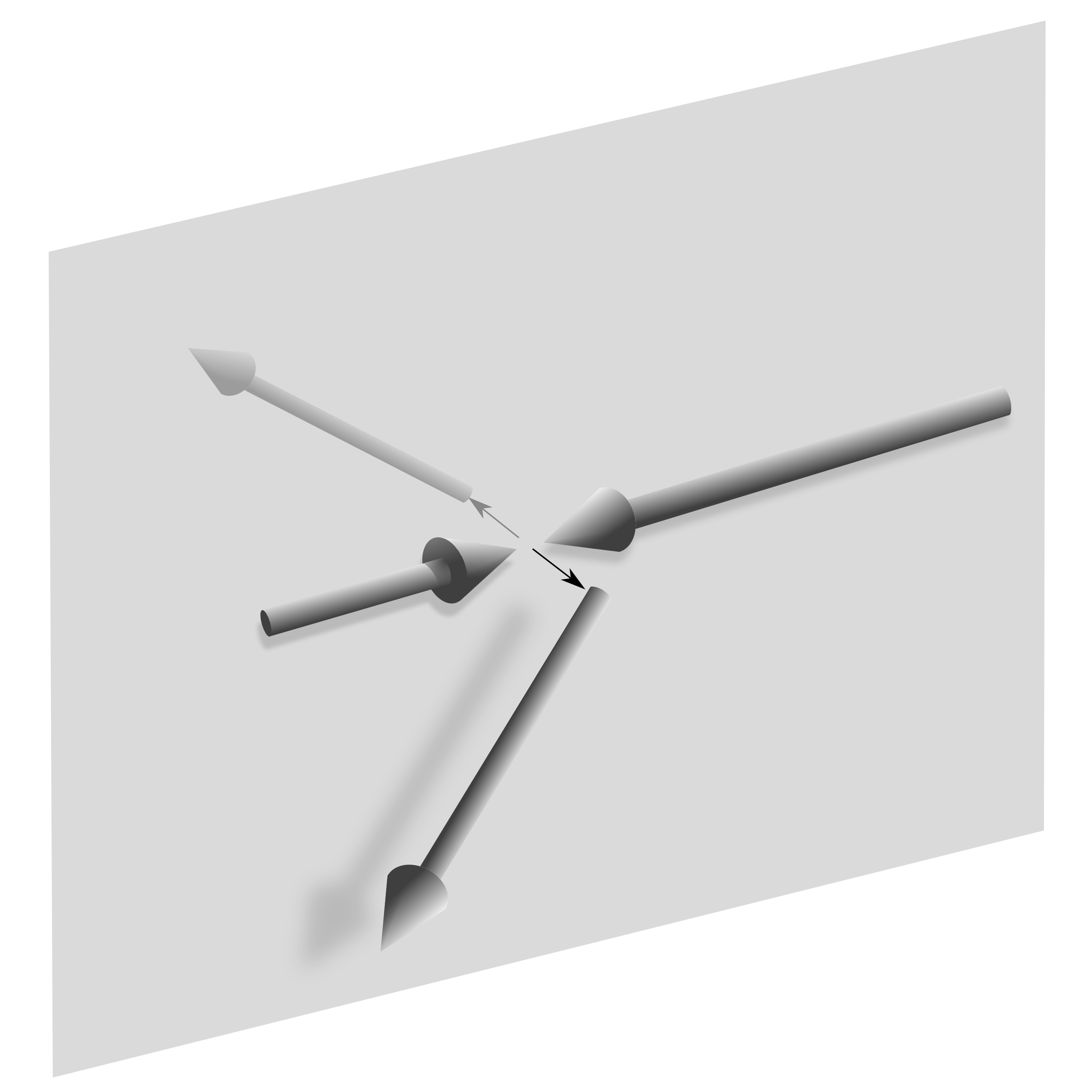}
\caption{A schematic illustration of the non-local collision of massless spinning particles in a non-center-of-mass frame.}
\label{collision}
\end{figure}

Recall that the a spinning massless particle's spin angular momentum is always locked with its momentum, $\vec{S}=s\hbar \hat{p}$, where $s=\pm 1/2$ for right-handed and left-handed Weyl fermions respectively. Let's suppose the particles can collide with each other. In particular, consider a head-on collision as illustrated in Figure \ref{head-on_collision}. In the center-of mass-frame of the collision, the incoming particles have zero total orbital angular momentum $\vec{L}$, zero total spin angular momentum $\vec{S}$ and hence zero total angular momentum $\vec{\mathcal{J}}$; so are the outgoing ones. Now we boost along the direction of the incoming particles. The $\vec{L}$, $\vec{S}$ and $\vec{\mathcal{J}}$ of the incoming particles are still zero as before. However, the outgoing particles now have non-zero $\vec{S}$ since their momenta are not in opposite directions. In order for $\vec{\mathcal{J}}$ to still be conserved, we must conclude that the outgoing particles have $\vec{L}=-\vec{S}\neq 0$, which means they do not fly out directly from the ``collision point'' of the incoming particles. Thus, a local collision viewed in the center-of-mass frame becomes a non-local collision in the boosted frame, schematically shown in Figure \ref{collision}. This enforces the idea that the notion of ``position'' must be frame dependent.

Clearly the collision here can be a fictitious one just to facilitate the argument that the position of a massless spinning particle must be frame dependent. The same must hold for non-colliding particles, for example photons (which we will mention later). A massive particle does not have this ambiguity because we can always go to its own center-of-mass frame to define its position; on contrary, there is no center-of-mass frame for a massless particle.

\section{Non-Trivial Frame Dependence}

To understand the Lorentz invariance in the semi-classical theory of Weyl fermion~\eqref{Weyl_Action}, we first use the intrinsic-time parameter formulation of the action introduced in Section \ref{Rev_Classical_Mech} to express \eqref{Weyl_Action} ``as Lorentz invariant as possible'':
\begin{eqnarray}
S=\int dS, \ \ \ \ dS[x, p, \lambda] = p_\mu dx^\mu + A_\mu(x) dx^\mu - a^\mu(\p) d\p_\mu - \mathcal{H}(p, x) \lambda d\tau
\label{Weyl_Action_rel}
\end{eqnarray}
where
\begin{eqnarray}
\p_\mu\equiv p_\mu + p_\nu n^\nu n_\mu, \ \ \ \ \ \ |\p| = \sqrt{\vec{p}_\mu p^\mu}
\end{eqnarray}
and $n^\mu$ is a normalized future time-like frame vector, which is taken to be $(1,0,0,0)$ in the non-relativistic formalism. But in the relativistic formalism we do not fix its components; in fact, the formalism we are presenting here works even if $n^\mu$ is spacetime dependent, and this fact will be useful when we consider the chiral vortical effect later. The Hamiltonian is given by
\begin{eqnarray}
\mathcal{H}(p, x) = p_\mu p^\mu - \hbar\, F_{\mu\nu}(x) \Sigma^{\mu\nu}(\p), \ \ \ \ \ \hbar\, \Sigma^{\mu\nu}(\p)\equiv -\frac{\hbar}{2}\frac{\epsilon^{\mu\nu\rho\sigma} p_\rho n_\sigma}{|\p|}
\label{Weyl_H_rel}
\end{eqnarray}
where $\hbar\Sigma^{\mu\nu}$ is the spin angular momentum of Weyl fermion (the Pauli-Lubanski pseudo-vector is proportional to $\epsilon_{\mu\nu\rho\lambda}\hbar\Sigma^{\mu\nu}p^\rho$); we can see it is orthogonal to the direction of momentum, which is a feature of massless spinning particle. Clearly the Weyl fermion's spin generates its magnetic dipole moment. The Berry curvature \eqref{Weyl_Berry_curv} now reads
\begin{eqnarray}
\hbar\, b^{\mu\nu}(\p) = -\frac{\hbar}{2}\frac{\epsilon^{\mu\nu\rho\sigma} p_\rho n_\sigma}{|\p|^3}.
\end{eqnarray}
$\lambda$ serves a Lagrange multiplier that demands $\mathcal{H}=0$, which leads to the ``on-shell condition''
\begin{eqnarray}
-p\cdot n \equiv -p_\mu n^\mu = |\p| - \hbar\, F_{\mu\nu} \Sigma^{\mu\nu}/2|\p|
\label{rel_dispersion}
\end{eqnarray}
(taking the $-p\cdot n>0$ solution is always understood) which is the $H_0$ in \eqref{Weyl_Action}. We can derive the EoM from the action. The EoM relating the physical time to intrinsic time is
\begin{eqnarray}
-n_\mu \ dx^\mu/(2\lambda d\tau) = -n_\mu p^\mu = |\p| - \hbar\, F_{\mu\nu} \Sigma^{\mu\nu}/2|\p|.
\end{eqnarray}
The other EoMs just reduce to \eqref{Weyl_EoM} when $n^\mu=(1,0,0,0)$ and are not reproduced here (the EoM for energy is to relate the $d/(2\lambda d\tau)$ of \eqref{rel_dispersion} to $d\p/(2\lambda d\tau)$ and $dx/(2\lambda d\tau)$).

In the form \eqref{Weyl_Action_rel} and \eqref{Weyl_H_rel}, Lorentz invariant is manifest as everything is written as contraction of Lorentz indices. The problem now becomes (the order $\hbar$ terms of) the action depends on a reference frame $n^\mu$.

\subsection{Infinitesimal Transformation}
\label{ssect_infsml_transf}

In fact, under an infinitesimal change of reference frame, if we let $x$ and $p$ transform by order $\hbar$ terms, the action can remain invariant to order $\hbar$. The symplectic part $p\cdot dx + A\cdot dx - a \cdot dp$ is invariant under infinitesimal change of frame
\begin{eqnarray}
n^\mu \rightarrow n^\mu+\beta^\mu, \ \ \ \ \ \ \beta \cdot n=0
\end{eqnarray}
accompanied with~\cite{Chen:2014cla}
\begin{eqnarray}
x^\mu \rightarrow x^\mu + \hbar\, b^{\mu\nu}(\p) \: \beta_\nu \ (p\cdot n), \ \ \ \ p_\mu \rightarrow p_\mu + F_{\mu\nu} \: \hbar\, b^{\nu\lambda}(\p) \: \beta_\lambda \ (p\cdot n).
\label{frame_dep}
\end{eqnarray}
This frame dependence of $x$ was originally found in \cite{Skagerstam:1992er} while that of $p$ is new. Substituting these into the Hamiltonian \eqref{Weyl_H_rel}, we find it transforms by an overall factor
\begin{eqnarray}
\mathcal{H} \rightarrow \mathcal{H} \left(1+ 2\hbar\, n^\lambda F_{\lambda\mu} b^{\mu\nu} \beta_\nu\right).
\end{eqnarray}
We can make the action invariant by letting $\lambda$ to transform with the opposite factor so that $\mathcal{H}\, \lambda$ remains invariant. (This is valid even when $n^\mu$ is $x$ dependent, because the $x$ dependence will only add to the transformation of $n$ an order $\hbar$ term, but $n$ itself appears in the action in order $\hbar$ terms only, so the difference will the $\mathcal{O}(\hbar^2)$.)

The invariance of the Hamiltonian part seems quite trivial, for we can always use $\lambda$ to cancel the transformation of $\mathcal{H}$. This is not the case; in fact, this is the most non-trivial part of the story. Note that $\mathcal{H}$ transforms by an overall factor, so the on-shell condition $\mathcal{H}=0$ is frame independent. Were it not transforming by an overall factor, the transformation of $\lambda$ must involve $1/\mathcal{H}$ which would be problematic. This condition is highly restrictive.

What is the most general Lorentz invariant semi-classical action with non-zero Berry curvature? Rotational invariance and point-like monopole dictates the Berry curvature to be that of Weyl fermion (the overall factor is fixed quantum mechanically). Then one can show frame independence dictates the most general dispersion relation to be given by \eqref{Weyl_H_rel}~\cite{Son:2012zy,Chen:2014cla} up to a mass term $m^2$ so small that $m^2 \hbar F/|\p|^4$ can be neglected. The allowance of such small mass term is compatible with the fact that a high speed massive Dirac or Majorana fermion can be approximately seen as a Weyl fermion. The exclusion of other possibilities reflects the fact that any non-interacting Lorentz invariant term that can be added to the Weyl Lagrangian must be at least order $\hbar^2 F^2$.

One can compute the commutation of two frame transformations by $\beta$ and $\beta'$. It is non-vanishing:
\begin{eqnarray}
x^\mu \rightarrow x^\mu + \hbar \frac{\epsilon^{\lambda\kappa\rho\sigma}n_\lambda p_\kappa \beta_\rho \beta'_\sigma}{|\p|^2} \frac{p^\mu}{|\p|}, \ \ \ \ \ p_\mu \rightarrow p_\mu + \hbar \frac{\epsilon^{\lambda\kappa\rho\sigma}n_\lambda p_\kappa \beta_\rho \beta'_\sigma}{|\p|^2} \: F_{\mu\nu}\frac{p^\nu}{|\p|}.
\end{eqnarray}
One can see the transformation is proportional to the EoM evolution of the variables~\cite{Chen:2014cla}. A transformation proportional to the EoM evolution always leaves the action invariant, as is obvious from \eqref{general_action_variation} letting $\delta \xi$ be proportional to \eqref{general_EoM}. Therefore, the frame transformation together with the EoM evolution transformation form a group. The mathematical origin of this algebra is discussed in the next section.

\subsection{Finite Transformation On-Shell}

For the frame transformation to be useful, we must not only have its infinitesimal form \eqref{frame_dep} but also the finite form. We would only need it when the ``on-shell condition'' $\mathcal{H}=0$ is satisfied. We can integrate the infinitesimal transformation imposing $\mathcal{H}=0$, and find, under $n\rightarrow n'$, \cite{Duval:2014cfa,Stone:2015kla,Chen:2015gta}
\begin{eqnarray}
x'^\mu = x^\mu + \Delta_{nn'}^\mu(p), \ \ \ \ \ p'_\mu = p_\mu + F_{\mu\nu}(x) \Delta_{nn'}^\nu(p), \ \ \ \ \ \Delta_{nn'}^\mu \equiv \frac{\hbar}{2} \frac{\epsilon^{\mu\nu\rho\sigma}p_\nu n_\rho n'_\sigma}{(p\cdot n) (p\cdot n')}.
\label{frame_dep_finite}
\end{eqnarray}
When $F_{\mu\nu}=0$, the total angular momentum
\begin{eqnarray}
\mathcal{J}^{\mu\nu} \equiv x^\mu p^\nu - x^\nu p^\mu + \hbar\, \Sigma^{\mu\nu}
\end{eqnarray}
is frame independent, but we are not aware of an analogy that is frame independent when $F_{\mu\nu}\neq 0$.

One can easily see Figures \ref{head-on_collision} and \ref{collision}, obtained based on physical arguments, agree with the mathematical form of the frame dependence.

\subsection{Current, Stress-Energy Tensor and Physical Interpretation}

If the position and momentum are frame dependent quantities, what are the physical observables of the Weyl fermion that are frame independent? They are the current and stress-energy tensor, for a single particle or for an ensemble. Understanding the frame independence of them gives us a clear physical interpretation for the frame dependence of position and momentum.

The single particle current, defined as
\begin{eqnarray}
J^\mu_{sp}(x)[z, p, \lambda] \equiv \left. \frac{\delta S[z, p, \lambda]}{\delta A_\mu(x)} \right|_\EoM,
\end{eqnarray}
is certainly frame independent, because both the action and the field $A_\mu(x)$ are (in this expression the particle's frame dependent position is $z$; $x$ is an arbitrary point in spacetime that has nothing to do with the particle). For an ensemble of particles, we need to integrate over the phase space with measure \eqref{Liouville_volume_intrinsic}. The measure is also frame independent -- the invariance of the symplectic part of the action leads to the invariance of $d^4 z \: d^4 p \: \sqrt{\det\omega}$ (the symplectic 2-form with time and energy is to expand \eqref{Berry_symplectic} so that the $i, j$ indices become $\mu, \nu$ indices), while the invariance of $\mathcal{H} \, \lambda$ leads to the invariance of $\delta(\mathcal{H})/\lambda$. Finally, we need the distribution function to depend on frame so that
\begin{eqnarray}
f'(\p', z') = f(\p, z), \ \ \ \ \ \mbox{i.e.} \ f'(\p', x) = f(\p, x) - \Delta_{nn'} \cdot \partial_x f(\p, x)
\label{f_transf}
\end{eqnarray}
(because here $z$ is the particle's position, while $x$ is just a generic point in the spacetime which does not depend on the frame). Then, the ensemble current
\begin{eqnarray}
J^\mu(x) = \int \frac{d^4 z \: d^4 p}{(2\pi\hbar)^4} \frac{\delta(\mathcal{H})\, \theta(-p\cdot n)}{2\lambda d\tau} \sqrt{\det\omega} \ f(\p, x) \ J^\mu_{sp}(x)[z, p, \lambda]
\end{eqnarray}
must be frame independent $J'^\mu(x)=J^\mu(x)$.

The current \eqref{BFG_Weyl_current} can be written as
\begin{eqnarray}
J^\mu(x) &=& \int \frac{d^3 \p}{(2\pi\hbar)^3 \, (-p\cdot n)} \left[\left(1+\hbar \: \frac{F_{\rho\sigma}b^{\rho\sigma}}{2}\right) p^\mu + \hbar \: b^{\mu\nu}F_{\nu\lambda} n^\lambda (-p\cdot n)  \right. \nonumber\\[.2cm]
&& \hspace{3.5cm} \left. \phantom{\frac{1}{1}} + \ \hbar\, \Sigma^{\mu\nu}\partial_{x^\nu}\right]f(\p, x).
\label{Weyl_current_rel}
\end{eqnarray}
One can manifestly check its frame independence under \eqref{frame_dep} with the aid of the collisionless Boltzmann equation.

The stress-energy tensor can be written as
\begin{eqnarray}
T^{\mu\nu} &=& \int \frac{d^3 \p}{(2\pi\hbar)^3 \, (-p\cdot n)} \left[ \left(1+\hbar \: \frac{F_{\rho\sigma}b^{\rho\sigma}}{2}\right) p^\mu p^\nu + \hbar \left( F_{\rho\sigma} p^{\{\mu} - F_\rho^{\: \{\mu} p_\sigma \right) b^{\nu\}\rho} n^\sigma (-p\cdot n) \right. \nonumber\\[.2cm]
&& \hspace{3.5cm} \left. \phantom{\frac{1}{1}} + \ \hbar\: p^{\{\mu} \Sigma^{\nu\}\lambda} \: \partial_{x^\lambda} \right] \: f(\p, x),
\label{Weyl_stress-tensor_rel}
\end{eqnarray}
It can be obtained by coupling the action to spacetime metric and then integrating over the phase space, therefore by the same argument it must also be frame independent $T'^{\mu\nu}(x)=T^{\mu\nu}(x)$.

Now we can give a physical interpretation to the frame dependence of the particle's position and momentum. As we did in Section \ref{ssect_BFG_Weyl_ensemble}, consider a single particle current where $f$ is localized in space and momentum space, i.e. a narrow wave packet; for concreteness one can take $f(\p, x)=(2\pi\hbar)^3 \delta^3(\x-\z(t))\delta^3(\p-\q(t))$ where the particle's position $\z(t)$ and momentum $\q(t)$ satisfy the EoM. The current is helical as illustrated in Figure \ref{Helical_Current_Pic}. It is easy to see the particle's ``position'' $\z$ must be interpreted as ``the spatial center of the helix''. While the helical current itself is frame independent, it is an elementary relativity fact that the notion of of ``spatial center'' of a rotating body is frame dependent, in a way qualitatively agree with \eqref{frame_dep_finite}. Similarly, $\p$ must be interpreted as the ``momentum carried at the spatial center of the helix'', and therefore is also frame dependent when there is EM force curving the helical trajectory. A more pictorial analysis of this interpretation is found in \cite{Stone:2015kla}. What we have seen is the massless, quantum version of the phenomenon that was first learned in the context of massive rotating bodies~\cite{Dixon:1970zza,Costa:2011zn}.

Clearly what we described here did not rely on the particle being a Weyl fermion. Similar frame dependence of the ``position'' or ``center'' occurs in photons too, except the helicity $\pm \hbar/2$ must be replaced by $\pm \hbar$. The associated experimental phenomenon for photon is that, when a circular polarized light beam (or photon) undergoes a refraction, the center of the beam (or photon wave packet) undergoes an order $\hbar$ slight shift, much like the scenario in Figure \ref{collision}. See \cite{bliokh2015transverse} for a review on this subject.

\section{Relation to Wigner Translation}
\label{sect_Wigner_transl}

In the previous section we have seen the frame dependence of the definition of a massless spinning particle's position and momentum. This ambiguity only arises for massless particle which does not have a natural frame, the rest (center-of-mass) frame, to define its position and momentum. In this section we see the mathematical origin of the frame dependence -- in particular, its relation to the Wigner translation~\cite{Wigner:1939cj} for massless particle. This relation can be seen via various methods~\cite{Duval:2014cfa,Stone:2015kla}. Here we demonstrate a method based on an alternative, group theoretic expression of the action.

First we briefly review the concept of little group and Wigner translation. Suppose we a have time-like four vector, say $(m, 0, 0, 0)$. Among the six dimensional Lorentz group, a three dimensional subgroup leaves this vector invariant; clearly this subgroup is just the $SO(3)$ rotation in the time-like vector's rest frame. Such Lorentz subgroup that leaves a four vector invariant is called the vector's \emph{little group}, or \emph{stablizer}. What if the four vector is null, say $k^a=(\kappa, 0, 0, \kappa)$ ($\kappa$ has the dimension of energy)? The little group is the Euclidean group $ISO(2)$ generated by \cite{weinberg1995quantum}
\begin{eqnarray}
J_3, \ \ \ \ \ A_3 \equiv J_2 - K_1, \ \ \ \ \ B_3 \equiv -J_1-K_2
\end{eqnarray}
where $J_i$ are rotation generators and $K_i$ are boost generators, whose non-zero components are
\begin{eqnarray}
(J_i)^j_{\ k} = (J_i)_{jk} = \epsilon_{ijk}, \ \ \ \ \ \ (K_i)^0_{\ j} = (K_i)^j_{\ 0} = -(K_i)_{0j}=(K_i)_{j0} = -\delta_{ij}.
\end{eqnarray}
One can easily check $J_3, A_3, B_3$ annihilate $k^a$. Clearly $J_3$ is the rotation along the spatial direction of the null vector. $A_3, B_3$ are certain combinations of rotation and boost; the transformation generated by them are called \emph{Wigner translation}.

Now we assume $A_\mu=0$ and consider the following action
\begin{eqnarray}
S[x, L] = \int \left(L_\mu^{\ a} k_a \ \frac{dx^\mu}{d\tau} - \frac{\hbar}{2} \frac{(J_3)^a_{\ b}}{2} \: L^\mu_{\ a} \: \frac{dL_\mu^{\ b}}{d\tau}\right) d\tau
\label{geo_action}
\end{eqnarray}
where $[x^\mu(\tau), L^\mu_{\ a}(\tau)]$ should be viewed as a worldline lying in the Poincar\'{e} group $\mathbb{R}^{3, 1}\rtimes SO(3,1)$. There is a natural group theoretical and geometrical way to construction of such actions from Lie groups; see \cite{Balachandran:1983pc,Skagerstam:1992er} and we will not review the construction here. We will very soon see this action is physically equivalent to the Weyl fermion action at $A_\mu=0$. We will interpret $L_\mu^{\ a} k_a$ as the momentum $p_\mu$ and $(\hbar/2) \, L^\mu_{\ a} L^\nu_{\ b} (J_3)_{ab}$ as the spin matrix $\hbar\, \Sigma^{\mu\nu}$. Note that the physical conditions $p_\mu p^\mu=0$ and $p_\mu \Sigma^{\mu\nu}=0$ are already built-in.

The action \eqref{geo_action} has two obvious properties. It is independent of the choice of $k_a$ as long as it stays null and $(J_3)^a_{\ b}$ stays the rotational generator annihilating $k_a$. This can be seen by performing a Lorentz transformation on the $ab$ indices (involving a redefinition of $L$). Moreover, it has the physical Poincar\'{e} invariance of adding a constant to $x$ and performing a Lorentz transformation on the $\mu\nu$ indices.

If we assert the action above describes the Weyl fermion, we must resolve the following: A Weyl fermion has six physical degrees of freedom $(\x, \p)$, while the action \eqref{geo_action} has ten degrees of freedom $(x^\mu, L^\mu_{\ a})$ of the Poincar\'{e} group. If they can match, the action \eqref{geo_action} must have four gauge (unphysical) degrees of freedom along the worldline. This is indeed the case. If we consider an generic infinitesimal Poincar\'{e} transformation
\begin{eqnarray}
x^\mu \rightarrow x^\mu + L^\mu_{\ a} \chi^a, \ \ \ \ \ \ L^\mu_{\ a} \rightarrow L^\mu_{\ b} (\delta^a_{\ b}+ \lambda^a_{\ b}), \ \ \ \ \lambda_{ab}=-\lambda_{ba}
\end{eqnarray}
we can easily find the following gauge transformations leave the action invariant:
\begin{eqnarray}
\chi^a &=& d(\tau) \ k^a + (\hbar/2) \left(a(\tau) \, (e_1)^a + b(\tau) \, (e_2)^a\right), \nonumber \\[.2cm]
\lambda^a_{\ b} &=& b(\tau) \, (A_3)^a_{\ b} - a(\tau) \, (B_3)^a_{\ b} + c(\tau) \, (J_3)^a_{\ b}
\end{eqnarray}
where $(e_1)^a = \delta_1^a$ and $(e_2)^a=\delta_2^a$ are the unit spatial vectors orthogonal to $k^a$, and $a, b, c, d$ are four arbitrary infinitesimal gauge parameters along the worldline. The $d$ transformation is a version of the transformation proportional to the EoM evolution that we mentioned at the end of Section \ref{ssect_infsml_transf}, and has nothing special to do with massless spinning particle. The $c$ transformation is an obvious one. Most interesting is the gauge Wigner translation parametrized by $a, b$ -- it involves a transformation of $x$, which means the ``position $x$'' is not a gauge invariant physical quantity. This hints the gauge Wigner transformation is related to the frame dependence \eqref{frame_dep}. Another evidence is that the momentum $p_\mu$ and the total angular momentum $\mathcal{J}^{\mu\nu}$ are gauge invariant under Wigner translation, in agreement with their frame independence when $A_\mu=0$. Yet another evidence is, one can check the commutation of two Wigner translations is a $d$ transformation, also in agreement with the algebra mentioned at the end of Section \ref{ssect_infsml_transf}. 

How to explicitly realize the relation? Note that the benefit of \eqref{geo_action} over \eqref{Weyl_Action_rel} is that the former is manifestly Lorentz invariant and frame independent (no frame vector $n^\mu$ is involved), but the price payed is it involves gauge degrees of freedom as mentioned above. To reduce \eqref{geo_action} to \eqref{Weyl_Action_rel}, we need to use a frame vector $n^\mu$ to fix the Wigner translation gauge. More particularly, note that in \eqref{Weyl_H_rel}, the spin matrix satisfies both $p_\mu \Sigma^{\mu\nu}=0$ as well as $n_\mu \Sigma^{\mu\nu}=0$, while in the present formalism of \eqref{geo_action}, we only have the former constraint but not the latter, because there is no frame $n^\mu$. To fix the Wigner gauge, we use an arbitrary frame vector $n^\mu$ (can be spacetime dependent), and impose the gauge fixing condition
\begin{eqnarray}
n_\mu \Sigma^{\mu\nu} = 0.
\end{eqnarray}
Equivalently, we are gauge fixing $L$ to take the form of ``first perform a boost (with respect to our frame $n$) along the direction of $\kk$ so that $\kappa$ is boosted to $|\p|$, then rotate the vector into the direction of $\p$''. Under such gauge fixing, the action \eqref{geo_action} reduces to the Weyl action \eqref{Weyl_Action_rel} with $\mathcal{H}=0$. If we pick another reference frame $n'^\mu$, the gauge fixing condition will change, so we must perform a Wigner translation to satisfy the new gauge fixing condition, and the transformation is precisely \eqref{frame_dep_finite} at $A_\mu=0$ (the easiest way to see this is to use the Wigner gauge invariance of $\mathcal{J}^{\mu\nu}$ \cite{Chen:2015gta}).

Everything said above did not rely on the smallness of $\hbar$, because there has been no external field $A_\mu$. How to extend the connection between Wigner translation and frame dependence to the case with $A_\mu\neq 0$? If we just add to \eqref{geo_action} the minimal coupling $A_\mu(x) dx^\mu$, the gauge invariance of Wigner translation is clearly broken at order $\hbar$, which is undesired. To remedy this to order $\hbar$, it can be shown that in addition to the minimal coupling, one must also replace 
\begin{eqnarray}
k_a \ \ \rightarrow \ \ k_a - \hbar F_{\rho\sigma}\Sigma^{\rho\sigma} \, \frac{\bar{k}_a}{4\kappa^2}
\label{k-replacement}
\end{eqnarray}
in the first term of \eqref{geo_action}, where $\bar{k}^a=(\kappa, 0, 0, -\kappa)$; moreover, the gauge Wigner transformation of $L$ must also be accompanied with a correction term to $\lambda^a_{\ b}$ by
\begin{eqnarray}
-\frac{\hbar}{4\kappa^2} \left[(K_3)^a_{\ b} \: \frac{F_{\rho\sigma} \: L^\rho_{\ c} L^\sigma_{\ d} (a \bar{A_3} + b \bar{B}_3)^{cd}}{2} + 2\left(b\bar{A}_3 - a\bar{B}_3\right)^a_{\ b} \: F_{\rho\sigma} \Sigma^{\rho\sigma} \right]
\end{eqnarray}
where $\bar{A}_3=J_2+K_1$ and $\bar{B_3}=-J_1+K_2$ are the Wigner translation generators for $\bar{k}$. The replacement \eqref{k-replacement} leads to the ``on-shell condition'' $\mathcal{H}=0$ with the magnetic dipole term in \eqref{Weyl_H_rel}. Performing the gauge fixing as before, one can show under a change of frame choice $n^\mu \rightarrow n'^\mu$, the Wigner translation needed to accommodate the new gauge fixing condition is precisely \eqref{frame_dep_finite}. 

This explains the mathematical original of the frame dependence of $x$ and $p$, valid to order $\hbar$. Furthermore, in the presence of $A_\mu$, one can show that no further correction can make the action \eqref{geo_action} Wigner translation invariant, or equivalently make the action \eqref{Weyl_Action_rel} frame independent, at order $\hbar^2$. This is the symmetry based argument towards our comment made in Section \ref{sect_BFG_PIderivation}, that the semi-classical theory generally cannot be extended to order $\hbar^2$, because at that order we have to consider the presence of other bands.

\section{Chiral Kinetic Theory with Collisions}
\label{sect_collisions}

We have already completed the discussion of Lorentz invariance in non-interacting chiral kinetic theory, with both the physical interpretation and mathematical structure explored. In this section we take one step further~\cite{Chen:2015gta}. We would like to consider chiral kinetic theory with collisions, and study what constraints Lorentz invariance places on it. We will construct an entropy current, and determine what the most general form of hydrodynamical limit (local equilibrium) is. As expected, it is the well known chiral hydrodynamics \cite{Son:2009tf,Neiman:2010zi,Loganayagam:2012pz}, from which we will compute the famous chiral vortical effect (CVE).

So far we have only been able to study chiral kinetic theory with collisions in the absence of EM field. Moreover, we have to point out the formalism presented here is solely based on symmetry principles, and lacks of a microscopic derivation. Microscopically, collisions are due to interactions between particles, so this section is beyond the non-interacting Berry Fermi gas theory. It is well-known that microscopic interactions lead to many effects in the kinetic theory other than collisions, for example an interaction potential energy; in this section we assume collisions is the dominating effect, while other effects can be ignored.

\subsection{Current}

Collisions in chiral kinetic theory must be non-trivial. As we have seen from Figures \ref{head-on_collision} and \ref{collision}, the collision must be non-local in general. However to construct a non-local collision kernel respecting Lorentz invariance is not obvious. Before we resolve this problem, though, we shall resolve another, perhaps more straightforward problem.

In the absence of EM field, the momentum $p_\mu$ is a frame independent variable. The current \eqref{Weyl_current_rel} reduces to
\begin{eqnarray}
J^\mu(x) = \int_{(p)} j^\mu(p, x) \equiv \int \frac{d^4 p}{(2\pi\hbar)^3} 2\delta(p\cdot p)\, \theta(-n\cdot p) \ j^\mu(p, x)
\end{eqnarray}
where the ``phase space current'' (the single particle current integrated over $z$) is
\begin{eqnarray}
j^\mu(p, x) \equiv p^\mu f(p, x) + \hbar\, \Sigma^{\mu\nu}(p) \partial_{x^\nu} f(p, x).
\label{phase_space_current_collisionless}
\end{eqnarray}
Clearly the integral $\int_{(p)}$ is frame independent. We expect $j^\mu$ to be frame independent too. But it is not. Recall that the distribution transforms as \eqref{f_transf}, i.e.
\begin{eqnarray}
f'(p, x)-f(p, x)=-\hbar\, \Delta_{nn'}\cdot \partial_x f(p, x).
\label{f_frame_dep_collisionless}
\end{eqnarray}
(recall that here $x$ is not ``a particle's position'' but a generic point in the spacetime, which is frame independent). Because $\hbar\, \Sigma^{\mu\nu}$ transforms as $-2x^{[\mu} p^{\nu]}$ does (invariance of total angular momentum), we find
\begin{eqnarray}
j'^\mu(p, x) - j^\mu(p, x) = \hbar, \Delta_{nn'}^\mu(p) \ \left(-p\cdot \partial_{x} f(p, x)\right).
\end{eqnarray}
When the ensemble is collisionless, the collisionless Boltzmann equation \eqref{Boltzmann_Eq} reads $p\cdot \partial_{x} f=0$, and therefore $j^\mu$ and hence $J^\mu$ are indeed frame independent.

But once we include collisions, the Boltzmann equation becomes
\begin{eqnarray}
p\cdot \partial_{x} f = \int_{BCD} C_{ABCD}[f] + \mathcal{O}(\hbar).
\label{Boltz_Eq_col_raw}
\end{eqnarray}
Here we are considering all possible collisions $AB\leftrightarrow CD$ with $p=p_A$, and integrating over the momenta of the other particles, $\int_B \equiv \int_{(p_B)}$ etc. The collision kernel is
\begin{eqnarray}
C_{ABCD} = W_{CD\rightarrow AB} - W_{AB\rightarrow CD}
\end{eqnarray}
where $W$ is the collision rate whose detailed form will be given later; it should be frame independent at zeroth order in $\hbar$. The Boltzmann equation at order $\hbar$ will also be given later. The issue now is, the collision term makes $j^\mu$ and hence $J^\mu(x)$ become frame dependent at order $\hbar$.

The physical observable $J^\mu$ being frame dependent is certainly unacceptable. But how can it be, given that we obtained $J^\mu$ from the action, which is itself frame independent? The problem is, the action is frame independent only up to boundary terms. For collisionless ensemble, the worldlines have no end points and therefore there would be no boundary terms. With collisions, however, the Weyl fermion action varies by a boundary term of $\left. A_\mu \hbar\Delta_{nn'}^\mu \right|_{initial}^{final}$, resulting in the frame dependence of $j^\mu$. This means, although our description of the particle's helical current is frame independent along the worldline, the description of the ends of the worldline depends on a frame, which causes the problem.

The problem is now reduced to, which frame should we choose to describe the end points? We do not need one frame for all the worldlines. The choice of frame can be different for each collision event. We assume, for each 2-to-2 collision event, there is a ``special frame'' $\bar{n}^\mu$ that, if we view the collision event in this special frame, then each particle's current is just the usual single particle current, ending (for incoming particles) or starting (for outgoing particles) at a common collision point, just as in the spinless case. If we view the event in another frame $n^\mu$, then we transform each single particle current according to the boundary term of the action, i.e. the single particle current transforms by $\hbar \Delta_{n\bar{n}}^\mu$ with a delta function localized at the collision point. In the context of 2-to-2 collision, there is a natural choice for the ``special frame'' $\bar{n}^\mu$ -- the center-of-mass (CoM) frame of the particular 2-to-2 collision:
\begin{eqnarray}
\bar{n}^\mu = \frac{(p+p_B)^\mu}{\sqrt{-(p+p_B)^2}} = \frac{(p+p_B)^\mu}{\sqrt{-2p\cdot p_B}} =  \frac{(p_C+p_D)^\mu}{\sqrt{-(p_C+p_D)^2}} = \frac{(p_C+p_D)^\mu}{\sqrt{-2p_C\cdot p_D}}.
\end{eqnarray}
In the following we will make this assumption.

This is just the idea behind Figure \ref{head-on_collision} made concrete. Based on this reasoning, the presence of collision modifies the collisionless phase space current \eqref{phase_space_current_collisionless} into
\begin{eqnarray}
j^\mu \equiv p^\mu f + \hbar\, \Sigma^{\mu\nu} \partial_{x^\nu} f + \int_{BCD} C_{ABCD} \: \hbar\, \Delta_{\bar{n}n}^\mu
\label{phase_space_current}
\end{eqnarray}
where the last term captures the frame dependent ``end point current'' localized at each collision point. From our reasoning, $j^\mu$ must be frame independent by construction. Then, in retrospect, it is in fact natural to ``define'' the phase space distribution by the time component of the phase space current:
\begin{eqnarray}
f = \frac{-n\cdot j}{-n\cdot p}.
\end{eqnarray}
Now that $j$ and $p$ are frame independent, we can find the frame dependence of $f$ is, in addition to \eqref{f_frame_dep_collisionless},
\begin{eqnarray}
f'-f = -\hbar\, \Delta_{nn'}\cdot \partial_{x} f + \int_{BCD} C_{ABCD} \frac{\hbar\, \Delta_{nn'} \cdot \bar{n}}{p \cdot \bar{n}},
\label{f_frame_dep}
\end{eqnarray}
where we have used the nice identity
\begin{eqnarray}
\Delta_{nn'}^\mu + \Delta_{n'n''}^\mu + \Delta_{n''n}^\mu = p^\mu \frac{\Delta_{nn'}\cdot n''}{p\cdot n''}
\label{Delta_identity}
\end{eqnarray}
that follows from the definition \eqref{frame_dep_finite} of $\Delta$. Now, one can compute $j'^\mu-j^\mu$ using \eqref{f_frame_dep} and find $j'^\mu-j^\mu=0$ as it should. This checks the consistency of our formalism.

Now that we have a frame independent phase space current $j^\mu$, we assert the physical current and the stress-energy tensor are give by
\begin{eqnarray}
J^\mu(x) = \int_{(p)} j^\mu(p, x), \ \ \ \ \ \ T^{\mu\nu}(x) = \int_{(p)} p^{\{\mu} j^{\nu\}}(p, x).
\label{current_stress-tensor_collision}
\end{eqnarray}
In the collisionless limit, they reduce to \eqref{Weyl_current_rel} and \eqref{Weyl_stress-tensor_rel} at $A_\mu=0$. We can check $\partial_{x^\mu} J^\mu=0$ and $\partial_{x^\mu} T^{\mu\nu}=0$ after we developed the Boltzmann equation.

\subsection{Boltzmann Equation}

Now we have found how to consistently express $J^\mu$ and $T^{\mu\nu}$ in terms of $f$ in the presence of collisions. But we still need to figure out how the distribution $f$ evolves in time, that is, what is the collisionful Boltzmann equation \eqref{Boltz_Eq_col_raw} to order $\hbar$. The collision kernel we find must reflect the non-local collision in Figure \ref{collision}.

In usual kinetic theory, i.e. the theory at zeroth order in $\hbar$, the collision rate for identical fermions is given by
\begin{eqnarray}
W_{AB\rightarrow CD}[f] &=& \frac{1}{2!} |\mathcal{M}(\mathrm{s}, \mathrm{t})|^2 \: (2\pi)^4 \delta^4(p_A+p_B-p_C-p_D) \: \times \nonumber \\[.2cm]
&& \hspace{1cm} f(p_A, x)  f(p_B, x) \left(1- f(p_C, x)\right) \left(1- f(p_D, x)\right)
\end{eqnarray}
where $\mathcal{M}$ is the scattering amplitude computed from quantum field theory, with Mandelstam variables $\mathrm{s} \equiv (p_A+p_B)^2$, $\mathrm{t} \equiv (p_A - p_C)^2$. But at order $\hbar$, the $f$'s are frame dependent, so the collision rate becomes ambiguous. Fortunately, again, for each collision event we have a natural choice of frame, the CoM frame $\bar{n}^\mu$. Therefore we are led to consider the distribution in the CoM frame for each collision event
\begin{eqnarray}
\bar{f} = \frac{-\bar{n}\cdot j}{-\bar{n}\cdot p},
\end{eqnarray}
and we propose the collision rate for each collision event should be
\begin{eqnarray}
W_{AB\rightarrow CD}[\bar{f}] &=& \frac{1}{2!} |\mathcal{M}(\mathrm{s}, \mathrm{t})|^2 \: (2\pi)^4 \delta^4(p_A+p_B-p_C-p_D) \: \times \nonumber \\[.2cm]
&& \hspace{1cm} \bar{f}(p_A, x)  \bar{f}(p_B, x) \left(1- \bar{f}(p_C, x)\right) \left(1- \bar{f}(p_D, x)\right).
\end{eqnarray}
With this proposition, we have a fix the right-hand-side of \eqref{Boltz_Eq_col_raw} to order $\hbar$. Since the right-hand-side does not depend on an arbitrarily chosen reference frame now, the left-hand-side must not be. The most natural choice of the left-hand-side would be $\partial_{x}\cdot j$, which reduces back to $p\cdot \partial_{x} f$ at zero order in $\hbar$. Therefore, we propose the Boltzmann equation with collisions at $A_\mu=0$ should be
\begin{eqnarray}
\partial_x \cdot j(p, x) = \int_{BCD} C_{ABCD}[\bar{f}] = \int_{BCD} \left(W_{AB\rightarrow CD}[\bar{f}] - W_{CD\rightarrow AB}[\bar{f}] \right)
\label{Boltz_Eq_col}
\end{eqnarray}
where $p=p_A$. Both side are now manifestly frame independent.

In practice, it is more convenient to use $f$ in some lab frame $n^\mu$, instead of having to consider $\bar{f}$ in the CoM frame of each collision. For this purpose, we may rewrite \eqref{Boltz_Eq_col} as
\begin{eqnarray}
p\cdot \partial_{x} f = \int_{BCD} C_{ABCD}[f] \ \left(1 - \int_{B'C'D'} \frac{dC_{AB'C'D'}[f]}{df} \ \frac{\hbar\, \Delta_{\bar{n}n} \cdot \bar{n}'}{p\cdot \bar{n}'}\right)
\label{Boltz_Eq_col_alt}
\end{eqnarray}
where $\bar{n}'$ is the COM frame of the collision $AB'\leftrightarrow C'D'$.

 As usual, integrating the Boltzmann equation \eqref{Boltz_Eq_col} with either $\int_{(p)}$ or $\int_{(p)} p^\mu$ yields zero. This reflects the conservation of charge and momentum under collisions. More precisely, this leads to $\partial_{x^\mu} J^\mu=0$ and $\partial_{x^\mu} T^{\mu\nu}=0$ (with the aid of \eqref{Boltz_Eq_col_alt}).

\subsection{Entropy Current}

Once collision is included, we are led to the most important concept of Boltzmann's kinetic theory -- the existence of an entropy current $S^\mu$ which is non-decreasing $\partial_x\cdot S \geq 0$. This is the famous Boltzmann's H-Theorem, revealing the fact the effects of collisions are irreversible and drive the systems towards disorder. When entropy is maximized, i.e. $\partial_x\cdot S \rightarrow 0$, the system has relaxed to local equilibrium, and this is the hydrodynamical limit of kinetic theory.

In our formalism, a general frame independent phase space current takes the form
\begin{eqnarray}
j_\phi^\mu = p^\mu \phi(f) + \hbar\, \Sigma^{\mu\nu} \partial_{x^\nu} \phi(f) + \int_{BCD} C_{ABCD} \: \hbar\, \Delta_{\bar{n}n}^\mu \frac{d\phi(f)}{df}
\end{eqnarray}
where $\phi(f)$ is an arbitrary smooth function of $f$; clearly the charge current $j$ is the $j_\phi$ when $\phi(f)=f$. Its frame independence ${j'}_\phi^\mu=j_\phi^\mu$ can be verified by \eqref{f_frame_dep} and \eqref{Delta_identity}. Moreover, its divergence is
\begin{eqnarray}
\partial_x \cdot j_\phi = \int_{BCD} C_{ABCD}[\bar{f}] \frac{d\phi(\bar{f})}{d\bar{f}}
\end{eqnarray}
using \eqref{Boltz_Eq_col_alt}.

With the above, if we know in usual kinetic theory (zeroth order in $\hbar$) what the $\phi$ corresponding to entropy current is, we immediately have the entropy current at first order in $\hbar$. It is elementary that in usual kinetic theory, the $\phi$ corresponding to entropy current is the phase space local entropy
\begin{eqnarray}
\mathfrak{s} = - f \ln f - (1-f) \ln (1-f).
\end{eqnarray}
Therefore, to first order in $\hbar$, the entropy current is
\begin{eqnarray}
S^\mu = \int_{(p)} j_\mathfrak{s}^\mu.
\end{eqnarray}
To see the entropy current is non-decreasing, we abbreviate
\begin{eqnarray}
r \equiv \frac{W_{CD\rightarrow AB}}{W_{AB\rightarrow CD}} = \frac{f_C f_D (1-f_A) (1-f_B)}{f_A f_B (1-f_C) (1-f_D)}.
\end{eqnarray}
Note that $C_{ABCD} = W_{AB\rightarrow CD} (r-1)$. Since $C_{ABCD}$ is even under $A\leftrightarrow B$ and $C\leftrightarrow D$ but odd under $AB\leftrightarrow CD$, we find
\begin{eqnarray}
\partial_{x} \cdot S = \frac{1}{4} \int_{ABCD} W_{AB\rightarrow CD} (r-1) \ln r \geq 0,
\end{eqnarray}
The above is vanishing only when $r=1$, i.e. when there is no net collision $C_{ABCD}=0$ (``detailed balance''). This is the local equilibrium state.

\subsection{Equilibrium and Chiral Vortical Effect}
\label{ssect_CVE}

Let's work out the most general local equilibrium state. We are going to show that the most general equilibrium distribution function, viewed in any frame $n^\mu$, is
\begin{eqnarray}
f(p, x) = \frac{1}{\exp\left( U^\mu(x)\: p_\mu - Y + \hbar\, \Sigma^{\mu\nu}(p) \: \partial_{x^\mu} U_\nu(x)/2 \right) + 1}
\label{f_eq_Weyl}
\end{eqnarray}
with $Y$ constant and $U^\mu(x)$ a future time-like vector satisfying
\begin{eqnarray}
\partial_{x^\mu} U_\nu(x) = \partial_{[x^\mu} U_{\nu]}(x) + \frac{\eta_{\mu\nu}}{4} \partial_{x^\lambda} U^\lambda(x).
\label{U_condition}
\end{eqnarray}
We will also show \eqref{f_eq_Weyl} is compatible with the frame transformation \eqref{f_frame_dep}. In usual notations one would write $U^\mu(x)=u^\mu(x)/ T(x)$ and $Y=\mu(x) / T(x)$ where $u$ is the normalized local fluid velocity, $T>0$ is the local temperature and $\mu$ is the local chemical potential.

As usual, it is convenient to define $g$ such that $f = 1/(e^g +1)$ so that $r$ defined above can be written as $r = \exp(\bar{g}_A + \bar{g}_B - \bar{g}_C - \bar{g}_D)$. To achieve local equilibrium, we need $r=1$, so $\bar{g}$ must be a linear combination of conserved quantities during a collision. The conserved quantities are particle number (charge), momentum $p^\mu$ and angular momentum $\mathcal{J}^{\mu\nu}$. Since we are viewing $g$ in the CoM frame, the orbital angular momentum vanishes, so the conservation of $\mathcal{J}^{\mu\nu}$ becomes the conservation of spin viewed in the CoM frame $\hbar \, \bar{\Sigma}^{\mu\nu}$. Therefore, in local equilibrium $\bar{g}$ must take the form
\begin{eqnarray}
\bar{g}_{eq}(p, x) = U^\mu(x)\: p_\mu - Y(x) + \hbar\, \bar{\Sigma}^{\mu\nu}(p) \: \Omega_{\mu\nu}(x)/2.
\label{g_eq_ansantz}
\end{eqnarray}
where $U^\mu, Y$ and $\Omega_{\mu\nu}$ are some coefficients. This is the constraint from $C_{ABCD}=0$ on $\bar{g}$.

But the coefficients $U, Y, \Omega$ are certainly not arbitrary. They need to satisfy two conditions:
\begin{itemize}
\item
We have expressed $g_{eq}$ in the CoM frame $\bar{n}$ of a particular collision $AB\leftrightarrow CD$. However, physically $f=1/(e^g+1)$ describes the distribution of the particle $A$, and is unrelated to the other particles $B, C, D$. This means, if we transform \eqref{g_eq_ansantz} into an arbitrary lab frame $n$ using \eqref{f_frame_dep}, in the lab frame the expression of $g_{eq}$ must be independent of $\bar{n}$, and hence independent of $p_B, p_C, p_D$. 

\item
The distribution $f=1/(e^g+1)$ must satisfy the Boltzmann equation \eqref{Boltz_Eq_col}. 
\end{itemize}
To satisfy the second condition, now that \eqref{g_eq_ansantz} guarantees $C_{ABCD}=0$, the Boltzmann's equation \eqref{Boltz_Eq_col_alt} implies $p\cdot \partial_x g =0$. This means
\begin{eqnarray}
\partial_{x^\mu} U^\nu \: p^\mu p_\nu = 0, \ \ \ \ \ \ p^\mu \partial_{x^\mu} Y = 0, \ \ \ \ \ \ \hbar \bar{\Sigma}^{\mu\nu} p^\lambda \partial_{x^\lambda} \Omega_{\mu\nu} = 0.
\label{U_Y_Omega_conditions}
\end{eqnarray}
The equations for $U$ and $Y$ reduce to \eqref{U_condition} (the antisymmetric part is generic, while the trace part relies on the Weyl fermion being massless) and $Y=const.$ respectively; we will leave the $\Omega$ equation for later. For the first condition, using \eqref{f_frame_dep} we find, in a lab frame $n$,
\begin{eqnarray}
g_{eq} = \left( U_\mu - \hbar\, \Delta_{\bar{n}n}^\nu \partial_{x^\nu} U_\mu + \hbar\, \Delta_{\bar{n}n}^\nu \Omega_{\nu\mu} \right) p^\mu - \left(Y - \hbar \, \Delta_{\bar{n}n}^\nu \partial_{x^\nu} Y \right) + \hbar\, \Sigma^{\mu\nu} \Omega_{\mu\nu}/2
\end{eqnarray}
Since we have already shown $Y$ is a constant and $U$ satisfies \eqref{U_condition}, the condition $g_{eq}$ being independent of $\bar{n}$ reduces to $\Sigma^{\mu\nu}\Omega_{\mu\nu} = \Sigma^{\mu\nu}\partial_{x^\mu} U_\nu$; this in turn solves the equation for $\Omega$ in \eqref{U_Y_Omega_conditions}. Finally, in order for $f$ to approach $0$ at large energy, $U$ must be future time-like. Thus, we have shown \eqref{f_eq_Weyl} with constant $Y$ and future time-like $U$ satisfying \eqref{U_condition} is the most general equilibrium distribution, and it is compatible with the change of frame \eqref{f_frame_dep}.

The phase space current \eqref{phase_space_current} in equilibrium is
\begin{eqnarray}
j_{eq}^\mu &=& p^\mu f_{eq}^0 + \frac{df_{eq}^0}{dg_{eq}^0} \hbar\, \frac{3}{2} \Sigma^{[\mu\nu} p^{\lambda]} \partial_{x^\nu} U_\lambda \nonumber \\[.2cm]
&=& p^\mu f_{eq}^0 + \frac{df_{eq}^0}{dg_{eq}^0} \frac{\hbar}{2} \frac{\epsilon^{\mu\nu\rho\sigma}}{2} p_\nu \: \partial_{x^\rho} U_\sigma
\label{j_eq_expression}
\end{eqnarray}
where $f_{eq}^0$ is $f_{eq}$ dropping the $\hbar$ term, i.e. the Fermi-Dirac distribution for spinless particle. $j_{eq}^\mu$ is explicitly frame independent. The phase space current $j_{\mathfrak{s}}^\mu$ is similar, but with $\mathfrak{s}(f_{eq}^0)$ in place of $f_{eq}^0$.

We would like to compute $J^\mu$, $T^{\mu\nu}$ and $S^\mu$ in equilibrium and see the chiral vortical effect. However we have a big missing piece. In this section we have been discussing the fermion only, but we also have to include anti-fermion contributions. The total current, stress-energy tensor and entropy current are
\begin{eqnarray}
J_{tot}^\mu = J^\mu - \wt{J}^\mu, \ \ \ \ \ T_{tot}^{\mu\nu} = T^{\mu\nu} + \wt{T}^{\mu\nu}, \ \ \ \ \ \ S_{tot}^\mu = S^\mu + \wt{S}^\mu
\end{eqnarray}
where, for anti-fermions, every $\hbar/2$ is replaced with $-\hbar/2$. The incorporation of anti-fermions into the collision kernel is also straightforward: now the $ABCD$ not only denote the particles' momenta, but also whether each particle is a fermion or anti-fermion. In equilibrium, anti-fermions should have $\wt{U}^\mu=U^\mu$ but $\wt{Y}=-Y$, because $Y$ is the coefficient for fermion number conservation.

Now we have all the recipes. In the $\int_{(p)}$ integration, it is natural to use the fluid velocity as the frame vector, i.e. $n^\mu = u^\mu$ (recall that $u^\mu$ is $U^\mu$ normalized). The energy in this frame is $E\equiv - u\cdot p$, while the spatial momentum is $\p^\mu \equiv p^\mu - E u^\mu$. Using the phase space current \eqref{j_eq_expression} (and also the phase space entropy current) and performing the $\p$ integrals, we find
\begin{eqnarray}
&& (J_{tot})_{eq}^\mu = \mathcal{N} u^\mu + \xi_J \, \omega^\mu, \ \ \ \ \ (S_{tot})_{eq}^\mu = \mathcal{S} u^\mu + \xi_S \, \omega^\mu, \nonumber \\[.2cm]
&& (T_{tot})_{eq}^{\mu\nu} = \mathcal{E} u^\mu u^\nu + \mathcal{P} \left( \eta^{\mu\nu}+u^\mu u^\nu \right) + 2\xi_T \, u^{\{\mu} \omega^{\nu\}}
\label{CVE_JST}
\end{eqnarray}
where $\omega^\mu \equiv \epsilon^{\mu\nu\rho\sigma} u_\nu \partial_{x^\rho} u_\sigma /2$ is the vorticity and the associated terms is the CVE. $\mathcal{N}$, $\mathcal{E}$, $\mathcal{P}$ and $\mathcal{S}$ are the usual density, energy density, pressure and entropy density for massless spinless particle:
\begin{eqnarray}
\mathcal{N} = \frac{4\pi}{(2\pi\hbar)^3} \frac{\mu^3 + \pi^2 \mu T^2}{3}, \ \ \ \ \ \mathcal{E} = \frac{4\pi}{(2\pi\hbar)^3} \frac{\mu^4+2\pi^2 \mu^2 T^2 + 7\pi^4 T^4/15}{4}
\end{eqnarray}
and $\mathcal{P}=\mathcal{E}/3$, $\mathcal{S} = \partial \mathcal{P}/ \partial T$. The CVE coefficients $\xi_J$, $\xi_T$ and $\xi_S$ are given by
\begin{eqnarray}
\xi_J &=& \frac{4\pi}{(2\pi\hbar)^3} \frac{\hbar}{2} \frac{1}{T} \int_0^\infty dE \: E^2 \left( \frac{\exp(E/T-Y)}{(\exp(E/T-Y) + 1)^2} - \frac{\exp(E/T+Y)}{(\exp(E/T+Y) + 1)^2} \right) \nonumber \\[.2cm]
&=& \frac{4\pi}{(2\pi\hbar)^3} \frac{\hbar}{2} \left( \mu^2 + \frac{\pi^2}{3} T^2\right),
\end{eqnarray}
\begin{eqnarray}
\xi_T &=& \frac{4\pi}{(2\pi\hbar)^3} \frac{\hbar}{2} \frac{2}{3T} \int_0^\infty dE \: E^3 \left( \frac{\exp(E/T-Y)}{(\exp(E/T-Y) + 1)^2} + \frac{\exp(E/T+Y)}{(\exp(E/T+Y) + 1)^2} \right) \nonumber \\[.2cm]
&=& \frac{4\pi}{(2\pi\hbar)^3} \frac{\hbar}{2} \frac{2}{3}\left( \mu^3 + \pi^2 \mu T^2\right),
\end{eqnarray}
\begin{eqnarray}
\xi_S &=& \frac{3}{2T} \xi_T - Y \xi_J = \frac{\hbar}{2} \frac{\mu T}{3}.
\end{eqnarray}
This is in agreement with the previous literature~\cite{Son:2009tf,Neiman:2010zi,Loganayagam:2012pz}, up to a ``change of hydrodynamic frame'', i.e. a redefinition of $u^\mu$ in \eqref{CVE_JST}. (The earliest computation \cite{PhysRevD.20.1807} of CVE had a different $\xi_T$ because the $\partial_x f$ contribution to $T^{\mu\nu}$ was missing.)

Usually in relativistic hydrodynamics one uses one of Eckart frame (define $u$ so that $J$ has no spatial component in the $u$ frame), Landau frame (define $u$ so that $T$ has no mixed temporal-spatial components in the $u$ frame) or entropy frame (define $u$ so that $S$ has no spatial component in the $u$ frame). However, our result \eqref{CVE_JST} is expressed in a frame of fluid velocity $u$ that does not admit any of these three usual conditions. What is the physical meaning of our $u$ frame? It is shown that our frame of fluid velocity $u$ is the ``no-drag frame''~\cite{Stephanov:2015roa}: if an impurity particle interacting with the fluid is moving in the fluid with the velocity $u$, it will experience no net drag force.

\section{Summary and Outlook}

In this chapter we clarified how Lorentz invariance is non-trivially realized in chiral kinetic theory, in the seemly frame dependent Berry curvature formalism. We provided both the physical interpretation and the mathematical origin of the non-trivial realization. Then we proposed a formalism of collisionful chiral kinetic theory in the absence of external force, and showed how the system is can be relaxed to the chiral hydrodynamics limit, in which we computed the chiral vortical effect. The collisionful chiral kinetic theory can be potentially applied to the study of quark matter in heavy ion collision experiment or early universe.

The collisionful chiral kinetic theory is worth further study. First, our proposed formalism is restricted to the absence external electromagnetic field; currently it is unclear how to write down a frame independent collisionful Boltzmann equation in the presence of external field. For application purpose it would be useful to make this extension. Second, we do not have a microscopic derivation for the collisionful case; we only write down the simplest possible formalism based on Lorentz symmetry considerations. It is desired to have a derivation from the Kadanoff-Baym method~\cite{kadanoff1962quantum} or other methods.

Another direction for further theoretical study is to couple the semi-classical Weyl fermion to curved spacetime. Within the group theoretical formalism in Section \ref{sect_Wigner_transl}, the coupling to a background metric can be introduced~\cite{Balachandran:1983pc}. For massless particles, the non-triviality is again to find quantum correction terms, this time to remedy the violation of Wigner translation gauge invariance due to the spacetime curvature. We have made certain progress in this direction, but there remains an unsatisfactory issue -- the gravitational chiral anomaly is beyond the reach of our current regime of validity. In this thesis we are not covering the relevant efforts. The completion towards this direction can be a future subject of study.

Although our discussion focuses on the issue of Lorentz invariance, through the discussion we gain a better understanding of the Berry curvature formalism in general. The $\partial_x f$ terms in the current and stress-energy tensor are generic. This means at order $\hbar$ the semi-classical particle's single particle current and stress-energy tensor generically have non-point-like feature. Thus, in general semi-classical systems, we always have to carefully interpret the ``position $\x$'' and ``momentum $\p$'' as the position of the ``center'' of the single particle current and the momentum carried at this ``center''. Moreover, if we would like to include a collision term in the Boltzmann equation, either collision between particles or collision off impurities, it is in general invalid to assume the collision is local; there would generally be some order $\hbar$ (same order as Berry curvature effects) non-locality in the collision kernel. For chiral kinetic theory we can use Lorentz symmetry to pin down this non-locality; for general systems perhaps a microscopic derivation is needed.

\chapter{Berry Fermi Liquid}
\label{chap_BFL}

From the previous chapters, we see the kinetic theory of Fermi gas with Berry curvature is a useful approach towards a wide range of physical systems. However, for most physical systems in reality, the assumption of free Fermi gas does not hold. The interaction between the fundamental fermions might be so strong that, if one excites a fermion with high energy, it very soon decays into low energy excitations and ``dissolves'' into the system; hence, the picture based on the notion ``particles'' becomes not so useful. In such scenarios, we shall wonder, how do we still define ``Berry phase effects''? How much of those we said about Berry Fermi gas still survives under interactions? Moreover, even if the interaction is sufficiently weak that ``particles'' can still be talked about and the picture of Fermi gas is a good approximation, we can still ask, does the interaction introduce any new and interesting effects? In this chapter we explore these problems.

In our discussion of Fermi gas, thanks to the single particle picture, we could discuss the physics of far-from-equilibrium systems. Once interaction is included, far-from-equilibrium systems become difficult to study in general. Therefore we will be less ambitious about interacting systems. We restrict our consideration to a large class of interacting fermionic systems called \emph{Fermi liquid}. A Fermi liquid is a system whose ground state has some non-zero fermion density specified by a Fermi surface similar to that of a non-interacting system, and more importantly, whose low energy (gapless) excitations behave as nearly-free fermionic particles near the Fermi surface, with decay rate suppressed by their low energy (while interactions are not necessarily weak); no particular assumption is made about high energy excitations. Due to these limitations in assumptions, we study the near-ground-state behaviors of the Fermi liquid under low energy, long wavelength external perturbations only.

The study of Fermi liquid started with the classic works of Landau~\cite{landau1957theory,landau1957oscillations} in 1957. Following his amazing physical intuition, Landau proposed that at low energy a Fermi liquid can be described by collisionless kinetic theory, and the leading effect (unsuppressed by the low energy) of interaction is a local interaction potential energy and nothing else; all other effects such as particle decay are suppressed by the low energy, even when interaction is not weak. Using his theory of Fermi liquid, Landau successfully explained the propagation of ``zero sound'' in low temperature Helium-3 liquid. Landau's Fermi liquid theory was also successful in explaining various properties of electrons in metals.

Originally Landau's theory was formulated based on physical intuition, the theory was later derived from quantum field theory by Landau himself and others~\cite{landau1959theory,Nozieres:1962zz,Luttinger:1962zz,kadanoff1962quantum,abrikosov1975methods}. In the 1990s, the remarkable intuition of Landau was concretely formulated in the language of low energy effective field theory~\cite{Polchinski:1992ed,Shankar:1993pf}.

Despite the remarkable success of the Landau Fermi liquid theory, the interesting anomalous Hall effect, chiral magnetic effect and other Berry phase physics are missed in Landau's framework. The reason is simple: Landau focused on the behaviors that are unsuppressed by the low energy / long wavelength, while these interesting physics are one order higher in the low energy / long wavelength expansion. One can see this easily: for instance, the longitudinal current $\delta J_L \sim \sigma_L E \sim const.\, A$ where $A$ is the electromagnetic connection, $E\sim \omega A$ is the electric field and $\sigma_L\sim const./\omega$ is the longitudinal conductivity, with $\omega$ being the small energy carried in $A$; on the other hand, the Hall current $\delta J_H\sim \sigma_H E \sim const.\, \omega A$. Therefore we see the longitudinal current is dominated by unsuppressed terms while the Hall current is suppressed by $\omega$. Thus, in order to study the anomalous Hall effect, chiral magnetic effect and other Berry phase physics in Fermi liquid, we must work to the first order in the low energy / long wavelength expansion. Moreover, for Berry curvature to appear, clearly the fermionic field must be multi-component. This is what we consider in this chapter, and the theory developed here would be called ``Berry Fermi liquid theory''.

We start with an Invitation section, demonstrating in the simple example of $(2+1)D$ Dirac fermion with a weak contact interaction that the anomalous Hall current is no longer given by the Berry curvature of the particles, but receives a correction due to the interaction. This will be explained by the emergent electric dipole moment as we develop the general framework of Berry Fermi liquid theory. Between the Invitation section and the presentation of the Berry Fermi liquid theory, we briefly review the Landau Fermi liquid theory.

This Chapter is mostly a reproduction of Ref.~\cite{chen2016berry}, with the example in Section \ref{sect_Invitation} added for demonstration purpose.

\section{Invitation: $(2+1)D$ Dirac Fermion with Weak Contact Interaction}
\label{sect_Invitation}

Let's start with a simple example. In this simple example, interaction is so weak that the decay or collision between the fermions can be neglected, so the very notion of particle is well-defined. Thus, we have a physical picture that is mostly analogous to a Fermi gas, except now there is some interaction between the fermions. We want to see if there is any interesting new effect arising from the interactions.

We consider the 2-component Dirac fermion field $\psi$ in $(2+1)D$. We consider $N$ identical copies (flavors) of the field and label the flavors as $\psi_a$. The Lagrangian is
\begin{eqnarray}
\mathcal{L} = -\bar{\psi}_a \, i\gamma^\mu \left(-i\partial_{x^\mu}-A_\mu-\epsilon_0\delta_\mu^0\right) \psi_a - m_0 \bar{\psi}_a \psi_a - \sigma \bar{\psi}_a\psi_a + \sigma^2 / 2g.
\end{eqnarray}
We choose the Dirac gamma matrices in the basis $\gamma^0=-i\sigma^3, \gamma^1=\sigma^2, \gamma^2=-\sigma^1$ (which leads to $\gamma^0\gamma^\mu=-\sigma^\mu$). Here $\bar{\psi}\equiv \psi^\dagger (i\gamma^0)$ as usual; $m_0$ is the mass parameter and $\epsilon_0$ is the chemical potential parameter. $\sigma$ is an auxiliary scalar field integrating out which yields the contact interaction $-(g/2)(\bar{\psi}_a\psi_a)^2$. We let $g$ to be small and keep it to first order; since particle decay / collision is order $g^2$, we have the picture of stable Dirac fermions. For this Lagrangian to be invariant under large gauge transformation, $N$ must be even, and that is why we did not set $N=1$ for simplicity.

We allow $m_0$ to take either sign, but for definiteness let's fix $\epsilon_0\geq |m_0|$. That is, in the ground state, the Dirac sea (negative energy band) is completely filled (no anti-fermion), while we have a Fermi sea of fermions in the positive energy band.

The Feynman rules are simple. The bare fermion propagator, renormalized with a filled Dirac sea and no Fermi sea (i.e. renormalized at $\epsilon_0=0$), is given by
\begin{eqnarray}
iG_0(p) \delta_{ab} &\equiv& \left.\langle \psi_a(-p) \psi_b^\dagger(p) \rangle \right|_{g=0} \nonumber \\[.2cm]
&=& \frac{i\delta_{ab} \ u_0(\p) u_0^\dagger(\p)}{p^0-(E_0(\p)-\epsilon_0)(1-i\epsilon)} + \frac{i\delta_{ab} \ w_0(\p) w_0^\dagger(\p)}{p^0-(-E_0(\p)-\epsilon_0)(1-i\epsilon)}
\end{eqnarray}
where $E_0(\p)\equiv \sqrt{\p^2+m_0^2}$ and the $i\epsilon$ prescription corresponds to the said ground state. The bare positive energy band spinor $u_0$ satisfies the Dirac equation $(\sigma^i p^i + \sigma^3 m)u_0(\p)=E_0(\p) u_0(\p)$ and the negative energy band spinor $w$ satisfies the Dirac equation $(\sigma^i p^i + \sigma^3 m)w_0(\p)=-E_0(\p) w_0(\p)$; alternatively,
\begin{eqnarray}
2u_0 u_0^\dagger = \sigma^0 + \frac{m_0}{E_0} \sigma^3 + \frac{p^i}{E_0} \sigma^i, \ \ \ \ \ 2w_0 w_0^\dagger = \sigma^0 - \frac{m_0}{E_0} \sigma^3 - \frac{p^i}{E_0} \sigma^i.
\end{eqnarray}
The explicit solutions to the bare spinors are
\begin{eqnarray}
u_0(\p)=\left[\begin{array}{c} \cos(\theta/2) \\ e^{i\phi}\sin(\theta/2) \end{array}\right], \ \ \ \ \ w_0(\p)=\left[\begin{array}{c} -e^{-i\phi}\sin(\theta/2) \\ \cos(\theta/2) \end{array}\right]
\end{eqnarray}
where $(E_0, \theta, \phi)$ are the spherical coordinates in the $(p^1, p^2, m_0)$ space. Note all these are the same as Weyl fermion but with $m_0$ in place of $p^3$. The other Feynman rules are: the $A_\mu\psi_a\psi_b^\dagger$ vertex is given by $i\sigma^\mu\delta_{ab}$; the ``propagator'' of the auxiliary $\sigma$ field is $ig/N$; the $\sigma\psi_a\psi_b^\dagger$ vertex is given by $-i\sigma^3\delta_{ab}$. 

If we include interaction effects and keep $g$ to first order, the bare propagator $iG_0$ is renormalized to full propagator $iG$, given by the following diagrams:
\begin{eqnarray}
\parbox{20mm}{
\begin{fmffile}{zzz-Dirac_G}
\begin{fmfgraph*}(20, 15)
\fmfleftn{l}{4}\fmfrightn{r}{4}
\fmf{fermion,width=2}{l2,r2}
\fmfdot{l2}
\fmfdot{r2}
\end{fmfgraph*}
\end{fmffile}
} \ \ \ \ \ = \ \ \ \ \ 
\parbox{20mm}{
\begin{fmffile}{zzz-Dirac_G0}
\begin{fmfgraph*}(20, 15)
\fmfleftn{l}{4}\fmfrightn{r}{4}
\fmf{fermion,width=0.2}{l2,r2}
\fmfdot{l2}
\fmfdot{r2}
\end{fmfgraph*}
\end{fmffile}
}
\ \ \ + \ \ \
\parbox{30mm}{
\begin{fmffile}{zzz-Dirac_G_1}
\begin{fmfgraph*}(30, 15)
\fmfleftn{l}{3}\fmfrightn{r}{3}
\fmf{fermion,width=0.2}{l1,m1}
\fmf{fermion,width=0.2}{m1,r1}
\fmfdot{l1}
\fmfdot{r1}
\fmf{phantom}{l2,m2,r2}
\fmf{phantom}{l3,m3,r3}
\fmffreeze
\fmf{dashes}{m1,m2}
\fmf{fermion,width=0.2,left}{m2,m3}
\fmf{fermion,width=0.2,left}{m3,m2}
\end{fmfgraph*}
\end{fmffile}
}
\ \ \ + \ \ \
\parbox{30mm}{
\begin{fmffile}{zzz-Dirac_G_2}
\begin{fmfgraph*}(30, 15)
\fmfleftn{l}{4}\fmfrightn{r}{4}
\fmf{plain,width=0.2}{l2,m1}
\fmf{fermion,width=0.2,tension=0.55}{m1,m2}
\fmf{plain,width=0.2}{m2,r2}
\fmfdot{l2}
\fmfdot{r2}
\fmffreeze
\fmf{dashes,left}{m1,m2}
\end{fmfgraph*}
\end{fmffile}
}
\end{eqnarray}
where the thick fermion line is the full propagator $iG$ while the thin fermion lines are the bare propagator $iG_0$; the dashed lines are the $\sigma$ field propagator. One can show that after including the 1-loop diagrams, $iG$ takes the same form as $iG_0$, but with the mass parameter $m_0$ renormalized to some physical mass $m$, and the chemical potential parameter $\epsilon_0$ renormalized to some physical chemical potential $\epsilon_F$; the renormalized energy $E(\p)$ and renormalized spinors $u(\p), w(\p)$ are defined with $m$ in place of $m_0$. Moreover, $p^0$ is shifted by a constant which has no physical effect and can be removed by a redefinition of $p^0$. (The details of the shifts from $m_0$ to $m$, $\epsilon_0$ to $\epsilon_F$ and the shift of $p^0$ are unimportant for the present discussion, but we include them here for completeness. To order $g$ the shift in $m$ is $m= m_0+(g/2\pi) m_0 \epsilon_0 \left(1-1/2N\right)$. The Fermi momentum $p_F=\sqrt{\epsilon_0^2-m_0^2}$ is a physical quantity and remains unchanged, and the physical chemical potential is given by $\epsilon_F = \sqrt{p_F^2+m^2}=\epsilon_0+(m_0/\epsilon_0)(m-m_0)$. The shift in $p^0$ is given by $p^0 + \epsilon_F-\epsilon_0 + (g/8\pi N)(\epsilon_0^2-m_0^2)$, and can be removed by a redefinition of $p^0$.)

Now we are ready to get to our main point: to see whether the presence of interaction alters the relation \eqref{semi-classical_AHE} between anomalous Hall effect and Berry curvature. Consider the linear response of $\delta J^\mu$ to an external $A_\nu$ field carrying momentum $q$. The relevant Feynman diagrams are
\begin{center}
\begin{fmffile}{zzz-Dirac-current-0}
\begin{fmfgraph*}(30, 25)
\fmfleftn{l}{3}\fmfrightn{r}{3}
\fmf{photon,tension=8}{l2,g}
\fmf{fermion,left,width=2}{G,g}
\fmf{fermion,left,width=2}{g,G}
\fmf{photon,tension=3,label.side=right,label.dist=10,label=$q$}{r2,G}
\fmflabel{$A_\nu$}{r2}
\fmflabel{$\mu$}{l2}
\momentumarrow{a}{up}{6}{r2,G}
\end{fmfgraph*}
\end{fmffile}
\hspace{1.5cm}
\begin{fmffile}{zzz-Dirac-current-2}
\begin{fmfgraph*}(35, 25)
\fmfleftn{l}{3}\fmfrightn{r}{3}
\fmf{photon,tension=8}{l2,g}
\fmf{fermion,left,width=2}{G1,g}
\fmf{fermion,left,width=2}{g,G1}
\fmf{fermion,left,width=2}{G2,G}
\fmf{fermion,left,width=2}{G,G2}
\fmf{dashes,tension=4}{G1,G2}
\fmf{photon,tension=3,label.side=right,label.dist=10,label=$q$}{r2,G}
\fmflabel{$A_\nu$}{r2}
\fmflabel{$\mu$}{l2}
\momentumarrow{a}{up}{6}{r2,G}
\end{fmfgraph*}
\end{fmffile}
\hspace{1.5cm}
\begin{fmffile}{zzz-Dirac-current-1}
\begin{fmfgraph*}(30, 25)
\fmfleftn{l}{3}\fmfrightn{r}{3}
\fmf{photon,tension=8}{l2,g}
\fmf{fermion,left,width=2}{G,g}
\fmf{fermion,left,width=2}{g,G}
\fmf{photon,tension=3,label.side=right,label.dist=10,label=$q$}{r2,G}
\fmflabel{$A_\nu$}{r2}
\fmflabel{$\mu$}{l2}
\momentumarrow{a}{up}{6}{r2,G}
\fmffreeze
\fmf{phantom}{l1,m1}
\fmf{phantom,tension=0.6}{r1,m1}
\fmf{phantom}{l3,m3}
\fmf{phantom,tension=0.6}{r3,m3}
\fmf{dashes,tension=0.5}{m1,m3}
\end{fmfgraph*}
\end{fmffile}
\end{center}
where the fermion lines are the full propagator $iG$. We further require $q^i=0$ and $q^0$ to be so small that can be kept to first order; the electric field is $E_j=iq^0 A_j$ and the magnetic field is zero. For the purpose of computing anomalous Hall current, we are only interested in $i\Pi^{[ij]}$ (at linear response $i\delta J^\mu(q) = i\Pi^{\mu\nu}(q) A_\nu(q)$), the antisymmetric part of these diagrams.

The first diagram is what would contribute when $g=0$, so we expect the first diagram to reproduce the Berry Fermi gas result \eqref{semi-classical_AHE}. Let's demonstrate this. 
With $q^i=0$, performing the loop integral and picking the $p^0$ poles in the loop yields
\begin{eqnarray}
&& -iN \int \frac{d^2 \p}{(2\pi)^2} \left[ \frac{\theta(\epsilon_F-E(\p))-1}{2E(\p)-q^0} \ \left(w^\dagger(\p) \sigma^{[i} u(\p) \right) \left(u^\dagger(\p) \sigma^{j]} w(\p) \right) \right. \nonumber \\[.2cm]
&& \hspace{3cm} + \left. \frac{\theta(\epsilon_F-E(\p))-1}{2E(\p)+q^0} \ \left(u^\dagger(\p) \sigma^{[i} w(\p) \right) \left(w^\dagger(\p) \sigma^{j]} u(\p) \right) \right] \nonumber \\[.2cm]
&=& iN \int \frac{d^2 \p}{(2\pi)^2} \frac{\left(1-\theta(\epsilon_F - E(\p)\right) q^0}{4E(\p)^2-(q^0)^2} \ \left(w^\dagger(\p) \sigma^{[i} u(\p) \right) \left(u^\dagger(\p) \sigma^{j]} w(\p) \right).
\end{eqnarray}
At leading order we can ignore the $(q^0)^2$ in the denominator. Moreover, note that $\sigma^i=\partial_{p^i} \left(\sigma^k p^k + \sigma^3 m\right)$, while $u, w$ are eigenstates of $\sigma^k p^k + \sigma^3 m$ with eigenvalues $\pm E(\p)$ respectively, therefore we have
\begin{eqnarray}
w^\dagger(\p) \sigma^i u(\p) = 2E(\p) \ w^\dagger(\p) \left(\partial_p^i u(\p)\right) = -2E(\p) \ \left(\partial_p^i w^\dagger(\p)\right) u(\p),
\end{eqnarray}
as well as complex conjugate version of it. Hence, using $\mathbf{1}=uu^\dagger+ww^\dagger$, we finally find
\begin{eqnarray}
-q^0 \frac{N}{2\pi} \int\frac{d^2 \p}{4\pi} \left[(-2i) \partial_p^{[i} w^\dagger \partial_p^{j]} w + \theta(\epsilon_F-E) (-2i) \partial_p^{[i} u^\dagger \partial_p^{j]} u\right].
\end{eqnarray}
This is indeed the Berry Fermi gas result \eqref{semi-classical_AHE}. The first term in the square bracket is the Berry curvature contribution from the filled Dirac sea, yielding $-\epsilon^{ij}\sgn(m)/2$, i.e. in the gapped phase $|\epsilon_F|\leq |m|$ the Chern number would be $\sgn(m) N /2$, which explains why we said $N$ must be even. The second term in the square bracket is the Berry curvature contribution from the Fermi sea, yielding $\epsilon^{ij}(1-|m|/\epsilon_F)\sgn(m)/2$.

To first order in $g$, we must also take into account the second and third Feynman diagrams. Unless those two diagrams vanish at first order in $q^0$, the extra contribution from those two diagrams would violate the Berry Fermi gas result \eqref{semi-classical_AHE}. It is easy to see the second Feynman diagram vanishes when $q^i=0$; however, the third has a non-trivial contribution, so \eqref{semi-classical_AHE} is indeed violated. The 2-loop integral in the third diagram is not as hard as it might look. Because the $\sigma$ propagator is momentum independent, we can carry our the momentum integral in each loop separately (and the two loops are clearly identical up to $q\rightarrow -q$), so we really only need to perform a 1-loop integral. We find
\begin{eqnarray}
g\: m\frac{q^0 \epsilon^{ij}}{16\pi^2} \left(1 - \frac{m^2}{\epsilon_F^2}\right).
\end{eqnarray}
Combining with the first diagram, we arrive at
\begin{eqnarray}
i\Pi^{[ij]}(q) = \frac{N}{2} \frac{q^0 \epsilon^{ij}}{2\pi} \left(\frac{m}{\epsilon_F}+\frac{g\: m}{4\pi N} \left(1 - \frac{m^2}{\epsilon_F^2}\right)\right), \ \ \ \ \ \ \delta J_H^i(q)=\Pi^{[ij]}(q) A_j(q).
\end{eqnarray}
The first term is the expected Berry curvature contribution from both bands. However, interaction introduces the second term, and thus violates the simple relation \eqref{semi-classical_AHE} between anomalous Hall conductivity and Berry curvature in Berry Fermi gas.

Through this simple example we see the violation of \eqref{semi-classical_AHE}. What is the physical nature of the new term? Inspecting the third Feynman diagram, it seems we shall view the dash line as a quantum correction to the bare $A_\nu \psi\psi^\dagger$ vertex, and this suggests the new term is related to the effective electric dipole moment arising from interactions. This is indeed the correct physical interpretation, as we will see along the development of the general formalism of Berry Fermi liquid.

What cannot be learned from this simple example -- in which the particles are stable since decay occurs at order $g^2$ which we neglect -- is whether Berry curvature effect is still worth talking about when the interaction is so strong that particles in the Fermi sea are unstable. The answer is positive -- more particularly, only the Berry curvature near the Fermi surface is involved. We will also show this along the development of our general formalism.

\section{Review of Landau Fermi Liquid Theory}

Before we present the kinetic theory of Berry Fermi liquid, let's start with a quick review of Landau Fermi liquid theory -- in particular, the computation of linear response in the theory. We assume that there is no external field violating spacetime translational symmetry except for the present external EM field. We assume the EM $U(1)$ charge conservation is not broken by the ground state. We do not assume the presence of any other symmetry.

Let us start with a system of fermions, interacting through a
finite-ranged interaction, in $d$ spatial dimensions with $d\geq 2$ (or
$(d+1)$ spacetime dimensions). We assume the ground state at chemical
potential $\epsilon_F$ is a Fermi liquid, with a sharp Fermi surface
(FS). (We assume that the Kohn-Luttinger instability
\cite{Kohn:1965zz} occurs at a temperature much smaller than any energy
scales of interest). The low energy excitations are fermionic
quasiparticles (or quasiholes) near the FS. For simplicity we assume
one, non-degenerate, FS, i.e., each momentum $\p$ near the FS corresponds
to only one quasiparticle.

We now perturb this system by a small external EM field $A_\mu$.
Physically, this causes a deformation of the FS, which can also be
viewed as creating quasiparticles and quasiholes, which in Landau's
theory are described by the quasiparticle distribution function
$\delta\!f(\p; x)$ with $\p$ near the FS. In the linear response
theory we keep $\delta\! f$ to linear order of $A_\mu$.

The Landau Fermi liquid theory matches with quantum field theory at
long wavelength. If we Fourier transform $-i\hbar\partial_{x^\mu}$ to
$q_\mu$, then in the long-wavelength limit under consideration,
$A_\mu$ and $\delta f$ only have $q$ modes with $q \ll p_F$ and $q \ll
\hbar/r_{int}$, where $p_F$ is the size scale of the FS (there is no notion
of ``Fermi momentum'' since we do not assume rotational symmetry), 
and $r_{int}$ is the range of interaction between quasiparticles (this is 
why we assumed finite-ranged interactions). In practice, we keep
$\hbar\,\partial_x$, or equivalent $q$, to leading order in Landau
Fermi liquid theory.

It can be shown that the collision (decay included) rate of quasiparticles is suppressed 
beyond leading order in $q$, due to the limited availability of decay channels. 
In particular, the suppression is by an extra order of $q$ for $d\geq 3$
\cite{landau1957oscillations,Luttinger:1961zz,abrikosov1975methods}, 
and by an extra $q \ln q$ for $d=2$ 
\cite{baym1978physics,chubukov2003nonanalytic,chubukov2005singular}. Thus, quasiparticle collision 
can be neglected in Landau Fermi liquid theory.

The computation of linear response in Landau Fermi liquid theory
proceeds in two steps. One first computes $\delta\!f$ as a linear
function of $A$ by solving the Boltzmann equation, and then expresses
(the quantum expectation of) the induced current $\delta J^\mu$ as a
linear function of $\delta\!f$, and hence of $A$. In Landau's Fermi liquid
theory, the energy of a single quasiparticle has the form
\begin{eqnarray}
\epsilon(\p; x) = E(\p) + \int_\kk \mathcal{U}(\p, \kk) \ \delta\! f(\kk; x)
\label{LFL_quasiparticle_energy}
\end{eqnarray}
where $\int_\kk \equiv \int d^d k / (2\pi\hbar)^d$. Here $E(\p)$ is the kinetic energy of the
quasiparticle, and $\mathcal{U}(\p, \kk)$, even under exchange of $\p$
and $\kk$, parameterizes the contact interaction between two
quasiparticles of momenta $\p$ and $\kk$. (If the system has rotational
symmetry, the Landau Fermi liquid parameters are obtained by putting
$\p$ and $\kk$ on the Fermi surface and expanding $\mathcal U$ in
angular harmonics in the angle between $\p$ and $\kk$.) Both $E$ and
$\mathcal{U}$ are microscopic inputs into Landau's theory.
Landau's Fermi liquid theory postulates the collisionless Boltzmann equation \eqref{Boltzmann_Eq}:
\begin{equation}
  \frac{\partial f(\p; x)}{\partial t} 
  + \frac{\partial\epsilon(\p;x)}{\partial p_i} \frac{\partial f(\p;x)}{\partial x^i}
  + \left( F_{i0}(x) + F_{ij}(x) \frac{\partial\epsilon(\p;x)}{\partial p_j} - \frac{\partial\epsilon(\p;x)}{\partial x^i}\right)
     \frac{\partial f(\p;x)}{\partial p_i} = 0.
\end{equation}
where $F_{i0}=\partial_{x^i} A_0-\partial_{t} A_i$ is the electric field, $F_{ij}=\partial_{x^i} A_j-\partial_{x^j} A_i$ is the magnetic field, and we have absorbed the electric charge into the field potential $A$. Writing $f(\p;x)=\theta(\epsilon_F-E(\p)) + \delta\!f(\p;x)$ and linearizing over $\delta\!f$ and $A$, one finds
\begin{eqnarray}
v^\mu(\p) \: \partial_{x^\mu} \delta\! f(\p; x) = \delta(\epsilon_F-E(\p)) \: v^i(\p) \: \left(F_{i0}(x) - \partial_{x^i} \epsilon(\p; x)\right)
\label{LFL_Boltzmann_Eq_coordinates}
\end{eqnarray}
where $v^0 \equiv 1$, $v^i(\p) \equiv \partial_p^i E(\p)$, and $x^0\equiv t$.
Notice that, due to the delta function on the right hand side, Eq.~\eqref{LFL_Boltzmann_Eq_coordinates} involves only the FS, but not, say, the whole Fermi sea. Performing the Fourier transformation $-i\partial_{x^\mu} \rightarrow q_\mu$, where, in our convention, $-q_0=q^0$ is the energy, while $q_i=q^i$ is the momentum, the Boltzmann equation then reads
\begin{eqnarray}
\delta\!f(\p; q) = \delta(\epsilon_F-E) \frac{v^i}{v^\mu q_\mu - i\epsilon} \left( -iF_{i0}(q) - q_i \int_\kk \mathcal{U}(\p, \kk) \: \delta\!f(\kk; q) \right)
\label{LFL_Boltzmann_Eq}
\end{eqnarray}
where $F_{\mu\nu}(q) = 2iq_{[\mu} A_{\nu]}$. This is an integral equation
from which one can find $\delta\!f$ in terms of $A$. It follows from
Eq.~(\ref{LFL_Boltzmann_Eq}) that the the coefficient of linear dependence
between $\delta\!f$ and $A$ is finite in the limit $q\to 0$ and $q^0/|{\bf q}|$ fixed.
In this thesis we count this as \emph{zeroth} order (leading order) in $q$. Note that we placed an $i\epsilon$ prescription in the denominator; its sign is such that $q^0$ appears as $q^0+i\epsilon$. This corresponds to the retarded boundary condition that at infinite past the system is in its ground state.

Now suppose we have solved for $\delta f$ as a linear function of $A$ from \eqref{LFL_Boltzmann_Eq}. Then the induced current in Landau Fermi liquid theory is given by
\begin{eqnarray}
\delta J^\mu(x) = \int_\p \left( v^\mu(\p) \: \delta\!f(\p; x) + \delta(\epsilon_F - E(\p)) \delta^\mu_i v^i(\p) \: (\epsilon(\p; x)-E(\p)) \right).
\label{LFL_current}
\end{eqnarray}
The first term is simply the current created by the quasiparticles that were
excited. The second term, by recognizing $\delta(\epsilon_F-E) \: v^i = -\partial_p^i \theta(\epsilon_F-E)$ and integrating by parts over $\p$, is the current due to quasiparticles in the Fermi sea
having their velocity perturbed by interactions with the excited quasiparticles
$\partial_p^i (\epsilon-E)$.
(Although ``quasiparticles in the Fermi sea'' are generally not well-defined far from the FS, from the expression \eqref{LFL_current} we clearly see only those quasiparticles near the FS are involved.) This is the procedure of computing linear response in Landau Fermi liquid theory.

\section{Berry Fermi Liquid Theory}
\label{sect_kinetic}

As introduced at the beginning of this chapter, we are to develop a kinetic theory of Fermi liquid incorporating Berry curvature. The most outstanding questions are how to define Berry curvature effects when high energy quasiparticles are unstable, which properties of Berry Fermi gas survive in the presence of interactions, and whether any new effects arise from interactions. The assumptions about the Fermi liquid is mostly the same as in Landau's theory. The differences are that here the fermionic field must be multi-component, and that we work to one order higher in low energy / long wavelength expansion (consequently, collisions cannot be neglected even for the meta-stable quasiparticles near the FS).

The kinetic formalism of Berry Fermi liquid theory, similar to Landau Fermi liquid theory,
consists of two
parts: the Boltzmann equation, and the expression of the current in
terms of the distribution function. We will also find that the
consistency of the theory requires certain relationships between the
chemical potential dependence of the Fermi velocity and the Landau interaction 
potential, and between the chemical potential dependence of the
Hall conductivity tensor (to be defined later) and the Berry curvature of
the fermionic quasiparticle. We will present this formalism in this
Section. In the next Section, we will show this kinetic formalism
exactly matches with quantum field theory (QFT) computation to all
orders in diagrammatic expansion, for a large class of QFTs.

We make a final remark. For simplicity, we again assume only one band crosses the Fermi level. However, the formalism below can be easily generalized to the cases of either i) multiple degenerate bands crossing the Fermi level, or ii) multiple bands crossing the Fermi level with disjoint FS; the generalizations are obvious in the QFT derivation. Therefore, with straightforward generalization, our formalism encompasses the example in Section \ref{sect_Invitation} in which $N$ degenerate bands cross the Fermi level.

\subsection{Boltzmann Equation}
\label{ssect_kinetic_BE}

In a Berry Fermi liquid, as in the usual Fermi liquid theory, the energy of a quasiparticle with momentum $\p$ near the FS depends on the occupation at other momenta. To first order in $A$ and first order in $\partial_x$, the energy is
\begin{eqnarray}
\epsilon(\p; x) = E(\p) - \mu^{\mu\nu}(\p) \frac{F_{\mu\nu}(x)}{2} + \int_\kk \left(\mathcal{U}(\p, \kk) \ \delta\!f(\kk; x) + \mathcal{V}^\nu(\p, \kk) \ \partial_{x^\nu} \delta\!f(\kk; x)\right).
\label{BFL_quasiparticle_energy}
\end{eqnarray}
Compared to \eqref{LFL_quasiparticle_energy}, here $\mu^{\mu\nu}$, antisymmetric in $\mu\nu$, is the EM dipole moment of the quasiparticles (the purely spatial components $\mu^{ij}$ correspond to the magnetic dipole moment and the mixed components $\mu^{i0}$ to the electric dipole moment), and $\mathcal{V}^\nu(\p, \kk)$, odd under exchange of $\p$ and $\kk$, is the gradient interaction potential between quasiparticles. The function $\mathcal V^\mu(\p,\kk)$ is the additional function parametrizing the dependence of the energy of the quasiparticle with momentum $\p$ on the gradient of the distribution function at $\kk$. Since we are performing a gradient expansion of the interaction between two quasiparticles, our assumption of interaction being finite-ranged is needed.

Extended to sub-leading order in spacetime derivative, the linearized (in $\delta f$ and $A$) Boltzmann equation now includes collision term. Although we need to include collision for completeness, we emphasize it is ``uninteresting'' towards the focus of this paper as it does not contribute to interesting physics such as the anomalous Hall effect, as we will show later in Section \ref{ssect_QFT_AppendixAB}.

The collision term is different from that in classical Boltzmann equation, and must be obtained quantum mechanically. The collisionful Boltzmann equation we find is to modify \eqref{LFL_Boltzmann_Eq_coordinates} by the replacement $v^i(\p) \partial_{x^i} \rightarrow v^i(\p) \partial_{x^i} - \delta(\epsilon_F-E(\p))\int_\kk \mathcal{C}(\p, \kk) \: \partial_{t}^2$ on both sides, yielding
\begin{eqnarray}
&& v^\mu(\p) \: \partial_{x^\mu} \delta\!f(\p; x) \ - \ \delta(\epsilon_F-E(\p))\int_\kk \mathcal{C}(\p, \kk) \ \partial_t^2 \delta\!f(\kk; x) \nonumber \\[.2cm]
&=& \delta(\epsilon_F-E(\p)) \: \left( v^i(\p) \: F_{i0}(x) - v^i(\p)\partial_{x^i} \epsilon(\p; x) \phantom{\int} \right. \nonumber \\[.1cm]
&& \hspace{3.5cm} \left. + \int_\kk \mathcal{C}(\p, \kk)\:\delta(\epsilon_F-E(\kk)) \ \partial_t^2 \epsilon(\kk; x)\right).
\label{BFL_Boltzmann_Eq}
\end{eqnarray}
Here $\mathcal{C}(\p, \kk)$, symmetric under exchange of $\p$ and $\kk$, is the effective collision kernel defined on the FS. It has the following properties (which we will show when we perform the QFT derivation in the next Section):
\begin{itemize}
\item
Collisions do not change the total number of fermionic excitations, i.e.
\begin{eqnarray}
\int_\p \delta(\epsilon_F-E(\p)) \ \mathcal{C}(\p, \kk) =0.
\label{Collision_conservation}
\end{eqnarray}
\item
$\mathcal{C}(\p, \kk)$ is not regular over the FS. It can be separated into a positive ``quasiparticle decay'' piece that is non-vanishing only when $\p=\kk$ on the FS, plus a piece that is non-vanishing for general values of $\p$ and $\kk$.
\end{itemize}
The $\partial_t^2$ in the collision term has been long known. Recall that in the ``thermal regime'' where temperature $T\gg \partial_t$, linearizing the classical Boltzmann collision term yields the scaling of $T^2$. But here we are in the ``quantum regime'' where temperature is negligible, $T\ll \partial_t^2$; according to Landau's semi-classical argument~\cite{landau1957oscillations}, in this regime the scaling should be replaced by $\partial_t^2$. Luttinger also has a field theory power counting argument \cite{Luttinger:1961zz}; we will adopt this method in Section \ref{ssect_QFT_AppendixAB}.

We have to emphasize that such parametrization of the collision term is only valid for $d\geq 3$. In $d=2$ the collision term cannot be parametrized in any simple form~\cite{chubukov2003nonanalytic,chubukov2005singular}, as we will discuss in Section \ref{ssect_QFT_AppendixAB}. Fortunately, our main focus -- the computation of the Hall current -- is not undermined by this failure of parametrizing collisions in $d=2$. In particular, we will show in Section \ref{ssect_QFT_AppendixAB} that at order $q$ and $q\ln q$, collisions only contributes to the longitudinal current but not the Hall current.

The Boltzmann equation can be solved in principle, order by order in $q$. First in \eqref{BFL_Boltzmann_Eq} we Fourier transform $-i\partial_{x^\mu}$ into $q_\mu$. Let us separate $\delta\!f=\delta\!f_0 + \delta\!f_1$, where the subscript labels the order in $q$. Then the Boltzmann equation \eqref{BFL_Boltzmann_Eq} reads (the collision term only holds for $d\geq 3$)
\begin{eqnarray}
\delta\!f_0(\p; q) = \delta(\epsilon_F-E) \frac{v^i}{v^\mu q_\mu - i\epsilon} \left( -iF_{i0}(q) - q_i \int_\kk \mathcal{U}(\p, \kk) \: \delta\!f_0(\kk; q) \right),
\label{BFL_Boltzmann_Eq_0}
\end{eqnarray}
\begin{eqnarray}
&& \delta\!f_1(\p; q) \nonumber \\[.2cm]
&=& \delta(\epsilon_F-E) \frac{v^i \: q_i}{v^\mu q_\mu - i\epsilon} \left( \mu^{\mu\nu} \frac{F_{\mu\nu}(q)}{2} - \int_\kk \left(\mathcal{U}(\p, \kk) \: \delta\!f_1(\kk; q) + \mathcal{V}^\nu(\p, \kk) \ iq_\nu\, \delta\!f_0(\kk; q) \right) \right) \nonumber \\[.2cm]
&& + \ \delta(\epsilon_F-E) \frac{i(q_0)^2}{v^\mu q_\mu - i\epsilon} \int_\kk \mathcal{C}(\p, \kk) \left(\delta\!f_0(\kk; q) + \delta(\epsilon_F-E(\kk)) \int_{\l}\mathcal{U}(\kk, \l) \: \delta\!f_0(\l; q) \right)
\label{BFL_Boltzmann_Eq_1}
\end{eqnarray}
at zeroth and first order in $q$ respectively. We will prove these two equations from QFT in Section \ref{ssect_QFT_BE}. Note that the zeroth order Boltzmann equation \eqref{BFL_Boltzmann_Eq_0} is that in Landau Fermi liquid theory.

One may have noticed that there is no reference to Berry curvature in the Boltzmann equation. Notably, the Berry curvature $b^{ij}$ should induce an anomalous velocity $b^{ij} F_{j\nu} v^\nu$~\cite{Sundaram:1999zz}. However, since \eqref{BFL_Boltzmann_Eq} is of order $A$, the effect of anomalous velocity will be order $A^2$, which we assumed to be negligible. (If one works beyond linear response, and assume stable quasiparticle, the anomalous velocity term would be present~\cite{shindou2008gradient}.) Other effects of Berry curvature are negligible in the Boltzmann equation for the same reason.

\subsection{Current}
\label{ssect_kinetic_current}

At equilibrium there is some equilibrium current $J_{eq.}^\mu$. In most systems at equilibrium only the charge density $J_{eq.}^0$ is non-zero, while $J_{eq.}^i=0$. As we perturb the system, extra current $\delta J^\mu$ of order $A$ will be induced. In our formalism, we propose
\begin{eqnarray}
\delta J^\mu(x) &=& \int_\p \left( v^\mu(\p) \: \delta\!f(\p; x) + \mu^{\mu\nu}(\p) \: \partial_{x_\nu} \delta\!f(\p; x) + \delta(\epsilon_F - E(\p)) \delta^\mu_i v^i(\p) \: (\epsilon(\p; x)-E(\p)) \right)\nonumber \\[.2cm]
&& + \: \sigma^{\mu\nu\lambda} \frac{F_{\nu\lambda}(x)}{2}\,.
\label{BFL_current}
\end{eqnarray}
Inside the integral on the right-hand side of \eqref{BFL_current} are three terms. The first term is the current due to the velocity of the deformation of the FS. The second term is the magnetization / electric polarization current~\cite{Chen:2014cla} due to the deformation of the FS. The third term, as in Landau Fermi liquid theory, is the current due to quasiparticles in the Fermi sea getting extra velocity $\partial_p^i (\epsilon-E)$ (rewritten by integrating $\p$ by parts). All these three terms involve only $\p$ near the FS, as desired.

In the last term of \eqref{BFL_current}, $\sigma^{\mu\nu\lambda}$, totally antisymmetric in $\mu\nu\lambda$, is the Hall conductivity tensor. As we will discuss in Section \ref{ssect_kinetic_sigma_EF}, it has very interesting relation to the FS, and that is how Berry curvature enters the formalism.

Now in \eqref{BFL_current} we Fourier transform $-i\partial_{x^\mu}$ into $q_\mu$. At zeroth and first order in $q$ respectively, the current reads
\begin{eqnarray}
\delta J_0^\mu (q) &=& \int_\p \left( v^\mu \: \delta\!f_0(\p; q) + \delta(\epsilon_F - E) \delta^\mu_i v^i \! \int_\kk \mathcal{U}(\p, \kk) \: \delta\!f_0(\kk; q) \right),
\label{BFL_current_0}
\end{eqnarray}
\begin{eqnarray}
\delta J_1^\mu (q) &=& \int_\p \left[ v^\mu \: \delta\!f_1(\p; q) + \mu^{\mu\nu} \: iq_\nu\, \delta\!f_0(\p; q) \phantom{\frac{F_\mu}{F_\nu}} \right. \nonumber \\[.2cm]
&& \hspace{.7cm} \left. + \: \delta(\epsilon_F - E) \delta^\mu_i v^i \left( -\mu^{\nu\lambda} \frac{F_{\nu\lambda}(q)}{2} \phantom{\int_\kk} \right. \right. \nonumber \\[.2cm]
&& \hspace{4.5cm} \left. \left. + \ \int_\kk \left(\mathcal{U}(\p, \kk) \: \delta\!f_1(\kk; q) + \mathcal{V}^\nu(\p, \kk) \ iq_\nu\, \delta\!f_0(\kk; q) \right) \right) \right] \nonumber \\[.2cm]
&& + \: \sigma^{\mu\nu\lambda} \frac{F_{\nu\lambda}(q)}{2}.
\label{BFL_current_1}
\end{eqnarray}
Notice $\delta J_0^\mu$ is that in Landau Fermi liquid theory. We will prove these two equations from QFT in Section \ref{ssect_QFT_current}. In the proof, we will also discuss the microscopic contributions to $\mu^{\mu\nu}$. The magnetic dipole moment is generally non-zero; in the presence of interactions~\cite{shindou2008gradient}, the electric dipole moment will also be non-zero in general, as we will see in the proof.

Although we call $\sigma^{\mu\nu\lambda}$ the Hall conductivity
tensor, it is not the full Hall conductivity as measured in linear
response. The full Hall conductivity also receives contributions from
the $\p$ integral, and depends on the ratio $|\q|/q^0$.
For example, in order to find the Hall conductivity for
spatially homogeneous electric field, we set $q_j=0$ and choose the
gauge $A_0=0$, so $F_{j0}=-iq_0 A_j$. From the Boltzmann equation we
have $\delta f_0 = -\delta(\epsilon_F-E) v^j A_j$,
which leads to the anomalous Hall current
\begin{align}
\delta J_H^i = \left(\sigma^{ij0} -\int_\p \delta(\epsilon_F-E) \ 2v^{[i} \mu^{j]0} \right) F_{j0} \hspace{1.5cm} (q^j=0, \ q^0 \mbox{ small})
\label{BFL_AHE_uniform}
\end{align}
(although $\delta f_1$ is 
non-zero due to collisions, we already mentioned that collisions do not contribute 
to the Hall current, as shown in Section \ref{ssect_QFT_AppendixAB}). Thus, the full
Hall conductivity, in the limit of taking $q_j= 0$ first and then taking $q_0$ small, receives 
contribution from both the $\sigma$ tensor and the electric dipole moment $\mu^{j0}$ of the quasiparticles -- the latter is generally 
non-zero in the presence of interaction~\cite{shindou2008gradient}, as we will show in the 
QFT derivation. Similarly, if we take the other order of limits, $q_0=0$ first and $q_j$ small 
so that $F_{j0}=i q_j A_0$, we will find
\begin{align}
\delta J_H^i = \left(\sigma^{ij0} + \int_\p \delta(\epsilon_F-E) \ \underline{\mu}^{ij}\right) F_{j0} \hspace{1.5cm} (q^0=0, \ q^j \mbox{ small})
\label{BFL_AHE_static}
\end{align}
where $\underline{\mu}^{ij}$ is related to the magnetic dipole moment $\mu^{ij}$ via the recursion relation
\begin{align}
\underline{\mu}^{ij}(\p) = \mu^{ij}(\p) - \int_\kk \delta(\epsilon_F-E(\kk)) \ \mathcal{U}(\p, \kk) \ \underline{\mu}^{ij}(\kk).
\end{align}
So again the $\sigma$ tensor does not give the measured Hall conductivity.

\subsection{Chemical Potential dependence of Kinetic Energy}
\label{ssect_kinetic_E_EF}

We have separated the quasiparticle distribution into an equilibrium
part $\theta(\epsilon_F-E)$ and an excitation part $\delta\!f$.
Within Fermi liquid theory, such a separation is ambiguous:
the same state may equally well be described either by starting with a
slightly lower chemical potential and exciting some quasiparticles
above the FS, or by starting with a slightly higher chemical potential
and exciting some quasiholes below the FS. Clearly, for the theory to
be self-consistent, all these different descriptions of the same state
must be equivalent. For this, the following relationship between $E$
at different chemical potentials must hold:
\begin{eqnarray}
\frac{\partial E(\p)}{\partial \epsilon_F} = \int_\kk \mathcal{U}(\p, \kk) \frac{\partial}{\partial \epsilon_F} \theta(\epsilon_F-E(\kk)) = \int_\kk \mathcal{U}(\p, \kk) \left(1- \frac{\partial E(\kk)}{\partial \epsilon_F} \right) \delta(\epsilon_F-E(\kk)).
\label{BFL_dE_dEF}
\end{eqnarray}
This can be physically understood from
\eqref{BFL_quasiparticle_energy}, setting $F_{\mu\nu}=0$ and
$\partial_x \delta f=0$. Furthermore, we will prove it from QFT in
Section \eqref{sssect_QFT_int_Prop_on_EF}. Taking $\partial_p^i$ of
\eqref{BFL_dE_dEF}, we obtain the chemical potential dependence of
$v^i(\p)$ on the FS.

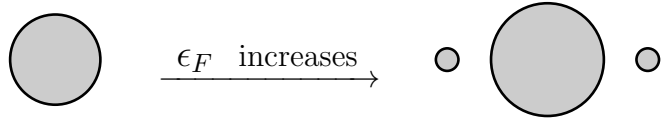
\begin{figure}
\centering
\begin{eqnarray*}
\parbox{15mm}{
\begin{fmffile}{zzz-FS_low_E_F}
\begin{fmfgraph*}(15, 15)
\fmfleftn{l}{3}\fmfrightn{r}{3}
\fmf{phantom}{l2,o,r2}
\fmfv{d.sh=circle,d.f=20,d.si=0.8h}{o}
\end{fmfgraph*}
\end{fmffile}
}
\hspace{.5cm} \xrightarrow{\ \mbox{\large $\epsilon_F$ \ } \mbox{increases} \ } \hspace{.1cm} 
\parbox{35mm}{
\begin{fmffile}{zzz-FS_high_E_F}
\begin{fmfgraph*}(40, 15)
\fmfleftn{l}{3}\fmfrightn{r}{3}
\fmf{phantom}{l2,ol,oll,o,orr,or,r2}
\fmfv{d.sh=circle,d.f=20,d.si=h}{o}
\fmfv{d.sh=circle,d.f=20,d.si=0.2h}{ol}
\fmfv{d.sh=circle,d.f=20,d.si=0.2h}{or}
\end{fmfgraph*}
\end{fmffile}
}
\end{eqnarray*}
\caption{Around some discrete values of chemical potential, the Fermi surface may develop new disconnected components, which may lead to a quantum phase transition. The behavior of the interacting system around such values of chemical potential remains unknown.}
\label{EF_increase_disconnect_comp}
\end{figure}

Strictly speaking, the reasoning above only applies when the FS
changes continuously with the chemical potential. If the system
undergoes a quantum phase transition at some $\epsilon_F$, around which the FS develops new
disconnected components, as illustrated in the Figure \ref{EF_increase_disconnect_comp}, 
then the formula \eqref{BFL_dE_dEF} not necessarily holds. 

\subsection{Chemical Potential dependence of Hall Conductivity Tensor}
\label{ssect_kinetic_sigma_EF}

The Hall conductivity tensor in \eqref{BFL_current} seem to have no
reference to the FS. But in fact the Hall conductivity tensor is
related to the FS via the Berry curvature in a very interesting
manner. We will distinguish two cases. In the first case, either $d=2$, or $d>2$ and
the Berry curvature is an exact 2-form on the FS, so that the system has no
anomaly-related transport effects. (The anomaly-related transport
effects include, e.g., the chiral magnetic effect in $(3+1)d$, but not
the anomalous Hall effect in $(2+1)d$.) Then we turn to the case in $d>2$ with
non-exact Berry curvature on the FS, so that the system exhibits
anomaly-related transport effects~\cite{Son:2012wh,Stephanov:2012ki}.

\subsubsection{Without Anomaly-Related Transport}
\label{sssect_kinetic_sigma_EF_no_anomaly}

Let's review the story in Berry Fermi gas. In Fermi gas, particles are stable, so one can define the Berry connection $a^j$ and Berry curvature $b^{ij}$ for all particles in the Fermi sea:
\begin{eqnarray}
a^j(\p) \equiv (-i\hbar) \: \mathfrak{u}^\dagger_\alpha(\p) \: \partial_p^j \mathfrak{u}^\alpha(\p),
\end{eqnarray}
\begin{eqnarray}
b^{ij}(\p) \equiv 2\partial_p^{[i} a^{j]}(\p) = (-2i\hbar) \: \partial_p^{[i} \mathfrak{u}^\dagger_\alpha(\p) \: \partial_p^{j]} \mathfrak{u}^\alpha(\p) 
\end{eqnarray}
where $\mathfrak{u}^\alpha(\p)$ is the spinor or Bloch state of the fermion. The Berry curvature induces an anomalous velocity~\cite{Sundaram:1999zz} and a change of the classical phase space measure~\cite{Xiao:2005qw, Duval:2005vn}, leading to Hall conductivity tensor of the form
\begin{eqnarray}
\sigma^{\mu\nu\lambda} = \sigma_{o}^{\mu\nu\lambda} + 3 \int_\p \theta(\epsilon_F-E(\p)) \: v^{[\mu}(\p) \: b^{\nu\lambda]}(\p)
\label{BFL_sigma_const_Fermi_sea}
\end{eqnarray}
where $a^0=0, \ b^{0\mu}=0$. Here $\sigma_{o}^{\mu\nu\lambda}$ is the contribution from valence bands / Dirac sea, and is independent of $\epsilon_F$. The second term seems like a Fermi sea property, but as observed by Haldane~\cite{Haldane:2004zz}, one can integrate $\p$ by parts and get
\begin{eqnarray}
\sigma^{\mu\nu\lambda} = \sigma_{o}^{\mu\nu\lambda} + 6 \int_\p \delta(\epsilon_F-E(\p)) \: \delta^{[\mu}_0 v^\nu(\p) \: a^{\lambda]}(\p),
\label{BFL_sigma_const_FS_1}
\end{eqnarray}
so that the kinetic part of the Hall conductivity tensor is actually a FS property; notice the kinetic part has no ${ijk}$ components, but only ${ij0}$ ones. For $d>2$ (recall $d$ is the number of spatial dimensions), there is another way to integrate \eqref{BFL_sigma_const_Fermi_sea} by parts, also promoted by Haldane~\cite{Haldane:2004zz}. Using $v^0\equiv 1 = \partial_p^k p^k/d$, we have
\begin{eqnarray}
\sigma^{ij0} &=& \sigma_{o}^{ij0} + \frac{3}{d-2} \int_\p \theta(\epsilon_F-E(\p)) \ b^{[ij}(\p) \ \partial_p^{k]} p_k \nonumber \\[.2cm]
&=& \sigma_{o}^{ij0} + \frac{3}{d-2}\int_\p \delta(\epsilon_F-E(\p)) \: b^{[ij}(\p) v^{k]}(\p) \: p_k \nonumber \\[.2cm]
&& \hspace{.85cm} + \: \frac{6}{d-2}\int_\p \partial_p^{[k} \left( \delta(\epsilon_F-E(\p)) \: v^i(\p) \: a^{j]}(\p) \: p_k\right)
\label{BFL_sigma_const_FS_2}
\end{eqnarray}
and $\sigma^{ijk}=\sigma_o^{ijk}$. The last line is a boundary term that is non-vanishing if the fermion is in a lattice and the FS intersects the boundary of our choice of first Brillouin zone~\cite{Haldane:2004zz} (because $p_k$ is not continuous when we identify the opposite boundaries of the first Brillouin zone). The advantage of \eqref{BFL_sigma_const_FS_2} over \eqref{BFL_sigma_const_FS_1} is that it involves $b^{ij}$ instead of the gauge dependent $a^i$ (except for the boundary term); as we will see later, this makes \eqref{BFL_sigma_const_FS_2} is more convenient for generalization to include anomaly-related transport effects.

Now we turn to the $\epsilon_F$ dependence of $\sigma^{\mu\nu\lambda}$ in Berry Fermi liquid. In the presence of interaction, the picture of quasiparticles is only valid near the FS, so whether the $\epsilon_F$ dependence of $\sigma^{\mu\nu\lambda}$ can be expressed as a FS property becomes important at conceptual level: It determines, in order to study linear response to EM field at long wavelength, whether knowing the system is a Fermi liquid at low energy is enough, or we have to know more beyond the low energy behaviors. Our conclusion is, the former is true -- the fact that the system is a Fermi liquid is enough. More exactly, we will show in Section \ref{sssect_QFT_dsigma_dEF} that if the FS changes continuously with the chemical potential, then
\begin{eqnarray}
\frac{d\sigma^{\mu\nu\lambda}}{d \epsilon_F} = 3\int_\p \delta(\epsilon_F-E(\p)) \left( v^{[\mu}(\p) \: b^{\nu\lambda]}(\p) - \frac{\partial E(\p)}{\partial \epsilon_F} \delta^{[\mu}_0 \: b^{\nu\lambda]}(\p)- 2\delta^{[\mu}_0 v^\nu(\p) \: b^{\lambda]F}(\p) \right).
\label{BFL_dsigma_dEF_detail}
\end{eqnarray}
Here $b^{\lambda F}$ is the mixed Berry curvature of momentum and chemical potential:
\begin{eqnarray}
b^{kF}(\p) \equiv (-i\hbar) \left( \partial_p^k \mathfrak{u}^\dagger_\alpha(\p) \: \frac{\partial \mathfrak{u}^\alpha(\p)}{\partial \epsilon_F} - \frac{\partial \mathfrak{u}^\dagger_\alpha(\p)}{\partial \epsilon_F} \: \partial_p^k \mathfrak{u}^\alpha(\p) \right)
\end{eqnarray}
and $b^{0F}=0$; it satisfies the Bianchi identity $\partial b^{\nu\lambda}/\partial \epsilon_F = 2 \partial_p^{[\lambda} b^{\nu] F}$. The spinor / Bloch state $\mathfrak{u}(\p)$ is understood as that of an on-shell quasiparticle near the FS. We will also show \eqref{BFL_dsigma_dEF_detail} is equivalent to
\begin{eqnarray}
\frac{d\sigma^{\mu\nu\lambda}}{d\epsilon_F} = \frac{d}{d\epsilon_F} \ 6 \int_\p \delta(\epsilon_F-E(\p)) \: \delta^{[\mu}_0 v^\nu(\p) \: a^{\lambda]}(\p),
\label{BFL_dsigma_dEF_1}
\end{eqnarray}
and, for $d>2$, also equivalent to
\begin{eqnarray}
\frac{d\sigma^{ij0}}{d\epsilon_F} &=& \frac{d}{d\epsilon_F} \ \frac{3}{d-2}\int_\p \delta(\epsilon_F-E(\p)) \: b^{[ij}(\p) v^{k]}(\p) \: p_k \nonumber \\[.2cm]
&& + \ \frac{d}{d\epsilon_F} \ \frac{6}{d-2}\int_\p \partial_p^{[k} \left( \delta(\epsilon_F-E(\p)) \: v^i(\p) \: a^{j]}(\p) \: p_k\right),
\label{BFL_dsigma_dEF_2e}
\end{eqnarray}
\begin{eqnarray}
\frac{d\sigma^{ijk}}{d\epsilon_F} &=& 0.
\label{BFL_dsigma_dEF_2b}
\end{eqnarray}
Thus, we conclude that in Berry Fermi liquid, \eqref{BFL_sigma_const_FS_1} and \eqref{BFL_sigma_const_FS_2} still hold as in Berry Fermi gas. Although we demonstrated in Section \ref{ssect_kinetic_current} that $\sigma^{\mu\nu\lambda}$ is not the full Hall conductivity, those remaining contributions are nevertheless always FS integrals. Therefore the full conductivity is always equal to a chemical potential independent part (as long as the FS changes continuously) plus a FS integral.

In Berry Fermi gas in $d=2$, $\sigma_{o}^{\mu\nu\lambda}$ is topological~\cite{Ishikawa:1986wx, Haldane:2004zz}. It would be interesting to study if it is still topological in Berry Fermi liquid. In particular, it is unknown whether $\sigma_{o}^{\mu\nu\lambda}$ can have a jump when the FS develops new disconnected components, as in the example of Figure \ref{EF_increase_disconnect_comp}.

\subsubsection{With Anomaly-Related Transport in $d>2$}
\label{sssect_kinetic_sigma_EF_with_anomaly}

For $d>2$, when the Berry curvature is not an exact 2-form on the FS, the system has anomaly-related transport effects.

Let's first review the effects in Berry Fermi gas. The expression \eqref{BFL_sigma_const_Fermi_sea} still holds, and we start from there. Now we have to take extra care when rewriting it via integration by parts. More precisely, $a^i$ cannot be continuously defined over the entire FS, so the expression \eqref{BFL_sigma_const_FS_1} is not so useful. The alternative expression \eqref{BFL_sigma_const_FS_2} promoted by \cite{Haldane:2004zz} is still useful as long as we take into account the ``Berry curvature defects'' where $\partial_p^{[k} b^{ij]} \neq 0$ (e.g. monopoles in $d=3$):
\begin{eqnarray}
\sigma^{ij0} &=& \sigma_{o}^{ij0} + \frac{3}{d-2}\int_\p \theta(\epsilon_F-E(\p)) \: b^{[ij}(\p) \ \partial_p^{k]} p_k \nonumber \\[.2cm]
&=& \sigma_o^{ij0} - \frac{3}{d-2}\int_\p \theta(\epsilon_F-E(\p)) \partial_p^{[k} b^{ij]}(\p) \: p_k  \nonumber \\[.2cm]
&& \hspace{.8cm} + \ \frac{3}{d-2}\int_\p \delta(\epsilon_F-E(\p)) \: b^{[ij}(\p) v^{k]}(\p) \: p_k \nonumber \\[.2cm]
&& \hspace{.8cm} + \ \frac{6}{d-2}\int_\p \partial_p^{[k} \left( \delta(\epsilon_F-E(\p)) \: v^i(\p) \: a^{j]}(\p) \: p_k\right).
\label{BFL_sigma_const_FS_anom_e}
\end{eqnarray}
The boundary term in the last line is explained below \eqref{BFL_sigma_const_FS_2}; although $a^i$ is not continuously defined over the FS, it can be continuously defined around where the FS intersects the boundary of the first Brillouin zone. The defects lie along where $\partial_p^{[k} b^{ij]} \neq 0$, and they are generically $d-3$ dimensional. In this thesis we assume there is no defect in the vicinity of the FS, and thus the second term is left unchanged under small continuous variation of $\epsilon_F$. In this spirit, we can combine the $\sigma_{o}^{ij0}$ term and the $\partial_p^{[k} b^{ij]}$ term and call their sum $\sigma_a^{ij0}$. A similar integration by parts can be carried out in the spatial components~\cite{Son:2012wh,Son:2012zy}:
\begin{eqnarray}
\sigma^{ijk} &=& \sigma_{o}^{ijk} + 3\int_\p \theta(\epsilon_F-E(\p)) \: b^{[ij}(\p) \ \partial_p^{k]} E(\p) \nonumber \\[.2cm]
&=& \sigma_o^{ijk} - 3\int_\p \theta(\epsilon_F-E(\p)) \partial_p^{[k} b^{ij]}(\p) \: E(\p) \nonumber \\[.2cm]
&& \hspace{.8cm} + \ 3\int_\p \delta(\epsilon_F-E(\p)) \: b^{[ij}(\p) v^{k]}(\p) \: \epsilon_F
\label{BFL_sigma_const_FS_anom_b}
\end{eqnarray}
(the second and third term separately vanish in the absence of Berry curvature defect), whose $\partial_p^{[k} b^{ij]}$ term is again independent of small continuous variation of $\epsilon_F$, and again we can combine the $\sigma_o^{ijk}$ term and the $\partial_p^{[k} b^{ij]}$ term and call their sum $\sigma_a^{ijk}$. The simplest example of \eqref{BFL_sigma_const_FS_anom_e} with Berry curvature defect is the anomalous Hall effect in Weyl metals~\cite{yang2011quantum}; the simplest example of \eqref{BFL_sigma_const_FS_anom_b} is the chiral magnetic effect~\cite{Son:2012wh,Stephanov:2012ki,Son:2012zy}.

For Berry Fermi liquid, \eqref{BFL_dsigma_dEF_detail} still holds when the Berry curvature is not exact on the FS, and we start from there. In Section \ref{sssect_QFT_dsigma_dEF} we will show \eqref{BFL_dsigma_dEF_detail} is equivalent to
\begin{eqnarray}
\frac{d\sigma^{ij0}}{d\epsilon_F} &=& \frac{d}{d\epsilon_F} \ \frac{3}{d-2}\int_\p \delta(\epsilon_F-E(\p)) \: b^{[ij}(\p) v^{k]}(\p) \: p_k \nonumber \\[.2cm]
&& + \ \frac{d}{d\epsilon_F} \ \frac{6}{d-2}\int_\p \partial_p^{[k} \left( \delta(\epsilon_F-E(\p)) \: v^i(\p) \: a^{j]}(\p) \: p_k\right),
\label{BFL_dsigma_dEF_e}
\end{eqnarray}
\begin{eqnarray}
\frac{d\sigma^{ijk}}{d\epsilon_F} &=& \frac{d}{d\epsilon_F} \ 3\int_\p \delta(\epsilon_F-E(\p)) \: b^{[ij}(\p) v^{k]}(\p) \: \epsilon_F
\label{BFL_dsigma_dEF_b}
\end{eqnarray}
as long as there is no Berry curvature defect near the FS; they reduce to \eqref{BFL_dsigma_dEF_2e}\eqref{BFL_dsigma_dEF_2b} if the Berry curvature is exact on the FS. Thus, for Berry Fermi liquid we can write
\begin{eqnarray}
\sigma^{ij\lambda} &=& \sigma_a^{ij\lambda} + 3\int_\p \delta(\epsilon_F-E(\p)) \: b^{[ij}(\p) v^{k]}(\p) \: P^\lambda_k (\p) \nonumber \\[.2cm]
&& \hspace{.85cm} + \ 6\int_\p \partial_p^{[k} \left( \delta(\epsilon_F-E(\p)) \: v^i(\p) \: a^{j]}(\p) \: P^\lambda_k\right),
\label{BFL_sigma_const_FS}
\end{eqnarray}
where $P^0_k\equiv p_k/(d-2)$ and $P^l_k\equiv \epsilon_F \delta^l_k$ (the second line vanishes if $\lambda$ is spatial). Here $\sigma_a^{\mu\nu\lambda}$ is independent of $\epsilon_F$ for generic values of $\epsilon_F$; but it depends on $\epsilon_F$ at special values of $\epsilon_F$ where some Berry curvature defect is brought across the Fermi level. Moreover, as before, it is unknown whether $\sigma_a^{\mu\nu\lambda}$ can have a jump in situations like Figure \ref{EF_increase_disconnect_comp}. In \eqref{BFL_sigma_const_FS_anom_e} and \eqref{BFL_sigma_const_FS_anom_b} for Fermi gas, we are able to separate $\sigma_a^{\mu\nu\lambda}$ into $\sigma_o^{\mu\nu\lambda}$ plus a Berry curvature defect term inside the Fermi sea. Such separation is generally impossible for Fermi liquid.

A final subtlety needs to be addressed. If we shift the definition of $\vec{p}$ by a constant vector, or shift the definitions of $E(\p)$ and $\epsilon_F$ together by a constant value, no physics should change. However, the kinetic term in \eqref{BFL_sigma_const_FS}, due to its $P^\lambda_k$ factor, does not necessarily satisfy this property in the presence of anomaly-related transport effects. There is no inconsistency here, as our starting point \eqref{BFL_dsigma_dEF_detail} does not have this problem. This just implies that, if we perform such shifts, $\sigma_a^{\mu\nu\lambda}$ also needs to be shifted such that $\sigma^{\mu\nu\lambda}$ remains unchanged. In Berry Fermi gas, this can be verified explicitly in \eqref{BFL_sigma_const_FS_anom_e} and \eqref{BFL_sigma_const_FS_anom_b}.

\subsection{Correspondence between Kinetic Theory and Field Theory}

In the next Section we will provide a diagrammatic derivation of the
Berry Fermi liquid theory. Here we summarize the identification of
the various quantities appearing the Berry Fermi liquid theory and
objects in the resummed perturbation theory.
\begin{itemize}
\item
$\delta f(\p; x)$, the quasiparticle distribution, is given by the FS singular part of the perturbed Wigner function, as introduced in Section \ref{ssect_QFT_BE}.
\item
$\delta J^\mu(x)$, the induced current, is the quantum expectation \eqref{current_QFT}.
\item
$E(\p)$, the kinetic energy of a quasiparticle, is defined by the full propagator \eqref{chiu_near_FS} near the FS. Its chemical potential dependence is given by \eqref{dE_dEF}.
\item
$\mathfrak{u}_\alpha(\p)$, the spinor / Bloch state of an quasiparticle, needed to define the Berry curvature, is defined in \eqref{G_expression} and above \eqref{u_derivative_on_shell}. Its chemical potential dependence is given below \eqref{du_dEF}.
\item
$\mu^{\mu\nu}(\p)$, the EM dipole moment of a quasiparticle, is given by \eqref{EM_dipole_def}, and discussed in detail in Section \ref{sssect_QFT_EM_dipole}.
\item
$\mathcal{U}(\p, \kk)$, the contact interaction energy between two quasiparticles, is given by \eqref{dE_dEF}.
\item
$\mathcal{V}^\mu(\p, \kk)$, the gradient interaction energy between two quasiparticles, is given by \eqref{grad_int_def}.
\item
$\mathcal{C}(\p, \kk)$, the near-FS effective collision kernel between two quasiparticles (in $d\geq 3$), is defined in \eqref{C_kernal_def}, whose details are discussed in Section \ref{sssect_QFT_Collision} and Section \ref{ssect_QFT_AppendixAB}.
\item
$\sigma^{\mu\nu\lambda}$, the Hall conductivity tensor, is defined in \eqref{sigma_def}. Its chemical potential dependence is given by the Berry curvature around the FS, as shown in Section \ref{sssect_QFT_dsigma_dEF}.
\end{itemize}
$E(\p)$ and $\mathcal{U}(\p, \kk)$ are familiar parameters in Landau Fermi liquid theory, while the other parameters $\mu^{\mu\nu}(\p), \mathcal{V}^\mu(\p, \kk), \mathcal{C}(\p, \kk)$ and $\sigma^{\mu\nu\lambda}$ are new.

\section{Derivation from Quantum Field Theory}
\label{sect_QFT}

In this Section we will prove the kinetic formalism presented above by analyzing the quantum field theory (QFT) to all orders in perturbation theory. Before we go into any details, we summarize the idea behind our proof as the following. Our goal is to compute linear response, i.e. induced current $\delta J$ as a linear function of electromagnetic (EM) connection $A$, to first order in the external momentum $q$ carried in $A$. The evaluation of $\delta J^\mu$ can be separated, technically, into two parts:
\begin{itemize}
\item
The first part is related to Cutkosky cut, and corresponds to the quasiparticle contributions to $\delta J$, that is, the first line of \eqref{BFL_current}. This part includes the excitation and collision of quasiparticles, described by the Boltzmann equation \eqref{BFL_Boltzmann_Eq}.
\item
The second part is the remaining contributions that are unrelated to Cutkosky cut, i.e. non-quasiparticle contributions. This second part gives rise to the Hall conductivity tensor in \eqref{BFL_current}, whose chemical potential dependence is given by the Berry curvature on the Fermi surface (FS).
\end{itemize}
The idea is simple and clear. Now we devote into the technical details.

First we state the assumptions about our QFT and its ground state and low energy spectrum.

Our QFT consists of a multi-component fermionic field $\psi^\alpha$, charged under EM. The index $\alpha$ can be a spinor index if $\psi$ is a Dirac spinor, or in general labels different bands. The fermionic field may interact via massive fields (generically denoted as $\phi$) or/and via finite-ranged self-interactions. But we assume any field other than $\psi$ to be EM neutral, and the EM couplings to $\psi$ only take place in the non-interacting terms of $\psi$ in the Lagrangian, but not in any interacting terms. Thus, there can be EM couplings such as $A\psi^\dagger \psi$ (including $F\psi^\dagger \psi$) and $AA\psi^\dagger \psi$, but there is no EM coupling like $A\phi\psi^\dagger \psi$ or $A\psi^\dagger \psi \psi^\dagger \psi$.

We assume the system is under chemical potential $\epsilon_F$ for the fermionic field and at negligible temperature (but high enough to avoid to Kohn-Luttinger instability~\cite{Kohn:1965zz}). We assume the EM $U(1)$ gauge invariance is not broken by the ground state. We assume there is no band degeneracy near the FS, and for simplicity, we assume the Fermi level crosses only one band of the spectrum of the fermionic field. We assume the only low energy excitations are quasiparticles of this band. Thereby the system is said to be a Fermi liquid at low energy.

We assume spacetime translational symmetry is not broken by anything except for the present external EM field. We do not assume any symmetry otherwise. In this Section we set $\hbar$ to $1$.

The proof is organized as the following. We first discuss the properties of a single full propagator and a pair of full propagators. Then we introduce the irreducible 2-particle interaction vertex, and discuss its relation to the chemical potential dependence of the full propagator (from which the chemical potential dependence \eqref{BFL_dE_dEF} of the kinetic energy follows). Then we introduce the properties of the bare EM coupling vertex. Next we present the recursion relation satisfied by the full EM coupling vertex; we will also extract the implications of the Ward-Takahashi identity. Having had all these preliminaries, we are ready to prove the main results. We first show the Boltzmann equation \eqref{BFL_Boltzmann_Eq_0}\eqref{BFL_Boltzmann_Eq_1} follows from the recursion relation satisfied by the full EM coupling vertex. Then we compute the quantum expectation of the current and show it takes the form \eqref{BFL_current_0}\eqref{BFL_current_1} given in the kinetic theory. Finally we study the chemical potential dependence \eqref{BFL_dsigma_dEF_detail} of the Hall conductivity tensor. Along the way, we will also discuss the microscopic ingredients of the EM dipole moment. As a bonus, we obtain an alternative diagrammatic proof to the Coleman-Hill theorem~\cite{Coleman:1985zi} for QFTs restricted to our assumptions.

\subsection{Propagator}
\label{ssect_QFT_prop}

\subsubsection{Single Propagator}

\begin{center}
\begin{fmffile}{zzz-G}
\begin{fmfgraph*}(35, 15)
\fmfleftn{l}{3}\fmfrightn{r}{3}
\fmf{fermion,label.side=left,label=$p$}{r2,l2}
\fmfdot{l2,r2}
\fmflabel{$\alpha$}{l2}
\fmflabel{$\beta$}{r2}
\end{fmfgraph*}
\end{fmffile}
\end{center}

The full propagator (all QFT quantities are time ordered unless otherwise specified) $iG^\alpha_{\ \beta}(p)$ is a matrix in the components of the fermionic field. We let the energy $p^0=-p_0=0$ on the FS. The assumption of Fermi liquid amounts to the assumptions of the form of $iG$ at small $p^0$. By general analytic properties of fermionic propagators \cite{Luttinger:1961zz, abrikosov1975methods} (in particular, the property that $G$ must be Hermitian at $p^0=0$), and the specific requirement that at low energy there is one species of stable quasiparticle, as $p^0\rightarrow 0$ the full propagator of our assumed Fermi liquid should take the form
\begin{eqnarray}
iG^\alpha_{\ \beta}(p) \simeq \frac{i u^\alpha(p) u^\dagger_\beta(p)}{\chi_u(p)} + \sum_w \frac{i w^\alpha(p) w^\dagger_\beta(p)}{\chi_w(p)}.
\label{G_expression}
\end{eqnarray}
The eigenvector $u$ is the band that crosses the Fermi level, with singular eigenvalue whose inverse is of the form
\begin{eqnarray}
\chi_u(p) = \frac{p^0 - \xi(\p) + i\epsilon\: \sgn\,\xi(\p)}{Z(\p)} \ + \ \cdots
\label{chiu_near_FS}
\end{eqnarray}
where $\xi(\p) \equiv E(\p) - \epsilon_F$ and $(\cdots)$ are terms of higher suppression in $p^0$; the quasiparticle renormalization factor $Z(\p)$ should be understood as the inverse of the coefficient of $p^0$ in $\chi_u$ (with rotational invariance, $Z$ can depend on $|\p|$; without rotational invariance, it may depend on all components of $\p$). The $w$'s are all other eigenvectors, and their eigenvalues $1/\chi_w$ are regular and nearly real.

For Landau Fermi liquid theory, \eqref{chiu_near_FS} is enough, but for Berry Fermi liquid we need to know one order higher in $p^0$, i.e. work up to $(p^0)^2$ order in the $(\cdots)$ terms. This will be handled later in Section \ref{sssect_QFT_Collision} and Section \ref{ssect_QFT_AppendixAB}. One should also worry about whether the diagonalization \eqref{G_expression} fails as we consider one order higher in $p^0$; using the method in Section \ref{ssect_QFT_AppendixAB} one can easily see this problem occurs only at two orders higher in $p^0$, so in this paper we do not need to worry about this.

In Section \ref{ssect_kinetic_sigma_EF}, we used the quasiparticle spinor / Bloch state $\mathfrak{u}(\p)$; it refers to $u(p)$ with $p$ on-shell and near the FS, i.e. $p^0=\xi(\p) \rightarrow 0$. The $p$ derivatives of $u(p)$ and $\mathfrak{u}(\p)$ are related by
\begin{eqnarray}
\left. (\partial_p^\mu+v^\mu \partial_{p^0}) u(p) \right|_{p \ on-shell} = \partial_p^\mu \mathfrak{u}(\p).
\label{u_derivative_on_shell}
\end{eqnarray}
Here $\partial_p^0 \equiv \partial / \partial p_0 = -\partial / \partial p^0 = \partial_{p_0} = - \partial_{p^0}$ while $\partial_p^i \equiv \partial / \partial p_i = \partial / \partial p^i  = \partial_p^i= \partial_{p^i}$, and recall that $v^0\equiv 1$ and $v^i\equiv \partial_p^i E$ as introduced in Section \ref{sect_kinetic}.

The following identity, which follows from the product rule of derivative, is useful in this paper:
\begin{eqnarray}
\partial_p^\mu (G^{-1})^\alpha_{\ \beta} \ u^\beta = - (G^{-1})^\alpha_{\ \beta} \ \partial_p^\mu u^\beta + \partial_p^\mu \chi_u u^\alpha + \chi_u \partial_p^\mu u^\alpha
\label{IBPtrick}
\end{eqnarray}
for $p$ near the FS; note that
\begin{eqnarray}
-Z \partial_p^\mu \chi_u = -Z\: u^\dagger_\alpha \ \partial_p^\mu (G^{-1})^\alpha_{\ \beta} \ u^\beta = v^\mu + \left(\mbox{terms vanish on the FS}\right).
\label{chiu_derivative}
\end{eqnarray}
There is a similar identity for $u^\dagger_\alpha \ \partial_p^\mu(G^{-1})^\alpha_{\ \beta}$.

Now we look at the momentum derivative of the full propagator:
\begin{eqnarray}
\partial_p^\nu iG^\alpha_{\ \beta}(p) &=& iG^\alpha_{\ \alpha'} \ \partial_p^\nu (iG^{-1})^{\alpha'}_{\ \beta'} \ iG^{\beta'}_{\ \beta} \ - \ iZ u^\alpha u^\dagger_\beta \ i\pi \delta(p^0-\xi) \ \partial_p^\nu \sgn\,\xi \nonumber \\[.2cm]
&=& iG^\alpha_{\ \alpha'} \ \partial_p^\nu (iG^{-1})^{\alpha'}_{\ \beta'} \ iG^{\beta'}_{\ \beta} - iZ u^\alpha u^\dagger_\beta \ i\delta_{FS} \ \delta^\nu_i v^i,
\label{G_derivative_indices}
\end{eqnarray}
where
\begin{eqnarray}
\delta_{FS}(p) \equiv 2\pi\delta(p^0) \: \delta(\xi(\p)).
\end{eqnarray}
The presence of the second term in \eqref{G_derivative_indices} is because as $p^i$ varies across the FS, the $p^0$ pole in $1/\chi_u$ moves across the real axis. This abrupt change is not captured by the first term. The expression of the second term can be obtained by principle function decomposition $(x \pm i\epsilon)^{-1}= \mathcal{P} x^{-1} \mp i\pi\delta(x)$.

To avoid having too many fermion component indices in equations in this thesis, we introduce two notations: single fermion linear space and double fermion linear space. Consider $G^\alpha_{\ \beta}$. In single fermion linear space, $^\alpha_{\ \beta}$ are viewed as two indices, so $G$ is viewed as a matrix in single fermion linear space. In double fermion linear space, $^\alpha_{\ \beta}$ together is viewed as one index, so $G$ is viewed as a vector in the the double fermion linear space. In our proof, only a few index contractions are to be understood in single fermion linear space, most are understood in double fermion linear space. To distinguish them, we will enclose objects contracted in single fermion linear space by curly brackets $\{ \ \}$, while do not enclose objects contracted in double fermion linear space by anything. For example, according to \eqref{chiu_derivative}, $v^\mu$ can be expressed as
\begin{eqnarray}
-\frac{v^\mu}{Z} = \left. u^\dagger_\alpha \: \partial_p^\mu (G^{-1})^\alpha_{\ \beta} \: u^\beta \right|_{p \ on \ FS}= \left. \left\{ u^\dagger \: \partial_p^\mu G^{-1} \: u \right\} \right|_{p \ on \ FS}= \left. (uu^\dagger)^T \partial_p^\mu G^{-1} \right|_{p \ on \ FS}
\label{v_in_QFT}
\end{eqnarray}
in explicit index notation, single fermion notation, and double fermion notation respectively. We will introduce more about double fermion notation in Section \ref{sssect_QFT_int_vertex}.

In the double fermion notation introduced above, \eqref{G_derivative_indices} can be expressed compactly as
\begin{eqnarray}
\partial_p^\nu iG = i\Delta_0 \ \partial_p^\nu iG^{-1} + (Zuu^\dagger) \delta_{FS} \delta^\nu_i v^i,
\label{G_derivative}
\end{eqnarray}
where
\begin{eqnarray}
i(\Delta_0)^{\alpha \ \gamma}_{\ \delta, \ \beta}(p) \equiv iG^\alpha_{\ \beta}(p) \ iG^\gamma_{\ \delta}(p)
\end{eqnarray}
is a matrix in double fermion notation. Since the momentum argument in both $iG$'s is the same, $i\Delta_0$ has a double pole in $p^0$ when all of its four indices are in the $u$ band.

\subsubsection{Double Propagator}

An important step towards the QFT foundation of the Landau Fermi liquid theory is the observation that, the semiclassical notion of ``deformation $\delta f$ of the FS'' originates from the pole structure of the double propagator $iG(p+q/2) iG(p-q/2)$, for $q$ small and $p$ near the FS~\cite{landau1959theory,abrikosov1975methods}. Now we make a similar analysis, but with non-trivial $u^\alpha(p)$, and work to first order in $q$.

\begin{center}
\begin{fmffile}{zzz-GG}
\begin{fmfgraph*}(35, 20)
\fmfleftn{l}{5}\fmfrightn{r}{5}
\fmf{fermion,label.side=left,label=$p-q/2$}{l4,r4}
\fmf{fermion,label.side=left,label=$p+q/2$}{r2,l2}
\fmfdot{l2,r2,l4,r4}
\fmflabel{$\delta$}{l4}
\fmflabel{$\alpha$}{l2}
\fmflabel{$\gamma$}{r4}
\fmflabel{$\beta$}{r2}
\end{fmfgraph*}
\end{fmffile}
\end{center}

Consider the product of two full fermionic propagators, as drawn above, with arbitrary $p$ and small $q$; more exactly, $q\ll p_F$ where $p_F$ is the size scale of the FS. To first order in $q$, we express the product in the form
\begin{eqnarray}
\!\!\!\!\!\! && iG^\alpha_{\ \beta}(p+q/2) \ iG^\gamma_{\ \delta}(p-q/2)\nonumber \\[.2cm]
\!\!\!\!\!\! &=& \left( i\Delta_0(p) + i\Delta'_0(p; q) + i\Delta^r_1(p; q) + i\Delta^s_1(p; q) + i\Delta'_1(p; q) \right)^{\alpha \ \gamma}_{\ \delta, \ \beta} - {D_1}^{\alpha \ \gamma}_{\ \delta, \ \beta}(p; q),
\label{doubleprop_parametrization}
\end{eqnarray}
where the subscripts $0$ or $1$ denote the order in $q$. Here $\Delta_0, \Delta^r_1, \Delta^s_1$ are regular as $q\rightarrow 0$. In particular, $\Delta_0$ has been introduced in \eqref{G_derivative}, and $\Delta^r_1$ and $\Delta^s_1$ follow from the two terms of \eqref{G_derivative_indices} when expanding $G(p\pm q/2)$ in $q$:
\begin{eqnarray}
(\Delta^r_1)^{\alpha \ \gamma}_{\ \delta, \ \beta}(p; q) \equiv -i \frac{q_\lambda}{2} \left( \left\{ G \ \partial_p^\nu G^{-1} \ G\right\}^\alpha_{\ \beta} G^\gamma_{\ \delta} - G^\alpha_{\ \beta} \left\{ G \ \partial_p^\nu G^{-1} \ G\right\}^\gamma_{\ \delta} \right),
\end{eqnarray}
\begin{eqnarray}
(\Delta^s_1)^{\alpha \ \gamma}_{\ \delta, \ \beta}(p; q) \equiv \frac{q_\lambda}{2} \delta^\lambda_k v^k \: Z \delta_{FS} \left(u^\alpha u^\dagger_\beta \: G^\gamma_{\ \delta} - G^\alpha_{\ \beta} \: u^\gamma u^\dagger_\delta\right).
\end{eqnarray}
Clearly both $\Delta^r_1$ and $\Delta^s_1$ vanish when all indices are projected onto one band.

When there is no FS, the expansion of the double propagator as $i\Delta_0+i\Delta^r_1+i\Delta^s_1$ is legitimate. When FS is present, such naive expansion in $q$ misses contributions that are related to the pole structure difference across the FS. These extra contributions are denoted by $\Delta'_0$ and $\Delta'_1$, which are singular as $q\rightarrow 0$. Explicitly, they are given by
\begin{eqnarray}
\Delta'_0 = \Delta' \ (uu^\dagger) (uu^\dagger)^T,
\label{Delta_prime}
\end{eqnarray}
\begin{eqnarray}
\Delta'_1 = \Delta' \ iq_\mu \mathcal{A}^\mu,
\end{eqnarray}
where we defined
\begin{eqnarray}
\Delta'(p; q) \equiv Z^2(\p) \: \delta_{FS}(p) \frac{v^i(\p) \ q_i}{v^\mu(\p) \ q_\mu - i \epsilon \: \sgn q^0},
\end{eqnarray}
which is familiar from Landau Fermi liquid theory~\cite{landau1959theory, abrikosov1975methods} when $u^\alpha$ is one-component (i.e. $u=1$ trivially). Also, we introduced the abbreviation
\begin{eqnarray}
(\mathcal{A}^\mu)^{\alpha \ \gamma}_{\ \delta, \ \beta} \equiv \frac{-i}{2}\left(\partial_p^\mu u^\alpha u^\dagger_\delta - u^\alpha \partial_p^\mu u^\dagger_\delta\right) (u^\gamma u^\dagger_\beta) - \frac{-i}{2}(u^\alpha u^\dagger_\delta) \left(\partial_p^\mu u^\gamma u^\dagger_\beta - u^\gamma \partial_p^\mu u^\dagger_\beta\right).
\end{eqnarray}
Below we present the derivation for $\Delta'_0$ and $\Delta'_1$.

Let's focus on the double $u$-band term in the double propagator:
\begin{eqnarray}
\frac{i\left(Zu^\alpha u^\dagger_\beta\right)(p+q/2)}{p^0+q^0/2-\xi(\p+\q/2)+i\epsilon\: \sgn\,\xi(\p+\q/2)} \ \frac{i\left(Zu^\gamma u^\dagger_\delta\right)(p-q/2)}{p^0-q^0/2-\xi(\p-\q/2)+i\epsilon\: \sgn\,\xi(\p-\q/2)}.
\label{ubandproduct}
\end{eqnarray}
What are missing in the naive expansion over $q$ are the contributions when the two $i\epsilon$ prescriptions in the denominators take opposite signs. To extract these missing pieces, we perform principle function decomposition $(x \pm i\epsilon)^{-1}= \mathcal{P} x^{-1} \mp i\pi\delta(x)$ and keep those terms which are non-vanishing only when $\sgn\, \xi(\p\pm\q/2)$ are opposite. Such terms are
\begin{eqnarray}
&& \left[ i\pi \ \frac{\sgn\,\xi(\p-\q/2) - \sgn\,\xi(\p+\q/2)}{-q^0 + \xi(\p+\q/2)-\xi(\p-\q/2)} \right. \nonumber \\[.2cm]
&& \hspace{3cm}\left. \times \frac{\delta(p^0+q^0/2-\xi(\p+\q/2)) + \delta(p^0-q^0/2-\xi(\p-\q/2))}{2} \right. \nonumber \\[.2cm]
&& \hspace{0cm} \left. - (i\pi)^2 2\theta(-\sgn\,\xi(\p-\q/2)\sgn\,\xi(\p+\q/2)) \right. \nonumber \\[.2cm]
&& \hspace{2.8cm}\left. \phantom{\frac{1}{1}} \times \delta(p^0+q^0/2-\xi(\p+\q/2))\delta(p^0-q^0/2-\xi(\p-\q/2)) \right] \nonumber \\[.2cm]
&& i\left(Zu^\alpha u^\dagger_\beta\right)(p+q/2) \ i\left(Zu^\gamma u^\dagger_\delta\right)(p-q/2).
\end{eqnarray}
Expanding the generalized functions in the square bracket in $q$, we have
\begin{eqnarray}
\!\!\!\!\!\!\! && \left[ i\pi \frac{-2 \delta(\xi(\p)) v^i q_i}{v^\mu(\p) q_\mu} \delta(p^0-\xi(\p)) - (i\pi)^2 2|v^i(\p) q_i| \delta(\xi(\p)) \delta(p^0-\xi(\p)) \delta(v^\mu(\p) q_\mu) \ + \ \mathcal{O}(q^2) \right] \nonumber \\[.2cm]
\!\!\!\!\!\!\! && i\left(Zu^\alpha u^\dagger_\beta\right)(p+q/2) \ i\left(Zu^\gamma u^\dagger_\delta\right)(p-q/2).
\end{eqnarray}
Now we recognize the square bracket is nothing but $-i\Delta'/Z^2$ expressed in principle function decomposition. Finally we expand the two $(Zuu^\dagger)$'s to zeroth and first order in $q$, we obtain the expression for $i\Delta'_0+i\Delta'_1$ presented above.

Developing along this line of thinking, one is led to the formalism of Cutkosky cut~\cite{Cutkosky:1960sp}, which we will discuss in Section \ref{ssect_QFT_AppendixAB}. In particular, see Eq.~\eqref{Delta_from_Cutkosky} for the derivation of $\Delta'$ from the Cutkosky cutting rule.

What is $D_1$ in \eqref{doubleprop_parametrization}? The step \eqref{ubandproduct} is not quite right, for it completely ignored the $(\cdots)$ terms in \eqref{chiu_near_FS}. While it is legitimate to do so at leading order in $q$ (in Landau's theory), at first order in $q$ there are missed contributions, which we call $D_1$. We will postpone its discussion to Section \ref{sssect_QFT_Collision}, when we discuss the quasiparticle decay term along with other quasiparticle collision terms.

\subsection{Interaction}
\label{ssect_QFT_int}

\subsubsection{$q$-2PI Interaction Vertex}
\label{sssect_QFT_int_vertex}

Let $i\wt{V}^{\alpha \ \gamma}_{\ \delta, \ \beta}(p, k; q)$ be the full $q$-2PI (defined below) interaction vertex, with two incoming fermions of momentum and index $(p-q/2, \delta)$ and $(k+q/2, \beta)$, and two outgoing fermions with momentum and index $(k-q/2, \gamma)$ and $(p+q/2, \alpha)$, as drawn below. \vspace{.4cm}
\begin{center}
\begin{fmffile}{zzz-wtV}
\begin{fmfgraph*}(45, 20)
\fmfleftn{l}{5}\fmfrightn{r}{5}
\fmf{phantom}{l4,lv4,rv4,r4}
\fmf{phantom}{r2,rv2,lv2,l2}
\fmf{fermion,tension=0.3,label.side=left,label=$p-q/2$}{l4,lv4}
\fmf{fermion,tension=0.3,label.side=left,label=$k+q/2$}{r2,rv2}
\fmf{fermion,tension=0.3,label.side=left,label=$k-q/2$}{rv4,r4}
\fmf{fermion,tension=0.3,label.side=left,label=$p+q/2$}{lv2,l2}
\fmflabel{$\delta$}{l4}
\fmflabel{$\alpha$}{l2}
\fmflabel{$\gamma$}{r4}
\fmflabel{$\beta$}{r2}
\fmf{phantom}{l3,o,r3}
\fmf{phantom,label.dist=0,label={\Large $i\wt{V}$}}{l3,r3}
\fmfv{d.sh=circle,d.f=empty,d.si=0.45w}{o}
\end{fmfgraph*}
\end{fmffile}
\end{center}
(In this thesis, external propagators without a solid dot at the end are always stripped off.) Here $q$-2PI means that $i\wt{V}$ is a sum of connected, 1PI (with respect to the fermion only) interaction diagrams, such that in each diagram there does not exist two internal fermion propagators that are dictated by momentum conservation to have momenta differing by $q$. Equivalently, for each diagram, one cannot find two internal fermion propagators cutting which will disconnect the diagram into two parts, such that the external lines of $(p-q/2, \delta)$ and $(p+q/2, \alpha)$ are on one part, while the external lines of $(k+q/2, \beta)$ and $(k-q/2, \gamma)$ are on the other part. For example, in the four diagrams below (fermionic propagators always mean full propagators), the two on the left are $q$-2PI, while the two on the right are not. \vspace{.4cm}
\begin{center}
\begin{fmffile}{zzz-q2PI-1}
\begin{fmfgraph*}(32, 20)
\fmfleftn{l}{2}\fmfrightn{r}{2}
\fmf{fermion}{l2,v2,v3,r2}
\fmf{fermion}{r1,v4,v1,l1}
\fmf{dashes,left=0.1,tension=0}{v4,v2}
\fmf{dashes,right=0.1,tension=0}{v1,v3}
\end{fmfgraph*}
\end{fmffile}
\begin{fmffile}{zzz-q2PI-2}
\begin{fmfgraph*}(32, 20)
\fmfleftn{l}{2}\fmfrightn{r}{2}
\fmf{fermion,tension=2}{l2,v1,l1}
\fmf{fermion,tension=1.5}{r1,v2,v3,r2}
\fmf{phantom,tension=2}{r1,v2}
\fmf{phantom,tension=2}{v3,r2}
\fmf{dashes,tension=2}{v1,v4}
\fmf{phantom}{v2,v5,v6,v4}
\fmf{phantom}{v3,v7,v8,v4}
\fmf{dashes,tension=0}{v2,v5}
\fmf{dashes,tension=0}{v3,v7}
\fmffreeze
\fmf{fermion,left=0.7}{v4,v7}
\fmf{fermion}{v7,v5}
\fmf{fermion,left=0.7}{v5,v4}
\fmf{dashes,left=0.15}{v7,v4}
\end{fmfgraph*}
\end{fmffile}
\hspace{1cm}
\begin{fmffile}{zzz-not-q2PI-1}
\begin{fmfgraph*}(32, 20)
\fmfleftn{l}{2}\fmfrightn{r}{2}
\fmf{fermion}{l2,v2,v3,v6,r2}
\fmf{fermion}{r1,v5,v4,v1,l1}
\fmf{dashes,left=0.1,tension=0}{v4,v2}
\fmf{dashes,right=0.1,tension=0}{v1,v3}
\fmf{dashes,tension=0}{v5,v6}
\end{fmfgraph*}
\end{fmffile}
\begin{fmffile}{zzz-not-q2PI-2}
\begin{fmfgraph*}(32, 20)
\fmfleftn{l}{2}\fmfrightn{r}{2}
\fmf{fermion,tension=2}{l2,v1,l1}
\fmf{fermion,tension=1.5}{r1,v2,v3,r2}
\fmf{phantom,tension=2}{r1,v2}
\fmf{phantom,tension=2}{v3,r2}
\fmf{dashes,tension=2}{v1,v4}
\fmf{phantom}{v2,v5,v6,v4}
\fmf{phantom}{v3,v7,v8,v4}
\fmf{dashes,tension=0}{v2,v5}
\fmf{dashes,tension=0}{v3,v7}
\fmffreeze
\fmf{fermion,left=0.7}{v4,v7}
\fmf{fermion}{v7,v5}
\fmf{fermion,left=0.7}{v5,v4}
\fmf{dashes,left}{v5,v7}
\end{fmfgraph*}
\end{fmffile}
\end{center}
More about $i\wt{V}$ is said in Section \ref{ssect_QFT_AppendixCD}.

For Fermi liquid, the limit $\wt{V}(p, k; 0)$ is generally regular and analytic in $q$. Keeping zeroth and first order in $q$, we write $\wt{V}(p, k; q) = \wt{V}_0(p, k) + \wt{V}_1(p, k; q)$. 

The $q$-2PI interaction vertex is the building block of full interaction vertex $iV$: The latter is a geometric series given by the recursion relation
\begin{eqnarray}
\parbox{30mm}{
\begin{fmffile}{zzz-V-recursion}
\begin{fmfgraph*}(25, 15)
\fmfleftn{l}{5}\fmfrightn{r}{5}
\fmf{phantom}{l4,lv4,rv4,r4}
\fmf{phantom}{r2,rv2,lv2,l2}
\fmf{fermion,tension=0}{l4,lv4}
\fmf{fermion,tension=0}{r2,rv2}
\fmf{fermion,tension=0}{rv4,r4}
\fmf{fermion,tension=0}{lv2,l2}
\fmf{phantom}{l3,o,r3}
\fmfv{d.sh=circle,d.f=empty,d.si=0.43w}{o}
\fmf{phantom,label.dist=0,label=$iV$}{l3,r3}
\end{fmfgraph*}
\end{fmffile}
}
= \hspace{.5cm}
\parbox{30mm}{
\begin{fmffile}{zzz-V-recursion-wtV}
\begin{fmfgraph*}(25, 15)
\fmfleftn{l}{5}\fmfrightn{r}{5}
\fmf{phantom}{l4,lv4,rv4,r4}
\fmf{phantom}{r2,rv2,lv2,l2}
\fmf{fermion,tension=0}{l4,lv4}
\fmf{fermion,tension=0}{r2,rv2}
\fmf{fermion,tension=0}{rv4,r4}
\fmf{fermion,tension=0}{lv2,l2}
\fmf{phantom}{l3,o,r3}
\fmfv{d.sh=circle,d.f=empty,d.si=0.43w}{o}
\fmf{phantom,label.dist=0,label=$i\wt{V}$}{l3,r3}
\end{fmfgraph*}
\end{fmffile}
}
+\hspace{.5cm}
\parbox{50mm}{
\begin{fmffile}{zzz-V-recursion-wtV-V}
\begin{fmfgraph*}(45, 15)
\fmfleftn{l}{5}\fmfrightn{r}{5}
\fmf{fermion}{l4,lm4,rm4,r4}
\fmf{fermion}{r2,rm2,lm2,l2}
\fmf{phantom,tension=2}{l4,lm4}
\fmf{phantom,tension=2}{rm4,r4}
\fmf{phantom,tension=2}{lm2,l2}
\fmf{phantom,tension=2}{r2,rm2}
\fmf{phantom}{l3,ol}
\fmf{phantom}{or,r3}
\fmf{phantom,tension=0.8}{ol,or}
\fmfv{d.sh=circle,d.f=empty,d.si=0.25w}{ol}
\fmfv{d.sh=circle,d.f=empty,d.si=0.25w}{or}
\fmfv{label.dist=0,label=$i\wt{V}$}{ol}
\fmfv{label.dist=0,label=$iV$}{or}
\end{fmfgraph*}
\end{fmffile}
}
\label{full_int_vertex}
\end{eqnarray}
The full interaction vertex is singular in the $q\rightarrow 0$ limit, due to the presence of $\Delta'$ in the double propagators, as well as collision factors to be discussed in Section \ref{sssect_QFT_Collision}.

Before we proceed, we say a bit more about the double fermion notation. Consider an object, perhaps with spacetime indices, $\left(X^{\alpha \ \gamma}_{\ \delta, \ \beta}\right)^{\mu\nu\rho\dots}(p, k; q)$. This object is a matrix in the double fermion linear space. We now introduce its transpose:
\begin{eqnarray}
\left(X^{\alpha \ \gamma}_{\ \delta, \ \beta}\right)^{\mu\nu\rho\dots}(p, k; q) = \left(\left(X^{\gamma \ \alpha}_{\ \beta, \ \delta}\right)^{\mu\nu\rho\dots}(k, p; -q) \right)^T.
\end{eqnarray}
Diagrammatically, the transpose in double fermion linear space corresponds to ``turning the diagram $180$ degrees''; note that the spacetime indices are unaffected by the transpose. Finally, we introduce the convention that, for objects like $X$ which involve two momenta $p$ and $k$, the contraction with another object implies a momentum integral, for example
\begin{eqnarray}
\left(X^{\mu\nu\rho\dots} Y\right)^\alpha_{\ \delta}(p; q) \equiv \int_k \left(X^{\alpha \ \gamma}_{\ \delta, \ \beta}\right)^{\mu\nu\rho\dots}(p, k; q) \ Y^\beta_{\ \: \gamma}(k; q)
\end{eqnarray}
where $\int_k \equiv \int d^{d+1} k / (2\pi)^{d+1}$.

Now, by definition of $i\wt{V}$, we see it satisfies $i\wt{V} = (i\wt{V} )^T$, and similarly for all the $\Delta$'s and $D_1$. We will need these transpose properties when we derive the current in Section \ref{ssect_QFT_current}.

\subsubsection{Full Interaction Vertex}

Using the double fermion notation introduced above, we expand the recursion relation \eqref{full_int_vertex} to zeroth and first order in $q$. At zeroth order,
\begin{eqnarray}
iV_0 = i\wt{V}_0 + i\wt{V}_0 \left(i\Delta_0+i\Delta'_0\right) iV_0 = i\bar{V}_0 + i\bar{V}_0 \ i\Delta'_0 \ iV_0
\end{eqnarray}
where we defined the geometric series $i\bar{V}_0$ via the recursion relation
\begin{eqnarray}
i\bar{V}_0 = i\wt{V}_0 + i\wt{V}_0 \ i\Delta_0 \ i\bar{V}_0,
\end{eqnarray}
$i\bar{V}_0$ can be understood as $iV$ in the limit $q^0 \rightarrow 0$, $q^i/q^0 \rightarrow 0$ (because $\Delta'$ vanishes in the $q^i/q^0 \rightarrow 0$ limit), and is closely related to Landau's contact interaction potential $\mathcal{U}$~\cite{landau1959theory,abrikosov1975methods}, as we will see later. 

At first order,
\begin{eqnarray}
iV_1 &=& i\wt{V}_1 + i\wt{V}_1\left(i\Delta_0+i\Delta'_0\right) iV_0  + i\wt{V}_0 \left(i\Delta^r_1+i\Delta^s_1+i\Delta'_1\right) iV_0 + i\wt{V}_0 \left(-C_1\right) iV_0 \nonumber \\[.2cm]
&& + \ i\wt{V}_0\left(i\Delta_0+i\Delta'_0\right) iV_1.
\label{full_int_vertex_1}
\end{eqnarray}
Here ${C_1}^{\alpha \ \gamma}_{\ \delta, \ \beta}(p, k; q)$ is the quasiparticle collision term. We explain this term below. 

\subsubsection{Quasiparticle Decay and Collision Vertices}
\label{sssect_QFT_Collision}

Let us emphasize our comment in Section \ref{ssect_kinetic_BE} again: Collision is a single band ($u$ band) effect that is ``uninteresting'', as our main interest is multi-component effects such as Berry curvature. More particularly, in Section \ref{ssect_QFT_AppendixAB} it is shown that collision has no contribution to the antisymmetric part of the current-current correlation (which include interesting physics such as anomalous Hall effect and chiral magnetic effect). Here we are including collision just for completeness.

The quasiparticle collision term ${C_1}^{\alpha \ \gamma}_{\ \delta, \ \beta}(p, k; q)$ is defined as
\begin{eqnarray}
-C_1(p, k; q) &\equiv& -D_1(p; q) \ (2\pi)^{d+1}\delta^{d+1}(p-k) \ - \ C^{ph}_1(p, k; q) \ - \ C^{pp}_1(p, k; q).
\end{eqnarray}
The decay term $D_1$ in $C_1$ is from \eqref{doubleprop_parametrization} but left unexplained there. Where do $D_1$, $C^{ph}_1$ and $C^{pp}_1$ come from? Recall that in \eqref{doubleprop_parametrization} we could not naively expand the two propagators in $q$ individually; there are terms non-analytic in $q$ to be carefully taken care of. Similarly, here in the recursion relation for $iV$, we cannot naively expand the $i\wt{V}$'s and the double propagators individually. The non-analytic contributions that are missed from such naive expansion are $D_1$, $C^{ph}_1$ and $C^{ph}_1$.
 
Formally, the three terms in the definition of $-C_1(p, k; q)$ correspond to the following three pairs of Cutkosky-cut sub-diagrams:
\begin{center}
\begin{fmffile}{zzz-Decay1}
\begin{fmfgraph*}(55, 40)
\fmfleftn{l}{9}\fmfrightn{r}{9}
\fmf{phantom}{r6,rm6,m6,lm6,l6}
\fmf{phantom}{l9,m9,r9}
\fmf{phantom}{r1,m1,l1}
\fmf{phantom,tension=2}{l1,m1}
\fmf{phantom,tension=2}{r9,m9}
\fmf{fermion,tension=0.3,label.side=left,label=$p+q/2$}{r3,l3}
\fmf{fermion,tension=0.3,label.side=left,label=$p-q/2$}{rm6,r6}
\fmf{fermion,tension=0.3,label.side=left,label=$p-q/2$}{l6,lm6}
\fmfdot{r3,l3,r6,l6}
\fmfv{d.sh=circle,d.f=20,d.si=0.12w}{rm6}
\fmfv{d.sh=circle,d.f=20,d.si=0.12w}{lm6}
\fmf{fermion,tension=0.2,label.side=right,label=$k+l$}{rm6,lm6}
\fmf{fermion,tension=0,right=0.6,label.side=left,label=$p+l$}{lm6,rm6}
\fmf{fermion,tension=0,left=0.6,label.side=left,label=$k-q/2$}{lm6,rm6}
\fmf{dots,tension=0}{m9,m1}
\end{fmfgraph*}
\end{fmffile}
\hspace{1cm}
\begin{fmffile}{zzz-Decay2}
\begin{fmfgraph*}(55, 40)
\fmfleftn{l}{9}\fmfrightn{r}{9}
\fmf{phantom}{r4,rm4,m4,lm4,l4}
\fmf{phantom}{l9,m9,r9}
\fmf{phantom}{r1,m1,l1}
\fmf{phantom,tension=2}{r1,m1}
\fmf{phantom,tension=2}{l9,m9}
\fmf{fermion,tension=0.3,label.side=left,label=$p-q/2$}{l7,r7}
\fmf{fermion,tension=0.3,label.side=left,label=$p+q/2$}{r4,rm4}
\fmf{fermion,tension=0.3,label.side=left,label=$p+q/2$}{lm4,l4}
\fmfdot{r4,l4,r7,l7}
\fmfv{d.sh=circle,d.f=20,d.si=0.12w}{rm4}
\fmfv{d.sh=circle,d.f=20,d.si=0.12w}{lm4}
\fmf{fermion,tension=0.2,label.side=right,label=$k+l$}{lm4,rm4}
\fmf{fermion,tension=0,right=0.6,label.side=left,label=$p+l$}{rm4,lm4}
\fmf{fermion,tension=0,left=0.6,label.side=left,label=$k+q/2$}{rm4,lm4}
\fmf{dots,tension=0}{m9,m1}
\end{fmfgraph*}
\end{fmffile}
\end{center}

\vspace{0.2cm}

\begin{center}
\begin{fmffile}{zzz-PHExchange1}
\begin{fmfgraph*}(45, 35)
\fmfleftn{l}{5}\fmfrightn{r}{5}
\fmf{phantom}{l4,m4,r4}
\fmf{phantom}{r2,m2,l2}
\fmf{fermion,tension=0.3,label.side=left,label=$p-q/2$}{l4,m4}
\fmf{fermion,tension=0.3,label.side=left,label=$k+q/2$}{r2,m2}
\fmf{fermion,tension=0.3,label.side=left,label=$k-q/2$}{m4,r4}
\fmf{fermion,tension=0.3,label.side=left,label=$p+q/2$}{m2,l2}
\fmfdot{r2,l2,r4,l4}
\fmfv{d.sh=circle,d.f=20,d.si=0.12w}{m4}
\fmfv{d.sh=circle,d.f=20,d.si=0.12w}{m2}
\fmf{fermion,tension=0,left=0.4,label.side=left,label=$k+l$}{m2,m4}
\fmf{fermion,tension=0,left=0.4,label.side=left,label=$p+l$}{m4,m2}
\fmf{dots}{r5,l1}
\end{fmfgraph*}
\end{fmffile}
\hspace{1cm}
\begin{fmffile}{zzz-PHExchange2}
\begin{fmfgraph*}(45, 35)
\fmfleftn{l}{5}\fmfrightn{r}{5}
\fmf{phantom}{l4,m4,r4}
\fmf{phantom}{r2,m2,l2}
\fmf{fermion,tension=0.3,label.side=left,label=$p-q/2$}{l4,m4}
\fmf{fermion,tension=0.3,label.side=left,label=$k+q/2$}{r2,m2}
\fmf{fermion,tension=0.3,label.side=left,label=$k-q/2$}{m4,r4}
\fmf{fermion,tension=0.3,label.side=left,label=$p+q/2$}{m2,l2}
\fmfdot{r2,l2,r4,l4}
\fmfv{d.sh=circle,d.f=20,d.si=0.12w}{m4}
\fmfv{d.sh=circle,d.f=20,d.si=0.12w}{m2}
\fmf{fermion,tension=0,left=0.4,label.side=left,label=$k+l$}{m2,m4}
\fmf{fermion,tension=0,left=0.4,label.side=left,label=$p+l$}{m4,m2}
\fmf{dots}{l5,r1}
\end{fmfgraph*}
\end{fmffile}
\end{center}

\vspace{0.2cm}

\begin{center}
\begin{fmffile}{zzz-PPExchange1}
\begin{fmfgraph*}(45, 35)
\fmfleftn{l}{5}\fmfrightn{r}{5}
\fmf{phantom}{l4,m4,r4}
\fmf{phantom}{r2,m2,l2}
\fmf{fermion,tension=0.3,label.side=left,label=$p-q/2$}{l4,m4}
\fmf{fermion,tension=0.3,label.side=right,label=$k-q/2$}{m2,r2}
\fmf{fermion,tension=0.3,label.side=right,label=$k+q/2$}{r4,m4}
\fmf{fermion,tension=0.3,label.side=left,label=$p+q/2$}{m2,l2}
\fmfdot{r2,l2,r4,l4}
\fmfv{d.sh=circle,d.f=20,d.si=0.12w}{m4}
\fmfv{d.sh=circle,d.f=20,d.si=0.12w}{m2}
\fmf{fermion,tension=0,right=0.4,label.side=right,label=$k-l$}{m4,m2}
\fmf{fermion,tension=0,left=0.4,label.side=left,label=$p+l$}{m4,m2}
\fmf{dots}{r5,l1}
\end{fmfgraph*}
\end{fmffile}
\hspace{1cm}
\begin{fmffile}{zzz-PPExchange2}
\begin{fmfgraph*}(45, 35)
\fmfleftn{l}{5}\fmfrightn{r}{5}
\fmf{phantom}{l4,m4,r4}
\fmf{phantom}{r2,m2,l2}
\fmf{fermion,tension=0.3,label.side=left,label=$p-q/2$}{l4,m4}
\fmf{fermion,tension=0.3,label.side=right,label=$k-q/2$}{m2,r2}
\fmf{fermion,tension=0.3,label.side=right,label=$k+q/2$}{r4,m4}
\fmf{fermion,tension=0.3,label.side=left,label=$p+q/2$}{m2,l2}
\fmfdot{r2,l2,r4,l4}
\fmfv{d.sh=circle,d.f=20,d.si=0.12w}{m4}
\fmfv{d.sh=circle,d.f=20,d.si=0.12w}{m2}
\fmf{fermion,tension=0,right=0.4,label.side=right,label=$k-l$}{m4,m2}
\fmf{fermion,tension=0,left=0.4,label.side=left,label=$p+l$}{m4,m2}
\fmf{dots}{l5,r1}
\end{fmfgraph*}
\end{fmffile}
\end{center}
The gray blobs represent full interaction vertices $iV$. The two cut sub-diagrams for $-D_1$ involve a quasiparticle decaying into two quasiparticles and a quasihole (or a hole decaying into two holes and a particle). The cut sub-diagrams for $-C^{ph}_1$ involve the exchange of an on-shell particle-hole pair, while the cut sub-diagrams for $-C^{pp}_1$ involve the exchange of an on-shell particle-particle (or hole-hole) pair. The computation of cut diagrams is explained in Section \ref{ssect_QFT_AppendixAB}; there, we will also argue that these three pairs are the only cut sub-diagrams that contribute at order $q$.

For $d\geq 3$ spatial dimensions, $C_1$ should scale as $\sim (q^0)^2$. This can be seen by counting the availability of collision channels constraint by energy and momentum conservation in the presence of FS~\cite{landau1959theory, abrikosov1975methods}. As we show in Section \ref{ssect_QFT_AppendixAB}, in $d\geq 3$ we can parametrize the cut sub-diagrams above by
\begin{eqnarray}
-D_1(p; q) = - \gamma(\p) \ \delta_{FS}(p) \ Z(\p)^3 \ (uu^\dagger)(p) \: (uu^\dagger)^T(p) \ \frac{|q^0| (q^0)^2}{(v^\mu(\p) q_\mu)^2},
\label{Decay_parametrization}
\end{eqnarray}
\begin{eqnarray}
-C^{ph}_1(p, k; q) = 2\lambda^{ph}(\p, \kk) \ \delta_{FS}(p) \ \delta_{FS}(k) \ (Z^2 uu^\dagger)(p) (Z^2 uu^\dagger)^T(k) \ \frac{|q^0|(q^0)^2}{(v(\p)^\mu q_\mu) (v(\kk)^\mu q_\mu)},
\label{Cph_parametrization}
\end{eqnarray}
\begin{eqnarray}
-C^{pp}_1(p, k; q) = - \lambda^{pp}(\p, \kk) \ \delta_{FS}(p) \ \delta_{FS}(k) \ (Z^2 uu^\dagger)(p) (Z^2 uu^\dagger)^T(k) \ \frac{|q^0|(q^0)^2}{(v(\p)^\mu q_\mu) (v(\kk)^\mu q_\mu)}.
\label{Cpp_parametrization}
\end{eqnarray}
(We have omitted the $i\epsilon$ prescription accompanying $v^\mu q_\mu$ in the denominator; in time-ordered correlation its sign should be $-\sgn(q^0)$, i.e. $\sgn(q_0)$, as usual.)
 In particular, the parameter $\gamma(\p)$, defined near the FS, is positive and regular, and is related to the imaginary part of the fermion self-energy via
\begin{eqnarray}
\chi_u(p) = \frac{p^0 - \xi(\p) + i\epsilon\: \sgn\,\xi(\p)}{Z(\p)} + i\frac{3}{2}\gamma(\p) \ p^0 |p^0| + (\mbox{higher orders in } p^0)
\label{chiu_near_FS_good}
\end{eqnarray}
as explained in Section \ref{ssect_QFT_AppendixAB}. The other two parameters, $\lambda^{ph}(\p, \kk)$ and $\lambda^{pp}(\p, \kk)$, defined near the FS, are both positive and regular, and symmetric under exchange of $\p$ and $\kk$. Moreover, from the computation in Section \ref{ssect_QFT_AppendixAB}, we have the relation
\begin{eqnarray}
\int_k \frac{-v^\mu(\kk) q_\mu}{Z(\kk)} (C^{ph}_R)_1(p, k; q) = 2\frac{v^\mu(\p) q_\mu}{Z(\p)}(D_R)_1(p; q) = 2\int_k \frac{v^\mu(\kk) q_\mu}{Z(\kk)} (C^{pp}_R)_1(p, k; q).
\label{C_D_relation}
\end{eqnarray}
In terms of the parameters $\gamma$, $\lambda^{ph}$ and $\lambda^{pp}$, this reads
\begin{eqnarray}
\int_{k} \ \delta_{FS}(k) \ Z(\kk) \ \lambda^{ph}(\p, \kk) = \gamma(\p) = \int_{k} \ \delta_{FS}(k) \ Z(\kk) \ \lambda^{pp}(\p, \kk).
\label{gamma_lambda_relation}
\end{eqnarray}
In particular, this relation implies
\begin{eqnarray}
\int_k C_1(p, k; q) \ \frac{uu^\dagger(k)}{Z(\kk)} v^\mu(\kk) q_\mu = 0.
\label{Collision_WTId}
\end{eqnarray}
This is related to the Ward-Takahashi identity, as we will see in Section \ref{sssect_QFT_WTId}. More physically, it is related to the fact that collisions do not change the total number of fermionic excitations, as discussed above \eqref{Collision_conservation}.

Piecing up the above, the collision factor $C_1$ can be written as
\begin{eqnarray}
-C_1(p, k; q) &=& -|q^0| \ \frac{(Zuu^\dagger)(p) \: q^0}{v^\mu(\p) q_\mu} \ \delta_{FS}(p) \ \mathcal{C}(\p, \kk) \ \delta_{FS}(k) \frac{(Zuu^\dagger)^T(k) \: q^0}{v^\mu(\kk) q_\mu}
\end{eqnarray}
where $\mathcal{C}(\p, \kk)$, symmetric in $\p, \kk$, is microscopically defined when both $\p$ and $\kk$ are on the FS:
\begin{eqnarray}
\delta_{FS}(p) \ \mathcal{C}(\p, \kk) \ \delta_{FS}(k) &\equiv& \delta_{FS}(p) Z(\p) \gamma(\p) \: (2\pi)^{d+1}\delta^{d+1}(p-k) \nonumber \\[.2cm]
&& + \ \delta_{FS}(p) Z(\p) \left(-2\lambda^{ph}(\p, \kk) + \lambda^{pp}(\p, \kk) \right) \delta_{FS}(k) Z(\kk)
\label{C_kernal_def}
\end{eqnarray}
(since $\mathcal{C}$ is defined only when $\p, \kk$ are on the FS, we can ``remove'' $\delta_{FS}(p)\delta_{FS}(k)$ from the first term on the right-hand-side unambiguously) and it satisfies
\begin{eqnarray}
\int_k \mathcal{C}(\p, \kk) \ \delta_{FS}(k) = 0.
\label{Collision_WTId_alt}
\end{eqnarray}
This is the collision effect $\mathcal{C}$ appearing in the kinetic theory in Section \ref{ssect_kinetic_BE}.

For $d=2$ spatial dimensions, $D_1$, $C^{ph}_1$ and $C^{pp}_1$ cannot be parametrized in any simple form, as discussed in Section \ref{ssect_QFT_AppendixAB}. Moreover, they are not order $q$; they also involve order $q \ln q$ terms, which are less suppressed than order $q$. The failure of the parametrization raises problem in the computation for e.g. the longitudinal current in $d=2$. But as shown in Section \ref{ssect_QFT_AppendixAB}, collisions do not contribute to the anomalous Hall current and the chiral magnetic current, so our main discussion about them is not undermined. Also, despite that there is no simple parametrization in $d=2$, \eqref{Collision_WTId} must still hold as it is dictated by the Ward-Takahashi identity.

\subsubsection{Chemical Potential dependence of Propagator}
\label{sssect_QFT_int_Prop_on_EF}

Having defined the $q$-2PI vertex $i\wt{V}$, we are ready to find the chemical potential dependence of the propagator. The procedure below is analogous to \cite{Nozieres:1962zz}, but allowing multi-component $u^\alpha$.

We define the notation
\begin{eqnarray}
\partial^F \equiv \partial/\partial \epsilon_F - \partial_{p^0}.
\end{eqnarray}
The subtraction of $\partial_{p^0}$ is because our $p^0$ is defined such that $p^0=0$ at the FS, and we want $\partial^F$ to extract the effects of physically shifting the FS; for example, $\partial^F (p^0-(E-\epsilon_F)) = \partial^F E = \partial E/ \partial \epsilon_F$. The FS dependence of the propagator can be derived in analogy to \eqref{G_derivative}, but with $\partial_p^\nu$ replaced with $\partial^F$:
\begin{eqnarray}
\partial^F iG = i\Delta_0 \ \partial^F iG^{-1} - (Zuu^\dagger) \left(1-\partial^F E \right) \delta_{FS}
\label{G_FSderivative}
\end{eqnarray}
where the expression of $\partial^F G^{-1}$, and hence $\partial^F E$, are to be derived below. The second term in \eqref{G_FSderivative} relied on the assumption that when the chemical potential changes, the FS changes continuously, so \eqref{G_FSderivative} (and hence all discussions below) does not apply to discrete values of $\epsilon_F$ around which the FS develops new disconnected components, as in the example Figure \ref{EF_increase_disconnect_comp}.

Let $G_{bare}$ be the bare fermion propagator and $G_{bare}^{-1}$ is its inverse ignoring the $i\epsilon$ pole structure. In the kinetic energy sector of the bare Lagrangian, $\epsilon_F$ always appears as $i\partial_{x^0} + \epsilon_F$, therefore $\partial^F G_{bare}^{-1} = 0$. Because $G^{-1}=G_{bare}^{-1} - \Sigma$ where $\Sigma$ is the self-energy, we get $\partial^F G^{-1} = -\partial^F \Sigma$. Diagrammatically, one can see in the presence of interaction, when the propagator is varied, the self-energy varies as $-\delta i\Sigma = i\wt{V}_0 \ \delta iG$. Therefore
\begin{eqnarray}
\partial^F iG^{-1} = -\partial^F i\Sigma = i\wt{V}_0 \ \partial^F iG.
\end{eqnarray}
Substituting \eqref{G_FSderivative} into the above yields
\begin{eqnarray}
\partial^F iG^{-1} = -\partial^F i\Sigma = - i\bar{V}_0 \ (Zuu^\dagger) \left(1-\partial^F E \right) \delta_{FS}.
\label{Ginv_FSderivative}
\end{eqnarray}
Recall that $\bar{V}_0$ is defined by the recursion relation $i\bar{V}_0 = i\wt{V}_0 + i\wt{V}_0 \ i\Delta_0 \ i\bar{V}_0$.

Let's focus on the change of the $u$-band eigenvalue of $G^{-1}$, given by
\begin{eqnarray}
\partial^F \chi_u = (uu^\dagger)^T \partial^F G^{-1}.
\end{eqnarray}
Now take $p$ near the FS. We can expand this in powers of $p^0$. In particular, by comparison with \eqref{chiu_near_FS}, we shall identify the coefficients at zeroth and first order in $p^0$ as
\begin{eqnarray}
\partial^F \chi_u = \left(-\frac{\partial^F E}{Z} - (E - \epsilon_F) \: \partial_{F} \frac{1}{Z} \right) + p^0 \: \partial^F \frac{1}{Z} + \mathcal{O}((p^0)^2).
\end{eqnarray}
To make this parallel with \eqref{chiu_derivative}, we shall define $v^F\equiv \partial^F E$. Notice $v^F$ has nothing to do with ``Fermi velocity'' (in this thesis there is no notion of Fermi velocity, as we did not assume rotational symmetry).

Now, for $p$ near the FS, we can read-off:
\begin{eqnarray}
\partial^F Z(\p) &=& \left.\partial_{p^0} \left( -Z^2 (uu^\dagger)^T \partial^F G^{-1} \right) \right|_{p \: on-shell} \nonumber \\[.2cm]
&=& \left. Z(\p) \ \partial_{p^0} \int_k (Zuu^\dagger)^T(p) \bar{V}_0(p, k) (Zuu^\dagger)(k) \ \left(1-\partial^F E(\kk) \right)\delta_{FS}(k) \right|_{p \: on-shell}
\end{eqnarray}
and
\begin{eqnarray}
&& \partial^F E(\p) = \left. -(Zuu^\dagger)^T \partial^F G^{-1} \right|_{p \: on-shell} = \int_k \mathcal{U}(\p, \kk) \ \left(1-\partial^F E(\kk) \right)\delta_{FS}(k), \nonumber \\[.2cm]
&& \mathcal{U}(\p, \kk) \equiv \left. (Zuu^\dagger)^T(p) \ \bar{V}_0(p, k) \ (Zuu^\dagger)(k) \right|_{p, k \: on-shell}.
\label{dE_dEF}
\end{eqnarray}
Thus we have proven \eqref{BFL_dE_dEF}. At the same time we found the microscopic expression for $\mathcal{U}$, which is the same as that in \cite{abrikosov1975methods} except here we need to contract with the four $u$'s. $\mathcal{U}$ is even under the exchange of $\p, \kk$, because $\bar{V}_0=(\bar{V}_0)^T$.

We can also find the change of the eigenvector $u$ for $p$ near the FS. Up to an unimportant complex phase, we have
\begin{eqnarray}
\partial^F u^\alpha(p) = \sum_w \frac{w^\alpha}{-\chi_w} \left\{ w^\dagger \partial^F G^{-1} u \right\}, \ \ \ \ \ \ \ \partial^F u^\dagger_\alpha(p) = \sum_w \left\{ u^\dagger \partial^F G^{-1} w \right\}\frac{w^\dagger_\alpha}{-\chi_w},
\label{du_dEF}
\end{eqnarray}
where $\partial^F G^{-1}$ is given by \eqref{Ginv_FSderivative}. $\partial^F \mathfrak{u}$ is related to $\partial^F u$ in a way similar to \eqref{u_derivative_on_shell}, with $v^\mu$ replaced by $v^F=\partial^F E$. It appears in the kinetic formalism through \eqref{BFL_dsigma_dEF_detail}, which we will prove in Section \ref{sssect_QFT_dsigma_dEF}.

\subsection{Electromagnetic Coupling}
\label{ssect_QFT_EM_WTId}

\subsubsection{Bare Electromagnetic Vertices}

By our assumptions about the QFT, the only bare EM coupling vertices take the form $A^n \psi^\dagger \psi$ for integer $n\geq 1$. Due to the smallness of $A$, we only need to concern about $n=1, 2$. In particular, we denote by $(i\wt{\Gamma}^\alpha_{\ \delta})^\mu(p; q)$ the bare $A\psi^\dagger\psi$ EM vertex with incoming fermion with momentum $p-q/2$ and index $\delta$, and outgoing fermion with momentum $p+q/2$ and index $\alpha$. We denote by $(i\wt{\Xi}^\alpha_{\ \delta})^{\mu\nu}(p; q, q')$ the bare $AA\psi^\dagger\psi$ vertex with incoming fermion with momentum $p-(q+q')/2$ and index $\delta$, outgoing fermion with momentum $p+(q+q')/2$ and index $\alpha$, and photons with incoming momenta $q$ and $q'$.
\begin{center}
\vspace{.5cm}
\hspace{1cm}
\begin{fmffile}{zzz-wtGamma}
\begin{fmfgraph*}(30,30)
\fmfleftn{l}{5}\fmfrightn{r}{5}
\fmf{fermion}{v,l1}
\fmf{fermion}{l5,v}
\fmf{photon,tension=1.5,label.dist=9,label=$q$}{r3,v}
\fmfv{label.dist=11,label={\large $i\wt{\Gamma}$}}{v}
\fmfv{label.dist=-12,label=$p+q/2$}{l2}
\fmfv{label.dist=-12,label=$p-q/2$}{l4}
\fmflabel{$\alpha$}{l1}
\fmflabel{$\delta$}{l5}
\fmflabel{$\mu$}{r3}
\momentumarrow{a}{up}{6}{r3,v}
\end{fmfgraph*}
\end{fmffile}
\hspace{4.5cm}
\begin{fmffile}{zzz-wtXi}
\begin{fmfgraph*}(30,30)
\fmfleftn{l}{5}\fmfrightn{r}{5}
\fmf{fermion}{v,l1}
\fmf{fermion}{l5,v}
\fmf{photon,tension=0.7,label.dist=9,label=$q' \hspace{-1.5mm}$}{r2,v}
\fmf{photon,tension=0.7,label.dist=9,label=$q \hspace{-1mm}$}{r4,v}
\fmfv{label.dist=11,label={\large $i\wt{\Xi}$}}{v}
\fmfv{label.dist=-10,label=$p+(q+q')/2$}{l2}
\fmfv{label.dist=-10,label=$p-(q+q')/2$}{l4}
\fmflabel{$\alpha$}{l1}
\fmflabel{$\delta$}{l5}
\fmflabel{$\mu$}{r4}
\fmflabel{$\nu$}{r2}
\momentumarrow{a}{down}{6}{r2,v}
\momentumarrow{b}{up}{6}{r4,v}
\end{fmfgraph*}
\end{fmffile}
\vspace{.5cm}
\end{center}
Since they are bare quantities, both of them are regular and analytic as $q\rightarrow 0$ (also $q'\rightarrow 0$ for $\wt{\Xi}$). To first order in $q$ (and $q'$ together for $\wt{\Xi}$) we separate them as $\wt{\Gamma}=\wt{\Gamma}_0+\wt{\Gamma}_1$ and $\wt{\Xi}=\wt{\Xi}_0+\wt{\Xi}_1$.

The EM $U(1)$ gauge invariance of the bare Lagrangian requires
\begin{eqnarray}
&& q_\mu \: i\wt{\Gamma}^\mu(p) = iG_{bare}^{-1}(p-q/2) - iG_{bare}^{-1}(p+q/2),
\end{eqnarray}
\begin{eqnarray}
&& q_\mu \: i\wt{\Xi}^{\mu\nu}(p; q, q') = i\wt{\Gamma}^\nu(p-q/2; q') - i\wt{\Gamma}^\nu(p+q/2; q'), \nonumber \\[.2cm]
&& q'_\nu \: i\wt{\Xi}^{\mu\nu}(p; q, q') = i\wt{\Gamma}^\mu(p-q'/2; q) - i\wt{\Gamma}^\mu(p+q'/2; q).
\end{eqnarray}
These lead to
\begin{eqnarray}
i\wt{\Gamma}_0^\mu = -i\partial_p^\mu G_{bare}^{-1}, \ \ \ \ \ i\wt{\Gamma}_1^\mu = i\hat{\mu}^{\mu\nu} \ iq_\nu,
\label{wtGamma_gaugeinv}
\end{eqnarray}
\begin{eqnarray}
i\wt{\Xi}_0^{\mu\nu}(p) = -\partial_p^\nu i\wt{\Gamma}_0^\mu(p) = -\partial_p^\mu i\wt{\Gamma}_0^\nu(p), \ \ \ \ \ \ i\wt{\Xi}_1^{\mu\nu}(p; q, q') = -\partial_p^\nu i\wt{\Gamma}_1^\mu(p; q) - \partial_p^\mu i\wt{\Gamma}_1^\nu(p; q').
\label{wtXi_gaugeinv}
\end{eqnarray}
Here $(\hat{\mu}^\alpha_{\ \delta})^{\mu\nu}(p)$ is the bare EM dipole matrix (such as that in the Pauli term) that is antisymmetric in $\mu\nu$ and Hermitian in $^\alpha_{\ \delta}$.

\subsubsection{Full Electromagnetic Vertex}

\begin{eqnarray*}
\parbox{30mm}{
\begin{fmffile}{zzz-Gamma-recursion}
\begin{fmfgraph*}(25, 15)
\fmfleftn{l}{3}\fmfrightn{r}{3}
\fmf{phantom}{r1,G1,l1}
\fmf{phantom}{l3,G2,r3}
\fmf{fermion}{G1,l1}
\fmf{fermion}{l3,G2}
\fmffreeze
\fmfpoly{empty,label=$i\Gamma$}{G2,G1,G3}
\fmf{photon,tension=4}{r2,G3}
\end{fmfgraph*}
\end{fmffile}
}
= \hspace{.5cm}
\parbox{20mm}{
\begin{fmffile}{zzz-Gamma-recursion-wtGamma}
\begin{fmfgraph*}(15, 15)
\fmfleftn{l}{3}\fmfrightn{r}{3}
\fmf{fermion}{v,l1}
\fmf{fermion}{l3,v}
\fmf{photon,tension=1.5}{r2,v}
\end{fmfgraph*}
\end{fmffile}
}
+\hspace{.5cm}
\parbox{45mm}{
\begin{fmffile}{zzz-Gamma-recursion-wtV-Gamma}
\begin{fmfgraph*}(40, 15)
\fmfleftn{l}{5}\fmfrightn{rr}{5}
\fmf{phantom}{l5,r5}\fmf{phantom,tension=3.5}{r5,rr5}
\fmf{phantom}{l4,r4}\fmf{phantom,tension=3.5}{r4,rr4}
\fmf{phantom}{l3,r3}\fmf{phantom,tension=3.5}{r3,rr3}
\fmf{phantom}{l2,r2}\fmf{phantom,tension=3.5}{r2,rr2}
\fmf{phantom}{l1,r1}\fmf{phantom,tension=3.5}{r1,rr1}
\fmffreeze
\fmf{phantom}{l4,lv4,rv4,r4}
\fmf{phantom}{r2,rv2,lv2,l2}
\fmf{fermion,tension=0}{l4,lv4}
\fmf{fermion,tension=0}{r2,rv2}
\fmf{fermion,tension=0}{rv4,r4}
\fmf{fermion,tension=0}{lv2,l2}
\fmf{phantom}{l3,o,r3}
\fmfv{d.sh=circle,d.f=empty,d.si=0.32w}{o}
\fmf{phantom,label.dist=0,label=$i\wt{V}$}{l3,r3}
\fmffreeze
\fmfpoly{empty,label=$i\Gamma$}{r4,r2,g}
\fmf{photon}{rr3,g}
\end{fmfgraph*}
\end{fmffile}
}
\end{eqnarray*}
Diagrammatically, one can see the full $A\psi^\dagger\psi$ EM vertex is given by the recursion relation
\begin{eqnarray}
i\Gamma^\mu = i\wt{\Gamma}^\mu + iV \ \left\{ iG \ i\wt{\Gamma}^\mu \ iG \right\} = i\wt{\Gamma}^\mu + i\wt{V} \ \left\{ iG \ i\Gamma^\mu \ iG \right\}
\end{eqnarray}
as drawn above. The recursion relation at zeroth order in $q$ is
\begin{eqnarray}
i\Gamma_0^\nu &=& i\wt{\Gamma}_0^\nu + i\wt{V}_0 \left(i\Delta_0 + i\Delta'_0 \right) i\Gamma_0^\nu \nonumber \\[.2cm]
&=& i\bar{\Gamma}_0^\nu + i\bar{V}_0 \ i\Delta'_0 \ i\Gamma_0^\nu,
\label{Gamma0_recursion}
\end{eqnarray}
where we defined
\begin{eqnarray}
i\bar{\Gamma}_0^\nu \equiv \left(\mathbf{1}+i\bar{V}_0 \ i\Delta_0\right) i\wt{\Gamma}_0^\nu.
\end{eqnarray}
The purpose of the second equality of \eqref{Gamma0_recursion} is that, now the effect of $\Delta'_0$, to be related to the deformation of the FS later, is singled out, and $i\bar{\Gamma}_0$ is independent of $q$.

The recursion relation at first order in $q$ is
\begin{eqnarray}
i\Gamma_1^\nu - i\bar{V}_0 \ i\Delta'_0 \ i\Gamma_1^\nu &=& \left(\mathbf{1}+i\bar{V}_0 \ i\Delta_0\right) i\wt{\Gamma}_1^\nu + i\bar{V}_0 \left(i\Delta'_1+i\Delta^r_1+i\Delta^s_1 - C_1\right) i\Gamma_0^\nu \phantom{,} \nonumber \\[.2cm]
&& + \left(\mathbf{1}+i\bar{V}_0 \ i\Delta_0\right) i\wt{V}_1 \left( i\Delta_0 + i\Delta'_0 \right) i\Gamma_0^\nu.
\label{Gamma1_recursion_1}
\end{eqnarray}
Of course the recursion \eqref{Gamma1_recursion_1} can be expressed in many equivalent ways; we have chosen to express it such that on the right-hand-side there is no $\Delta'_0$ (including those hidden in $\Gamma_0^\nu$) to the left of any quantity of order $q$. For the purpose of deriving the Boltzmann equation, we want to further rewrite \eqref{Gamma1_recursion_1} so that each $\Gamma_0^\nu$ has $\Delta'_0$ or $\Delta'_1$ or $C_1$ on its immediate left. We can achieve so by substituting \eqref{Gamma0_recursion} for those $\Gamma_0$'s in \eqref{Gamma1_recursion_1} whose immediate left are not yet $\Delta'_0$ or $\Delta'_1$ or $C_1$. The result is
\begin{eqnarray}
i\Gamma_1^\nu = i\bar{\Gamma}_1^\nu + i\bar{V}_1 \ i\Delta'_0 \ i\Gamma_0^\nu + i\bar{V}_0 \ i\Delta'_1 \ i\Gamma_0^\nu + + i\bar{V}_0 \ (-C_1) \ i\Gamma_0^\nu + i\bar{V}_0 \ i\Delta'_0 \ i\Gamma_1^\nu,
\label{Gamma1_recursion}
\end{eqnarray}
where
\begin{eqnarray}
i\bar{V}_1 \equiv \left(\mathbf{1}+i\bar{V}_0 \ i\Delta_0\right) i\wt{V}_1 \left(\mathbf{1}+i\bar{V}_0 \ i\Delta_0\right)^T + i\bar{V}_0 \left(i\Delta^r_1+i\Delta^s_1\right) i\bar{V}_0,
\label{barV1}
\end{eqnarray}
\begin{eqnarray}
i\bar{\Gamma}_1^\nu &\equiv& \left(\mathbf{1}+i\bar{V}_0 \ i\Delta_0\right) i\wt{\Gamma}_1^\nu + i\bar{V}_1 \ i\Delta_0 \ i\wt{\Gamma}_0^\nu + i\bar{V}_0 \left(i\Delta^r_1+i\Delta^s_1\right) i\wt{\Gamma}_0^\nu \nonumber \\[.2cm]
&=& \left(\mathbf{1}+i\bar{V}_0 \ i\Delta_0\right) \left(i\wt{\Gamma}_1^\nu + i\wt{V}_1 \ i\Delta_0 \ i\bar{\Gamma}_0^\nu \right) + i\bar{V}_0 \left(i\Delta^r_1+i\Delta^s_1\right) i\bar{\Gamma}_0^\nu
\label{barGamma1}
\end{eqnarray}
are partial sums at first order in $q$ that involve no factor of $\Delta'$ or $C_1$. By construction, $i\bar{V}_1$ and $i\bar{\Gamma}_1$ are analytic in $q$ as $q\rightarrow 0$. Thus, in \eqref{Gamma1_recursion} we singled out the $\Delta'_0, \Delta'_1$ and $C_1$ effects, which are to be related to the quasiparticle excitations.

We do not need to consider the ``full $AA\psi^\dagger \psi$ vertex''. In fact, the only place $\wt{\Xi}$ shows up in our proof is the expression of the current, in which we will immediately use \eqref{wtXi_gaugeinv} to eliminate $\wt{\Xi}$.

\subsubsection{Ward-Takahashi Identity}
\label{sssect_QFT_WTId}

Later in Section \ref{ssect_QFT_BE} we will show how the Boltzmann equation \eqref{BFL_Boltzmann_Eq_0}\eqref{BFL_Boltzmann_Eq_1} follow exactly from \eqref{Gamma0_recursion} and \eqref{Gamma1_recursion}. Before that, we need to answer a question: In QFT, it is the matrix $\Gamma^\nu$ governing the coupling to $A$, while in the kinetic formalism, it is the velocity $v^\nu$ (plus order $q$ couplings such as EM dipole). How to relate $\Gamma^\nu$ to $v^\nu$? The answer is the generalized Ward-Takahashi identity~\cite{takahashi1957generalized}:
\begin{eqnarray}
\left\{iG(p+q/2) \ i\Gamma^\nu(p; q) \ iG(p-q/2) \right\} q_\nu = iG(p-q/2) - iG(p+q/2).
\label{WTId}
\end{eqnarray}
We want to extract its implications at leading and sub-leading orders in $q$ in the presence of FS.

At leading order in $q$, the Ward-Takahashi identity reads
\begin{eqnarray}
\left( i\Delta_0 + i\Delta'_0 \right) \Gamma_0^\nu \ q_\nu = -i\Delta_0 \ \partial_p^\nu G^{-1} \ q_\nu + i(Zuu^\dagger) \delta_{FS} \delta^\nu_i v^i q_\nu
\label{WTId_1_raw}
\end{eqnarray}
using \eqref{G_derivative}. This is equivalent to
\begin{eqnarray}
\Gamma_0^\nu \ q_\nu = -\partial_p^\nu G^{-1} \ q_\nu.
\label{WTId_1}
\end{eqnarray}
One can easily verify the equivalence by contracting $i\Delta_0+i\Delta'_0$ on the left of \eqref{WTId_1}, with the aid of \eqref{v_in_QFT}, to recover \eqref{WTId_1_raw}. We will see the result \eqref{WTId_1} is related to the gauge invariance of $\delta f_0$.

We can extract more detailed information from \eqref{WTId_1} -- we gain an identity similar to the original Ward identity~\cite{ward1950identity}, but in the presence of FS. For this purpose let's treat $|\q|/q^0$ as an independent small expansion parameter, and expand \eqref{WTId_1} to its zeroth and first order. This gives us two equations, about $\bar{\Gamma}^0_0$ and $\bar{\Gamma}^i_0$ respectively. Solving them with the help of \eqref{Gamma0_recursion} and the explicit expression for $\Delta'_0$, we find the Ward identity in the presence of FS:
\begin{eqnarray}
\bar{\Gamma}_0^\nu = -\partial_p^\nu G^{-1} + \bar{V}_0 \ (Zuu^\dagger) \delta_{FS} \delta^\nu_i v^i.
\label{Gamma0_WTId}
\end{eqnarray}
We can equivalently express \eqref{Gamma0_WTId} as
\begin{eqnarray}
i\Delta_0 \ i\bar{\Gamma}_0^\nu = -\partial_p^\nu iG + \left(\mathbf{1} + i\Delta_0 \ i\bar{V}_0 \right) (Zuu^\dagger) \delta_{FS} \delta^\nu_i v^i
\label{Gamma0_WTId_1}
\end{eqnarray}
using \eqref{G_derivative}. As we will see later, this result will help us relate the EM vertex in QFT to the velocity in the kinetic formalism.

At sub-leading order in $q$, the Ward-Takahashi identity reads
\begin{eqnarray}
\left(\Delta_0+\Delta'_0\right) \Gamma_1^\nu \ q_\nu + \left(\Delta^r_1+\Delta^s_1+\Delta'_1+iC_1\right) \Gamma_0^\nu \ q_\nu = 0.
\end{eqnarray}
The $C_1 \Gamma_0^\nu q_\nu$ term vanishes on its own, due to \eqref{WTId_1}, \eqref{v_in_QFT} and \eqref{Collision_WTId}. For the remaining terms, we can conclude
\begin{eqnarray}
\Delta_0 \ \Gamma_1^\nu \ q_\nu + \Delta^r_1 \ \Gamma_0^\nu \ q_\nu = 0 = \Delta'_0 \ \Gamma_1^\nu \ q_\nu + \left(\Delta^s_1+\Delta'_1\right) \Gamma_0^\nu \ q_\nu.
\label{WTId_2}
\end{eqnarray}
The two sides must vanish separately because the right-hand-side involves the singular factor $\delta(\xi(\p))$, while the left-hand-side does not. Later we will see \eqref{WTId_2} is related to the gauge invariance of $\delta f_1$.

\subsection{Boltzmann Equation}
\label{ssect_QFT_BE}

Having extracted \eqref{Gamma0_WTId} from the Ward-Takahashi identity, we are ready to prove the Boltzmann equation \eqref{BFL_Boltzmann_Eq_0}\eqref{BFL_Boltzmann_Eq_1} from the recursion relations \eqref{Gamma0_recursion} and \eqref{Gamma1_recursion}. The distribution of excitations $\delta f$ will be defined in terms of QFT quantities, and as one should expect, our definition agrees with the Wigner function approach. Our derivation also provides the microscopic expressions for $\mathcal{V}^\nu$ and $\mu^{\mu\nu}$.

\subsubsection{Zeroth Order in $q$}

When an external EM field of small $q$ is present, the propagation of a quasiparticle is no longer translationally invariant -- the two-point propagator now depends on both $p$ and $q$. More precisely, 
\begin{eqnarray}
iG(p) \ \ \ \longrightarrow \ \ \ iG(p) + \left\{iG(p+q/2) \ i\Gamma^\nu(p; q) \ iG(p-q/2) \right\} A_\nu(q)
\label{prop_shift_by_A}
\end{eqnarray}
at linear response. We will focus on the shifted piece.

At zeroth order in $q$, using the identity \eqref{Gamma0_WTId}, we can express the recursion relation \eqref{Gamma0_recursion} as
\begin{eqnarray}
i\Gamma_0^\nu A_\nu = -i\partial_p^\nu G^{-1} A_\nu - i\bar{V}_0 (Zuu^\dagger) \ \delta W_0,
\label{Gamma0_in_W0}
\end{eqnarray}
where $\delta W_0$ is a quantity restricted on the FS:
\begin{eqnarray}
\delta W_0 \equiv \frac{(uu^\dagger)^T}{Z} \ \Delta'_0 \ \Gamma_0^\nu A_\nu - \delta_{FS} v^i A_i.
\label{W0_in_Gamma0}
\end{eqnarray}
We can see $\delta W_0$ is gauge invariant from \eqref{WTId_1} and \eqref{v_in_QFT}. Substituting \eqref{Gamma0_in_W0} into \eqref{W0_in_Gamma0}, we find the recursion relation for $\delta W_0$:
\begin{eqnarray}
\delta W_0 = \delta_{FS} \ \frac{v^i}{v^\mu q_\mu-i\epsilon\: \sgn(q^0)} \left(-iF_{i0} - q_i \ \mathcal{U} \ \delta W_0 \right).
\label{Boltzmann_Eq_0}
\end{eqnarray}
This proves the Boltzmann equation \eqref{BFL_Boltzmann_Eq_0} at zeroth order in $q$, if we make the identification
\begin{eqnarray}
2\pi \delta(p^0-(E(\p)-\epsilon_F)) \ \delta f (\p; q) \equiv \delta W(p; q)
\end{eqnarray}
to factor out the on-shell condition. Note that the computation above is time-ordered, therefore the $i\epsilon$ prescription depends on $\sgn(q^0)$; when computing the physical quasiparticle distribution in kinetic theory, retarded boundary condition should be used, which corresponds to removing the $\sgn(q^0)$ factor in the $i\epsilon$ prescription. This proof is a generalization to that in \cite{abrikosov1975methods}, with multi-component spinor / Bloch state $u^\alpha$ and the presence of external EM field, and without rotational symmetry.

The definition \eqref{W0_in_Gamma0} of $\delta W_0$ agrees with the quasiparticle Wigner function to first order in $A$ and zeroth order in $q$. The first term of \eqref{W0_in_Gamma0} corresponds to the singular part of \eqref{prop_shift_by_A} projected onto the $u$ band (at zeroth order in $q$), which we identify as the distribution of excited quasiparticles; the factor of $Z$ difference is the quasiparticle wave function renormalization. The second term of \eqref{W0_in_Gamma0} is due to the Peierl's substitution in the equilibrium part $\theta(\epsilon_F-E)$ of the Wigner function; Fourier transforming to the position space, it corresponds to the Wilson loop at first order in $A$ in the Wigner function.

\subsubsection{First Order in $q$}

At first order in $q$, we assert we should define
\begin{eqnarray}
\delta W_1 \equiv \frac{(uu^\dagger)^T}{Z} \left(\Delta'_0 \ \Gamma_1^\nu A_\nu + \Delta'_1 \ \Gamma_0^\nu A_\nu + iC_1 \ \Gamma_0^\nu A_\nu\right).
\label{W1_in_Gamma}
\end{eqnarray}
Its gauge invariance follows from \eqref{WTId_2} and \eqref{Collision_WTId}. It also agrees with the order $q$ singular part of the Wigner function -- as can be seen from \eqref{prop_shift_by_A} -- projected onto the $u$ band. In particular, the projection onto the $u$ band should be done by the momentum space Wilson line
\begin{eqnarray}
\lim_{n\rightarrow \infty} \ \left(u^\alpha u_{\alpha_1 \phantom{\beta\!\!\!\!}}^\dagger\right)\left(p+q/2\right) \ \left(u^{\alpha_1} u_{\alpha_2 \phantom{\beta\!\!\!\!}}^\dagger\right)\left(p+q(n-1)/2n\right) \cdots \left(u^{\alpha_{2n}} u_\beta^\dagger\right)\left(p-q/2\right).
\end{eqnarray}
It equals $u^\alpha(p) u^\dagger_\beta(p) + \mathcal{O}(q^2)$, so we can just use $(uu^\dagger)^T(p)$ at first order in $q$.

Now we derive the kinetic recursion relation for $\delta W_1$. Substituting \eqref{Gamma0_recursion} and \eqref{Gamma1_recursion} into \eqref{W1_in_Gamma}, we have
\begin{eqnarray}
\!\!\!\!\! \delta W_1 &=& \frac{(uu^\dagger)^T}{Z} \left[-\Delta'_0 \ \bar{V}_1 \ \Delta'_0 \ \Gamma_0^\nu - \Delta'_0 \ \bar{V}_0 \ \Delta'_0 \ \Gamma_1^\nu + iC_1 \ \Gamma_0^\nu - \Delta'_0 \ \bar{V}_0 \ iC_1 \ \Gamma_0^\nu \right. \nonumber \\[.2cm]
&& \hspace{2cm} \left. - \ \left(\Delta'_0 \ \bar{V}_0 \ \Delta'_1 + \Delta'_1 \ \bar{V}_0 \ \Delta'_0\right) \Gamma_0^\nu + \left(\Delta'_0 \ \bar{\Gamma}_1^\nu + \Delta'_1 \ \bar{\Gamma}_0^\nu \right) \right] \ A_\nu.
\end{eqnarray}
We use the identity
\begin{eqnarray}
\Delta'_1 &=& (uu^\dagger) (uu^\dagger)^T \Delta'_1 + \Delta'_1 (uu^\dagger) (uu^\dagger)^T \nonumber \\[.2cm]
&=& (uu^\dagger) (uu^\dagger)^T \Delta'_1 + iq_\lambda \mathcal{A}^\lambda \ \Delta'_0 = \Delta'_0 \ iq_\lambda \mathcal{A}^\lambda + \Delta'_1 (uu^\dagger) (uu^\dagger)^T
\label{Delta_prime_1_Id}
\end{eqnarray}
and the fact $(uu^\dagger)^T \Delta'_1 (uu^\dagger) = 0$ to rewrite $\delta W_1$ as
\begin{eqnarray}
\delta W_1 &=& \frac{\Delta' (uu^\dagger)^T}{Z} \left[-\left( \bar{V}_1 + iq_\lambda \mathcal{A}^\lambda \ \bar{V}_0 + \bar{V}_0 \ iq_\lambda \mathcal{A}^\lambda \right) \Delta'_0 \ \Gamma_0^\nu + \left(\bar{\Gamma}_1^\nu + iq_\lambda \mathcal{A}^\lambda \ \bar{\Gamma}_0^\nu \right) \right. \nonumber \\[.2cm]
&& \hspace{4.3cm} \left. - \ \bar{V}_0 (uu^\dagger) (uu^\dagger)^T \left(\Delta'_0 \ \Gamma_1^\nu + \Delta'_1 \ \Gamma_0^\nu + iC_1 \ \Gamma_0^\nu\right) \right] A_\nu \nonumber \\[.2cm]
&& + \ \frac{(uu^\dagger)^T}{Z} iC_1 \ \Gamma_0^\nu A_\nu
\end{eqnarray}
The second line can be easily identified as $(1/Z^2) \Delta' \ \mathcal{U} \ \delta W_1$. In the first line, we substitute \eqref{W0_in_Gamma0} for $\Delta'_0 \Gamma_0^\nu A_\nu$. Then we define the gradient interaction potential via
\begin{eqnarray}
&& \hspace{-1.5cm} iq_\mu \: \mathcal{V}^\mu(\p, \kk) \equiv \nonumber \\[.2cm]
&& \hspace{-1cm} \left. (Zuu^\dagger)^T(p) \left[ \bar{V}_1(p, k; q) +iq_\mu \left( \mathcal{A}^\mu(p) \bar{V}_0(p, k) + \bar{V}_0(p, k) \mathcal{A}^\mu(k) \right) \right] (Zuu^\dagger)(k) \right|_{p, k \ on \ FS}
\label{grad_int_def}
\end{eqnarray}
(note that even if the microscopic interaction is contact interaction, in kinetic theory $\mathcal{V}^\mu$ is still non-zero) and define the EM dipole moment via
\begin{eqnarray}
iq_\mu \: \mu^{\mu\nu}(\p) \equiv \left. (Zuu^\dagger)^T \left( \bar{\Gamma}_1^\nu + iq_\mu \mathcal{A}^\mu \ \bar{\Gamma}_0^\nu \right) \right|_{p \ on \ FS} - iq_\mu \: \mathcal{V}^\mu \ \delta_{FS} \delta^\nu_i v^i.
\label{EM_dipole_def}
\end{eqnarray}
As we will show explicitly below, $\mu^{\mu\nu}$ is antisymmetric in $\mu\nu$. With these definitions, the recursion relation for $\delta W_1$ becomes
\begin{eqnarray}
\delta W_1 &=& \delta_{FS} \ \frac{v^i \: q_i}{v^\mu q_\mu - i\epsilon \: \sgn(q^0)} \left(\mu^{\nu\lambda} \frac{F_{\nu\lambda}}{2} - \mathcal{U} \ \delta W_1 - \mathcal{V}^\nu \ iq_\nu \ \delta W_0 \right) \nonumber \\[.2cm]
&& + \ \frac{(uu^\dagger)^T}{Z} iC_1 \ \Gamma_0^\nu A_\nu.
\label{Boltzmann_Eq_1_raw}
\end{eqnarray}
(The gauge invariance of $\delta W_1$ also implicitly requires the antisymmetry of $\mu^{\mu\nu}$.)

The last step is to rewrite the $C_1$ term:
\begin{eqnarray}
\frac{(uu^\dagger)^T}{Z} iC_1 \ \Gamma_0^\nu A_\nu &=& \frac{i|q^0| q^0}{v^\mu q_\mu} \ \delta_{FS} \ \mathcal{C} \ \delta_{FS} \left(\frac{v^i q_i}{v^\mu q_\mu} -1\right) (Z uu^\dagger)^T \ \Gamma_0^\nu A_\nu \nonumber \\[.2cm]
&=& \frac{i|q^0| q^0}{v^\mu q_\mu} \ \delta_{FS} \ \mathcal{C} \left[(\delta W_0 + \delta_{FS} v^i A_i) \phantom{(Z uu^\dagger)^T} \right. \nonumber \\[.2cm] 
&& \hspace{2cm} \left. - \delta_{FS} (Z uu^\dagger)^T \left(-\partial_p^\nu G^{-1} A_\nu - \bar{V}_0 (Zuu^\dagger) \ \delta W_0\right) \right] \nonumber \\[.2cm]
&=& \frac{i|q^0| q^0}{v^\mu q_\mu} \ \delta_{FS} \ \mathcal{C} \left(\delta W_0 + \delta_{FS} \ \mathcal{U} \ \delta W_0 \right)
\end{eqnarray}
where in the second equality we used \eqref{W0_in_Gamma0} and \eqref{Gamma0_in_W0}, and in the third equality we used \eqref{v_in_QFT} and \eqref{Collision_WTId}.

Now we have
\begin{eqnarray}
\delta W_1 &=& \frac{\delta_{FS}}{v^\mu q_\mu - i\epsilon \: \sgn(q^0)} \left[ v^i q_i \left(\mu^{\nu\lambda} \frac{F_{\nu\lambda}}{2} - \mathcal{U} \ \delta W_1 - \mathcal{V}^\nu \ iq_\nu \ \delta W_0 \right) \right. \nonumber \\[.2cm]
&& \hspace{5cm} \left. + \ i|q^0| q^0 \ \mathcal{C} \left(\delta W_0 + \delta_{FS} \ \mathcal{U} \ \delta W_0 \right) \phantom{\frac{1}{1}} \right].
\label{Boltzmann_Eq_1}
\end{eqnarray}
The computation done here is time-ordered. When computing physical quasiparticle distribution, we should use retarded boundary condition, which corresponds to using the retarded versions of $\Delta'$ and $C$ -- that is, to remove the $\sgn(q^0)$ on the $i\epsilon$ prescription, and remove the absolute value on $|q^0|$ in the collision term. This proves \eqref{BFL_Boltzmann_Eq_1}.

\subsubsection{Electromagnetic Dipole Moment}
\label{sssect_QFT_EM_dipole}

The definition \eqref{EM_dipole_def} of $\mu^{\mu\nu}$ is unusual, and its antisymmetry in $\mu\nu$ is not manifest. Now we present it in a more familiar form that is explicitly antisymmetric. Using \eqref{Gamma0_WTId}, \eqref{Gamma0_WTId_1} and the explicit expressions of $\bar{\Gamma}_1$ and $\bar{V}_1$, we can express the EM dipole moment as
\begin{eqnarray}
\mu^{\mu\nu} = \mu_{bare}^{\mu\nu} + \mu_{band}^{\mu\nu} + \mu_{anom.}^{\mu\nu}
\end{eqnarray}
which we explain term by term below.

The bare EM dipole moment is due to the bare EM dipole matrix (e.g. the Pauli term):
\begin{eqnarray}
\mu_{bare}^{\mu\nu} \equiv \left. (Zuu^\dagger)^T \left(\mathbf{1} - \bar{V}_0 \Delta_0 \right) \hat{\mu}^{\mu\nu} \right|_{p \ on \ FS}
\end{eqnarray}
where $\hat{\mu}^{\mu\nu}$ has been introduced in \eqref{wtGamma_gaugeinv} and is antisymmetric in $\mu\nu$.

The band EM dipole moment, due to the $p$ dependence of $u$, is
\begin{eqnarray}
\mu_{band}^{\mu\nu} \equiv -\left. (Zuu^\dagger)^T \mathcal{A}^\mu \ \partial_p^\nu G^{-1} \right|_{p \ on \ FS} = \left. -i Z \ \left\{ \partial_p^{[\mu} u^\dagger \ G^{-1} \ \partial_p^{\nu]} u \right\} \right|_{p \ on \ FS}.
\end{eqnarray}
In the second equality we used the trick \eqref{IBPtrick}. It is explicitly antisymmetric in $\mu\nu$. In non-interacting theory, $u$ depends only on $\p$ but not $p^0$, so $\mu_{band}$ would be purely magnetic (e.g. the $g=2$ magnetic dipole of free Dirac fermion). In interacting theory, $u$ may or may not depend on $p^0$, so $\mu_{band}$ may or may not have electric dipole components.

The anomalous EM dipole moment, due to interactions, is defined via
\begin{eqnarray}
iq_\mu \: \mu_{anom.}^{\mu\nu} &\equiv& -\left. (Zuu^\dagger)^T \bar{V}_0 \left( -\left(\Delta^r_1 + \Delta^s_1\right) \partial_k^\nu G^{-1} + iq_\mu \mathcal{A}^\mu \ (Zuu^\dagger) \delta_{FS} \delta^\nu_i v^i \right) \right|_{p \ on \ FS} \nonumber \\[.2cm]
&& -\left. (Zuu^\dagger)^T \left(\mathbf{1} - \bar{V}_0 \Delta_0 \right) \wt{V}_1 \ \partial_k^\nu iG \right|_{p \ on \ FS}.
\end{eqnarray}
To get a better understanding of $\mu_{anom.}^{\mu\nu} $, we do the following. For the term with $\Delta^r_1$, we use the explicit expression of $\Delta^r_1$. For the term with $\Delta^s_1$, we use the identity
\begin{eqnarray}
\Delta^s_1 \ \partial_k^\nu G^{-1} = \mathcal{A}^\nu \ (Zuu^\dagger) \delta_{FS} \delta^\mu_i v^i \ iq_\mu
\label{Deltas1_dGinv}
\end{eqnarray}
which again follows from the trick \eqref{IBPtrick}. Now, the anomalous EM dipole moment reads
\begin{eqnarray}
\mu_{anom.}^{\mu\nu} &=& -\left. (Zuu^\dagger)^T \bar{V}_0 \left( \left\{G (\partial_p^{[\mu} G^{-1}) G (\partial_p^{\nu]} G^{-1}) G\right\} + 2\mathcal{A}^{[\mu} \ (Zuu^\dagger) \delta_{FS} \delta^{\nu]}_i v^i \right) \right|_{p \ on \ FS} \nonumber \\[.2cm]
&& + \left. (Zuu^\dagger)^T \left(\mathbf{1} - \bar{V}_0 \Delta_0 \right) \partial_q^\mu \left( i\wt{V}_1 \ \partial_k^\nu iG \right) \right|_{p \ on \ FS}.
\label{EM_dipole_anom}
\end{eqnarray}
The antisymmetry in $\mu\nu$ is manifest in the first line. Gauge invariance of \eqref{Boltzmann_Eq_1_raw} requires the second line above to be antisymmetric in $\mu\nu$ too; more explicitly we show this from diagrams in Section \ref{ssect_QFT_AppendixCD}.

In general, $\mu_{anom.}^{i0}\neq 0$, so even when there is no bare electric dipole matrix, the quasiparticle will still acquire an electric dipole moment due to interactions. This gives rise to the second term in \eqref{BFL_AHE_uniform} which is absent in usual Fermi gas.

\subsection{Current}
\label{ssect_QFT_current}

We now prove the expression of the current \eqref{BFL_current_0}\eqref{BFL_current_1}. Previously we have defined $\delta W$, $\mathcal{U}$, $\mathcal{V}^\nu$ and $\mu^{\mu\nu}$ from QFT, but we have not shown they are real in the position space. But these immediately follow once we have \eqref{BFL_current}, because in position space the quantum expectation of the current must be real for arbitrary $A$, $q$, interaction strength and initial / boundary conditions of $\delta W$. (However, we note that $\mu_{band}^{\mu\nu}$ and $\mu_{anom.}^{\mu\nu}$ are not separately real in general -- due to interactions, the $\chi_w(p)$'s are generally complex near the FS, and hence is $\mu_{band}^{\mu\nu}$.)

\begin{center}
\begin{fmffile}{zzz-current-1}
\begin{fmfgraph*}(45, 25)
\fmfleftn{l}{3}\fmfrightn{r}{3}
\fmf{photon,tension=8}{l2,g}
\fmf{fermion,left=0.7}{G1,g}
\fmf{fermion,left=0.7}{g,G2}
\fmfpoly{empty,label=$i\Gamma$,tension=1}{G2,G1,G3}
\fmf{photon,tension=3,label.side=right,label.dist=10,label=$q$}{r2,G3}
\fmflabel{$A_\nu$}{r2}
\fmflabel{$\mu$}{l2}
\momentumarrow{a}{up}{6}{r2,G3}
\end{fmfgraph*}
\end{fmffile}
\hspace{2cm}
\begin{fmffile}{zzz-current-2}
\begin{fmfgraph*}(25, 25)
\fmfleftn{l}{6}\fmfrightn{r}{6}
\fmf{phantom}{l6,m,r6}
\fmf{phantom_arrow,left=1.22}{l3,r3}
\fmffreeze
\fmf{photon,left=0.3}{o,l2}
\fmf{photon,left=0.3}{r2,o}
\fmf{plain,left,tension=0.1}{o,m}
\fmf{plain,left,tension=0.1}{m,o}
\fmf{phantom}{l2,lm,l3}
\fmf{phantom}{r2,rm,r3}
\fmflabel{$A_\nu$}{rm}
\fmflabel{$\mu$}{lm}
\fmf{phantom,label.side=left,label.dist=11,label=$q$}{r2,l2}
\momentumarrow{a}{down}{7}{r2,l2}
\end{fmfgraph*}
\end{fmffile}
\end{center}

As drawn above, the expectation of the current induced by $A$ at linear response is given by
\begin{eqnarray}
i\delta J^\mu(q) &=& - \int_p \tr \left\{ i\wt{\Gamma}^\mu(p; -q) \ iG(p+q/2) \ i\Gamma^\nu(p; q) \ iG(p-q/2) \right\} A_\nu(q). \nonumber \\[.2cm]
&& - \int_p \tr \left\{ {i\wt{\Xi}^{\mu\nu}}(p; -q, q) \ iG(p) \right\} A_\nu(q)
\label{current_QFT}
\end{eqnarray}
where the negative sign is due to the fermion loop. In the second line, we use \eqref{wtXi_gaugeinv} and integrate $p$ by parts to eliminate $\wt{\Xi}$. Below we work in double fermion notation, at zeroth and first order in $q$ separately.

We emphasize that here we are computing the time-ordered correlation of $\delta J$ and $A$, while in linear response we should compute the retarded correlation. This difference only shows up in the recursion relation that $\delta W$ satifies, i.e. the Boltzmann equation, and there we have already handled this difference. The expression of $\delta J$ in terms of $\delta W$ is the same for time-ordered and retarded correlation.

\subsubsection{Zeroth Order in $q$}

At zeroth order in $q$, 
\begin{eqnarray}
i\delta J_0^\mu = -(i\wt{\Gamma}_0^\mu)^T \left(i\Delta_0 \ i\Gamma_0^\nu + i\Delta'_0 \ i\Gamma_0^\nu + \partial_p^\nu iG\right) A_\nu
\end{eqnarray}
where the integration over $p$ is understood. For the $i\Gamma_0^\nu$ in the first term, whose immediate left is not $\Delta'_0$, we apply the recursion relation \eqref{Gamma0_recursion}, and get
\begin{eqnarray}
i\delta J_0^\mu = -(i\wt{\Gamma}_0^\mu)^T \left(\mathbf{1} + i\Delta_0 \ i\bar{V}_0\right) i\Delta'_0 \ i\Gamma_0^\nu A_\nu - (i\wt{\Gamma}_0^\mu)^T \left( i\Delta_0 \ i\bar{\Gamma}_0^\nu + \partial_p^\nu iG\right) A_\nu
\end{eqnarray}
Due to \eqref{Gamma0_WTId_1} and the facts $(\bar{V}_0)^T = \bar{V}_0$, $(\Delta_0)^T = \Delta_0$, the above reduces to
\begin{eqnarray}
\delta J_0^\mu = (\bar{\Gamma}_0^\mu)^T \left( \Delta'_0 \ \Gamma_0^\nu \ A_\nu - (Zuu^\dagger) \ \delta_{FS} \delta^\nu_i v^i\right) A_\nu = (\bar{\Gamma}_0^\mu)^T (Zuu^\dagger) \ \delta W_0.
\end{eqnarray}
Finally, applying \eqref{Gamma0_WTId}, we obtain
\begin{eqnarray}
\delta J_0^\mu = (v^\mu)^T \ \delta W_0 + \left(\delta^\mu_i v^i \delta_{FS}\right)^T \mathcal{U} \ \delta W_0.
\end{eqnarray}
The transpose on the left implies integration over $p$. This is \eqref{BFL_current_0}.

\subsubsection{First Order in $q$}

At first order in $q$, 
\begin{eqnarray}
i\delta J_1^\mu &=& -(i\wt{\Gamma}_1^\mu(-q))^T \left(i\Delta_0 + i\Delta'_0\right) i\Gamma_0^\nu A_\nu - (i\wt{\Gamma}_0^\mu)^T \left(i\Delta_0 + i\Delta'_0\right) i\Gamma_1^\nu(q) A_\nu \nonumber \\[.2cm]
&& - \ (i\wt{\Gamma}_0^\mu)^T \left(i\Delta'_1 + i\Delta^r_1 + i\Delta^s_1 - C_1 \right)(q) \ i\Gamma_0^\nu A_\nu \nonumber \\[.2cm]
&& - \ (i\wt{\Gamma}_1^\mu(-q))^T \ \partial_p^\nu iG \ A_\nu - (i\wt{\Gamma}_1^\nu(q))^T \ \partial_p^\mu iG \ A_\nu.
\end{eqnarray}
We rewrite this according to the following: If the immediate left of an $i\Gamma_0^\nu$ is not $\Delta'_0$ or $\Delta'_1$ or $C_1$, we apply the recursion relations \eqref{Gamma0_recursion} to it; similarly, if the immediate left of an $i\Gamma_1^\nu$ is not $\Delta'_0$, we apply \eqref{Gamma1_recursion} to it. We find
\begin{eqnarray}
i\delta J_1^\mu &=& - (i\bar{\Gamma}_0^\mu)^T \left(i\Delta'_0 \ i\Gamma_1^\nu(q) + i\Delta'_1(q) \ i\Gamma_0^\nu - C_1(q) \ i\Gamma_0^\nu\right) A_\nu - (i\bar{\Gamma}_1^\mu(-q))^T \ i\Delta'_0 \ i\Gamma_0^\nu A_\nu \nonumber \\[.2cm]
&& + \ \mbox{(terms regular in $q$)}.
\label{deltaJ1_mid}
\end{eqnarray}
We will take care of the terms in the second line of \eqref{deltaJ1_mid} later. To terms in the first line, we apply the identity \eqref{Delta_prime_1_Id}, and get
\begin{eqnarray}
(i\bar{\Gamma}_0^\mu)^T (Zuu^\dagger) \ \delta W_1 + \left( (i\bar{\Gamma}_1^\mu(-q))^T + (i\bar{\Gamma}_0^\mu)^T iq_\lambda \mathcal{A}^\lambda \right) (Zuu^\dagger) \left(\delta W_0 + \delta_{FS} v^j A_j \right) 
\end{eqnarray}
Now use \eqref{Gamma0_WTId} in the first term, and \eqref{EM_dipole_def} and the facts $\mathcal{A}^\lambda=-(\mathcal{A}^\lambda)^T$, $\mathcal{V}^\nu(\p, \kk)=-\mathcal{V}^\nu(\kk, \p)$ in the second term, the first line of \eqref{deltaJ1_mid} becomes
\begin{eqnarray}
&& i(v^\mu)^T \delta W_1 + \left(\delta^\mu_i v^i \delta_{FS}\right)^T i\mathcal{U} \ \delta W_1 \nonumber \\[.2cm]
&& + \ \left((i\mu^{\mu\nu})^T+ (\delta_{FS}\delta^\mu_i v^i)^T i\mathcal{V}^\nu \right) iq_\nu \left(\delta W_0 + \delta_{FS} v^j A_j \right).
\label{deltaJ1_mid_1st}
\end{eqnarray}
Notice that the $\delta W$ dependence agrees with \eqref{BFL_current_1}.

The second line of \eqref{deltaJ1_mid} -- terms regular in $q$ -- can be read-off diagrammatically:
\begin{eqnarray}
&& - (i\wt{\Gamma}_1^\mu(-q))^T \left(i\Delta_0 \ i\bar{\Gamma}_0^\nu + i\partial_p^\nu G\right) A_\nu - \left((i\Delta_0 \ i\bar{\Gamma}_0^\mu)^T + (i\partial_p^\mu G)^T \right) i\wt{\Gamma}_1^\nu(q) A_\nu \nonumber\\[.2cm]
&& -(i\bar{\Gamma}_0^\mu)^T \left(i\Delta^r_1+i\Delta^s_1\right)(q) \ i\bar{\Gamma}_0^\nu A_\nu - \left( i\Delta_0 \ i\bar{\Gamma}_0^\mu \right)^T i\wt{V}_1(q) \left( i\Delta_0 \ i\bar{\Gamma}_0^\nu \right) A_\nu
\end{eqnarray}
where the $\partial_p G$ terms follow from the $\wt{\Xi}$ terms in \eqref{current_QFT}. There are many equivalent expressions; we have chosen to express it so that it appears ``symmetric'' to read from left to right and from right to left. Now, substitute \eqref{Gamma0_WTId} into the $(\Delta^r_1+\Delta^s_1)$ term, and substitute \eqref{Gamma0_WTId_1} into the rest; next, for each of the two terms in the second line above, we expand like $-a \, c \, b = -a \, c\, b_2 - a_2 \, c \, b + a_2 \, c\, b_2 - a_1 \, c\, b_1$ for $a=a_1+a_2$, $b=b_1+b_2$. The result is
\begin{eqnarray}
&& - (i\bar{\Gamma}_1^\mu(-q))^T \ (Zuu^\dagger) \delta_{FS} \delta^\nu_j v^j A_\nu \nonumber \\[.2cm]
&& - (\delta_{FS}\delta^\mu_i v^i)^T (Zuu^\dagger)^T \ i\bar{\Gamma}_1^\nu(q) \ A_\nu + (\delta_{FS}\delta^\mu_i v^i)^T (Zuu^\dagger)^T \ i\bar{V}_1(q) \ (Zuu^\dagger) \delta_{FS} \delta^\nu_j v^j A_\nu \nonumber \\[.2cm]
&& - (i\partial_p^\mu G^{-1})^T \left(i\Delta^r_1+i\Delta^s_1\right)(q) \ i\partial_p^\nu G^{-1} \ A_\nu - (i\partial_p^\mu G)^T \ i\wt{V}_1(q) \ i\partial_k^\nu G \ A_\nu.
\label{sigma_Y_ref}
\end{eqnarray}
The last term is to be expressed using \eqref{tildeY}; it vanishes, as we show diagrammatically and combinatorially in Section \ref{ssect_QFT_AppendixCD}. The remaining terms, inspecting the definitions of $\mu^{\mu\nu}$ and $\mathcal{V}^\nu$, can be expressed line by line as
\begin{eqnarray}
&& -\left((i\mu^{\mu\lambda})^T+ (\delta_{FS}\delta^\mu_i v^i)^T i\mathcal{V}^\lambda - (i\bar{\Gamma}_0^\mu)^T \mathcal{A}^\lambda \ (Zuu^\dagger) \right) iq_\lambda \ \delta_{FS} \delta^\nu_j v^j A_\nu \nonumber \\[.2cm]
&& - (\delta_{FS}\delta^\mu_i v^i)^T \left( i\mu^{\lambda\nu} - (Zuu^\dagger)^T \mathcal{A}^\lambda \ i\bar{\Gamma}_0^\nu \phantom{\left( \mathcal{A}^\lambda i\bar{V_0} - i\bar{V_0} \mathcal{A}^\lambda \right)} \right. \nonumber \\[.2cm]
&& \hspace{3cm} + \left. (Zuu^\dagger)^T \left( \mathcal{A}^\lambda i\bar{V_0} - i\bar{V_0} \mathcal{A}^\lambda \right) (Zuu^\dagger) \delta_{FS} \delta^\nu_j v^j \right) iq_\lambda A_\nu \nonumber \\[.2cm]
&& - (i\partial_p^\mu G^{-1})^T \left(i\Delta^r_1+i\Delta^s_1\right)(q) \ i\partial_p^\nu G^{-1} \ A_\nu.
\end{eqnarray}
Substituting \eqref{Gamma0_WTId} for $\bar{\Gamma}_0$, and using the definition of $\mu_{band}^{\mu\nu}$, we find
\begin{eqnarray}
&& -\left((i\mu^{\mu\lambda})^T+ (\delta_{FS}\delta^\mu_i v^i)^T i\mathcal{V}^\lambda\right) iq_\lambda \ \delta_{FS} \delta^\nu_j v^j A_\nu - (\delta_{FS} \delta^\mu_i v^i)^T i\mu^{\lambda\nu} iq_\lambda A_\nu \nonumber \\[.2cm]
&& + \: (i\mu_{band}^{\mu\lambda})^T\delta_{FS} \delta^\nu_j v^j \ iq_\lambda A_\nu + (\delta_{FS} \delta^\mu_i v^i)^T i\mu_{band}^{\lambda\nu} \ iq_\lambda A_\nu \nonumber \\[.2cm]
&& - \: (i\partial_p^\mu G^{-1})^T \left(i\Delta^r_1+i\Delta^s_1\right)(q) \ i\partial_p^\nu G^{-1} \ A_\nu.
\end{eqnarray}
Finally, for the $\Delta^r_1$ term, use its explicit expression, and for the $\Delta^s_1$ term, use \eqref{Deltas1_dGinv} and the definition of $\mu_{band}^{\mu\nu}$. We arrive at
\begin{eqnarray}
&& -\left((i\mu^{\mu\nu})^T+ (\delta_{FS} \delta^\mu_i v^i)^T i\mathcal{V}^\nu \right) iq_\nu \ \delta_{FS} v^j A_j - (\delta_{FS} \delta^\mu_i v^i)^T i\mu^{\nu\lambda} F_{\nu\lambda}/2 \nonumber \\[.2cm]
&& + \ i\sigma^{\mu\nu\lambda} F_{\nu\lambda}/2,
\label{deltaJ1_mid_2nd}
\end{eqnarray}
where the Hall conductivity tensor $\sigma^{\mu\nu\lambda}$, totally antisymmetric in $\mu\lambda\nu$, is defined by $\sigma^{\mu\nu\lambda} \equiv \sigma_r^{\mu\nu\lambda}+\sigma_s^{\mu\nu\lambda}$, with
\begin{eqnarray}
&& \sigma_r^{\mu\nu\lambda} \equiv \int_p \: \tr\left\{ (\partial_p^{[\mu} iG^{-1}) \: iG \: (\partial_p^\nu iG^{-1}) \: iG \: (\partial_p^{\lambda]} iG^{-1}) \: iG \right\}, \nonumber \\[.2cm]
&& \sigma_s^{\mu\nu\lambda} \equiv 3 \int_p \delta_{FS} \ v^i \delta^{[\mu}_i \mu_{band}^{\nu\lambda]}.
\label{sigma_def}
\end{eqnarray}
One may notice the similarity between the definitions of $\sigma^{\mu\nu\lambda}$ and $\mu_{anom.}^{\nu\lambda}$ (except in $\sigma^{\mu\nu\lambda}$, the $\wt{V}_1$ term vanishes due to the proof in Section \ref{ssect_QFT_AppendixCD}).

After combining \eqref{deltaJ1_mid_1st} and \eqref{deltaJ1_mid_2nd} into \eqref{deltaJ1_mid}, we arrive at
\begin{eqnarray}
\delta J_1^\mu &=& (v^\mu)^T \delta W_1 + (\mu^{\mu\nu})^T iq_\nu \delta W_0 + (\delta_{FS} \delta^\mu_i v^i)^T \left( - \mu^{\nu\lambda} F_{\nu\lambda}/2 + \mathcal{U} \ \delta W_1 + \mathcal{V}^\nu \ iq_\nu \delta W_0 \right) \nonumber \\[.2cm]
&& + \ \sigma^{\mu\nu\lambda} F_{\nu\lambda}/2.
\end{eqnarray}
This is \eqref{BFL_current_1}.

\subsubsection{Detour: Coleman-Hill Theorem}

Interestingly, our derivation for $\delta J_1^\mu$ above, most crucially the cancellation in Section \ref{ssect_QFT_AppendixCD}, provides an alternative diagrammatic proof to the Coleman-Hill theorem, for QFTs restricted to our assumptions (which are less general than in the original proof). The theorem states that in a gapped fermionic system, the Hall conductivity is unaffected by the interactions. When the system is gapped, i.e. in the absence of FS, our result reduces to $\delta J_1^\mu = \sigma_r^{\mu\nu\lambda} F_{\nu\lambda}/2$, that is, the full Hall conductivity is equal to $\sigma_r^{\mu\nu\lambda}$. Let $g$ be some interaction strength, we have (denoting $\partial^g \equiv \partial / \partial g$)
\begin{eqnarray}
\partial^g \sigma_r^{\mu\nu\lambda} &=& - \int_p \: \partial^g \ \tr\left\{ (\partial_p^{[\mu} G^{-1}) \: G \: (\partial_p^\nu G^{-1}) \: G \: (\partial_p^{\lambda]} G^{-1}) \: G \right\} \nonumber \\[.2cm]
&=& - \int_p \: 4\partial^{[g} \ \tr\left\{ (\partial_p^\mu G^{-1}) \: G \: (\partial_p^\nu G^{-1}) \: G \: (\partial_p^{\lambda]} G^{-1}) \: G \right\}.
\end{eqnarray}
In the second equality we added some total derivative terms so to antisymmetrize the $\partial^g$ altogether with the three $\partial_p$'s. But because of the antisymmetrization, the integrand actually vanishes. This means the full Hall conductivity is independent of interaction strength. This proves the Coleman-Hill theorem, for QFTs restricted to our assumptions. In asserting ``the integrand vanishes'', we implicitly made use of the fact that $\partial iG = \{ iG \ \partial iG^{-1} \ iG \}$ in the absence of FS, the fact that $G_{bare}^{-1}$ by definition is independent of $g$, and the physical assumption that the dependence of the self-energy $\Sigma$ on $g$ is non-singular.

\subsubsection{Chemical Potential Dependence of the Hall Conductivity Tensor}
\label{sssect_QFT_dsigma_dEF}

Now we prove \eqref{BFL_dsigma_dEF_detail}, the important result relating the Hall conductivity to the Berry curvature on the FS.

We first consider the $\epsilon_F$ dependence of $\sigma_r^{\mu\nu\lambda}$:
\begin{eqnarray}
\partial^F \sigma_r^{\mu\nu\lambda} = - \int_p \: 4\partial^{[F} \ \tr\left\{ (\partial_p^\mu G^{-1}) \: G \: (\partial_p^\nu G^{-1}) \: G \: (\partial_p^{\lambda]} G^{-1}) \: G \right\}.
\end{eqnarray}
The integrand is non-vanishing because in the presence of FS, the derivative outside the trace acts on the $p^0$ pole structure of the $iG$'s, and the pole structure depends on $p_i$ and $\epsilon_F$. In fact, by similar reasoning that led to the FS term in \eqref{G_derivative_indices}, here we are led to
\begin{eqnarray}
\partial^F \sigma_r^{\mu\nu\lambda} &=& 12 \int_p \: i\pi \delta(p^0-\xi) \ \partial^{[F}\sgn\, \xi \: \times \nonumber \\[.2cm]
&& \hspace{2cm} \sum_{w} \sum_{w'} \frac{Z}{\chi_w \chi_{w'}} \left\{ u^\dagger \: (\partial_p^\mu G^{-1}) \: w w^\dagger \: (\partial_p^\nu G^{-1}) \: w' w'^\dagger \: (\partial_p^{\lambda]} G^{-1}) \: u \right\} \nonumber \\[.2cm]
&& + \ 12 \int_p \: i\pi \left(-\partial_{p^0} \delta(p^0-\xi) \right) \ \partial^{[F}\sgn\, \xi \: \times \nonumber \\[.2cm]
&& \hspace{2.2cm} \sum_w \frac{Z^2}{\chi_w} \left\{ w^\dagger \: (\partial_p^\mu G^{-1}) \: uu^\dagger \: (\partial_p^\nu G^{-1}) \: uu^\dagger \: (\partial_p^{\lambda]} G^{-1}) \: w \right\}.
\end{eqnarray}
The first term arises from the single pole (appearing as $i\pi \delta(p^0-\xi) \: \sgn\, \xi$ in principle function decomposition) when one of the three $G$'s is in the $u$ band; the second term arises from the double pole (appearing as $i\pi (-\partial_{p^0} \delta(p^0-\xi)) \: \sgn\, \xi$) when two of the $G$'s are in the $u$ band. The triple pole contribution when all three $G$'s are in the $u$ band vanishes under the total antisymmetrization. We can evaluate the above using the trick \eqref{IBPtrick}. We find
\begin{eqnarray}
\!\!\!\!\!\! && \partial^F \sigma_r^{\mu\nu\lambda} \ = \nonumber \\[.2cm]
\!\!\!\!\!\! && 12 \int_p \: i\pi \delta(p^0) \ \partial^{[F} \sgn\,\xi \: \left\{ \partial_p^\mu u^\dagger \: (\mathbf{1}- uu^\dagger) \: Z(\partial_p^\nu G^{-1}) \: (\mathbf{1}- uu^\dagger) \: \partial_p^{\lambda]} u \right\} \nonumber \\[.2cm]
\!\!\!\!\!\! && - \ 12 \int_p i\pi \delta(p^0) \ \partial^{[F}\sgn\,\xi \: \partial_{p^0} \left\{ \partial_p^\mu u^\dagger \ Z^2\partial_p^\nu \chi_u \sum_w \left(\chi_u - \chi_w\right)^2 \frac{ww^\dagger}{\chi_w} \ \partial_p^{\lambda]} u \right\}.
\end{eqnarray}
In the first line, $\mathbf{1}- uu^\dagger$ can be further replaced by $\mathbf{1}$ thanks to \eqref{IBPtrick} and the antisymmetrization. In the second line, we need the following relevant terms, according to \eqref{chiu_near_FS_good}:
\begin{eqnarray}
\left. -Z^2\partial_p^\nu \chi_u \ (\chi_u-\chi_w)^2 \right|_{p \ on \ FS} = Zv^\nu \chi_w^2
\end{eqnarray}
\begin{eqnarray}
&& \left. \partial_{p^0} \left(-Z^2\partial_p^\nu \chi_u \ (\chi_u-\chi_w)^2\right) \right|_{p \ on \ FS} \nonumber \\[.2cm] &=& Zv^\nu \partial_{p^0} \chi_w^2 + \left(\partial_p^\nu Z + \, 2iZ\gamma \: \partial_p^\nu |p^0| \right) \chi_w^2 - 2v^\nu\chi_w.
\end{eqnarray}
Similar results hold when $\partial_p^\nu \chi_u$ is replaced with $\partial^F \chi_u$; recall that $v^F\equiv \partial^F E$. The remaining problem is, how to understand $\delta(p^0) \partial_{p^0} |p^0|$? We should understand it as $0$, because in this thesis, the generalized function $\delta(p^0)$ always arises as the approximation to a narrow rectangular function over an interval centered at $p^0=0$ (more precisely, the rectangular function is $\theta(p^0-q^0/2) - \theta(p^0+q^0/2)$, see Section \ref{ssect_QFT_AppendixAB}), and $\delta(p^0) \partial_{p^0} |p^0|$ corresponds to taking the difference of $|p^0|$ between the two sides of the interval, which is obviously $0$. (Note that in $d=2$, the parametrization \eqref{chiu_near_FS_good} does not apply; however, general analytic properties of $\delta\Sigma$ still requires it to be odd in $p^0$~\cite{Luttinger:1961zz, abrikosov1975methods}, and therefore the corresponding contribution here must still vanish.) Thus, at the end, we have
\begin{eqnarray}
\!\!\!\!\!\! \partial^F \sigma_r^{\mu\nu\lambda} &=& 12 \int_p \: i 2\pi \delta(p^0) \ \partial^{[F}\theta(\epsilon_F-E) \: \times \nonumber \\[.2cm]
&& \hspace{.5cm} \left( -2v^\mu \left\{ \partial_p^\nu u^\dagger \: \partial_p^{\lambda]} u \right\} + \left(\partial_p^\mu + v^\mu \partial_{p^0} \right) \left\{ \partial_p^\nu u^\dagger \: ZG^{-1} \: \partial_p^{\lambda]} u \right\} \right).
\label{dsigma_r}
\end{eqnarray}
Note that the derivatives of $-2\theta(\epsilon-E)$ are always the same as those of $\sgn\, \xi$; we choose to express as the former because it admits the intuition as the ``Fermi sea'', at least near the FS.

Next, from the expression of $\mu_{band}^{\nu\lambda}$, we observe $\sigma_s^{\mu\nu\lambda}$ can be expressed as
\begin{eqnarray}
\sigma_s^{\mu\nu\lambda} = 3 \int_p \: 2\pi \delta(p^0-(E-\epsilon_F)) \ \partial_p^{[\mu} \theta(\epsilon_F-E) \ i \left\{ \partial_p^\nu u^\dagger \: ZG^{-1} \: \partial_p^{\lambda]} u \right\}.
\label{sigma_s_alt}
\end{eqnarray}
Now we take $\partial^F$ and find
\begin{eqnarray}
\partial^F \sigma_s^{\mu\nu\lambda} &=& 12 \int_p \partial^{[F} \left( 2\pi \delta(p^0-(E-\epsilon_F)) \ \partial_p^\mu \theta(\epsilon_F-E) \ i \left\{ \partial_p^\nu u^\dagger \: ZG^{-1} \: \partial_p^{\lambda]} u \right\} \right) \nonumber \\[.2cm]
&=& -12 \int_p \: i 2\pi \ \partial^{[F} \theta(\epsilon_F-E) \left( \delta(p^0) \partial_p^\mu - v^\mu \partial_{p^0} \delta(p^0) \right) \left\{ \partial_p^\nu u^\dagger \: ZG^{-1} \: \partial_p^{\lambda]} u \right\} \nonumber \\[.2cm]
&=& -12 \int_p \: i 2\pi \delta(p^0) \ \partial^{[F} \theta(\epsilon_F-E) \left( \partial_p^\mu + v^\mu \partial_{p^0} \right) \left\{ \partial_p^\nu u^\dagger \: ZG^{-1} \: \partial_p^{\lambda]} u \right\}
\label{dsigma_s}
\end{eqnarray}
In the second equality we used
\begin{eqnarray}
\partial_p^\mu \delta(p^0-(E-\epsilon_F)) = - v^\mu \partial_{p^0} \delta(p^0-(E-\epsilon_F))
\end{eqnarray}
and likewise for $\partial^F$.

Finally we combine \eqref{dsigma_r} and \eqref{dsigma_s} and obtain
\begin{eqnarray}
\partial^F \sigma^{\mu\nu\lambda} = 12 \int_p \: 2\pi \delta(p^0) \ \partial^{[F} \theta(\epsilon_F-E) \: v^\mu \: (-2i) \left\{ \partial_p^\nu u^\dagger \: \partial_p^{\lambda]} u \right\}.
\end{eqnarray}
Due to \eqref{u_derivative_on_shell} (and the similar version for $\partial^F$) and the antisymmetrization, we can replace $u$ with $\mathfrak{u}$, and perform the $p^0$ integral to obtain
\begin{eqnarray}
\partial^F \sigma^{\mu\nu\lambda} = 12 \int_\p \partial^{[F} \theta(\epsilon_F-E) \: v^\mu \: b^{\nu\lambda]}.
\label{sigma_FSderivaitve}
\end{eqnarray}
Expanding the antisymmetrization explicitly, this is \eqref{BFL_dsigma_dEF_detail}.

If $d=2$, or if $d>2$ and the Berry curvature is an exact 2-form on the FS, we can continuously define $\mathfrak{u}(\p)$ over the FS, and derive \eqref{BFL_dsigma_dEF_1} via integration by parts:
\begin{eqnarray}
\partial^F \sigma^{\mu\nu\lambda} &=& 24 \int_\p \partial^{[F} \left(\partial_p^\mu \theta(\epsilon_F-E) \: v^\nu \: a^{\lambda]} \right) \nonumber \\[.2cm]
&=& \partial^F \ 6 \int_\p \delta(\epsilon_F-E) \: \delta^{[\mu}_0 v^\nu \: a^{\lambda]}
\end{eqnarray}
where in the first equality we used the fact that $v^\mu$ and $v^F$ are respectively $-\partial_p^\mu$ and $-\partial^F$ acted on $p^0-(E-\epsilon_F)$, and in the second equality, total $\p$ derivatives vanish due to the $\p$ integral, while total $p_0$ derivative vanishes trivially as the integrand has no $p^0$ dependence. 

If $d>2$ and the Berry curvature is not an exact 2-form on the FS, we need to integrate by parts in another way. We can rewrite \eqref{sigma_FSderivaitve} using the $P^\lambda_k$ introduced below \eqref{BFL_sigma_const_FS}:
\begin{eqnarray}
\partial^F \sigma^{ij\lambda} = 12 \int_\p \partial^{[F} \theta(\epsilon_F-E) \: b^{ij} \: \partial_p^{k]} P^\lambda_k
\end{eqnarray}
where the total antisymmetrization is in indices $[Fijk]$. Using the assumption that there is no band degeneracy near the FS and hence $\partial_p^{[k}b^{ij]}=0$ near the FS, we have
\begin{eqnarray}
\partial^F \sigma^{ij\lambda} = 12 \int_\p \partial^{[F} \left( -\partial_p^k \theta(\epsilon_F-E) \: b^{ij]} \: P^\lambda_k \right).
\end{eqnarray}
When we proceed further, note that if the fermions are in a lattice and if the FS intersects the boundary of our choice of first Brillouin zone, then for $\lambda=0$ we cannot drop the total $p$-derivative terms because $p_k$ is not continuous when we identify the opposite boundaries of the first Brillouin zone~\cite{Haldane:2004zz}. We have
\begin{eqnarray}
\partial^F \sigma^{ij\lambda} = \partial^F \left( 3 \int_\p \delta(\epsilon_F-E) b^{[ij} v^{k]} \: P^\lambda_k + 6\int_\p \partial_p^{[k} \left( \delta(\epsilon_F-E) \: v^i \: a^{j]} \: P^\lambda_k\right) \right)
\end{eqnarray}
(we used the Bianchi identity $\partial^F \partial b^{\nu\lambda} = 2 \partial_p^{[\lambda} b^{\nu] F}$). This proves \eqref{BFL_dsigma_dEF_e} and \eqref{BFL_dsigma_dEF_b}, and hence \eqref{BFL_sigma_const_FS}.

We remind that the derivation above fails at discrete values of $\epsilon_F$ around which the FS develops new disconnected components (e.g. Figure \ref{EF_increase_disconnect_comp}). Across those values of $\epsilon_F$ it is unclear whether $\sigma^{\mu\nu\lambda}$ may have a jump, as commented in Section \ref{ssect_kinetic_sigma_EF}.

\subsection{Cutkosky Cut and Quasiparticle Collision}
\label{ssect_QFT_AppendixAB}

In the proof present above, we have left behind the technical discussion of the Cutkosky formalism and the computation of quasiparticle collisions. In this section we complete the relevant discussions. More exactly, we present the followings:
\begin{itemize}
\item
First we present the Cutkosky cutting rule for our fermionic system.
\item
Then we introduce how to count the power of $q$ in a cut diagram, and show $D_1$, $C^{ph}_1$ and $C^{pp}_1$ are the only cut sub-diagrams that contribute at order $q$. Then restrict to $d\geq 3$ and justify their parametrization \eqref{Decay_parametrization}, \eqref{Cph_parametrization} and \eqref{Cpp_parametrization}, and show the relation \eqref{Collision_WTId}.
\item
Then we discuss the difficulty of parametrizing quasiparticle collision in $d=2$, and explain why \eqref{Collision_WTId} still holds in $d=2$ despite the difficulties of parametrization.
\item
Finally, we show (regardless of $d\geq 3$ and $d=2$) that collisions have no contribution to the antisymmetric part $\Pi^{[\mu\nu]}$ of the current-current correlation. So collision are ``uninteresting'' to the main focus of this thesis.
\end{itemize}

\subsubsection{Review of Cutkosky Cut}

The formalism of Cutkosky cut is a formal procedure to compute discontinuities (with respect to external momenta) in Feynman diagrams. A Feynman diagram has a discontinuity if, in an intermediate step of the process described the diagram, all intermediate particles go on-shell. We have seen an example of this in the derivation for $\Delta'$ following \eqref{ubandproduct}. From the derivation for $\Delta'$, we have the sense that all it matters is the $i\epsilon$ prescription whose effects can be extracted by principle function decomposition; other details of the propagator are irrelevant. This suggests the method can be generalized to other diagrams. Indeed, Landau made pioneering contribution in this direction~\cite{Landau:1959fi}, and the full formalism was established by Cutkosky~\cite{Cutkosky:1960sp}. The Cutkosky cutting rule was originally developed for bosons. Here, based on our needs, we present the rule for fermions; the modifications have clear physical meanings as long as one is familiar with some basic analytic properties of fermion propagators~\cite{Luttinger:1961zz, abrikosov1975methods}.

Consider a two point correlation $\Pi$ with momentum $q$ (for definiteness, we let $q$ run from the right to the left of the diagram). In this section, unless otherwise specified, $\Pi$ refers to the current-current correlation $\Pi^{\mu\nu}(q)$, and $\Re\Pi$ and $\Im\Pi$ really mean the Hermitian and anti-Hermitian parts of $\Pi^{\mu\nu}$. The Cutkosky cutting rule gives the difference between the retarded correlation and the advanced correlation:
\begin{eqnarray}
\Pi_{cut} &\equiv& -i(\Pi_R-\Pi_A) = \Im\Pi_R-\Im\Pi_A = 2\Im\Pi_R = -2\Im\Pi_A \nonumber\\[.2cm]
&=& \Pi_{cut-} - \Pi_{cut+}.
\label{def_Cut}
\end{eqnarray}
The three equalities in the first line follow from general analytic properties of two-point correlations~\cite{Luttinger:1961zz, abrikosov1975methods}. In the second line, $\Pi_{cut+}$ is defined as the following. Consider a certain Feynman diagram in $\Pi$, with a certain Cutkosky cut -- a cut through a number of internal fermion propagators such that the Feynman diagram is disconnected into two parts, with one current insertion (or other operators, depending on what $\Pi$ is) contained in each part. Clearly the total momentum running from right to left across the cut is $q$. For those fermion propagators that are being cut, we place the fermions on-shell, which, according to the Cutkosky cutting rule, means to replace each cut propagator by
\begin{eqnarray}
iG(p) \ \ \ \longrightarrow \ \ 2\pi Z(\p) \ \delta(p^0-\xi_\p) \ \sgn(p^0) \theta(\mp p^0),
\end{eqnarray}
where $\theta(\mp p^0)$ is taken when the fermion runs across the cut from right to left / from left to right (given we have chosen $q$ to run from right to left). Then we sum over all possible ways of cutting over all Feynman diagrams, and the result is defined as $\Pi_{cut-}$. And $\Pi_{cut+}$ is defined in a similar manner, but with $\theta(\pm p^0)$ taken when the fermion runs across the cut from right to left / from left to right.

An important consequence is, by energy conservation, all these cut propagators must have energies between $\pm|q^0|$. This is because, those step functions require the on-shell quasiparticles' energies to appear in energy conservation in the form ``the sum of positive energies minus the sum of negative energies is equal to $\mp q^0$'' respectively in $\Pi_{cut \mp}$ (so $\Pi_{cut\mp}$ is non-vanishing only for negative / positive $q^0$ respectively, hence our $\mp$ subscript). In retrospect, this justifies why we could restrict to the $u$ band in a multi-band system, and why we could ignore the possibility of cutting through an interaction mediator: Because by the assumptions about our QFT, neither the other bands of the fermion nor the short-ranged interaction mediator(s) have any low energy on-shell excitation.

\subsubsection{Computation of Quasiparticle Collision $d\geq 3$}

Below we discuss how to count the power of $q$ in a cut sub-diagram. We have to note that the power counting introduced below has missing piece. In $d=2$ the missing piece is order $q \ln q$ (less suppressed than order $q$ when $q$ is small) and lead to complications to be discussed later. For now we work with $d\geq 3$, where this missing piece is neglected as they are beyond order $q$. After introducing the power counting, we will argue the only cut sub-diagram that contributes at zeroth order in $q$ is a cut through double propagator, leading to $\Delta'$, and the only ones that contribute at first order in $q$ are those three pairs of cut sub-diagrams for $D_1$, $C^{ph}_1$ and $C^{pp}_1$.

Consider a cut through $n>2$ internal fermion propagators. Generically they all have different internal momenta. As discussed above, all their energies are restricted by $|q^0|$, hence the integration over the $n$ internal energies yields a suppression of order $(q^0)^n$; on the other hand, the argument of the delta function of energy conservation is of order $q^0$. Therefore, the contribution of the $n$ cut propagators is of order $(q^0)^{n-1}$.

The case of $n=2$ has a difference. For $n=2$, when one on-shell fermion is low energy, and hence near the FS, the other one, due to spatial momentum conservation and the smallness of $\q$, is automatically near the FS too, and hence low energy too (because of on-shell). This means the smallness of their energies provides only one constraint, instead of two independent constraints. This lowers the power counting of $q^0$ by one. Thus, the cut through two propagator is not first order but zeroth order in $q$. In particular, let's compute the cut through the double propagator of momenta $p\pm q/2$. According to the cutting rule, the cut sub-diagram is equal to
\begin{eqnarray}
&& 2\Im\Delta'_R(q) \nonumber \\[.2cm]
&=& 2\pi Z(\p+\q/2) \delta(p^0+q^0/2-\xi(\p+\q/2)) \ 2\pi Z(\p-\q/2) \delta(p^0-q^0/2-\xi_(\p-\q/2)) \nonumber\\[.2cm]
&& (-1) \left(\theta(-(p^0+q^0/2)) \theta(p^0-q^0/2) - \theta(p^0+q^0/2) \theta(-(p^0-q^0/2)) \right) \nonumber \\[.2cm]
&=& (2\pi Z(\p))^2 \ q^0 \ \delta(p^0) \ \delta(p^0-\xi(\p)) \ \delta(v^\mu(\p) q_\mu) \ + \ \mathcal{O}(q^2).
\label{Delta_from_Cutkosky}
\end{eqnarray}
To relate this to the time-ordered $\Delta'$, we use the Kramers-Kronig dispersion relation of a general two-point correlation~\cite{Luttinger:1961zz, abrikosov1975methods}:
\begin{eqnarray}
i\Pi(q) = \frac{i}{2\pi} \int d\omega \frac{\Pi_{cut}(\omega, \q)}{-q^0+\omega-i\epsilon \: \sgn\, q^0} \ + \ i\, (\mbox{real terms unrelated to Cutkosky cut}).
\label{dispersion_relation}
\end{eqnarray}
(The $\Pi$ here is time-ordered; if retarded or advanced, the $\sgn\, q^0$ should be replaced with $\pm 1$.) Performing the integration yields the time ordered $i\Delta'(q)$ in \eqref{Delta_prime}, as desired. (The integration generally involves $\Pi_{cut}$ at non-small values of $\omega$. But $\Im\Delta'$ in particular is non-vanishing only when $\omega$ equals the small value $v^i q_i$.)

For $n>2$, the $n$ cut propagators contribute order $(q^0)^{n-1}$, and for current-current correlation $n$ must be even (with $n/2$ cut propagators running across the cut from right to left, and the other $n/2$ from left to right). So it seems the corrections from Cutkosky cut beyond $\Delta'$ (beyond $n=2$) are at least of order $q^3$. (This justifies our analytic expansion of the $q$-2PI interaction vertex to zeroth and first order in $q$.) How can there be order $q$ sub-diagrams? Consider the following situation. Given that all cut propagators are on-shell, if there is a pair of propagators, one cut and one uncut, whose momenta are dictated by momentum conservation to differ by $q$, then that uncut propagator will be nearly-on-shell (due to the smallness of $q$), and contributes a factor of order $1/q$. There can be at most one such nearly-on-shell propagator on either the left or the right of the cut, so there can be at most two of them in total. Therefore, there exist cut sub-diagrams at order $q$: Such cut sub-diagrams have four cut propagators, and two nearly-on-shell propagators, one on each side of the cut. These propagators can be organized in six different ways, which are the three pairs of cut sub-diagrams for $D_1$, $C^{ph}_1$ and $C^{pp}_1$ respectively, presented in Section \ref{sssect_QFT_Collision}.

Now we evaluate the sub-diagrams for $D_1$, $C^{ph}_1$ and $C^{pp}_1$ according to the cutting rule. First,
\begin{eqnarray}
&& 2(D_R)_1(p; q) \nonumber \\[.2cm]
&=& \int_{k, l} \left(-\frac{1}{2}\right) \left(\frac{iZ(\p-\q/2)}{-q^0-\xi(\p-\q/2) + \xi(\p+\q/2)}\right)^2 \nonumber \\[.2cm]
&& \hspace{1cm} i^2 \left|V(p-q/2, k+l \rightarrow k-q/2, p+l)\right|^2 \nonumber \\[.2cm]
&& \hspace{1cm} (2\pi)^4 Z(\p+\q/2) Z(\kk-\q/2) Z(\p+\l) Z(\kk+\l) \nonumber \\[.2cm]
&& \hspace{1cm} \delta(p^0+q^0/2-\xi(\p+\q/2)) \ \delta(k^0-q^0/2-\xi(\kk-\q/2)) \nonumber \\[.2cm]
&& \hspace{1cm} \delta(p^0+l^0-\xi(\p+\l)) \ \delta(k^0+l^0-\xi(\kk+\l)) \nonumber \\[.2cm]
&& \hspace{1cm}\left[ \: \theta(-(p^0+q^0/2)) \theta(k^0-q^0/2) \theta(-(k^0+l^0)) \theta(p^0+l^0) \right. \nonumber \\[.2cm]
&& \hspace{2cm} \left. - \theta(p^0+q^0/2) \theta(-(k^0-q^0/2)) \theta(k^0+l^0) \theta(-(p^0+l^0)) \: \right] \nonumber \\[.2cm]
&& \ - \ (\mbox{with } q\leftrightarrow -q).
\label{DR1}
\end{eqnarray}
The $-1/2$ is due to fermionic statistics. The products of step functions restricts the energies $p^0, k^0$ and $l^0$ to order $q^0$; for example, the first product of step functions restricts $q^0/2<k^0<-l^0<p^0<-q^0/2$. We already argued that the $q$ suppression in the cut sub-diagrams is dictated by the step functions, so in the $Z$'s, the $V$'s and the on-shell delta functions, we can neglect the $q$-dependences, as well as the $p^0, k^0, l^0$ dependences. We are then led to
\begin{eqnarray}
&& 2(D_R)_1(p; q) \nonumber \\[.2cm]
&=& -\frac{1}{2}\int_{k, l} \frac{i^2 Z(\p)^3 Z(\kk) Z(\p+\l) Z(\kk+\l)}{(v^\mu(\p) q_\mu)^2} \ i^2 \left|V(p, k+l \rightarrow k, p+l)\right|^2 \nonumber \\[.2cm]
&& \hspace{.8cm} (2\pi)^4 \delta(\xi(\p)) \delta(\xi(\kk)) \delta(\xi(\p+\l)) \delta(\xi(\kk+\l)) \nonumber \\[.2cm]
&& \hspace{.8cm} \left[ \phantom{+} \theta(-(p^0+q^0/2)) \theta(k^0-q^0/2) \theta(-(k^0+l^0)) \theta(p^0+l^0) \right. \nonumber \\[.2cm]
&& \hspace{1cm} - \theta(p^0+q^0/2) \theta(-(k^0-q^0/2)) \theta(k^0+l^0) \theta(-(p^0+l^0)) \nonumber \\[.2cm]
&& \hspace{1cm} + \theta(-(k^0+q^0/2)) \theta(p^0-q^0/2) \theta(-(p^0+l^0)) \theta(k^0+l^0) \nonumber \\[.2cm]
&& \hspace{1cm} \left. - \theta(k^0+q^0/2) \theta(-(p^0-q^0/2)) \theta(p^0+l^0) \theta(-(k^0+l^0)) \: \right].
\end{eqnarray}
Inspecting the $p, q$ dependence, together with power counting, we justify the parametrization \eqref{Decay_parametrization} with non-negative $\gamma$ (what is remained to be shown is that the $\gamma$ here is the same $\gamma$ that appears in $\Im\Sigma$). In particular, the sign of $(D_R)_1$ is given by the sign of $q^0$, and so the time-ordered $D_1=(D_R)_1 \sgn\, q^0$ is non-negative. Next,
\begin{eqnarray}
&& 2(C^{ph}_R)_1(p, k; q) \nonumber \\[.2cm]
&=& \int_l \ (-1) \ \frac{iZ(\p-\q/2)}{-q^0-\xi(\p-\q/2) + \xi(\p+\q/2)} \ \frac{iZ(\kk+\q/2)}{q^0-\xi(\kk+\q/2) + \xi(\kk-\q/2)} \nonumber \\[.2cm]
&& \hspace{.5cm} iV(p-q/2, k+l \rightarrow k-q/2, p+l) \ iV(k+q/2, p+l \rightarrow p+q/2, k+l) \nonumber \\[.2cm]
&& \hspace{.5cm} (2\pi)^4 Z(\p+\q/2) Z(\kk-\q/2) Z(\p+\l) Z(\kk+\l) \nonumber \\[.2cm]
&& \hspace{.5cm} \delta(p^0+q^0/2-\xi(\p+\q/2)) \ \delta(k^0-q^0/2-\xi(\kk-\q/2)) \nonumber \\[.2cm]
&& \hspace{.5cm} \delta(p^0+l^0-\xi(\p+\l)) \ \delta(k^0+l^0-\xi(\kk+\l)) \nonumber \\[.2cm]
&& \hspace{.5cm}\left[ \: \theta(-(p^0+q^0/2)) \theta(k^0-q^0/2) \theta(-(k^0+l^0)) \theta(p^0+l^0) \right. \nonumber \\[.2cm]
&& \hspace{1.5cm} \left. - \theta(p^0+q^0/2) \theta(-(k^0-q^0/2)) \theta(k^0+l^0) \theta(-(p^0+l^0)) \: \right] \nonumber \\[.2cm]
&& \ + \ (\mbox{with } p\leftrightarrow k, \mbox{ except the arguments of the $V$'s kept unchanged}).
\label{CphR1}
\end{eqnarray}
Making the small $q$ approximations we made for $(D_R)_1$, we are led to
\begin{eqnarray}
&& 2(C^{ph}_R)_1(p, k; q) \nonumber \\[.2cm]
&=& - \int_{l} \frac{i^2 Z(\p)^2 Z(\kk)^2 Z(\p+\l) Z(\kk+\l)}{-(v^\mu(\p) q_\mu)(v^\mu(\kk) q_\mu)} \ i^2 \left|V(p, k+l \rightarrow k, p+l) \right|^2 \nonumber \\[.2cm]
&& \hspace{.8cm} (2\pi)^4 \delta(\xi(\p)) \delta(\xi(\kk)) \delta(\xi(\p+\l)) \delta(\xi(\kk+\l)) \nonumber \\[.2cm]
&& \hspace{.8cm} \left[ \phantom{+} \theta(-(p^0+q^0/2)) \theta(k^0-q^0/2) \theta(-(k^0+l^0)) \theta(p^0+l^0) \right. \nonumber \\[.2cm]
&& \hspace{1cm} - \theta(p^0+q^0/2) \theta(-(k^0-q^0/2)) \theta(k^0+l^0) \theta(-(p^0+l^0)) \nonumber \\[.2cm]
&& \hspace{1cm} + \theta(-(k^0+q^0/2)) \theta(p^0-q^0/2) \theta(-(p^0+l^0)) \theta(k^0+l^0) \nonumber \\[.2cm]
&& \hspace{1cm} \left. - \theta(k^0+q^0/2) \theta(-(p^0-q^0/2)) \theta(p^0+l^0) \theta(-(k^0+l^0)) \: \right].
\end{eqnarray}
This justifies the parametrization \eqref{Cph_parametrization} with non-negative $\lambda^{ph}$. And last,
\begin{eqnarray}
&& 2(C^{pp}_R)_1(p, k; q) \nonumber \\[.2cm]
&=& \int_l \left(-\frac{1}{2}\right) \frac{iZ(\p-\q/2)}{-q^0-\xi(\p-\q/2) + \xi(\p+\q/2)} \ \frac{iZ(\kk-\q/2)}{-q^0-\xi(\kk-\q/2) + \xi(\kk+\q/2)} \nonumber \\[.2cm]
&& \hspace{.5cm} iV(p-q/2, k+q/2 \rightarrow k-l, p+l) \ iV(k-l, p+l \rightarrow p+q/2, k-q/2) \nonumber \\[.2cm]
&& \hspace{.5cm} (2\pi)^4 Z(\p+\q/2) Z(\kk+\q/2) Z(\p+\l) Z(\kk-\l) \nonumber \\[.2cm]
&& \hspace{.5cm} \delta(p^0+q^0/2-\xi(\p+\q/2)) \ \delta(k^0+q^0/2-\xi(\kk+\q/2)) \nonumber \\[.2cm]
&& \hspace{.5cm} \delta(p^0+l^0-\xi(\p+\l)) \ \delta(k^0-l^0-\xi(\kk-\l)) \nonumber \\[.2cm]
&& \hspace{.5cm}\left[ \: \theta(-(p^0+q^0/2)) \theta(-(k^0+q^0/2)) \theta(k^0-l^0) \theta(p^0+l^0) \right. \nonumber \\[.2cm]
&& \hspace{1.5cm} \left. - \theta(p^0+q^0/2) \theta(k^0+q^0/2) \theta(-(k^0-l^0)) \theta(-(p^0+l^0)) \: \right] \nonumber \\[.2cm]
&& \ - \ (\mbox{with } q\leftrightarrow -q, \mbox{ except the arguments of the $V$'s kept unchanged}).
\label{CppR1}
\end{eqnarray}
Making the small $q$ approximations we made for $(D_R)_1$, we are led to
\begin{eqnarray}
&& 2(C^{pp}_R)_1(p, k; q) \nonumber \\[.2cm]
&=& -\frac{1}{2} \int_{l} \frac{i^2 Z(\p)^2 Z(\kk)^2 Z(\p+\l) Z(\kk-\l)}{(v^\mu(\p) q_\mu)(v^\mu(\kk) q_\mu)} \ i^2 |V(p, k \rightarrow k-l, p+l)|^2 \nonumber \\[.2cm]
&& \hspace{.8cm} (2\pi)^4 \delta(\xi(\p)) \delta(\xi(\kk)) \delta(\xi(\p+\l)) \delta(\xi(\kk-\l)) \nonumber \\[.2cm]
&& \hspace{.8cm} \left[ \phantom{+} \theta(-(p^0+q^0/2)) \theta(-(k^0+q^0/2)) \theta(k^0-l^0)) \theta(p^0+l^0) \right. \nonumber \\[.2cm]
&& \hspace{1cm} - \theta(p^0+q^0/2) \theta(k^0+q^0/2) \theta(-(k^0-l^0)) \theta(-(p^0+l^0)) \nonumber \\[.2cm]
&& \hspace{1cm} + \theta(k^0-q^0/2)) \theta(p^0-q^0/2) \theta(-(p^0+l^0)) \theta(-(k^0-l^0)) \nonumber \\[.2cm]
&& \hspace{1cm} \left. - \theta(-(k^0-q^0/2)) \theta(-(p^0-q^0/2)) \theta(p^0+l^0) \theta(k^0-l^0) \: \right].
\end{eqnarray}
This justifies the parametrization \eqref{Cpp_parametrization} with non-negative $\lambda^{pp}$. 

Inspecting the exact expressions \eqref{DR1}, \eqref{CphR1} and \eqref{CppR1} (these expressions are before we apply power counting, and hence hold in $d=2$ as well), we can observe the important relation \eqref{C_D_relation}. In particular, to see the second equality in \eqref{C_D_relation}, we shift $k\rightarrow k+l \mp q/2$ in the two cut sub-diagrams contributing to $(C^{pp}_R)_1$. This leads to \eqref{Collision_WTId}, which is required by the Ward-Takahashi identity.

We have computed the order $q$ contributions to $\Im\Pi_R$. Is there an associated part in $\Re\Pi$ through the Kramers-Kronig dispersion relation \eqref{dispersion_relation}? For $\Im\Sigma_R$, and hence the $(D_1)_R$ contribution to $\Im\Pi_R$, it is known that the associated real part amounts to a correction to $Z(\p)$; but $Z(\p)$ itself appeared in our cutting rule to start with, so this just means we must use the self-consistent, i.e. physical, value of $Z(\p)$. As long as we have done so, there is no further contribution to $\Re\Pi$ from $(D_1)_R$. From Ward identity we know $(C^{ph})_R$ and $(C^{pp})_R$ must have no further contribution to $\Re\Pi$ either. This differs from the scenario in the zeroth order contribution, where $\Delta'$ in \eqref{Delta_prime} has both real and imaginary parts.

It remains to show the $\gamma$ in the parametrization of $D_1$ is the same $\gamma$ that appears in $\Im\Sigma$. Now we apply the Cutkosky cutting rule to $-\Sigma(p)$ with small $p^0$. The leading cut diagram is order $(p^0)^2$, involving three cut propagators~\cite{Luttinger:1961zz}. More explicitly, at this order we have
\begin{eqnarray}
2\Im(-\Sigma_R)(p) &=& \int_{k, l} \left(-\frac{1}{2} \right) i^2 |V(p, k+l \rightarrow k, p+l)|^2 \ (2\pi)^3 Z(\kk) Z(\kk+\l) Z(\p+\l) \nonumber \\[.2cm]
&& \hspace{.75cm} \delta(k^0-\xi(\kk)) \delta(k^0+l^0-\xi(\kk+\l)) \delta(p^0+l^0-\xi(\p+\l)) \nonumber \\[.2cm]
&& \hspace{.75cm} \left[ \: (-1)^2 \ \theta(-k^0) \theta(k^0+l^0) \theta(-(p^0+l^0)) \phantom{{X^T}^T} \right. \nonumber \\[.2cm]
&& \hspace{1.5cm} \left. \phantom{{X^T}^T} - (-1) \ \theta(k^0) \theta(-(k^0+l^0)) \theta(p^0+l^0) \: \right].
\end{eqnarray}
We can see $-\Im\Sigma_R$ is explicitly positive, as it should~\cite{Luttinger:1961zz, abrikosov1975methods}. Combined with power counting, we justify the parametrization $-\Im\Sigma=-\Im\Sigma_R \, \sgn\, p^0 = \gamma' p^0 |p^0|$ at small $p^0$ with positive $\gamma'$. To verify $\gamma'=(3/2)\gamma$, we compare the evaluations of $-\Im\Sigma_R$ and $(D_R)_1$ and restrict to order $q$, we find
\begin{eqnarray}
&& 2(D_R)_1(p; q) \nonumber \\[.2cm]
&=& \frac{i^2 Z^2}{(v^\mu q_\mu)^2} \ 2\pi Z \delta(\xi) \left(\theta(-(p^0+q^0/2)) \theta(p^0-q^0/2) + \theta(p^0+q^0/2) \theta(-(p^0-q^0/2)) \right) \nonumber \\[.2cm]
&& \left[ \: \sgn(p^0-q^0/2) (-2\Im\Sigma_R(p+q/2)) - \sgn(p^0+q^0/2) (-2\Im\Sigma_R(p-q/2)) \: \right] \nonumber \\[.2cm]
&=& 2\frac{Z^3 \: 2\pi \delta(\xi)}{(v^\mu q_\mu)^2}\: \sgn(q^0)\theta(|q^0|/2-|p^0|) \left(-\Im\Sigma_R(p-q/2) - \Im\Sigma_R(p+q/2) \right).
\label{D_Sigma_relation}
\end{eqnarray}
The last factor is equal to $\gamma'(\p) \left(2(p^0)^2 + (q^0)^2/2\right)$ to leading order. Now, we average $p^0$ between $\pm|q^0|/2$ before replacing the step function with $\delta(p^0) q^0$; this is equivalent to making the approximation
\begin{eqnarray}
\sgn(q^0)\: \theta(|q^0|/2-|p^0|) = 2\frac{\delta(p^0)}{1!}\: \frac{q^0}{2} + 2 \frac{\partial_{p^0}^2 \delta(p^0)}{3!} \left(\frac{q^0}{2}\right)^3 + \cdots,
\end{eqnarray}
where the second term is needed to take into account the $(p^0)^2$ from $\Im\Sigma_R$. This leads to \eqref{Decay_parametrization}, and in particular, verifies $\gamma'=(3/2)\gamma$.

\subsubsection{Difficulties in $d=2$}

As we mentioned previously, the power counting of $q$ presented there has missing piece. Here it is: The power counting relied on the assumption that the internal momenta carried by the cut propagators are generically not close to each other. However, in the integration over the internal momenta, there must be some region where this assumption does not hold -- the internal momenta, restricted near the FS, can appear collinear with one another. To estimate the scale of the the contribution from the collinear regime, one views this regime as a quasi-one-dimensional system~\cite{baym1978physics}. Take the self-energy $\Sigma$ for example. In $d=1$ dimension, the self-energy $\Sigma\sim p^0 \ln p^0$ -- this leads to the well-known non-Fermi liquid behavior. In higher dimensions, quasi-one-dimensional power counting estimates the contribution from the collinear regime to be $\sim (p^0)^d \ln p^0$. For $d=2$ this is less suppressed than the usual $(p^0)^2$.

Denote the collinear regime contribution to $\Sigma$ as $\delta\Sigma$; the leading contribution has three intermediate collinear on-shell fermions. Based on the quasi-one-dimensional power counting, along with the general analyticity requirements that $-\Sigma_R(\omega, \p)$ is analytic in the $\omega$ upper-half-plane and $-\Im\Sigma_R>0$~\cite{Luttinger:1961zz, abrikosov1975methods}, it is tempting to parametrize the collinear regime contribution as
\begin{eqnarray}
-\delta\Sigma_R(\omega, \p) = ia(\p) \, \omega^2 \ln(-i\omega/p_F) + (\mbox{higher order contributions})
\end{eqnarray}
for small $\omega$ in the upper-half-plane. Here $p_F$ is the size scale of the FS, and $a$ is positive; the branch cut of $\ln$ is placed along the negative real axis. Taking $\omega=p^0+i\epsilon$ gives $-\delta\Sigma_R(p)$. Unfortunately, when $\xi(\p)$ is of the same order as $p^0$, such parametrization is wrong. It is known~\cite{chubukov2003nonanalytic,chubukov2005singular} that $-\delta\Sigma_R$ has complicated dependence on $p^0$ and $p^0-\xi(\p)$, but still scales as $(p^0)^2 \ln p^0$ when $\xi(\p)$ and $p^0$ are of the same order. 

Now that there is no simple way to parametrize $-\delta\Im\Sigma$, there is no simple way to parametrize the decay factor $-D_1$ (despite the $1$ subscript, here it also involves terms of order $q \ln q$), because the latter can be expressed in terms of the former, with $p^0$ being order $q^0$ and $p^0-\xi$ being order $v^\mu q_\mu$. The collinear regime contributions to $-C^{ph}_1$ and $-C^{pp}_1$ are terms of order $q \ln q$ proportional to $\delta^{d+1}(p\mp k)$ (respectively corresponding to forward and back scattering); this terms have no simple parametrization either.

Although the collision term in $d=2$ is not parametrized, we know \eqref{C_D_relation} must still hold, as it is required by the Ward-Takahashi identity.

\subsubsection{Collision being Uninteresting}

Finally we want to show that collisions are ``uninteresting" to the focus of this thesis -- the collision term $C_1$ has no contribution to the antisymmetric part $\Pi^{[\mu\nu]}$ of the current-current correlation, and therefore has no contribution to the anomalous Hall effect or the chiral magnetic effect. First, we note that $C_1$ (despite the $1$ subscript, in $d=2$ it also involves terms of order $q\ln q$) has a special property
\begin{eqnarray}
C_1(p, k; q)=C_1(k, p; q).
\label{C1_pk_symm}
\end{eqnarray}
In particular, the $D_1$ term in $C_1$ has this property simply because it is proportional to $\delta^{d+1}(p-k)$. On the other hand, inspecting the exact expressions \eqref{CphR1} and \eqref{CppR1}, we see $C^{ph}_1$ and $C^{pp}_1$ are symmetric under $p\leftrightarrow k$ up to the $q$-dependences in the $V$'s. But the $q$-dependences in the $V$'s can be neglected, for theirs effects would be further suppressed by order $q$, i.e. contribute to order $q^2$ (and order $q^2 \ln q$ in $d=2$) which are beyond order $q$. Therefore \eqref{C1_pk_symm} holds to order $q$ -- including in $d=2$, since the argument here did not rely on power counting. The collision contribution to $i(i\Im\Pi^{\mu\nu})$ is given by
\begin{eqnarray}
&& -\int_{p, k} i\Re\Gamma_0^{\mu}(p; -q) \ (-C_1(p, k; q)) \ i\Re\Gamma_0^{\nu}(k; q) \nonumber \\[.2cm]
&& \hspace{2cm} -\int_{p, k} i \left(i\Im\Gamma_0^{\mu} \right)(p; -q) \ (-C_1(p, k; q)) \ i\left(i\Im\Gamma_0^{\nu}\right)(k; q).
\end{eqnarray}
The $q$-dependence in $\Gamma_0^\mu$ comes from $\Delta'$ which is even in $q$, so $\Gamma_0^{\mu}$ is also even in $q$. Thus, due to \eqref{C1_pk_symm}, the expression above is symmetric in $\mu\nu$. Now that $C_1$ does not contribute to $\Im\Pi^{[\mu\nu]}$, by the Kramers-Kronig dispersion relation it has no associated contribution to $\Re\Pi^{[\mu\nu]}$ either, and thus our claim is proven.

\subsection{Some Cancellation of Diagrams}
\label{ssect_QFT_AppendixCD}

In the presentation of the proof, there are two technical diagrammatic cancellations that are left unproven. One is \eqref{tildeY} in the EM dipole moment, and the other is the vanishing of the $\wt{V}_1$ term in \eqref{sigma_Y_ref} in the Hall conductivity tensor. Here we complete these technical steps. 

\subsubsection{In Electromagnetic Dipole Moment}

We explicitly show from Feynman diagrams the second line of \eqref{EM_dipole_anom} is antisymmetric in $\mu\nu$. Consider diagrams in the $q$-2PI sum $i\wt{V}^{\alpha \ \gamma}_{\ \delta, \ \beta}(p, k; q)$. We can separate these diagrams into two types:
\begin{itemize}
\item
Type I: The $q$-2PI diagram has a fermion line at the top, running in with momentum and index $(p-q/2, \delta)$ and running out with $(k-q/2, \gamma)$, and a fermion line at the bottom, running in with momentum and index $(k+q/2, \beta)$ and running out with $(p+q/2, \alpha)$. Type I diagrams are summed in $\wt{V}$ with plus sign. Some examples are shown below.

\begin{center}
\begin{fmffile}{zzz-A-TypeI-1}
\begin{fmfgraph*}(40, 20)
\fmfleftn{l}{2}\fmfrightn{r}{2}
\fmf{fermion}{r1,v1,l1}
\fmf{fermion}{l2,v2,v3,r2}
\fmf{dashes, tension=0}{v1,v2}
\fmf{dashes, tension=0}{v1,v3}
\end{fmfgraph*}
\end{fmffile}
\begin{fmffile}{zzz-A-TypeI-2}
\begin{fmfgraph*}(40, 20)
\fmfleftn{l}{2}\fmfrightn{r}{2}
\fmf{fermion}{l2,v3,v4,r2}
\fmf{fermion}{r1,v2,v1,l1}
\fmf{dashes,tension=0}{v1,v3}
\fmffreeze
\fmf{phantom}{v2,v7,v8,v4}
\fmf{dashes,tension=0}{v2,v7}
\fmf{dashes,tension=0}{v4,v8}
\fmf{fermion,left,tension=0}{v7,v8}
\fmf{fermion,left,tension=0}{v8,v7}
\fmffreeze
\fmf{phantom}{r1,v2,v1,v5,v6,l1}
\fmf{dashes, right=0.7, tension=0}{v2,v5}
\end{fmfgraph*}
\end{fmffile}
\begin{fmffile}{zzz-A-TypeI-3}
\begin{fmfgraph*}(40, 20)
\fmfleftn{l}{2}\fmfrightn{r}{2}
\fmf{fermion}{l2,v1,v2,r2}
\fmf{fermion}{r1,v3}
\fmf{fermion,tension=1.5}{v3,v4,v5,l1}
\fmffreeze
\fmf{phantom}{l1,v6,v3}
\fmffreeze
\fmf{phantom}{v1,v7,v9,v8,v6}
\fmf{fermion,left,tension=0.15,tag=1}{v7,v8}
\fmf{fermion,left,tension=0.15}{v8,v7}
\fmffreeze
\fmf{dashes,tension=1.5}{v2,v10,v3}
\fmf{dashes}{v1,v7}
\fmf{dashes}{v8,v4}
\fmf{dashes}{v8,v5}
\fmffreeze
\fmfposition
\fmfipath{p[]}
\fmfiset{p1}{vpath1(__v7,__v8)}
\fmfi{dashes}{point length(p1)/2 of p1 -- vloc(__v10)}
\fmfdot{v10}
\end{fmfgraph*}
\end{fmffile}
\end{center}

For Type I diagrams, we can assign momenta on the internal propagators so that $-q/2$ runs through the top fermion line, $+q/2$ runs through the bottom fermion line, and all other internal propagators are independent of $q$.

\item
Type II: The $q$-2PI diagram has a fermion line on the left, running in with momentum and index $(p-q/2, \delta)$ and running out with $(p+q/2, \alpha)$, and a fermion line on the right, running in with momentum and index $(k+q/2, \beta)$ and running out with $(k-q/2, \gamma)$. Type II diagrams are summed in $\wt{V}$ with minus sign, because of fermionic statistics. Some examples are shown below.

\begin{center}
\begin{fmffile}{zzz-A-TypeII-1}
\begin{fmfgraph*}(40, 20)
\fmfleftn{l}{2}\fmfrightn{r}{2}
\fmf{fermion,tension=2}{r1,v1,v2,r2}
\fmf{fermion,tension=2}{l2,v4,v3,l1}
\fmf{dashes,left=0.3,tension=0.4}{v1,v3}
\fmf{dashes,left=0.3,tension=0.4}{v4,v2}
\end{fmfgraph*}
\end{fmffile}
\begin{fmffile}{zzz-A-TypeII-2}
\begin{fmfgraph*}(40, 20)
\fmfleftn{l}{2}\fmfrightn{r}{2}
\fmf{fermion,tension=2}{l2,v1,l1}
\fmf{fermion,tension=1.5}{r1,v2,v3,r2}
\fmf{phantom,tension=2}{r1,v2}
\fmf{phantom,tension=2}{v3,r2}
\fmf{dashes,tension=2}{v1,v4}
\fmf{phantom}{v2,v5,v6,v4}
\fmf{phantom}{v3,v7,v8,v4}
\fmf{dashes,tension=0}{v2,v5}
\fmf{dashes,tension=0}{v3,v7}
\fmffreeze
\fmf{fermion,left=0.7}{v4,v7}
\fmf{fermion}{v7,v5}
\fmf{fermion,left=0.7}{v5,v4}
\fmf{dashes,left=0.15}{v7,v4}
\end{fmfgraph*}
\end{fmffile}
\begin{fmffile}{zzz-A-TypeII-3}
\begin{fmfgraph*}(40, 20)
\fmfleftn{l}{2}\fmfrightn{r}{2}
\fmf{fermion,tension=1}{r1,v1,v12,v8,v2,r2}
\fmf{fermion,tension=1}{l2,v4,v9,v3,l1}
\fmf{dashes,tension=2}{v1,v6}
\fmf{dashes,tension=2}{v7,v3}
\fmf{fermion,left,tension=1.2}{v6,v7}
\fmf{fermion,left,tension=1.2}{v7,v6}
\fmf{phantom,tension=0.7}{v2,v4}
\fmf{phantom,tension=2}{r1,v1}
\fmf{phantom,tension=2}{r2,v2}
\fmf{phantom,tension=2}{l1,v3}
\fmf{phantom,tension=2}{l2,v4}
\fmffreeze
\fmf{dashes,tension=1.2}{v8,v11}
\fmf{dashes,tension=1}{v9,v10}
\fmf{dashes,tension=1}{v4,v10}
\fmf{fermion,left,tension=1}{v11,v10}
\fmf{fermion,left,tension=1}{v10,v11}
\fmf{dashes,left=0.5,tension=0.4}{v12,v2}
\end{fmfgraph*}
\end{fmffile}
\end{center}

For Type II diagrams, however we assign the internal momenta, $q$ will in general appear in some propagator(s) on both the left and right fermion lines, as well as on some internal propagators in the middle.

\end{itemize}
We want to show $\int_k \left(i\wt{V}_1\right)^{\alpha \ \gamma}_{\ \delta, \ \beta}(p, k; q) \ \partial_k^\nu iG^\beta_{\ \gamma}(k)$ is equal to $q_\mu$ times a quantity antisymmetric in $\mu\nu$. We consider the Type I and Type II contributions separately.

For Type I diagrams in $i\wt{V}(p, k; q)$, when expanded to linear order in $q$, we pick one propagator $iG(l\pm q/2)$ on the bottom (top) fermion line and replace it with $(\pm q_\mu/2) \partial_l^\mu iG(l)$, and in all other propagators set $q$ to zero; we sum up all possible ways of such expansion, and sum up all possible Type I diagrams. (One may wonder why the interaction vertices are independent of $q$. We explain this at the end of this section.) As a result, Type I contribution to $\int_k \left(i\wt{V}_1\right)^{\alpha \ \gamma}_{\ \delta, \ \beta}(p, k; q) \ \partial_k^\nu iG^\beta_{\ \gamma}(k)$ can be summarized as
\begin{eqnarray}
\frac{q_\mu}{2} \int_k \int_l \left(i\wt{Y}\right)^{\alpha \: \ \xi \: \ \gamma}_{\ \delta, \ \zeta, \ \beta}(p, l, k) \ \partial_l^\mu iG^\zeta_{\ \xi}(l) \ \partial_k^\nu iG^\beta_{\ \gamma}(k),
\label{tildeY}
\end{eqnarray}
where $i\wt{Y}$ is a sum of connected diagrams:
\begin{itemize}
\item
Diagrams satisfying the following conditions are summed in $i\wt{Y}$ with plus sign:

There is a fermion line running in with momentum and index $(p, \delta)$ and running out with $(k, \gamma)$, a fermion line running in with momentum and index $(k, \beta)$ and running out with $(l, \xi)$, and a fermion line running in with momentum and index $(l, \zeta)$ and running out with $(p, \alpha)$.

Moreover, among the internal fermion propagators on these three fermion lines, none of them is dictated by momentum conservation to have momentum $p, k$ or $l$, and no pair of them is dictated by momentum conservation to have same momenta. Equivalently, among those fermion propagators, one cannot cut any one or two of them to disconnect the diagram.
\vspace{.2cm}
\begin{center}
\begin{fmffile}{zzz-wtY}
\begin{fmfgraph*}(35, 35)
\fmfsurroundn{v}{24}
\fmflabel{$\alpha$}{v12}
\fmflabel{$\delta$}{v10}
\fmflabel{$\xi$}{v20}
\fmflabel{$\zeta$}{v18}
\fmflabel{$\gamma$}{v4}
\fmflabel{$\beta$}{v2}
\fmf{fermion,label.side=left,label=$k$}{pk2,v4}
\fmf{fermion,label.side=left,label=$p$}{v10,pk1}
\fmf{fermion}{pk1,pk2}
\fmf{fermion,label.side=left,label=$p$}{lp3,v12}
\fmf{fermion,label.side=left,label=$l$}{v18,lp1}
\fmf{fermion}{lp1,lp2}
\fmf{fermion}{lp2,lp3}
\fmf{fermion,label.side=left,label=$l$}{kl1,v20}
\fmf{fermion,label.side=left,label=$k$}{v2,kl1}
\fmf{dashes,left=0.2,tension=0.3}{pk2,lp2}
\fmf{phantom,left=0.3,tension=0.3}{kl1,pk1}
\fmf{dashes,left,tension=0}{lp3,lp1}
\fmf{phantom,left,tension=0.5}{lp1,lp2,lp3}
\fmffreeze
\fmf{dashes,tension=1}{pk1,o1}
\fmf{dashes,tension=2.5}{o2,kl1}
\fmf{fermion,left}{o1,o2}
\fmf{fermion,left}{o2,o1}
\end{fmfgraph*}
\end{fmffile}
\hspace{1cm}
\begin{fmffile}{zzz-not-wtY-1}
\begin{fmfgraph*}(28, 28)
\fmfsurroundn{v}{24}
\fmf{fermion}{pk2,v4}
\fmf{fermion}{v10,pk1}
\fmf{fermion,tension=2}{pk1,pk2}
\fmf{fermion}{lp3,v12}
\fmf{fermion}{v18,lp1}
\fmf{fermion}{lp1,lp2}
\fmf{fermion}{lp2,lp3}
\fmf{fermion}{kl1,v20}
\fmf{fermion}{v2,kl1}
\fmf{dashes,left=0.3,tension=0.3}{pk1,lp2}
\fmf{dashes,left=0.3,tension=0.3}{kl1,pk2}
\fmf{dashes,left,tension=0}{lp3,lp1}
\fmf{phantom,left,tension=0.5}{lp1,lp2}
\end{fmfgraph*}
\end{fmffile}
\hspace{0.1cm}
\begin{fmffile}{zzz-not-wtY-2}
\begin{fmfgraph*}(28, 28)
\fmfsurroundn{v}{24}
\fmf{fermion}{pk2,v4}
\fmf{fermion}{v10,pk1}
\fmf{fermion}{pk1,pk2}
\fmf{fermion}{lp2,v12}
\fmf{fermion}{v18,lp1}
\fmf{fermion}{lp1,lp2}
\fmf{fermion}{kl2,v20}
\fmf{fermion}{kl1,kl2}
\fmf{fermion}{v2,kl1}
\fmf{dashes,left=0.3,tension=0.3}{pk2,lp2}
\fmf{dashes,left=0.3,tension=0.3}{kl1,pk1}
\fmf{dashes,left=0.2,tension=0}{pk2,kl1}
\fmf{dashes,tension=0.3}{kl2,lp1}
\end{fmfgraph*}
\end{fmffile}
\hspace{0.1cm}
\begin{fmffile}{zzz-not-wtY-3}
\begin{fmfgraph*}(28, 28)
\fmfsurroundn{v}{24}
\fmf{fermion}{v2,kk,v4}
\fmf{fermion}{v10,pp,v12}
\fmf{fermion}{v18,ll1,ll2,v20}
\fmf{dashes,tension=1.3}{kk,o}
\fmf{dashes,tension=1.3}{pp,o}
\fmf{dashes,right=0.2}{ll2,o}
\fmfdot{o}
\fmf{dashes,left=0.2,tension=0.4}{pp,ll1}
\end{fmfgraph*}
\end{fmffile}
\end{center}
\vspace{.5cm}
For example, the diagram on the left contributes to $i\wt{Y}$, while the three on the right do not.

\item
Diagrams satisfying the conditions above, but with $(k, {}^\gamma_{\ \beta})$ and $(l, {}^\xi_{\ \zeta})$ switched, are summed in $i\wt{Y}$ with minus sign.
\end{itemize}
Notice that $i\wt{Y}$ is a totally antisymmetric 3-tensor in the double fermion linear space, i.e. it is antisymmetric under the exchange of any two of $(p, {}^\alpha_{\ \delta})$, $(k, {}^\gamma_{\ \beta})$ and $(l, {}^\xi_{\ \zeta})$. Thus, Type I contribution is antisymmetric in $\mu\nu$.

For Type II diagram contribution, we use a diagrammatic technique developed by Ward~\cite{Ward:1950xp}. Consider, for instance, the diagram below.
\begin{center}
\begin{fmffile}{zzz-A-prototype-no-symm}
\begin{fmfgraph*}(50, 20)
\fmfleftn{l}{5}\fmfrightn{r}{5}
\fmf{fermion,tension=5}{l4,v4,v2,l2}
\fmf{phantom}{v2,r1}
\fmf{phantom}{v4,r5}
\fmf{phantom}{v2,v3,v4}
\fmffreeze
\fmf{phantom}{v3,v6}
\fmf{phantom,tension=1}{r3,v6}
\fmffreeze
\fmf{fermion,left,tension=0,tag=1}{v6,r3}
\fmf{plain,left,tension=0,tag=2}{r3,v6}
\fmf{dashes,tension=2.5}{v7,v2}
\fmf{fermion,left,tension=2}{v7,v8}
\fmf{fermion,left,tension=2}{v8,v7}
\fmf{phantom,tension=1}{v8,r2}
\fmffreeze
\fmfipath{p[]}
\fmfiset{p1}{vpath1(__v6,__r3)}
\fmfiset{p2}{vpath2(__r3,__v6)}
\fmfi{dashes}{point length(p1)/3 of p1 .. vloc(__v4)}
\fmfi{dashes}{point 8length(p2)/8 of p2 .. vloc(__v8)}
\fmfi{dashes}{point 5length(p2)/7 of p2 .. vloc(__v8)}
\fmfi{dashes}{point 4length(p1)/7 of p2 .. point length(p1)/5 of p1}
\end{fmfgraph*}
\end{fmffile}
\end{center}
Let us call this a ``prototype diagram''. Let us call the loop on the right the $k$-loop, whose loop momentum is assigned $k$; it consists of five fermion propagators. A Type II diagram contribution to $\int_k i\wt{V}_1(p, k; q) \ \partial_k^\nu iG(k)$ is obtained by picking one propagator $iG$ on the $k$-loop -- in the example above there are five ways to do so -- and replacing it with $\partial_k^\nu iG$, and then letting the external momentum $q$ flow-in through it. The sum of all these five resulting diagrams forms a ``prototype class'' associated with the above prototype diagram. For each prototype class, we can fix one interaction propagator (it maybe an auxiliary propagator), through which $q$ in the $k$-loop flows out towards the left; for instance, in the example above, we may assign internal momenta so that $q$ always flows out the $k$-loop along the dashed line on the top. Now, for each Type II diagram in this prototype class, we expand $q$ at first order (since we are looking at $\wt{V}_1$). This corresponds to picking one internal propagator (fermion or interaction) that has $q$ in its argument, replacing it with $q_\mu$ times its momentum derivative, and then in all other internal propagators set $q$ to zero. Now:
\begin{itemize}
\item
If our picked $q$-dependent propagator is on the $k$-loop, then we have a $\partial_k^\nu iG(k)$ and a $\partial_k^\mu iG(k)$ on the $k$-loop, and summing up all such possibilities in the prototype class yields a quantity antisymmetric in $\mu\nu$, in a manner similar to the Type I diagram contribution.
\item
If our picked $q$-dependent propagator is not on the $k$-loop, then there is only one $\partial_k^\nu iG(k)$ on the $k$-loop, and summing up all such possibilities is equivalent to taking a total $k$-derivative on the $k$-loop (and $k$ is integrated over later). So the sum of such possibilities vanishes.
\end{itemize}
Thus, Type II diagram contribution to $\int_k (\partial_q^\mu i\wt{V}_1(p, k; q)) \ \partial_k^\nu iG(k)$ is also antisymmetric in $\mu\nu$. (There is a small caveat in the use of prototype diagrams. For example the prototype below
\begin{center}
\begin{fmffile}{zzz-A-prototype-symm}
\begin{fmfgraph*}(50, 20)
\fmfleftn{l}{5}\fmfrightn{r}{5}
\fmf{fermion,tension=5}{l4,v4,v2,l2}
\fmf{phantom}{v2,r1}
\fmf{phantom}{v4,r5}
\fmf{phantom}{v2,v3,v4}
\fmffreeze
\fmf{phantom}{v3,v6}
\fmf{phantom,tension=1}{r3,v6}
\fmffreeze
\fmf{fermion,left,tension=0,tag=1}{v6,r3}
\fmf{plain,left,tension=0,tag=2}{r3,v6}
\fmffreeze
\fmfposition
\fmfipath{p[]}
\fmfiset{p1}{vpath1(__v6,__r3)}
\fmfiset{p2}{vpath2(__r3,__v6)}
\fmfi{dashes}{point length(p1)/5 of p1 -- vloc(__v4)}
\fmfi{dashes}{point length(p1)/5 of p1 -- vloc(__v2)}
\fmfi{dashes}{point 4length(p1)/5 of p2 -- vloc(__v2)}
\fmfi{dashes}{point 4length(p1)/5 of p2 -- vloc(__v4)}
\end{fmfgraph*}
\end{fmffile}
\end{center}
has a symmetry of exchanging the two propagators on the $k$-loop. So when relating it to Type II contribution by replacing one $iG$ on the $k$-loop with $\partial_k^\nu iG$, we need an extra factor of $1/2$. Clearly this does not affect the final antisymmetry in $\mu\nu$.)

A left-over subtlety has to be addressed: When we were expanding the momentum running along a fermion line, we did not have contribution from the interaction vertices on the fermion line. Why is that? Recall our assumption (for simplicity, not for principle) about the QFT that any bare interaction vertex has no coupling to $A$, i.e. there is no e.g. $A\phi\psi^\dagger \psi$ bare vertex or $A\psi^\dagger \psi \psi^\dagger \psi$ bare vertex. By EM $U(1)$ gauge invariance, this also means the bare interaction vertices cannot depend on the momentum running along the charged fermion line. So a bare interaction vertex at most depends on the momentum running along the neutral interaction lines (which maybe auxiliary), for example in $(\psi^\dagger \psi) \: \partial_x^2 \phi$ or in $(\psi^\dagger \psi)^2\partial_x^2(\psi^\dagger \psi)$.

\subsubsection{In Hall Conductivity Tensor}

In this section we show
\begin{eqnarray}
\sigma_Y^{\mu\nu\lambda} \equiv (i\partial_p^\mu G)^T \ \partial_q^\nu i\wt{V}_1(q) \ i\partial_k^\lambda G,
\end{eqnarray}
which appears in \eqref{sigma_Y_ref}, vanishes. According to the previous discussion, we separate the contributions of Type I diagrams and Type II diagrams in $i\wt{V}_1$. In $\sigma_Y$, since $\partial iG$ is contracted on both sides, it is easy to see Type II contribution vanishes using Ward's method presented previously -- at least one of the $k$-loop and the $p$-loop involves a total derivative. We are left with Type I contribution to $\sigma_Y$. By \eqref{tildeY} we can express it as
\begin{eqnarray}
\sigma_Y^{\mu\nu\lambda} = \frac{1}{2} \int_k \int_l \int_p \left(i\wt{Y}\right)^{\alpha \: \ \xi \: \ \gamma}_{\ \delta, \ \zeta, \ \beta}(p, l, k) \ \partial_p^\mu iG^\delta_\alpha(p) \ \partial_l^\nu iG^\zeta_{\ \xi}(l) \ \partial_k^\lambda iG^\beta_{\ \gamma}(k).
\end{eqnarray}
We can describe diagrams in $\sigma_Y^{\mu\nu\lambda}$ as the following:
\begin{itemize}
\item
The diagram has a fermion loop, which we call the outer loop (formed by connecting the three fermion lines in $i\wt{Y}$ with the three differentiated propagators). We redefine $p$ so that it is now the loop momentum running around the outer loop. Interaction vertices separate the outer loop into $n\geq 3$ segments. Three of the segments are differentiated propagators $\partial_p^\mu iG$, $\partial_p^\nu iG$ and $\partial_p^\lambda iG$, the remaining $n-3$ segments are propagators $iG$.

Moreover, the interaction lines inside make the outer loop 2PI; that is, among all segments on the outer loop, no pair of them are dictated by momentum conservation to have the same momentum.

If $\partial_p^\mu$, $\partial_p^\nu$, $\partial_p^\lambda$ appear on the outer loop in the cyclic order against the fermion arrow, then the diagram is summed in $\sigma_Y^{\mu\nu\lambda}$ with coefficient $-1/2$ (the minus sign is because now we have an extra fermion loop -- the outer loop -- compared to $\wt{Y}$). If they appear on the outer loop in the cyclic order along the fermion arrow, then the diagram is summed with coefficient $+1/2$.
\end{itemize}
To show $\sigma_Y^{\mu\nu\lambda}=0$, below we introduce four notions.

First, let us be blind between $iG$ and $\partial_p iG$ on the outer loop. Then we are led to consider prototype diagrams like this one
\begin{center}
\begin{fmffile}{zzz-B-prototype}
\begin{fmfgraph*}(30, 30)
\fmfleftn{l}{1}\fmfrightn{r}{1}
\fmf{fermion,left,tag=1}{l1,r1}
\fmf{plain,left,tag=2}{r1,l1}
\fmffreeze
\fmfsurroundn{v}{24}
\fmf{phantom}{v12,ml1,mr1,v3}
\fmfdot{mr1}
\fmf{phantom}{v14,ml2,mr2,v2}
\fmffreeze
\fmf{phantom}{mr2,mm,v20}
\fmf{fermion,left,tension=0}{ml2,ml1}
\fmf{fermion,left,tension=0}{ml1,ml2}
\fmf{fermion,left,tension=0,tag=3}{mr2,mm}
\fmf{fermion,left,tension=0,tag=4}{mm,mr2}
\fmffreeze
\fmfposition
\fmfipath{p[]}
\fmfiset{p1}{vpath1(__l1,__r1)}
\fmfiset{p2}{vpath2(__r1,__l1)}
\fmfiset{p3}{vpath3(__mr2,__mm)}
\fmfiset{p4}{vpath4(__mm,__mr2)}
\fmfi{dashes}{point 1length(p1)/5 of p1{right} .. vloc(__mr1)}
\fmfi{dashes}{point 5length(p1)/7 of p1{left} .. vloc(__mr1)}
\fmfi{dashes}{point 0length(p3)/7 of p3 -- vloc(__mr1)}
\fmfi{dashes}{point 1length(p1)/3 of p1.. {down}vloc(__ml1)}
\fmfi{dashes}{point 4length(p2)/5 of p2{right} .. vloc(__ml2)}
\fmfi{dashes}{point 1length(p4)/8 of p4{down} .. point 1length(p2)/2 of p2}
\fmfi{dashes}{point 7length(p3)/9 of p3{down} .. point 1length(p2)/4 of p2}
\end{fmfgraph*}
\end{fmffile}
.
\end{center}
A prototype diagram defines a prototype class: Diagrams contributing to $\sigma_Y$ are in the same prototype class if, after ignoring the distinction between $\partial_p iG$ and $iG$ on the outer loop, they reduce to the same prototype diagram. In fact, the sum $S^{\mu\nu\lambda}$ (we drop the $\mu\nu\lambda$ indices from here on) of diagrams (with coefficients $\pm 1/2$ assigned as before) within a prototype class vanishes, as we will show later. This leads to $\sigma_Y=0$, because clearly diagrams in $\sigma_Y$ are partitioned into prototype classes.

Second, let us fix a prototype class, and consider the placement of the three $\partial_p iG$'s on the outer loop. For simplicity, in the below we will restrict to prototype classes with no symmetry factor (there will be a symmetry factor of $1/n_s$ if the prototype diagram has a $\mathbb{Z}_{n_s}$ cyclic symmetry with respect to the outer loop, where $n_s$ divides $n$); we will return to the case with symmetry factor later. Now consider for example the diagram
\begin{center}
\begin{fmffile}{zzz-B-prototype-antisymmetrization}
\begin{fmfgraph*}(30, 30)
\fmfleftn{l}{1}\fmfrightn{r}{1}
\fmf{fermion,left,tag=1}{l1,r1}
\fmf{plain,left,tag=2}{r1,l1}
\fmffreeze
\fmfsurroundn{v}{24}
\fmf{phantom}{v12,ml1,mr1,v3}
\fmfdot{mr1}
\fmf{phantom}{v14,ml2,mr2,v2}
\fmffreeze
\fmf{phantom}{mr2,mm,v20}
\fmf{fermion,left,tension=0}{ml2,ml1}
\fmf{fermion,left,tension=0}{ml1,ml2}
\fmf{fermion,left,tension=0,tag=3}{mr2,mm}
\fmf{fermion,left,tension=0,tag=4}{mm,mr2}
\fmffreeze
\fmfposition
\fmfipath{p[]}
\fmfiset{p1}{vpath1(__l1,__r1)}
\fmfiset{p2}{vpath2(__r1,__l1)}
\fmfiset{p3}{vpath3(__mr2,__mm)}
\fmfiset{p4}{vpath4(__mm,__mr2)}
\fmfi{dashes}{point 1length(p1)/5 of p1{right} .. vloc(__mr1)}
\fmfi{dashes}{point 5length(p1)/7 of p1{left} .. vloc(__mr1)}
\fmfi{dashes}{point 0length(p3)/7 of p3 -- vloc(__mr1)}
\fmfi{dashes}{point 1length(p1)/3 of p1.. {down}vloc(__ml1)}
\fmfi{dashes}{point 4length(p2)/5 of p2{right} .. vloc(__ml2)}
\fmfi{dashes}{point 1length(p4)/8 of p4{down} .. point 1length(p2)/2 of p2}
\fmfi{dashes}{point 7length(p3)/9 of p3{down} .. point 1length(p2)/4 of p2}
\fmffreeze
\fmfi{plain,width=4}{subpath (5length(p1)/7,length(p1)) of p1}
\fmfi{plain,width=4}{subpath (0length(p2)/4,1length(p2)/4) of p2}
\fmfi{plain,width=4}{subpath (1length(p2)/4,1length(p2)/2) of p2}
\fmfi{plain,width=4}{subpath (4length(p2)/5,5length(p2)/5) of p2}
\fmfi{plain,width=4}{subpath (0length(p1)/5,1length(p1)/5) of p1}
\end{fmfgraph*}
\end{fmffile}
.
\end{center}
This diagram represents the sum (with coefficients $\pm 1/2$ assigned as before) of all diagrams in the given prototype class such that the three $\partial_p iG$'s appear on the three thickened segments. This sum is manifestly antisymmetric in $\mu\nu\lambda$. Let us call the set of diagrams contributing to such sum an ``antisymmetrization class''. Obviously a prototype class can be partitioned into antisymmetrization classes. The purpose of introducing this notion is only for introducing the next notion.

The third notion to introduce is a partitioning finer than a prototype class (still, we restrict to those without symmetry factor) but coarser than an antisymmetrization class. In an antisymmetrization class, the three $\partial_p iG$'s are separated by a number of $iG$'s, for example, in the previous antisymmetrization diagram, the three $\partial_p iG$'s are separated by $0, 1$ and $2$ $iG$'s. But there are other antisymmetrization diagrams whose three $\partial_p iG$'s are also separated by $0, 1$ and $2$ $iG$'s. Let us introduce the notation $(012)$, which represents the sum of them:
\begin{center}
\begin{fmffile}{zzz-B-prototype-separation-1}
\begin{fmfgraph*}(20, 20)
\fmfleftn{l}{1}\fmfrightn{r}{1}
\fmf{plain,left,tag=1}{l1,r1}
\fmf{plain,left,tag=2}{r1,l1}
\fmffreeze
\fmfsurroundn{v}{24}
\fmf{phantom}{v12,ml1,mr1,v3}
\fmfdot{mr1}
\fmf{phantom}{v14,ml2,mr2,v2}
\fmffreeze
\fmf{phantom}{mr2,mm,v20}
\fmf{fermion,left,tension=0}{ml2,ml1}
\fmf{fermion,left,tension=0}{ml1,ml2}
\fmf{fermion,left,tension=0,tag=3}{mr2,mm}
\fmf{fermion,left,tension=0,tag=4}{mm,mr2}
\fmffreeze
\fmfposition
\fmfipath{p[]}
\fmfiset{p1}{vpath1(__l1,__r1)}
\fmfiset{p2}{vpath2(__r1,__l1)}
\fmfiset{p3}{vpath3(__mr2,__mm)}
\fmfiset{p4}{vpath4(__mm,__mr2)}
\fmfi{dashes}{point 1length(p1)/5 of p1{right} .. vloc(__mr1)}
\fmfi{dashes}{point 5length(p1)/7 of p1{left} .. vloc(__mr1)}
\fmfi{dashes}{point 0length(p1)/7 of p3 -- vloc(__mr1)}
\fmfi{dashes}{point 1length(p1)/3 of p1.. {down}vloc(__ml1)}
\fmfi{dashes}{point 4length(p1)/5 of p2{right} .. vloc(__ml2)}
\fmfi{dashes}{point 1length(p1)/8 of p4{down} .. point 1length(p1)/2 of p2}
\fmfi{dashes}{point 7length(p1)/9 of p3{down} .. point 1length(p1)/4 of p2}
\fmffreeze
\fmfi{plain,width=4}{subpath (5length(p1)/7,length(p1)) of p1}\fmfi{plain,width=4}{subpath (0length(p2)/4,1length(p2)/4) of p2}
\fmfi{plain,width=4}{subpath (1length(p2)/4,1length(p2)/2) of p2}
\fmfi{plain,width=4}{subpath (4length(p2)/5,5length(p2)/5) of p2}\fmfi{plain,width=4}{subpath (0length(p1)/5,1length(p1)/5) of p1}
\end{fmfgraph*}
\end{fmffile}
\hspace{0cm}
\begin{fmffile}{zzz-B-prototype-separation-2}
\begin{fmfgraph*}(20, 20)
\fmfleftn{l}{1}\fmfrightn{r}{1}
\fmf{plain,left,tag=1}{l1,r1}
\fmf{plain,left,tag=2}{r1,l1}
\fmffreeze
\fmfsurroundn{v}{24}
\fmf{phantom}{v12,ml1,mr1,v3}
\fmfdot{mr1}
\fmf{phantom}{v14,ml2,mr2,v2}
\fmffreeze
\fmf{phantom}{mr2,mm,v20}
\fmf{fermion,left,tension=0}{ml2,ml1}
\fmf{fermion,left,tension=0}{ml1,ml2}
\fmf{fermion,left,tension=0,tag=3}{mr2,mm}
\fmf{fermion,left,tension=0,tag=4}{mm,mr2}
\fmffreeze
\fmfposition
\fmfipath{p[]}
\fmfiset{p1}{vpath1(__l1,__r1)}
\fmfiset{p2}{vpath2(__r1,__l1)}
\fmfiset{p3}{vpath3(__mr2,__mm)}
\fmfiset{p4}{vpath4(__mm,__mr2)}
\fmfi{dashes}{point 1length(p1)/5 of p1{right} .. vloc(__mr1)}
\fmfi{dashes}{point 5length(p1)/7 of p1{left} .. vloc(__mr1)}
\fmfi{dashes}{point 0length(p1)/7 of p3 -- vloc(__mr1)}
\fmfi{dashes}{point 1length(p1)/3 of p1.. {down}vloc(__ml1)}
\fmfi{dashes}{point 4length(p1)/5 of p2{right} .. vloc(__ml2)}
\fmfi{dashes}{point 1length(p1)/8 of p4{down} .. point 1length(p1)/2 of p2}
\fmfi{dashes}{point 7length(p1)/9 of p3{down} .. point 1length(p1)/4 of p2}
\fmffreeze
\fmfi{plain,width=4}{subpath (1length(p2)/4,1length(p2)/2) of p2}
\fmfi{plain,width=4}{subpath (1length(p2)/2,4length(p2)/5) of p2}
\fmfi{plain,width=4}{subpath (1length(p1)/5,1length(p1)/3) of p1}
\end{fmfgraph*}
\end{fmffile}
\hspace{0cm}
\begin{fmffile}{zzz-B-prototype-separation-3}
\begin{fmfgraph*}(20, 20)
\fmfleftn{l}{1}\fmfrightn{r}{1}
\fmf{plain,left,tag=1}{l1,r1}
\fmf{plain,left,tag=2}{r1,l1}
\fmffreeze
\fmfsurroundn{v}{24}
\fmf{phantom}{v12,ml1,mr1,v3}
\fmfdot{mr1}
\fmf{phantom}{v14,ml2,mr2,v2}
\fmffreeze
\fmf{phantom}{mr2,mm,v20}
\fmf{fermion,left,tension=0}{ml2,ml1}
\fmf{fermion,left,tension=0}{ml1,ml2}
\fmf{fermion,left,tension=0,tag=3}{mr2,mm}
\fmf{fermion,left,tension=0,tag=4}{mm,mr2}
\fmffreeze
\fmfposition
\fmfipath{p[]}
\fmfiset{p1}{vpath1(__l1,__r1)}
\fmfiset{p2}{vpath2(__r1,__l1)}
\fmfiset{p3}{vpath3(__mr2,__mm)}
\fmfiset{p4}{vpath4(__mm,__mr2)}
\fmfi{dashes}{point 1length(p1)/5 of p1{right} .. vloc(__mr1)}
\fmfi{dashes}{point 5length(p1)/7 of p1{left} .. vloc(__mr1)}
\fmfi{dashes}{point 0length(p1)/7 of p3 -- vloc(__mr1)}
\fmfi{dashes}{point 1length(p1)/3 of p1.. {down}vloc(__ml1)}
\fmfi{dashes}{point 4length(p1)/5 of p2{right} .. vloc(__ml2)}
\fmfi{dashes}{point 1length(p1)/8 of p4{down} .. point 1length(p1)/2 of p2}
\fmfi{dashes}{point 7length(p1)/9 of p3{down} .. point 1length(p1)/4 of p2}
\fmffreeze
\fmfi{plain,width=4}{subpath (1length(p2)/2,4length(p2)/5) of p2}
\fmfi{plain,width=4}{subpath (4length(p2)/5,5length(p2)/5) of p2}\fmfi{plain,width=4}{subpath (0length(p1)/5,1length(p1)/5) of p1}
\fmfi{plain,width=4}{subpath (1length(p1)/3,5length(p1)/7) of p1}
\end{fmfgraph*}
\end{fmffile}
\hspace{0cm}
\begin{fmffile}{zzz-B-prototype-separation-4}
\begin{fmfgraph*}(20, 20)
\fmfleftn{l}{1}\fmfrightn{r}{1}
\fmf{plain,left,tag=1}{l1,r1}
\fmf{plain,left,tag=2}{r1,l1}
\fmffreeze
\fmfsurroundn{v}{24}
\fmf{phantom}{v12,ml1,mr1,v3}
\fmfdot{mr1}
\fmf{phantom}{v14,ml2,mr2,v2}
\fmffreeze
\fmf{phantom}{mr2,mm,v20}
\fmf{fermion,left,tension=0}{ml2,ml1}
\fmf{fermion,left,tension=0}{ml1,ml2}
\fmf{fermion,left,tension=0,tag=3}{mr2,mm}
\fmf{fermion,left,tension=0,tag=4}{mm,mr2}
\fmffreeze
\fmfposition
\fmfipath{p[]}
\fmfiset{p1}{vpath1(__l1,__r1)}
\fmfiset{p2}{vpath2(__r1,__l1)}
\fmfiset{p3}{vpath3(__mr2,__mm)}
\fmfiset{p4}{vpath4(__mm,__mr2)}
\fmfi{dashes}{point 1length(p1)/5 of p1{right} .. vloc(__mr1)}
\fmfi{dashes}{point 5length(p1)/7 of p1{left} .. vloc(__mr1)}
\fmfi{dashes}{point 0length(p1)/7 of p3 -- vloc(__mr1)}
\fmfi{dashes}{point 1length(p1)/3 of p1.. {down}vloc(__ml1)}
\fmfi{dashes}{point 4length(p1)/5 of p2{right} .. vloc(__ml2)}
\fmfi{dashes}{point 1length(p1)/8 of p4{down} .. point 1length(p1)/2 of p2}
\fmfi{dashes}{point 7length(p1)/9 of p3{down} .. point 1length(p1)/4 of p2}
\fmffreeze
\fmfi{plain,width=4}{subpath (5length(p1)/7,length(p1)) of p1}\fmfi{plain,width=4}{subpath (0length(p2)/4,1length(p2)/4) of p2}
\fmfi{plain,width=4}{subpath (4length(p2)/5,5length(p2)/5) of p2}\fmfi{plain,width=4}{subpath (0length(p1)/5,1length(p1)/5) of p1}
\fmfi{plain,width=4}{subpath (1length(p1)/5,1length(p1)/3) of p1}
\end{fmfgraph*}
\end{fmffile}
\hspace{0cm}
\begin{fmffile}{zzz-B-prototype-separation-5}
\begin{fmfgraph*}(20, 20)
\fmfleftn{l}{1}\fmfrightn{r}{1}
\fmf{plain,left,tag=1}{l1,r1}
\fmf{plain,left,tag=2}{r1,l1}
\fmffreeze
\fmfsurroundn{v}{24}
\fmf{phantom}{v12,ml1,mr1,v3}
\fmfdot{mr1}
\fmf{phantom}{v14,ml2,mr2,v2}
\fmffreeze
\fmf{phantom}{mr2,mm,v20}
\fmf{fermion,left,tension=0}{ml2,ml1}
\fmf{fermion,left,tension=0}{ml1,ml2}
\fmf{fermion,left,tension=0,tag=3}{mr2,mm}
\fmf{fermion,left,tension=0,tag=4}{mm,mr2}
\fmffreeze
\fmfposition
\fmfipath{p[]}
\fmfiset{p1}{vpath1(__l1,__r1)}
\fmfiset{p2}{vpath2(__r1,__l1)}
\fmfiset{p3}{vpath3(__mr2,__mm)}
\fmfiset{p4}{vpath4(__mm,__mr2)}
\fmfi{dashes}{point 1length(p1)/5 of p1{right} .. vloc(__mr1)}
\fmfi{dashes}{point 5length(p1)/7 of p1{left} .. vloc(__mr1)}
\fmfi{dashes}{point 0length(p1)/7 of p3 -- vloc(__mr1)}
\fmfi{dashes}{point 1length(p1)/3 of p1.. {down}vloc(__ml1)}
\fmfi{dashes}{point 4length(p1)/5 of p2{right} .. vloc(__ml2)}
\fmfi{dashes}{point 1length(p1)/8 of p4{down} .. point 1length(p1)/2 of p2}
\fmfi{dashes}{point 7length(p1)/9 of p3{down} .. point 1length(p1)/4 of p2}
\fmffreeze
\fmfi{plain,width=4}{subpath (1length(p2)/4,1length(p2)/2) of p2}
\fmfi{plain,width=4}{subpath (1length(p1)/5,1length(p1)/3) of p1}
\fmfi{plain,width=4}{subpath (1length(p1)/3,5length(p1)/7) of p1}
\end{fmfgraph*}
\end{fmffile}
\hspace{0cm}
\begin{fmffile}{zzz-B-prototype-separation-6}
\begin{fmfgraph*}(20, 20)
\fmfleftn{l}{1}\fmfrightn{r}{1}
\fmf{plain,left,tag=1}{l1,r1}
\fmf{plain,left,tag=2}{r1,l1}
\fmffreeze
\fmfsurroundn{v}{24}
\fmf{phantom}{v12,ml1,mr1,v3}
\fmfdot{mr1}
\fmf{phantom}{v14,ml2,mr2,v2}
\fmffreeze
\fmf{phantom}{mr2,mm,v20}
\fmf{fermion,left,tension=0}{ml2,ml1}
\fmf{fermion,left,tension=0}{ml1,ml2}
\fmf{fermion,left,tension=0,tag=3}{mr2,mm}
\fmf{fermion,left,tension=0,tag=4}{mm,mr2}
\fmffreeze
\fmfposition
\fmfipath{p[]}
\fmfiset{p1}{vpath1(__l1,__r1)}
\fmfiset{p2}{vpath2(__r1,__l1)}
\fmfiset{p3}{vpath3(__mr2,__mm)}
\fmfiset{p4}{vpath4(__mm,__mr2)}
\fmfi{dashes}{point 1length(p1)/5 of p1{right} .. vloc(__mr1)}
\fmfi{dashes}{point 5length(p1)/7 of p1{left} .. vloc(__mr1)}
\fmfi{dashes}{point 0length(p1)/7 of p3 -- vloc(__mr1)}
\fmfi{dashes}{point 1length(p1)/3 of p1.. {down}vloc(__ml1)}
\fmfi{dashes}{point 4length(p1)/5 of p2{right} .. vloc(__ml2)}
\fmfi{dashes}{point 1length(p1)/8 of p4{down} .. point 1length(p1)/2 of p2}
\fmfi{dashes}{point 7length(p1)/9 of p3{down} .. point 1length(p1)/4 of p2}
\fmffreeze
\fmfi{plain,width=4}{subpath (5length(p1)/7,length(p1)) of p1}\fmfi{plain,width=4}{subpath (0length(p2)/4,1length(p2)/4) of p2}
\fmfi{plain,width=4}{subpath (1length(p2)/2,4length(p2)/5) of p2}
\fmfi{plain,width=4}{subpath (1length(p1)/3,5length(p1)/7) of p1}
\end{fmfgraph*}
\end{fmffile}
.
\end{center}
In general, we call the set of diagrams contributing to $(abc)$ a ``cyclic class''. The name is because $(abc)$ is by definition the same object as $(bca)$ and $(cab)$. Clearly, a prototype class is partitioned into cyclic classes, and a cyclic class is partitioned into antisymmetrization classes. It is easy to see the sum $S$ of diagrams in the prototype class can be expressed as
\begin{eqnarray}
S=\sum'_{a+b+c=n-3} n \cdot s_{abc} \cdot (abc), \ \ \ \ \ \ \ \ s_{abc}=\left\{ \begin{array}{cl} 1/3 \ \ & \mbox{ if } a=b=c \\ 1 \ \ & \mbox{ otherwise} \end{array} \right.
\label{B_S_and_abc}
\end{eqnarray}
where $a,b,c$ are non-negative integers, and the prime on the sum means we only count $(abc)$, $(bca)$ and $(cab)$ once because they are the same object. We will see soon that by introducing the notion of cyclic class, we boil the Feynman diagram cancellation problem to a combinatorial problem.

In general $(abc)$, the sum of diagrams in a cyclic class, is not equal to zero. We need to introduce the fourth notion that bridges between prototype class and cyclic class; but unlike the three notions above, this fourth notion is not a partitioning. Given the prototype class, consider the following diagrams with two $\partial_p iG$'s: along the fermion arrow on the outer loop, we have $\partial_p^\mu iG$, then $m$ $iG$'s, then $\partial_p^\nu iG$, and then the remaining $(n-2-m)$ $iG$'s. We assume $m\leq n-2-m$. Sum up all such diagrams. But before we integrate over the outer loop momentum $p$, we take a total $\partial_p^\lambda$ derivative. Then we totally antisymmetrize between $\mu\nu\lambda$ and multiply by $3$. We denote the result by $\langle m \rangle$. By construction, $\langle m \rangle=0$ due to the total derivative. But on the other hand, it is easy to see $\langle m \rangle$ is a sum of cyclic classes:
\begin{eqnarray}
&& \hspace{4cm} \langle m \rangle = \sum'_{a+b+c=n-3} s^m_{abc} \cdot (abc), \label{B_m_and_abc} \\[.2cm]
&& s^m_{abc}=\left\{ \begin{array}{ll} 1 \ \ & \mbox{ if only one of $a, b, c$ is $m$,} \\[-.2cm] & \mbox{ and the other two do not sum up to $m-1$} \\ 1+1=2 \ \ & \mbox{ if two of $a, b, c$ are $m$, the other is not $m$} \\ 1 \ \ & \mbox{ if $a=b=c=m$} \\ -1 \ \ & \mbox{ if two of $a, b, c$ sum up to $m-1$, and the other is not $m$} \\ 1-1=0 & \mbox{ if one of $a, b, c$ is $m$, and the other two sum up to $m-1$} \\ 0 \ \ & \mbox{ otherwise}. \end{array} \right. \nonumber 
\end{eqnarray}
We used the fact that the vertices on the outer loop are independent of $p$, whose reason is explained at the end of the previous section. Diagrams contributing to a fixed $\langle m \rangle$ do not form an equivalence class, because clearly a given $(abc)$ can appear in several different $\langle m \rangle$'s.

Now that every $\langle m \rangle$ is equal to $0$, it would be desirable to express $S$ as a linear combination of $\langle m \rangle$'s. Indeed, now we shall show that
\begin{eqnarray}
S = \sum_{m=0}^{m\leq (n-2)/2} \left(n-2-2m\right) \cdot \langle m \rangle = 0.
\label{B_S_and_m}
\end{eqnarray}
This is a simple combinatorial problem that can be shown by matching the coefficient of each $(abc)$ on both sides, using equations \eqref{B_S_and_abc} and \eqref{B_m_and_abc}. Due to the cyclic property of $(abc)$, we can assume $a\leq b$, $a\leq c$. Then we discuss over 8 possibilities:
\begin{enumerate}

\item $a<b<c<(n-2)/2$: On the right-hand-side, $(abc)$ appears in three different $\langle m \rangle$'s:
\begin{itemize}
\item $m=a$: The coefficient of $(abc)$ is $(n-2-2a)\cdot 1$.
\item $m=b$: The coefficient of $(abc)$ is $(n-2-2b)\cdot 1$.
\item $m=c$: The coefficient of $(abc)$ is $(n-2-2c)\cdot 1$.
\end{itemize}
The sum of these coefficients is $n$, matches with the coefficient on the left-hand-side.

The case $a<c<b<(n-2)/2$ works in the same manner.

\item $a<b<c=(n-2)/2$: On the right-hand-side, $(abc)$ appears in three different $\langle m \rangle$'s:
\begin{itemize}
\item $m=a$: The coefficient of $(abc)$ is $(n-2-2a)\cdot 1$.
\item $m=b$: The coefficient of $(abc)$ is $(n-2-2b)\cdot 1$.
\item $m=c$: The coefficient of $(abc)$ is $0$.
\end{itemize}
The sum of these coefficients is $n$, matches with the coefficient on the left-hand-side.

The case $a<c<b=(n-2)/2$ works in the same manner.

\item $a<b<(n-2)/2<c$: On the right-hand-side, $(abc)$ appears in three different $\langle m \rangle$'s:
\begin{itemize}
\item $m=a$: The coefficient of $(abc)$ is $(n-2-2a)\cdot 1$.
\item $m=b$: The coefficient of $(abc)$ is $(n-2-2b)\cdot 1$.
\item $m-1=a+b$: The coefficient of $(abc)$ is $(n-2-2(a+b+1))\cdot(-1)$.
\end{itemize}
The sum of these coefficients is $n$, matches with the coefficient on the left-hand-side.

The case $a<c<(n-2)/2<b$ works in the same manner.

\item $a=b<c<(n-2)/2$: On the right-hand-side, $(abc)$ appears in two different $\langle m \rangle$'s:
\begin{itemize}
\item $m=a=b$: The coefficient of $(abc)$ is $(n-2-a-b)\cdot 2$.
\item $m=c$: The coefficient of $(abc)$ is $(n-2-2c)\cdot 1$.
\end{itemize}
The sum of these coefficients is $n$, matches with the coefficient on the left-hand-side.

The case $a=c<b<(n-2)/2$ works in the same manner.

\item $a=b<c=(n-2)/2$: On the right-hand-side, $(abc)$ appears in two different $\langle m \rangle$'s:
\begin{itemize}
\item $m=a=b$: The coefficient of $(abc)$ is $(n-2-a-b)\cdot 2$.
\item $m=c$: The coefficient of $(abc)$ is $0$.
\end{itemize}
The sum of these coefficients is $n$, matches with the coefficient on the left-hand-side.

The case $a=c<b=(n-2)/2$ works in the same manner.

\item $a=b<(n-2)/2<c$: On the right-hand-side, $(abc)$ appears in two different $\langle m \rangle$'s:
\begin{itemize}
\item $m=a=b$: The coefficient of $(abc)$ is $(n-2-a-b)\cdot 2$.
\item $m-1=a+b$: The coefficient of $(abc)$ is $(n-2-2(a+b+1))\cdot(-1)$.
\end{itemize}
The sum of these coefficients is $n$, matches with the coefficient on the left-hand-side.

The case $a=c<(n-2)/2<b$ works in the same manner.

\item $a<b=c<(n-2)/2$: On the right-hand-side, $(abc)$ appears in two different $\langle m \rangle$'s:
\begin{itemize}
\item $m=a$: The coefficient of $(abc)$ is $(n-2-2a)\cdot 1$.
\item $m=b=c$: The coefficient of $(abc)$ is $(n-2-b-c)\cdot 1$.
\end{itemize}
The sum of these coefficients is $n$, matches with the coefficient on the left-hand-side.

\item $a=b=c$: On the right-hand-side, $(abc)$ appears only when $m=a=b=c$, with coefficient $(n-2-2(n-3)/3)\cdot 1 = n/3$. This matches with the coefficient on the left-hand-side.

\end{enumerate}
This completes our proof, for prototype classes that have no symmetry factor.

For a prototype class whose prototype diagram has a $\mathbb{Z}_{n_s}$ symmetry with respect to the outer loop, we can pick one segment on the outer loop to be ``the special segment''; in the sum of diagrams, this leads to over-counting by a factor of $n_s$. But now that there is no $\mathbb{Z}_{n_s}$ symmetry any more, we can show the sum is zero as before. The factor of $n_s$ has no effect on the zero.

\section{Summary and Outlook}

In this chapter, we have extended Landau's Fermi liquid theory to incorporate Berry curvature effects. Among other effects, we showed the anomalous Hall conductivity receives two contributions: the non-quasiparticle Hall conductivity tensor, plus the contribution of electric dipole moment of the excited quasiparticles. The latter can be viewed a new effect due to interactions, because a non-interacting fermion usually has no intrinsic electric dipole. As for the former, it can be further separated into a chemical potential dependent part and a chemical potential independent part. Remarkably, we showed the chemical potential dependent part is a Fermi surface property given by the Berry curvature around the Fermi surface, as in non-interacting Fermi gas. On the other hand, some puzzles about chemical potential independent part remain, as we will comment about below.

One can see several directions to
extend our theory. First, we assumed the Fermi level crosses only one band; 
one can generalize this to multiple bands. 
Two scenarios are of particular physical interest: either that the multiple bands crossing the Fermi level are 
completely degenerate~\cite{shindou2008gradient}, or that these multiple bands 
have completely disjoint Fermi surfaces. The generalization of our theory to both
 scenarios is straightforward. Second, our discussion is limited to linear response. It
would be interesting to extend the scope of the kinetic theory to
include also nonlinear response, so that important effects such as the $(3+1)d$ chiral anomaly can be captured. 
Third, we assumed that the quantum field
theory describing the fermions does not have couplings of the type
$A\phi \psi^\dagger \psi$. It would be interesting to see if the 
kinetic theory can be extended to include couplings of this type in
the QFT. Also, one may try to understand if long-ranged
interactions can be included, to the extent that these interactions do
not destroy the Fermi liquid ground state.

Some interesting questions are raised in the context of our Berry
Fermi liquid theory. We have found that, beside a
$\epsilon_F$-dependent piece, the Hall conductivity contains a constant
piece $\sigma_o^{\mu\nu\lambda}$ in Section \ref{sssect_kinetic_sigma_EF_no_anomaly}.
Is this contribution topological and not renormalized by interactions?
Does it receive jumps at discrete values of $\epsilon_F$ around which
the FS develops new disconnected components (e.g. in Figure \ref{EF_increase_disconnect_comp})? In gapped system,
$\sigma_o^{\mu\nu\lambda}=\sigma_r^{\mu\nu\lambda}$ is topological
\cite{Ishikawa:1986wx}.

More broadly, one may ask: Is it possible to
have a notion of topologically equivalent / distinct Fermi liquids?
For example, are the Fermi liquids in normal metal and in Weyl metal
topologically distinct under some notion? In this thesis, we see both a puzzle 
and a hint regarding such notion. The puzzle is the possible jump of
$\sigma^{\mu\nu\lambda}$ mentioned previously. The hint is the manifestation 
of anomaly-related transport effects in the distinction between \eqref{BFL_dsigma_dEF_b} 
and \eqref{BFL_dsigma_dEF_2b}. It would be interesting to study if such 
problems can be covered under a coherent framework.

We note also that the matching between the microscopic theory and the 
Fermi liquid theory is done here at the level of dynamical
equations. If there is a way to do the matching at the level of
action and path integral measure, like that in Berry Fermi gas 
\cite{Chen:2014cla} (see Section \ref{sect_BFG_PIderivation}), it would provide a much more transparent derivation 
of the Berry Fermi liquid theory. It may also help to extend the kinetic 
theory beyond linear response.

Finally, given the generality of the assumptions, the formalism should
have broad applications in physical systems. It would be interesting
if predictions of the Berry Fermi liquid theory can be directly
compared to experiments.

\chapter{Conclusion}

In this thesis we first studied general Berry Fermi gas theory and its microscopic justification. Then we studied the chiral kinetic theory of Weyl fermions as a specific example, and discussed how Lorentz invariance is non-trivially realized in chiral kinetic theory. Finally we studied general Berry Fermi liquid theory and its microscopic justification. The Berry phase physics being non-trivial and interesting in these systems should now be evident. At the end of each chapter, we have a ``Summary and Outlook'' section, summarizing the important ideas and results in that chapter, and more importantly, discussing future directions of study relevant to that chapter. We are not going to repeat those summarizations and discussions here. Here we discuss some more general possibilities.

From our study of Berry Fermi liquid, it is clear that to define momentum space Berry curvature, all it needs is the eigenvector $u$ of the full propagator; the eigenvalue part of the propagator is irrelevant to the definition. This suggests we should be able to define the momentum space Berry curvature even for non-Fermi liquids. Then, we should wonder what we can say about Berry curvature in such systems. For example, at least for non-Fermi liquids that are ``sufficiently similar'' to a normal Fermi liquid (for instance those whose $p^0$ in the denominator of the propagator is renormalized to the form $p^0 \ln(p^0)$), can we still explore the relation between anomalous Hall effect and Berry phase? If something can be said, it will not only be theoretically satisfactory, but may also have potential application value to important systems such as the composite Fermi liquid at half-filled Landau level.

Another problem worth thinking about is a more theoretical one. In $d$ spatial dimensions the Berry curvature defect where $\partial_p^{[i} b^{jk]}\neq 0$ is generally $d-3$ dimensional. So for $d>3$ the Berry curvature defect would be an extended object. Then there may be non-trivial topological phenomena, arising from the interplays between the path of the particle in the momentum space, the Berry curvature defect, and the topology of the Brillouin zone.

Yet another interesting thought would be to study Berry phase effects on non-point-like objects. In particular, one can consider a ``gluon string'' with ``quarks'' (chiral fermions) attached to its ends, and study if there is any non-trivial interplay between the Berry phase and the string.

And there must be many other hidden aspects of Berry phase physics awaiting for us to uncover.


\end{document}